%% file: Thesis.tex
\documentclass[
	paper=A4,					
	twoside=true,				
	openright,					
	parskip=full,				
	chapterprefix=true,			
	12pt,						
	headings=normal,			
	bibliography=totoc,			
	listof=totoc,				
	titlepage=on,				
	index=totoc,				
	captions=tablebelow,			
	draft=false,				
]{scrreprt}

\newcommand{\thesisTitle}{Amplitudes de Transition Holomorphes Semi-Classiques en Gravit\'{e} Quantique \`{a} Boucles}
\newcommand{\thesisName}{Fabio D'Ambrosio}
\newcommand{\thesisDate}{23 Septembre 2019}

\newcommand{\thesisDirecteur}{Carlo Rovelli}
\newcommand{\thesisCoDirecteur}{Simone Speziale}

\newcommand{\thesisFirstRapporteur}{Karim Noui}
\newcommand{\thesisSecondRapporteur}{Francesca Vidotto}

\newcommand{\thesisFirstExaminateur}{Marc Geiller}
\newcommand{\thesisSecondExaminateur}{Alejandro Perez}
\newcommand{\thesisThirdExaminateur}{Federico Piazza}
\newcommand{\thesisFourthExaminateur}{Carlo Rovelli}

\newcommand{\thesisUniversity}{\protect{Aix-Marseille Universit\'{e}}}
\newcommand{\thesisUniversityDepartment}{Centre de Physique Th\'{e}orique}

\usepackage[utf8]{inputenc}
\usepackage[english]{babel}    	
\usepackage[				   	
	figuresep=colon,           	
	sansserif=false,           	
	hangfigurecaption=false,	   	
	hangsection=true,          	
	hangsubsection=true,       	
	colorize=full,             	
	colortheme=bluemagenta    	
]{cleanthesis}
\usepackage{MyMath}

\hypersetup{					
	pdftitle={\thesisTitle},		
	pdfauthor={\thesisName},		
	plainpages=false,			
	colorlinks=true,			
	pdfborder={0 0 0},			
	breaklinks=true,			
	bookmarksnumbered=true,		%
	bookmarksopen=true			%
}
\usepackage[belowskip=-20pt,aboveskip=20pt]{caption}
\usepackage{adjustbox}
\usepackage{multirow}
\arrayrulecolor[rgb]{0.85,0.85,0.85}
\definecolor{myGray}{rgb}{0.85,0.85,0.85}
\begin{document}

\renewcaptionname{english}{\figurename}{Fig.}
\renewcaptionname{english}{\tablename}{Table}

\let\oldcite=\cite                                                              
\renewcommand{\cite}[1]{\textcolor{blue}{\oldcite{#1}}}
\makeatletter
\renewcommand*{\bibleftbracket}{\blx@postpunct\textcolor{black}{[}}
\renewcommand*{\bibrightbracket}{\blx@postpunct\textcolor{black}{]}\midsentence}
\makeatother

\let\oldref=\ref
\renewcommand{\ref}[1]{{\fontsize{12}{12}\ToCFont\oldref{#1}}}

\let\oldurl=\url
\renewcommand{\url}[1]{{\ToCFont\color{blue}\oldurl{#1}}}

\begin{titlepage}
	\pdfbookmark[0]{Titlepage}{Titlepage}
	\tgherosfont
	\centering

\begin{figure}[!htb]
	\begin{minipage}{.5\textwidth}
		\raggedright
		\includegraphics[height=2cm]{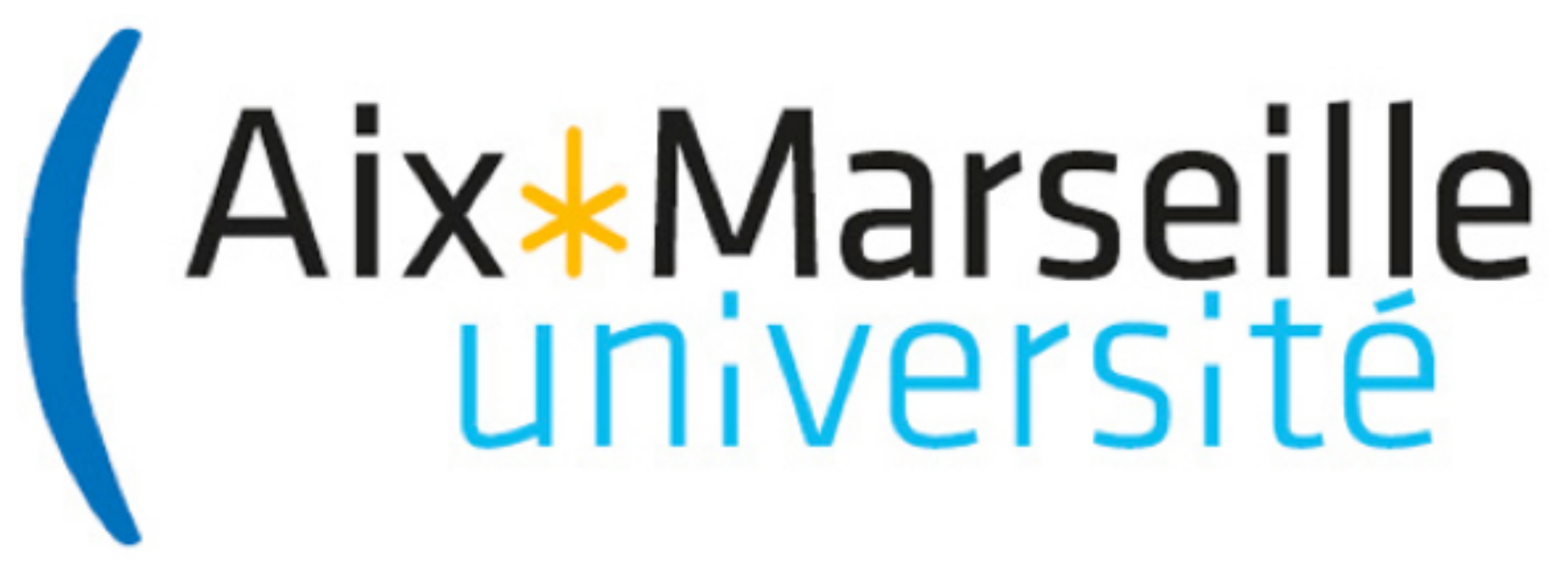} 
	\end{minipage}
	\begin{minipage}{.5\textwidth}
		\raggedleft
		\includegraphics[height=2cm]{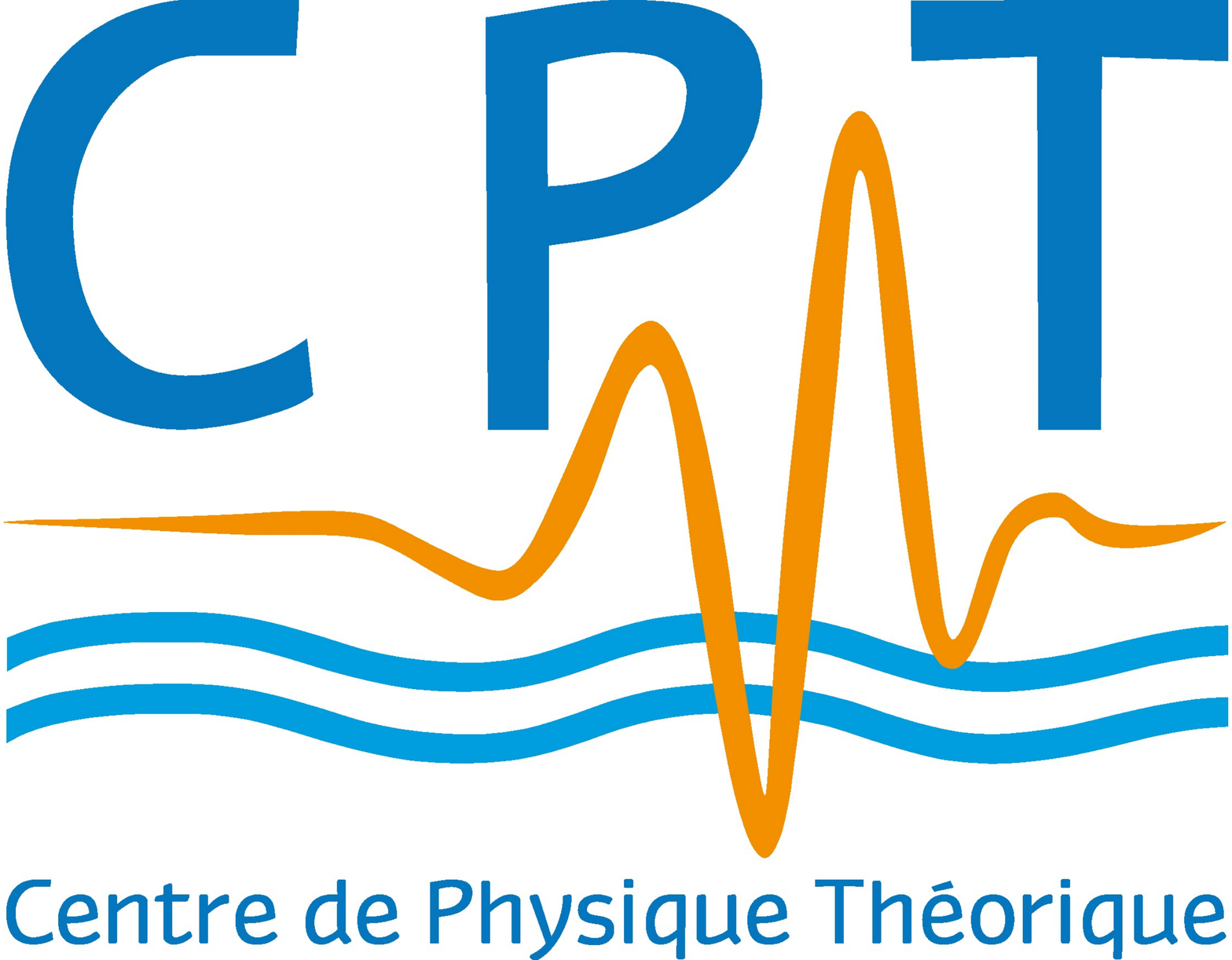}
	\end{minipage}
\end{figure}
\vspace{0.5cm}

\centering{\Large{Aix-Marseille Universit\'{e}}}\\[1mm]
\centering{ED352 Physique et Sciences de la Mati\`{e}re}\\[6mm]
\centering{\Large{Centre de Physique Th\'{e}orique}}\\[1mm]
\centering{\'{E}quipe de Gravit\'{e} Quantique}

\centering{Th\`{e}se de Doctorat pres\'{e}nt\'{e}e par}\\[3mm]
\centering{\Large{Fabio D'Ambrosio}}\\[3mm]
\centering{pour obtenir le grad universitaire de}\\[3mm]
\centering{\Large{Docteur en Physique Th\'{e}orique et Math\'{e}matique}}\\[3mm]

\par\noindent\rule{\textwidth}{0.4pt}\\[3mm]
	{\LARGE \color{MyBlue}\textbf{\thesisTitle}}\\[3mm]
\par\noindent\rule{\textwidth}{0.4pt}

	\begin{minipage}[t]{1\textwidth}
		\centering
		Soutenue le \thesisDate\,  devant le jury compos\'{e} de
	\end{minipage}\\[10mm]

	\begin{minipage}[t]{.3\textwidth}
		\raggedleft
		{\large Directeur de Th\`{e}se:}
	\end{minipage}
	\hspace*{15pt}
	\begin{minipage}[t]{.65\textwidth}
		{\large \thesisDirecteur} 
	\end{minipage}\\[5mm]

	\begin{minipage}[t]{.3\textwidth}
		\raggedleft
		\large{Rapporteurs:}
	\end{minipage}
	\hspace*{15pt}
	\begin{minipage}[t]{.65\textwidth}
		{\large \thesisFirstRapporteur} \\[2mm]
	  	{\large \thesisSecondRapporteur} \\[2mm]
	\end{minipage} \\[5mm]
	
	\begin{minipage}[t]{.3\textwidth}
		\raggedleft
		\large{Examinateurs:}
	\end{minipage}
	\hspace*{15pt}
	\begin{minipage}[t]{.65\textwidth}
		{\large \thesisFirstExaminateur}\\[2mm]
	  	{\large \thesisSecondExaminateur}\\[2mm]
		{\large \thesisThirdExaminateur}\\[2mm]
		{\large \thesisFourthExaminateur}\\[2mm]
	\end{minipage} \\[10mm]

\end{titlepage}
\cleardoublepage

\begin{titlepage}
	\textcolor{white}{empty line}\\[30mm]
	\centering{\Large{Fabio D'Ambrosio}}\\[10mm]
	{\LARGE \color{MyBlue}\textbf{Semi-Classical Holomorphic Transition Amplitudes in Covariant Loop Quantum Gravity}}\\[3mm]
	\par\noindent\rule{\textwidth}{0.4pt}
\end{titlepage}

\cleardoublepage

\pagenumbering{roman}			
\cleardoublepage

\cleardoublepage

\pagestyle{plain}				

\input{Content/French_Summary}	
\cleardoublepage

\input{Content/Summary}			
\cleardoublepage


\setcounter{tocdepth}{2}			
{\hypersetup{hidelinks}
\tableofcontents
}				
\cleardoublepage

\addcontentsline{toc}{chapter}{\listfigurename}
{\hypersetup{hidelinks}
\listoffigures	
}					
\cleardoublepage

\raggedbottom 
\pagenumbering{arabic}			
\setcounter{page}{1}			
\pagestyle{maincontentstyle} 	

\input{Content/Chapter_1} 		
\input{Content/Chapter_2}		
\input{Content/Chapter_3} 		
\input{Content/Chapter_4}		
\input{Content/Chapter_5}		
\input{Content/Chapter_6}		
\input{Content/Appendices}		
\cleardoublepage

\colorlet{arxivc}{blue}
\def\arxivtourl#1{http://arxiv.org/abs/#1}      
\DeclareFieldFormat{eprint:arxiv}{%
    \ifhyperref
        {\href{\arxivtourl#1}{[{\color{arxivc}arXiv:\nolinkurl{#1}}]}}
        {\nolinkurl{#1}}}

{\hypersetup{colorlinks=true,urlcolor=blue}
\urlstyle{rm}
\setstretch{1.1}
\renewcommand{\bibfont}{\normalfont\small}
\setlength{\biblabelsep}{0.3cm}
\setlength{\bibitemsep}{0.5\baselineskip plus 0.5\baselineskip}
\printbibliography[title={References},nottype=online]
\printbibliography[heading=subbibliography,title={Websites},type=online,prefixnumbers={@}]
}

\cleardoublepage

\end{document}

%% file: Content/French_Summary.tex
\pdfbookmark[0]{R\'{e}sum\'{e}}{R\'{e}sum\'{e}}
\chapter*{R\'{e}sum\'{e}}
\label{sec:Resume}
\vspace*{-10mm}
\addcontentsline{toc}{chapter}{R\'{e}sum\'{e}} 
La gravitation quantique \`{a} boucle covariante (CLQG) est une th\'{e}orie sp\'{e}culative de la gravitation quantique qui s'est d\'{e}velopp\'{e}e \`{a} partir de plusieurs directions de recherche diff\'{e}rentes. Depuis sa cr\'{e}ation, il a \'{e}t\'{e} \'{e}tabli que sa limite classique est li\'{e}e au calcul de Regge, le propagateur du graviton et \`{a} la fonction de corr\'{e}lation \`{a} trois points poss\`{e}dent la structure tensorielle attendue de la gravitation quantique perturbative, et la th\'{e}orie n'a pas de divergence ultraviolette. De plus, la CLQG a \'{e}t\'{e} \'{e}tendue aux couplages de mati\`{e}re avec les fermions et les champs de Yang-Mills, les groupes quantiques permettent l'introduction d'une constante cosmologique positive (qui rend d'ailleurs la th\'{e}orie finie aussi dans l'infrarouge), et cette th\'{e}orie a \'{e}t\'{e} utilis\'{e}e pour \'{e}tudier la cosmologie quantique.\\
Plus r\'{e}cemment, elle a \'{e}galement \'{e}t\'{e} appliqu\'{e}e \`{a} la transition trou noir/trou blanc -- un mod\`{e}le particulier d'effondrement stellaire sans singularit\'{e}, r\'{e}solvant l'\'{e}nigme de l'information et conduisant potentiellement \`{a} des effets observables. Cependant, plusieurs obstacles ont emp\^{e}ch\'{e} de progresser dans l'enqu\^{e}te sur ce sc\'{e}nario physique. \\
Ces obstacles vont de probl\`{e}mes conceptuels, tels que la question de savoir comment extraire les pr\'{e}dictions physiques d'une th\'{e}orie quantique de la gravit\'{e} ind\'{e}pendante du background \`{a} des probl\`{e}mes de calcul en raison de l'absence de m\'{e}thodes syst\'{e}matiques pour \'{e}valuer les amplitudes de transition de la CLQG.\\
Cette th\`{e}se aborde directement certaines de ces questions. Apr\`{e}s un chapitre introductif, nous passerons en revue la th\'{e}orie canonique de la LQG et travaillerons \`{a} la d\'{e}finition d'\'{e}tats semi-classiques coh\'{e}rents. Ces \'{e}tats joueront un r\^{o}le important dans les chapitres suivants o\`{u} ils faciliteront les calculs et les interpr\'{e}tations physiques. De plus, nous d\'{e}rivons une mesure pour les \'{e}tats du noyau de la chaleur coh\'{e}rents dans la param\'{e}trisation de la g\'{e}om\'{e}trie twist\'{e}e par rapport \`{a} laquelle ils satisfont \`{a} une r\'{e}solution d'identit\'{e}. Cette mesure entre directement dans la d\'{e}finition des observables physiques. \\
Dans le chapitre suivant, nous pr\'{e}sentons les id\'{e}es principales des mod\`{e}les de mousse de spin et fournissons le cadre math\'{e}matique n\'{e}cessaire pour discuter des espaces-temps discr\'{e}tis\'{e}s. En particulier, nous d\'{e}veloppons un algorithme de triangulation simplicial pour des vari\'{e}t\'{e}s de topologie $I\times\Sigma$ qui peut facilement \^{e}tre impl\'{e}ment\'{e} sur un ordinateur. L'exemple de $I^2\times\mathcal S^2$, qui peut \^{e}tre utilis\'{e} pour d\'{e}crire des r\'{e}gions compactes dans l'espace-temps de Schwarzschild, est discut\'{e} plus en d\'{e}tail.\\
Le chapitre quatre est enti\`{e}rement consacr\'{e} \`{a} la CLQG. Apr\`{e}s une br\`{e}ve d\'{e}finition de la th\'{e}orie, nous pr\'{e}sentons une m\'{e}thode de r\'{e}\'{e}criture de l'amplitude de transition de CLQG sous une forme rappelant l'int\'{e}grale du chemin de Feynman et nous introduisons l'amplitude de transition dite holomorphique. Ce sera le point de d\'{e}part du calcul de CLQG de la transition trou noir/trou blanc et de la d\'{e}termination du temps de rebond. Plusieurs questions conceptuelles et informatiques seront abord\'{e}es. \\
Enfin, au chapitre cinq, nous utilisons la m\'{e}canique quantique comme guide pour d\'{e}velopper une nouvelle m\'{e}thode d'approximation des amplitudes de transition holomorphes en l'absence de points critiques. Ces techniques sont ensuite appliqu\'{e}es \`{a} la CLQG, o\`{u} elles peuvent \^{e}tre comprises comme un d\'{e}veloppement semi-classique de l'amplitude de la transition holomorphe autour d'un espace-temps classique. Cette m\'{e}thode jette un nouvel \'{e}clairage sur la question dite du cosinus et reproduit le r\'{e}sultat obtenu pour la transition trou noir/trou blanc, obtenu au chapitre quatre.

%% file: Content/Summary.tex
\pdfbookmark[0]{Summary}{Summary}
\chapter*{Summary}
\label{sec:Summary}
\vspace*{-10mm}
\addcontentsline{toc}{chapter}{Summary} 
Covariant Loop Quantum Gravity (CLQG) is a tentative theory of quantum gravity which has emerged from a number of different research directions. In the years since its inception it has been established that its classical limit is related to (area) Regge calculus, the graviton propagator and the three-point function possess the tensorial structure expected from perturbative quantum gravity, and the theory is ultraviolet-finite. Moreover, CLQG has been extended to matter couplings with fermions and Yang-Mills fields, quantum groups allow the introduction of a positive cosmological constant (which incidentally renders the theory also infrared-finite), and the theory has been used to study quantum cosmology.\\	
More recently, it has also been applied to the so-called black hole to white hole transition -- a particular model of stellar collapse which is singularity-free, resolves the information puzzle and potentially leads to observable effects. However, several obstacles have impeded progress in the investigation of this physical scenario.\\
These obstacles range from conceptual issues, such as the question how to extract physical predictions from a background independent theory of quantum gravity, to computational problems due to a lack of systematic methods to evaluate CLQG transition amplitudes.\\
This thesis addresses some of these issues directly. After an introductory chapter, we will review the theory of canonical LQG and work toward a definition of coherent semi-classical states. These states will play an important role in subsequent chapters where they facilitate computations and physical interpretations. Moreover, we derive a measure for coherent heat kernel states in the twisted geometry parametrization with respect to which they satisfy a resolution of identity. This measure directly enters in the definition of physical observables.\\
In the next chapter, we present the main ideas of spin foam models and we provide the necessary mathematical framework to discuss discretized spacetimes. In particular, we develop a simplicial triangulation algorithm for manifolds of topology $I\times\Sigma$ which can easily be implemented on a computer. The example of $I^2\times\mathcal S^2$, which can be used to describe compact regions in the Schwarzschild spacetime, is discussed in more detail.\\
Chapter four is entirely devoted to CLQG. After a brief definition of the theory, we proceed to present a method to recast the CLQG transition amplitude in a form reminiscent of Feynman's path integral and we introduce the so-called holomorphic transition amplitude. This will be the starting point for the CLQG computation of the black hole to white hole transition and the determination of the bounce time. Several conceptual and computational issues will be discussed.\\
Finally, in chapter five we use quantum mechanics as a guide line to develop a new approximation method for holomorphic transition amplitudes in the absence of critical points. These techniques are then applied to CLQG where they can be understood as a semi-classical expansion of the holomorphic transition amplitude around a classical background spacetime. This method sheds new light on the so-called cosine issue and it reproduces the result for the black hole to white hole transition obtained in chapter four.

%% file: Content/Chapter_1.tex
\chapter{Introduction}
\label{chap:intro}
Covariant Loop Quantum Gravity (CLQG) is a tentative theory of quantum gravity which has emerged from a number of different research directions. In the years since its inception it has been established that its classical limit is related to (area) Regge calculus, the graviton propagator and the three-point function possess the tensorial structure expected from perturbative quantum gravity, and the theory is ultraviolet-finite. Moreover, CLQG has been extended to matter couplings with fermions and Yang-Mills fields, quantum groups allow the introduction of a positive cosmological constant (which incidentally renders the theory also infrared-finite), and the theory has been used to study quantum cosmology.\\	
More recently, it has also been applied to the so-called black hole to white hole transition -- a particular model of stellar collapse which is singularity-free, resolves the information puzzle and potentially leads to observable effects. However, several obstacles have impeded progress in the investigation of this physical scenario.\\
These obstacles range from conceptual issues, such as the question how to extract physical predictions from a background independent theory of quantum gravity, to computational problems due to a lack of systematic methods to evaluate CLQG transition amplitudes.\\
This thesis addresses some of these issues directly and contributes to the development of the field of CLQG. Th structure of this manuscript is as follows:

In chapter~\ref{chap:2} we review the canonical theory of Loop Quantum Gravity (LQG). The main purpose is to work toward a definition of coherent and semi-classical states, which will play an important role in subsequent chapters. In the last section of chapter~\ref{chap:2}, we obtain a first small result: We propose a measure for coherent heat kernel states in the twisted geometry parametrization and show that these states satisfy a resolution of identity with respect to this integration measure. With this, we close a gap in the existing literature. Moreover, this result is important when it comes to the computation of observables from CLQG transition amplitudes, as we discuss in chapter~\ref{chap:CLQG}. The resolution of identity in the twisted geometry parametrization could also play a role in reformulating certain CLQG transition amplitudes in terms of auxiliary variables which have a clear geometric interpretation. Work in this direction is currently on going and we will briefly discuss this possibility in chapter~\ref{chapter_5}.

Chapter~\ref{chap:MathI} introduces the heuristic picture of spin foams and illuminates the relation between LQG on one hand side and Feynman's path integral on the other hand side. We also introduce some basic notions of simplicial discretizations. This serves the purpose to introduce important terminology which is needed in CLQG and it leads us to our second result: A triangulation algorithm for manifolds of topology~$I\times\Sigma$.\\
The need for such an algorithm arises from current efforts to describe the black hole to white hole transition in terms of CLQG amplitudes. In this context, one naturally encounters spacetime regions of topology $I^2\times\mathcal{S}^2$. While it is easy to find a triangulation of $I\times\mathcal{S}^2$, which we also discuss in detail in chapter~\ref{chap:MathI}, it is extremely difficult to generalize the construction to $I^2\times\mathcal{S}^2$. That is where the algorithm comes into play. It has been implemented on a computer and a consistent simplicial triangulation of $I^2\times\mathcal{S}^2$ has been determined.\\
However, knowing the triangulation of a spacetime region is only the first step. Without a systematic method to evaluate CLQG transition amplitudes, the algorithm is of little use. This leads us then to chapters~\ref{chap:CLQG} and~\ref{chapter_5}.

In chapter~\ref{chap:CLQG}, Covariant Loop Quantum Gravity is properly introduced. We refrain from giving a ``derivation'' of the theory and instead just define it. In particular, we show how to recast the CLQG propagator in a form which resembles a Feynman path integral and we define the so-called holomorphic transition amplitude. \\
The main goal is then to determine the black hole to white hole transition amplitude based on a very simple spin foam and compute the bounce time -- a particular observable which is of physical relevance for this transition scenario. To that end, we first discuss the classical spacetime which is used to model the black hole to white hole transition in CQLG and we clarify certain aspects of the construction and the definition of bounce time. This represents joint work with Marios Christodoulou. \\
Then we present the actual computation of the transition amplitude for a simple spin foam, we estimate the bounce time and we clarify several conceptual issues. These represent results that have emerged from a collaboration with Marios Christodoulou. 

The aim of chapter~\ref{chapter_5} is then to generalize the computation method presented in chapter~\ref{chap:CLQG} and to put it on a more solid mathematical foundation. To achieve that, we take two steps back and first consider quantum mechanical holomorphic transition amplitudes in the continuum theory.\\
The reason for studying quantum mechanics first is its simplicity. It allows us to introduce new concepts, understand the meaning of various equations and formal manipulations and in particular it allows us to develop a new approximation method for holomorphic amplitudes.\\
It is found that holomorphic transition amplitudes posses a unique classical limit, unlike the usual Feynman path integral, and the aforementioned approximation method can be understood as a semi-classical expansion around a classical solution. In particular, it quantitatively predicts how holomorphic transition amplitudes decay when we move away from the classical solution.  Moreover, it is shown in Appendix~\ref{App:HolomorphicAmplitudes} that the approximation method developed in this thesis reproduces the exact analytical result for the holomorphic amplitude of the free particle and the harmonic oscillator.\\
Before proceeding to CLQG amplitudes, we add a layer of complexity and study the discretized holomorphic transition amplitude in quantum mechanics. It is shown that most results of the continuum theory are also valid in the discrete theory. This is an important and non-trivial observation which also sheds some new light on issues that appear in CLQG.\\
Finally, we emulate the approximation strategy  developed for quantum mechanics also in the CLQG case. This provides us with a more systematic way to compute CLQG transition amplitudes in a semi-classical regime and it is shown that the new method reproduces the result of the black hole to white hole computation presented in chapter~\ref{chap:CLQG}. It also provides adequate answers to critique that has previously been raised with the computation performed in chapter~\ref{chap:CLQG}.

We then conclude in chapter~\ref{chap:conclusion} with a summary of all results and a discussion of open questions and future research directions. 

%% file: Content/Chapter_2.tex
\chapter{Loop Quantum Gravity Basics}
\label{chap:2}
Progress in Quantum Gravity was impeded for a long time due to the problems discussed in the previous chapter: Perturbative techniques are in conflict with GR's background independence and lead to a non-renormalizable quantum theory while the program of canonical quantization and the path integral approach are ridden with mathematical difficulties.\\
In the mid eighties, however, canonical quantum gravity enjoyed a revival and underwent a phase of rapid development. Conceptual issues were resolved, the theory was given a solid mathematical foundation, and interesting physical consequences were worked out. All of this resulted in a genuine background independent and non-perturbative candidate theory of quantum gravity: Loop Quantum Gravity.\\
While there are still open problems, this theory is a major step forward and it greatly improved our understanding of how to do background independent quantum physics. And it all started with the introduction of new variables.

\section{On the Choice of Variables}

From early on we learn that an adequate choice of variables can drastically simplify the solution of a physical problem. This is especially true in quantum mechanics where an unfortunate choice of variables can complicate the definition of a Hilbert space or the construction of an operator algebra.\\
In field theories there is the additional complication that the choice of variables (i.e. of fundamental fields) directly impacts the action principle because the admissible actions are determined by the theory's field content and symmetries. In metric GR, in absence of matter, the only restriction on the action principle is general covariance and any gauge invariant functional of the metric $g_{\mu\nu}$ can be used to construct an action. Therefore, the most general action for GR in metric variables is formally given by
\begin{equation}\label{eq:MostGeneralAction}
	S[g_{\mu\nu}] = \frac{1}{2\kappa}\int_{\mathcal{M}}\dd^4x\sqrt{-g}\left(\Lambda+R+\sum_{n=2}^{\infty}\alpha_n R^n + \beta_1 R_{\mu\nu\rho\sigma}R^{\mu\nu\rho\sigma}+\dots\right),
\end{equation}
where the $\alpha$'s and $\beta$'s stand for infinitely many coupling constants that need to be determined through experiments and observations. It is true that so far all tests of GR are consistent with $\alpha=\beta=0$ but it is also true that higher order curvature terms could make an appearance in future observations. Even if this is not the case, we find ourselves in the awkward situation of having to dismiss an infinite number of admissible actions before quantizing the theory.\\
According to \cite{Weinberg:1980, PerezBHReview}, it could be that the Wilsonian renormalization group flow \cite{Wilson:1973} selects only a finite number of terms in \eqref{eq:MostGeneralAction}. This possibility, known as the asymptotic safety scenario, is currently under active investigation \cite{Niedermaier:2006}.\\
A different sieving mechanism which drastically reduces the terms needed to take into account for quantization is provided by considering matter couplings. Insisting on a classical theory of gravity which is torsionless in conjunction with the fact that we observe fermions in our universe forces us to give up the metric variables. Instead, we need to introduce a so-called tetrad field which allows to couple fermionic matter to the gravitational field.

\subsection{Fermions and Gravity: The Tetrad Formalism}
Let $(\mathcal{M},g_{\mu\nu})$ be a four-dimensional Lorentzian spacetime manifold coordinatized by~$x$. A frame field, or tetrad field, is then defined as the diffeomorphism $e:T\mathcal{M}\to\mathbb R^4$ from the tangent bundle $T\mathcal{M}$ to a fixed vector space $\mathbb R^4$ equipped with the Minkowski metric $\eta_{IJ} = \text{diag}(-1,1,1,1)$\footnote{Greek lower-case letters $\mu,\nu,\dots = 0,1,2,3$ denote spacetime indices while roman upper-case letters $I,J,\dots = 0,1,2,3$ refer to (internal) Minkowski space indices.}. This map is explicitly given by 
\begin{equation}\label{eq:TetradMap}
	g_{\mu\nu}(x) e^{\mu}_I(x) e^{\nu}_J(x) = \eta_{IJ}.
\end{equation}
At every point $x$ there exist four spacetime vector fields $e^\mu_I$ which provide an orthonormal frame in which the metric reduces to $\eta_{IJ}$. This definition beautifully captures Einstein's intuition that locally gravitation can be transformed away and spacetime looks like Minkowski space. As an added benefit, the local reduction to Minkowski spacetime allows us to define fermions using standard techniques.\\
However, there is more to this formalism. In fact, the Minkowski indices do more than just distinguish between the four vector fields $e^\mu_I$. We may think of spacetime as being equipped with two distinct vectorial structures. Its tangent bundle on one hand side and a collection of Minkowski spaces attached to every spacetime point $x$ on the other hand side \cite{BojowaldBook} (technically speaking this is a vector bundle \cite{BaezBook}). In this context, Minkowski space is often referred to as the internal (vector) space. Such a terminology is often encountered in particle physics where internal spaces are used as representation spaces for gauge groups. \\
As we will now show, the tetrad formalism introduces a new gauge symmetry which acts only on the internal space. From definition \eqref{eq:TetradMap} it follows easily that there exists a field $e^{I}_\mu=\eta^{IJ} e^\nu_J g_{\mu\nu}$ with the properties
\begin{equation}
	e^{I}_\mu e^{\mu}_J = \delta^{I}_J\quad\text{and}\quad e^{I}_\mu e^{\nu}_I = \delta^\mu_\nu.
\end{equation}
Hence, the field $e^{I}_\mu$ can be regarded as the inverse of the tetrad $e^\mu_I$ and we call it the \textit{co-tetrad}. With the help of the co-tetrad we can recast equation \eqref{eq:TetradMap} as
\begin{equation}
	g_{\mu\nu}(x) = e^{I}_\mu(x) e^{J}_\nu(x) \eta_{IJ}.
\end{equation}
This equation shows that knowing all the components of the co-tetrad field allows us to reconstruct the spacetime metric. Hence, we may also shift our perspective and regard the co-tetrad as the fundamental field and the metric as a derived object. Notice, however, that the co-tetrad has $16$ independent components while the metric has, due to its symmetry, only ten components.\\
The additional six degrees of freedom are easily recognized to pertain to an internal gauge symmetry. In fact, we can perform a Lorentz transformation in the internal space by
\begin{equation}
	e^{I}_\mu(x)\longrightarrow \Lambda^{I}_{\ J}(x) e^J_\mu(x),
\end{equation}
but by the very definition of a Lorentz transformation we find that
\begin{equation}
	\Lambda^{I}_{\ K}(x) \Lambda^{J}_{\ L}(x) e^K_\mu(x) e^L_\nu(x) \eta_{IJ} = e^K_\mu(x) e^L_\nu(x) \eta_{KL} = g_{\mu\nu}(x).  
\end{equation}
For the reconstruction of the metric it does therefore not matter which Lorentz frame we pick at $x$. Under local $SO(3,1)$ gauge transformations, all choices are equivalent.\\
Given that besides general covariance we also have an internal gauge symmetry, and given that there are two distinct vectorial spaces associated to the manifold $\mathcal M$, it is not surprising that there are also two distinct notions of covariant differentiation \cite{BojowaldBook, MercuriReview}. For spacetime tensor fields we can still employ the usual covariant derivative $\nabla_\mu$ in terms of the Levi-Civita connection $\Gamma_{\mu\nu}^\lambda$. For tensor fields with Minkowski indices, however, we need to introduce a connection $1$-form $\omega^{IJ}_{\mu}$. For an internal vector field we can then define the covariant derivative as
\begin{equation}
	\mathcal D_\mu v^{I}(x) := \nabla_\mu v^{I}(x) + \omega_{\mu\,\ J}^{\ I}(x) v^{J}(x).
\end{equation}
This covariant derivative also applies to tensors with mixed indices, such as the co-tetrad $e^{I}_\mu$ for instance. Its generalization to tensors with several (mixed) indices is straightforward \cite{BojowaldBook}, but it will not be need in the following. What is more important for us is the transformation behavior of the connection $1$-form under position-dependent Lorentz transformations $\Lambda^{I}_{\ J}(x)$. For $\mathcal D_\mu$ to be a covariant derivative, mapping internal vectors to internal vectors, we must require
\begin{equation}
	\tilde{\mathcal D}_\mu\tilde{v}^{I} = \Lambda^{I}_{\ J}\mathcal D_\mu v^{J} \quad \text{for}\quad \tilde{v}^{I} = \Lambda^{I}_{\ J}(x) v^{J}. 
\end{equation}
One can show \cite{BojowaldBook} that this requirement translates into the following transformation behavior for the connection $1$-form:
\begin{equation}
	\tilde{\omega}^{\ I}_{\mu\,\ J} = (\Lambda^{-1})^{I}_{K}\, \omega^{\ K}_{\mu\,\ L} \, \Lambda^L_J\, +\, (\Lambda^{-1})^{I}_{\ K}\partial_\mu\Lambda^{K}_{\ J}.
\end{equation}
Another natural requirement is that the connection $\mathcal D_\mu$ annihilates the Minkowski metric. A straight forward computation then shows that the connection $1$-form has to be antisymmetric in its internal indices, i.e.
\begin{equation}
	\omega^{\,\ IJ}_\mu = -\omega^{\,\ JI}_\mu.
\end{equation}
This condition guarantees that parallel transporting a vector along a curve simply amounts to a Lorentz transformation and the Minkowski space structure is therefore preserved. Moreover, the antisymmetry implies $\omega^{\,\ IJ}_\mu\in\mathfrak{so}(3,1)$.\\
A further advantage of tetrads is that they lend themselves easily to the differentialform notation. By introducing the $1$-forms
\begin{equation}
	e^{I}:= e^{I}_\mu \dd x^\mu\quad\text{and}\quad \omega^{IJ} := \omega^{IJ}_\mu \dd x^\mu
\end{equation}
we can define a geometry in the sense of Cartan, which includes torsion and contains Riemannian geometry as a special case. The basic geometrical objects are the so-called torsion $2$-form and the curvature $2$-form:
\begin{align}\label{eq:CartanForms}
	T^{I} &= \dd e^{I} + \omega^{I}_{\ J}\wedge e^{I} & &\text{(Torsion $2$-form)}\notag\\
	F^{I}_{\ J} &=  \dd\omega^{I}_{\ J} + \omega^{I}_{\ K}\wedge\omega^{K}_{\ J} & &\text{(Curvature $2$-form)}.
\end{align}
To recover Riemannian geometry we require vanishing torsion which is tantamount to imposing
\begin{equation}\label{eq:TorsionlessCondition}
	\dd_\omega e^{I}:=\dd e^{I} + \omega^{I}_{\ J}\wedge e^{I} = 0
\end{equation}
It can be shown that this equation has a unique solution $\omega[e]$ for the connection form which is called the \textit{(torsionless) spin connection} and it is explicitly given by
\begin{equation}\label{eq:TorsionlessSpinConnection}
	\omega_{\mu\,\ J}^{\,\ I} = e^{\nu\, I}\nabla_\mu e_{\nu\, J}.
\end{equation}
One can further show that the Riemann curvature tensor is related to the curvature $2$-form by
\begin{equation}
	F^{IJ} = e^{I}_\mu\, e^{J}_\nu\, R^{\mu\nu}_{\,\,\rho\sigma}\,\dd x^{\rho}\wedge\dd x^{\sigma}.
\end{equation}
Together with $e^{I}\wedge e^{J}\wedge e^{K}\wedge e^{L} = \varepsilon^{IJKL}|e|\dd^4 x = \varepsilon^{IJKL}\sqrt{-g}\dd^4 x$ this suffices to show that the Einstein-Hilbert action can be rewritten as \cite{SpezialeDonaReview, RovelliBook}
\begin{equation}\label{eq:EHequalsT}
	S_\text{EH}[g(e)] = -\frac{1}{2\kappa} \int_\mathcal{M} \varepsilon_{IJKL}\, e^{I}\wedge e^{J}\wedge F^{KL} =: S_\text{T}[e].
\end{equation}
It is worth noting that, despite the above equality, the Einstein-Hilbert action and the tetrad action $S_T[e]$ are not equivalent. This can be seen by performing a time-reversal $T\in SO(3,1)$ gauge transformation or a parity reversal $P\in SO(3,1)$ in the internal space \cite{Rovelli:2012b, RovelliBook}. As the metric is invariant under internal gauge transformations, the Einstein-Hilbert action remains unchanged. However, the tetrad action $S_\text{T}[e]$ flips sign. The difference of the two actions becomes manifest when written in terms of the tetrad and the Ricci scalar:
\begin{align}\label{eq:EH_vs_Palatini}
	S_\text{EH}[e] &= \frac{1}{\kappa}\int_\mathcal{M} |\det e|\, R[e]\, \dd^4x\notag\\
	S_\text{T}[e] &= \frac{1}{\kappa}\int_\mathcal{M} (\det e)\, R[e]\, \dd^4 x.	
\end{align}
The tetrad action is evidently sensitive to the orientation of the frame field. While this sign difference is irrelevant in the pure gravity sector, it has potentially observable consequences for fermionic matter, as it couples directly to the tetrad~\cite{Christodoulou:2012}. This sign will also make an appearance in the classical limit of the spin foam model discussed in chapter~\ref{chap:CLQG}.\\
For the remainder of this chapter, the relative sign difference will be irrelevant and instead we will be concern with a special type of action.

\subsection{First Order Formalism}
Consider the so-called (tetradic) Palatini action
\begin{equation}
	S_\text{P}[e,\omega] := \frac{1}{2\kappa}\int_\mathcal{M}\varepsilon_{IJKL}e^{I}\wedge e^{J}\wedge F^{KL}[\omega].
\end{equation}
This action looks deceitfully similar to \eqref{eq:EHequalsT}, but there is an important difference: The tetrad $e$ and the connection $\omega$ are treated as independent variables and the action is therefore \textit{polynomial}. This is often called the \textit{first order formulation} and one can easily show that it reproduces the Einstein field equations\footnote{Formulations where the metric or the tetrad are the only variables are generally referred to as \textit{second order formulations}.}. To that end we compute the variation of $S_\text{P}[e,\omega]$ with respect to the connection:
\begin{align}
	\delta_\omega S_\text{P}[e,\omega] &= \frac{1}{2\kappa} \int_\mathcal{M}\varepsilon_{IJKL}\, e^{I}\wedge e^{J}\wedge \dd_\omega(\delta\omega^{IJ})\notag\\
	&= -\frac{1}{2\kappa}\int_\mathcal{M}\varepsilon_{IJKL}\, \dd_\omega\left(e^{I}\wedge e^{J}\right)\wedge \delta\omega^{IJ}.
\end{align}
In the first line we used the Palatini identity $\delta_\omega F^{IJ} = \dd_\omega(\delta\omega^{IJ})$ (a direct consequence of Cartan's second structure equation \eqref{eq:CartanForms}) and in the second line we performed a partial integration. Imposing the vanishing of the variation and using the invertibility of the tetrad finally results in
\begin{equation}
	\dd_\omega e^{I} = 0.
\end{equation}
We recognize this purely algebraic equation to be the torsionless condition~\eqref{eq:TorsionlessCondition} and this in turn implies that $\omega$ has to be the torsionless spin connection given by\footnote{The presence of a non-dynamical equation which simply fixes one of the variables should not come as a surprise. After all, the Palatini action is simply the Einstein-Hilbert action with twice the variables and there has to be a relation between them.} \eqref{eq:TorsionlessSpinConnection}. The Palatini action hence reduces to the tetrad action \eqref{eq:EHequalsT} and the Einstein field equations follow from the variation with respect to $e$.\\
For gravity coupled to matter fields we find again that the first order formulation and the second order formulation are not completely equivalent. Fermions couple in a slightly different way in the two formulations and translating between them requires the use of a four-fermion interaction term \cite{Freidel:2005, Perez:2005, Mercuri:2006}.\\
The Palatini action can easily be extended to include a cosmological term
\begin{equation}\label{eq:CosmologicalTerm}
	S_\Lambda[e] := \frac{\Lambda}{2\kappa}\int_\mathcal{M}\varepsilon_{IJKL}\, e^{I}\wedge e^{J}\wedge e^{K}\wedge e^{L}.
\end{equation}
But this is not the only possible extension. In fact, there are four more $4$-forms that can be expressed with $e$ and $\omega$ and which are compatible with general covariance and internal gauge invariance~\cite{PerezBHReview}:
\begin{align}\label{eq:TopologicalTerms}
	S_\text{top.}[e,\omega] := \frac{1}{2\kappa}\int_\mathcal{M}&\overset{\text{Holst}}{\overbrace{\alpha_1 e_{I}\wedge e_{J}\wedge F^{IJ}}} + \overset{\text{Nieh-Yan}}{\overbrace{\alpha_2\left(\dd_\omega e^{I}\wedge \dd_\omega e_{I} - e_I\wedge e_J\wedge F^{IJ}\right)}}\notag\\
	&   + \underset{\text{Pontrjagin}}{\underbrace{\alpha_3F_{IJ}\wedge F^{IJ}}} + \underset{\text{Euler}}{\underbrace{\alpha_4 \varepsilon_{IJKL}F^{IJ}\wedge F^{KL}}}
\end{align}
In the pure gravity sector, and independently of the values of the coupling constants $\alpha_1,\alpha_2,\alpha_3,\alpha_4$, these terms do not modify the Einstein field equations \cite{PerezBHReview}. Instead, they describe topological invariants of the manifold $\mathcal M$. A canonical analysis of the Palatini action supplemented by the topological terms \eqref{eq:TopologicalTerms} reveals that they induce a canonical transformation on the phase space of GR \cite{Mercuri:2006, Date:2008, Rezende:2009}. In particular, it turns out that the coupling constants of the Pontrjagin and Euler term can be chosen freely \cite{Rezende:2009} while the coupling constants of the Holst and Nieh-Yan terms appear in the definition of the canonical momentum as
\begin{equation}
	\gamma = \frac{1}{\alpha_1 + 2 \alpha_2}.
\end{equation}
In the literature, $\gamma$ is usually referred to as the Barbero-Immirzi parameter\footnote{Here it is assumed that $\gamma$ is real and non-zero. Complex values can also be considered, see \cite{RovelliBook,ThiemannBook,AshtekarBook} and references therein.} \cite{Barbero:1994, Immirzi:1996} and the Nieh-Yan term is neglect such that one usually has $\alpha_1 = \frac{1}{\gamma}$. Since the canonical analysis of the most general first order action only depends on $\gamma$ and not on the individual coupling constants, we will set $\alpha_2=\alpha_3=\alpha_4=0$ and $\alpha_1=\frac{1}{\gamma}$. Moreover, we will not consider matter couplings and also set $\Lambda=0$. Hence, in the canonical analysis and the consequent quantization we will be concerned with the so-called Holst action \cite{Holst:1995, Hojman:1980}
\begin{equation}\label{eq:HolstAction}
		S_\text{Holst}[e,w] := \frac{1}{2\kappa}\int_\mathcal{M}(\varepsilon_{IJKL}+\frac{1}{\gamma}\eta_{IK}\, \eta_{JL})\left(e^{I}\wedge e^{J}\wedge F^{KL}[\omega]\right).
\end{equation}
Before tackling the canonical analysis, we emphasize the drastic reduction in admissible action principles we need to consider for the quantization. Our insistence on a torsionless classical theory of gravity which can be coupled to fermionic matter led us to introduce the tetrad as fundamental field. In terms of tetrads, there are only six $4$-forms that are compatible with general covariance and internal gauge invariance: The Palatini action, the cosmological term, and the four topological invariants \eqref{eq:TopologicalTerms}. We choose $\Lambda = 0$ for simplicity and concern ourselves with pure gravity, i.e. we disregard the presence of matter. The sole effect of the topological invariants is then to induce canonical transformations on the phase space of the theory. These transformations depend on the parameter $\gamma$ and as we will see, this induces a one-parameter quantization ambiguity reminiscent of the $\theta$-vacuum ambiguity in QCD. Remember that in QCD it is possible to add a topological term to the action which is compatible with the field content and the symmetries of the theory,
\begin{equation}
	S_\textsf{QCD} = \int \Tr\left(F\wedge(*+\theta) F\right).
\end{equation}
This additional term leads however to quantum theories which are unitarily inequivalent and the $\theta$ parameter makes an explicit appearance.\\
In the next subsection, we will discover a further connection between gravity and non-Abelian Yang-Mills theory.

\subsection{Canonical Analysis and the Ashtekar-Barbero Connection}\label{ssec:CanonicalAnalysis}
There are many different ways to perform the canonical analysis of the Holst action and introduce the Ashtekar-Barbero connection \cite{Ashtekar:1987,Barbero:1994,PerezBHReview,AshLewReview}. Every approach has its advantages and disatvantages and can be used to emphasize different aspects of the analysis. Here we opt for simplicity and follow \cite{PerezBHReview,AshLewReview}.\\
A general strategy to identify canonically conjugate variables is to perform the variation of the action without imposing the boundary conditions. This even works for discrete systems \cite{DAmbrosio:2019} and in the case of Newtonian mechanics one finds
\begin{equation}
	\delta S[p] = \int_{t_i}^{t_f} \left(\PD{L}{q}-\frac{\dd}{\dd t}\PD{L}{\dot q}\right)\,\delta q\, \dd t + \left.\PD{L}{\dot q}\,\delta q\right\vert_{t_i}^{t_f}.
\end{equation}
The boundary term vanishes after imposing the boundary conditions and requiring $\delta S\overset{!}{=}0$ leads to the equations of motion. However, if we do not impose any of these conditions we can simply read off the momentum canonically conjugate to $q$ from the boundary term.\\
Applying this strategy to the Holst action we find, with $p_{IJKL}:=\varepsilon_{IJKL} + \frac{1}{\gamma}\eta_{IK}\,\eta_{JL}$,
\begin{align}\label{eq:VariationHolst}
	\delta S_\text{Holst}[e,\omega] &= \frac{1}{2\kappa}\int_\mathcal{M} \left(2\,p_{IJKL} \delta e^{I}\wedge e^{J}\wedge F^{KL}[\omega] + p_{IJKL} e^{I}\wedge e^{J}\wedge \dd_\omega(\delta \omega^{KL})\right)\notag\\
	&= \frac{1}{2\kappa}\int_\mathcal{M}\left(2\,p_{IJKL} \delta e^{I}\wedge e^{J}\wedge F^{KL}[\omega]-p_{IJKL}\dd_\omega(e^{I}\wedge e^{J})\wedge\delta\omega^{KL}\right)\notag\\
	&+ \frac{1}{2\kappa}\int_{\partial\mathcal{M}}p_{IJKL} e^{I}\wedge e^{J}\wedge\delta\omega^{KL}.
\end{align}
In the first line we used the Palatini identity $\delta_\omega F^{KL}[\omega] = \dd_\omega(\delta\omega^{KL})$ and then performed a partial integration to obtain the second line. From the bulk integral we can simply read off the algebraic equations for the connection and the equations of motion for the tetrad. At this point it is possible to check that $\omega$ still reduces to the torsionless spin connection and that the equations of motion for the tetrad remain unchanged because the term containing the Barbero-Immirzi parameter vanishes. From the boundary term we infer that the canonically conjugate momentum to $\omega$ is given by
\begin{equation}
	\frac{1}{2\kappa}(\varepsilon_{IJKL}+\frac{1}{\gamma}\eta_{IK}\,\eta_{JL})\, e^{I}\wedge e^{J}.
\end{equation}
It is convenient, and very common, to introduce a partial gauge-fixing in the internal space. This so-called \textit{time-gauge} amounts to choosing the $e^{0}_a$ component of the tetrad to be aligned with the normal $n_a$ of the boundary $\partial\mathcal{M}$, i.e.
\begin{equation}
	e^{0}_a = n_a.
\end{equation}
In other words, we choose the time axis of the frame field to coincide with the time axis singled out by the spacelike boundary \cite{PerezBHReview}. This partial gauge-fixing is always possible due to the $SO(3,1)$ gauge-freedom and, once the time-axis has been fixed, reduces the internal gauge group to $SO(3)$.\\
This partial gauge-fixing also allows us to rewrite the last term in~\eqref{eq:VariationHolst} in a convenient way by separating the $e^{0}$ component from the spatial components $e^{a}$ with $a=1,2,3$.
\begin{align}
	&\frac{1}{\kappa}\int_{\Sigma}\left(\varepsilon_{0abc}e^{0}\wedge e^{a}\wedge \delta\omega^{bc} + \frac{1}{\gamma}e^0\wedge e^{a}\wedge \delta\omega_{0a}\right) - \frac{1}{\kappa} \int_{\Sigma} \left(\varepsilon_{0abc}\, e^{a}\wedge e^{b}\wedge\delta \omega^{c 0} + \frac{1}{\gamma}e^{a}\wedge e^{b}\wedge \delta\omega_{ab}\right)	\notag\\
	& = -\frac{1}{\gamma\kappa}\int_{\partial\mathcal M}\left[\varepsilon_{abc}\, e^{a}\wedge e^b\right]\wedge \delta \left(\gamma\omega^{c0}+\varepsilon^{cdf} \omega_{df}\right)
\end{align}
We defined $\Sigma:=\partial\mathcal M$ and used the fact that $e^0$ is normal to $\Sigma$ and therefore its pull-back to $\Sigma$ vanishes, making the first term in the first line disappear. From the last line we can now read off a new configuration space variable
\begin{equation}
	A^{a} := \gamma\omega^{c0}+\varepsilon^{cdf} \omega_{df}.
\end{equation}
This is the celebrated Ashtekar-Barbero connection \cite{Ashtekar:1987, Barbero:1994}. It transforms as an $SO(3)$ connection under internal gauge transformations and it is canonically conjugate to the momentum variable
\begin{equation}
	E^{a} := \varepsilon^{a}_{\,\ bc}\, e^{b}\wedge e^{c},
\end{equation}
which is often referred to as \textit{electric field} for reasons that will become clear in the sequel. The associated Poisson brackets to the phase space couple $(A^{i}_a, E^{a}_i)$ read
\begin{align}
	\{A^{i}_a(x), A^{j}_b(y)\} & = 0\notag\\
	\{E^{a}_i(x), E^{b}_j(y)\} &= 0\notag\\
	\{E^{a}_i(x), A^{j}_b(y)\} &= \gamma\kappa\, \delta^{a}_b \delta^{i}_j \delta^{(3)}(x,y).
\end{align}
This phase space structure is equivalent to a non-Abelian $SO(3)$ Yang-Mills theory, as partially anticipated in the previous subsection. Having worked out the phase space structure and the fact that it is similar to Yang-Mills theory puts us in a good position for quantization. However, to achieve a background independent quantization we need to take into account general covariance. This will be the content of the next subsection.\\
To conclude the present subsection we briefly discuss the constraints in the phase space variables $(A^{i}_a, E^{a}_i)$, as they will be important for the quantum theory presented in section~\ref{sec:Quantization}. Through rather long and tedious computations it can be shown that there are seven constraints: The Gauss constraints ($3$), the vector constraints ($3$) and the scalar constraint ($1$). They are explicitly given by
\begin{align}
	G_i(A^{j}_a, E^{a}_j) &= D_a E^{a}_j \approx 0	\notag\\
	V_b(A^{j}_a, E^{a}_j) &= E^{a}_j F_{ab} - (1+\gamma^2) K^{i}_a G_i(A^{j}_a, E^{a}_j) \approx 0\notag\\
	S(A^{j}_a, E^{a}_j) &=\frac{E^{a}_i E^{a}_j}{\sqrt{\det E}}\left(\varepsilon^{ij}_{\,\,\ k} F^k_{\,\ ab}- 2(1+\gamma^2) K^{i}_{[a}K{j}_{b]}\right) \approx 0,
\end{align}
where $D_a E^{a}_j = \partial_a E^{a}_j + \varepsilon_{ij}^{\,\,\ k} A^{j}_a E^{a}_k$ is the covariant divergence of the electric field, analogous to $\nabla\cdot E=0$ from vacuum electrodynamics, and $F_{ab} = \partial_a A^{i}_b - \partial_b A^{i}_a + \varepsilon^{i}_{\,\ jk} A^{j}_a A^{k}_b$ is the curvature tensor of the Ashtekar-Barbero connection.\\
These constraints reduce the $18$-dimensional phase space spanned by $(A^{i}_a, E^{a}_i)$ since they impose seven conditions among these variables and they generate seven gauge transformations on the constraint surface. Hence, one finds $18-7-7 = 4$ degrees of freedom in the reduced phase space which correspond to the two propagating degrees of freedom of GR. Just as expected.\\
The aforementioned gauge transformations can be studied by smearing the constraints with suitable test function. For the Gauss constraint we may define the smeared version as 
\begin{equation}
	G[\alpha] := \int_{\mathcal S} \alpha^{i} G_i(A^{j}_a, E^{a}_j)\, \dd^3 x.
\end{equation}
This constraint then generates a well-defined gauge transformation by virtue of the Poisson bracket applied to the canonical variables \cite{Perez2004Review}:
\begin{equation}
	\delta_G A^{i}_a := \{A^{i}_a, G[\alpha]\} = -D_a \alpha^{i}\quad\text{and}\quad \delta_G E^{a}_i := \{E^{a}_i, G[\alpha]\} = [E,\alpha]_i
\end{equation}
These equations are again familiar from non-Abelian Yang-Mills theory and furthermore one can show that, upon defining $A_a := A^{i}_a \tau_i\in\mathfrak{su}(2)$ and $E^{a} := E^{a}_i \tau^{i}\in\mathfrak{su}(2)$ and exponentiating the above quations, these gauge transformations amount to
\begin{equation}
	\tilde{A}_a = U A_a U^\dagger + U \partial_a U^\dagger\quad\text{and}\quad \tilde{E}^{a} = U E^{a} U^\dagger\quad\text{with}\quad U\in SU(2),
\end{equation}
which is the standard way under which the connection and the electric field transform in non-Abelian Yang-Mills theory. We can proceed in a similar fashion for the vector constraints and define
\begin{equation}
	V[N^{a}] := \int_{\mathcal V} N^{a} V_a(A^{j}_a, E^{a}_j)\, \dd^3 x,
\end{equation} 
where $\mathcal V$ is some $3$-dimensional hypersurface in $\mathcal M$. The Poisson brackets of this constraint with the canonical variables are given by \cite{Perez2004Review}
\begin{equation}
	\delta_V A^{i}_a := \{A^{i}_a, V[N^{a}]\} = \mathcal L_{\vec{N}} A^{i}_a\quad\text{and}\quad \delta_V E^{a}_i := \{E^{a}_i, V[N^{a}]\} = \mathcal L_{\vec{N}} E^{a}_i,
\end{equation}
where $\mathcal L_{\vec{N}}$ is the Lie derivative in direction of the vector $N^{a}$. Hence, the gauge transformations generated by the vector constraint are spatial diffeomorphisms. Similarly, one can show that the scalar constraint generates coordinate time evolution.

\subsection{The Holonomy-Flux Variables}\label{ssec:HolonomyFlux}
The Poisson algebra of the previous subsection is not suitable for quantization. The reason is the presence of the Dirac distribution and the explicit appearance of coordinates $x$ and $y$. To obtain a well-defined, regularized, Poisson algebra it is only natural to integrate the connection and the electric field against test functions. Since our objective is to obtain a background independent quantum theory, we have to build integrals of the $(A^{i},E_i)$ variables which do not involve any background structures. We may furthermore exploit the fact that the connection and the electric field are $1$-forms and $2$-forms, respectively, which can naturally be integrated over $1$- and $2$-dimensional manifolds\footnote{It is worth noting that different integration manifolds can be considered and integrals different from the ones studied here can be built \cite{ThiemannBook}. These constructions also lead to a regularization, but they introduce complications elsewhere. The regularization considered here is the standard one and so far the simplest that has been found.}. With this in mind, we regularize the electric field as
\begin{equation}\label{eq:Flux}
	E[\mathcal S] := \int_{\mathcal S}n_a(\star E)^a = \int_{\mathcal S} \dd^2 \sigma\, n_a E^{a}\in\mathfrak{su}(2),
\end{equation}
where $\mathcal S\subset\mathcal M$ is a two dimensional surface, $E^{a}:=E^{a}_i \tau^{i}$, and $n_a = \varepsilon_{abc} \PD{x^b}{\sigma^1}\PD{x^c}{\sigma^2}$ is the normal to that surface. The quantity \eqref{eq:Flux} is called the \textit{flux} of the electric field. This new variable depends, as indicated above, on the choice of surface and it is manifestly coordinate-independent.\\
In the case of the Ashtekar-Barbero connection, which is a $1$-form, it is more natural to consider a path $\gamma:[0,1]\to\mathcal M$ as integration manifold. Given such a connection $A^{i}_a$, we can associate to it an element of $SU(2)$ by defining $A_a:= A^{i}_a \tau_i$ where $\tau_i$ are the generators of $SU(2)$. Then we define the \textit{holonomy} by
\begin{equation}
	h_\gamma[A] := \textsf P \exp\left(\int_\gamma A^{i}_a(\gamma(s)) \frac{\dd \gamma^{a}(s)}{\dd s}\, \tau_i\, \dd s\right) \in SU(2),
\end{equation}
where $\textsf{P} \exp$ denotes the path-ordered exponential function. As can easily be shown \cite{BaezBook, SpezialeDonaReview}, the holonomy satisfies the first order differential equation
\begin{equation}
	\frac{\dd}{\dd t} h_\gamma[A(t)] - h_\gamma[A(t)] A(\gamma(t)) = 0\quad\text{with}\quad h_\gamma[A(0)] = \id_{2\times 2}.
\end{equation} 
From the definition it also follows that the holonomy satisfies the following properties \cite{SpezialeDonaReview}:
\begin{itemize}
	\item[a)] The holonomy of two composed paths is the product of the individual holonomies: $h_{\gamma_1\circ \gamma_2}[A] = h_{\gamma_1}[A] h_{\gamma_2}[A]$. This also implies $h_{\gamma^{-1}}[A] = h_\gamma^{-1}[A]$.
	\item[b)] Under local $SU(2)$ transformations $U(x)$, the holonomy transforms as $h_{\gamma}[A]\rightarrow U(s(\gamma)) h_{\gamma}[A] U^\dagger(t(\gamma))$, where $s(\gamma)$ and $t(\gamma)$ denote the source and target points of the path $\gamma$. 
	\item[c)] Under diffeomorphisms $\phi$, the holonomy transforms as $h_\gamma[\phi^* A] = h_{\phi^{-1}\circ\gamma}[A]$.
	\item[d)] The functional derivative with respect to the connection gives
	\begin{equation*}
		\frac{\delta h_\gamma[A]}{\delta A^{i}_a(x)} = \begin{cases}
			\frac12 \dot x^{a} \delta^{(3)}(\gamma(s), x) \tau_i h_\gamma[A] & \text{if } x \text{ is the source of } \gamma\\
			\frac12 \dot x^{a} \delta^{(3)}(\gamma(s), x) h_\gamma[A] \tau_i  & \text{if } x \text{ is the target of } \gamma\\
			\dot x^{a} \delta^{(3)}(\gamma(s), x) h_\gamma(0,s)\tau_i h_\gamma(s;1) & \text{if } x \text{ is inside } \gamma
		\end{cases}
	\end{equation*}
\end{itemize}
What makes the holonomy particularly interesting is its simple transformation under the action of $SU(2)$ elements. This facilitates the construction of gauge-invariant quantities which solve the Gauss constraint and, moreover, these variables are similar to the Wilson loops used in lattice QCD.\\
As it will turn out, the holonomy-flux variables are also well-suited for quantization.

\section{Canonical Quantization \`a la Dirac}\label{sec:Quantization}
With the introduction of new variables we have come a big step closer to kneading GR into a form more suitable for quantization. The transition from metric variables to tetrads and connection variables greatly restricted the number of admissible action principles we need to take into account for quantization. The subsequent introduction of the Ashtekar-Barbero and its canonically conjugate momentum as phase space variables was a very natural step: They simply appeared during the canonical analysis of the Holst action and their presence can be traced back to a canonical transformation induced by the topological terms.\\
These variables allow us to think about GR as a $SU(2)$ Yang-Mills theory with polynomial constraints. This suggests the use of a canonical quantization \`{a} la Dirac \cite{Perez2004Review} (for a more rigorous treatment see \cite{ThiemannBook, AbhayPullinBook}):
\begin{my_list_item}
	\item[1)] Find a representation of the phase space variables of the theory as operators in an auxiliary, so-called \textit{kinematical}, Hilbert space $\mathcal H_\textsf{kin}$ on which the Poisson brackets can be promoted to commutators. Schematically:
	\begin{equation}
		[\hat A,\hat B] = i \hbar\, \widehat{\{A,B\}}.
	\end{equation}
	\item[2)] Promote the constraints to self-adjoint operators on $\mathcal H_\textsf{kin}$.
	\item[3)] Characterize the space of solutions to the constraints and define the physical inner product. This defines the physical Hilbert space $\mathcal H_\textsf{phys}$.
	\item[4)] Find a complete set of gauge invariant observables, i.e. operators that commute with the constraints. 
\end{my_list_item}
In the next subsections we will follow these steps as far as possible and clarify some of the terminology.

\subsection{The Kinematical Hilbert Space}
\begin{center}
	\begin{figure}[htb]
		\centering
		\includegraphics[width=0.8\textwidth]{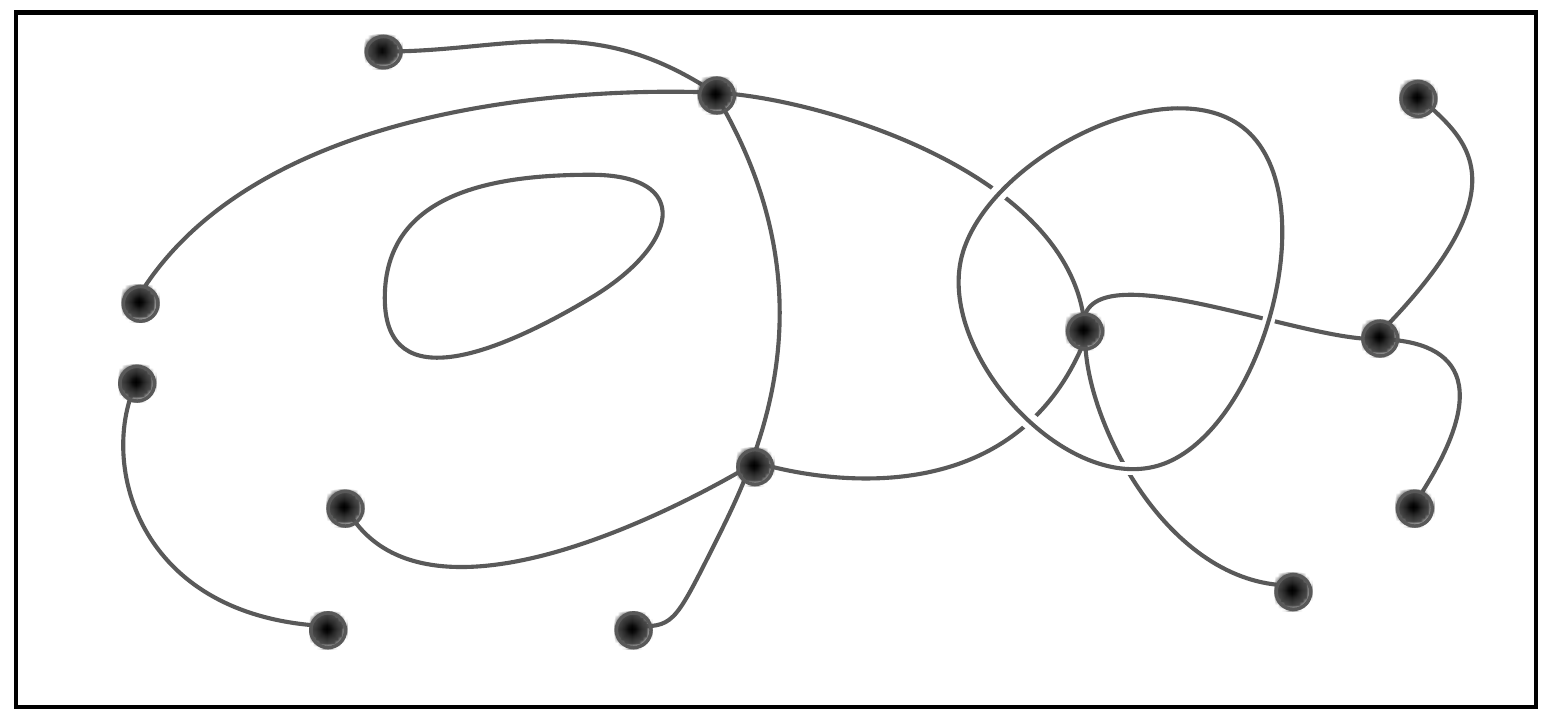}
		\caption{An example of a graph $\Gamma$.}
		\label{fig:GraphExample}
	\end{figure}
\end{center}

In subsection \ref{ssec:HolonomyFlux} we introduced a regularized version of the Poisson algebra. This suggests the use of the holonomy-flux algebra as starting point for quantization. However, we cannot simply promote the Poisson brackets to commutators. First we need to find an adequate Hilbert space on which the holonomy-flux variables can be represented as operators.\\
To that end, consider the space $\mathcal A$ of real Ashtekar-Barbero connections $A^{i}_a$ defined on a three-dimensional hypersurface $\Sigma$ of fixed topology. The $SU(2)$ holonomies introduced in \ref{ssec:HolonomyFlux} are functions on the space $\mathcal A$ and we can use them as basic building blocks to introduce a new space on which to represent the holonomy-flux variables \cite{AbhayPullinBook}. The construction proceeds as follows:\\
Let $\Gamma$ be a \textit{graph}, i.e. an ordered collection of smooth oriented paths $\gamma_\ell\subset\Sigma$, with $\ell=1,...,L$. These paths meet at most at their endpoints (see Figure~\ref{fig:GraphExample}) and they will be referred to as \textit{links} while their endpoints are called \textit{nodes}.
Let $f:SU(2)^L\to\mathbb C$ be a smooth function of $L$ group elements. The couple $(\Gamma, f)$ then defines a functional of the connection $A$ through
\begin{equation}
	\Psi_{\Gamma,f}[A] := f(h_{\gamma_1}[A]\dots, h_{\gamma_L}[A]).
\end{equation}
These functions are called \textit{cylindrical function} and they belong to a linear space denoted by $\textsf{Cyl}_\Gamma$. The space of cylindrical function can be turned into a Hilbert space $\mathcal H_\Gamma$ by equipping it with a scalar product. The shift from connections as fundamental variables to holonomies becomes crucial at this point: Since the holonomies are $SU(2)$ group elements, there is a natural candidate for a scalar product $\innerp{\,\cdot\,}{\,\cdot\,}_\textsf{kin}:\textsf{Cyl}_\Gamma\times\textsf{Cyl}_\Gamma\to\mathbb C$ which is
\begin{equation}\label{eq:KinematicalScalarProduct}
	\innerp{\Psi_{\Gamma,f}}{\Psi_{\Gamma,g}}_\textsf{kin} := \int_{SU(2)^L}\dd h_1\cdots\dd h_L\, \overline{f(h_1,\dots, h_L)}\, g(h_1,\dots, h_L).
\end{equation}
We introduced the shorthand notation $h_\ell := h_{\gamma_\ell}[A]$ for the holonomies and $\dd h_\ell$ denotes the $SU(2)$ Haar measure. This measure is  gauge-invariant in the sense that
\begin{equation}
	\dd h_\ell = \dd(g h_\ell) = \dd(h_\ell g) = \dd h_\ell^{-1}\quad\forall g\in SU(2)
\end{equation}
and normalized as
\begin{equation}
	\int_{SU(2)}\dd h_\ell = \id_{SU(2)}.
\end{equation}
A direct consequence of the gauge-invariance of the Haar measure is that the scalar product as a whole is gauge invariant under $SU(2)$ transformations, as can easily be verified.\\
The inner product \eqref{eq:KinematicalScalarProduct} is also invariant under extended spatial diffeomorphisms\footnote{Extended diffeomorphisms are continuous and invertible maps $\phi:\Sigma\to\Sigma$ such that the maps and their inverses are smooth everywhere except, possibly, in a finite number of isolated points \cite{RovelliBook}.} $\phi\in\textsf{Diff}^*(\Sigma)$. This follows directly from the transformation behavior of the holonomy under diffeomorphisms (see subsection~\ref{ssec:HolonomyFlux}) which amounts to moving around $\Gamma$ on $\Sigma$ and the fact that the right hand side of the scalar product is insensitive to the embedding of $\Gamma$ into the manifold $\Sigma$.\\
Notice that so far we considered only a single graph $\Gamma$. Since we regularized the continuous connection on a fixed structure, this is tantamount to truncating the degrees of freedom of the theory. A natural and necessary step is therefore to extend the fixed-graph space $\textsf{Cyl}_\Gamma$ to
\begin{equation}
	\textsf{Cyl} := \bigcup_{\Gamma\subset\Sigma} \textsf{Cyl}_\Gamma.
\end{equation}
The scalar product \eqref{eq:KinematicalScalarProduct} can be extended in a natural way to functionals on $\textsf{Cyl}$. In fact, observe that $(\Gamma, f)$ and $(\Gamma', f')$ may define the same functional under certain circumstances. For instance, if $\Gamma$ shares $L'$ links with $\Gamma'$ and contains $L''$ other links but the functions satisfy $f(h_1,\dots, h_{L'}, h_{L'+1},\dots, h_{L''}) = f'(h_1,\dots, h_{L'})$, then we clearly have $\Psi_{\Gamma,f}[A] = \Psi_{\Gamma',f'}[A]$.\\
This observation suggests to extend the scalar product in the following way: Given two functionals, $\Psi_{\Gamma, f}$ and $\Psi_{\Gamma', f'}$, trivially extend the functions $f$, $f'$ to $\Gamma\cup\Gamma'$ and then define
\begin{equation}\label{eq:InnerProduct}
	\innerp{\Psi_{\Gamma,f}}{\Psi_{\Gamma',f'}}_\textsf{kin} := \innerp{\Psi_{\Gamma\cup\Gamma',f}}{\Psi_{\Gamma\cup\Gamma',f'}}_\textsf{kin}.
\end{equation} 
The extended scalar product inherits the $SU(2)$ and $\textsf{Diff}^*(\Sigma)$ gauge invariance from the scalar product \eqref{eq:KinematicalScalarProduct}. The kinematical Hilbert space is now defined as the Cauchy completion of the space of cylindrical functions $\textsf{Cyl}$ with respect to the inner product \eqref{eq:InnerProduct}. This can formally be written as
\begin{equation}
	\mathcal H_\textsf{kin} = \overline{\bigoplus_{\Gamma\subset\Sigma}\mathcal H_\Gamma}.
\end{equation}
and it means that in addition to cylindrical functions we add the limits of all Cauchy convergent sequences $\|\Psi_m-\Psi_n\|$ in the norm induced by the scalar product \eqref{eq:InnerProduct}. \\
A key result due to Ashtekar and Lewandowski~\cite{Ashtekar:1994} states that the kinematical Hilbert space $\mathcal H_\textsf{kin}$ can be realized as a $L_2$ Hilbert space of (generalized, distributional) connections, i.e.
\begin{equation}
	\mathcal H_\textsf{kin} = L_2[A, \dd\mu_\text{AL}],
\end{equation}
where $\dd\mu_\text{AL}$ is the Ashtekar-Lewandowski measure on the space of generalized connections. This means in particular that \eqref{eq:InnerProduct} can be seen as a scalar product between cylindrical functionals of the connection $A$ with respect to the measure $\dd\mu_\text{AL}$,
\begin{equation}
	\int\dd\mu_\text{AL} \overline{\Psi_{\Gamma,f}[A]}\,\Psi_{\Gamma',f'}[A] \equiv \innerp{\Psi_{\Gamma,f}}{\Psi_{\Gamma',f'}}_\textsf{kin}.
\end{equation}
Moreover, Ashtekar and Lewandowski showed that this inner product gives a faithful representation of the algebra of cylindrical functions -- without the need to introduce any background structure.

\subsection{An orthogonal Basis for $\mathcal H_\textsf{kin}$}
There is a particularly useful basis of $\mathcal H_\textsf{kin}$, called the \textit{gauge-variant spin-network basis} \cite{ThiemannBook}, which will play an important role in the next subsection.\\
By construction, we have $\mathcal H_\Gamma = L_2[SU(2)^L, \dd h]$ and we can use the Peter-Weyl theorem to expand any function of $\mathcal H_\Gamma$ into an orthogonal basis. To see how to do that in practice, we start with the simpler space $L_2[SU(2), \dd h]$.\\
Let $j\in\frac12\mathbb N\setminus\{0\}$ label inequivalent irreducible representations of $SU(2)$, denote the carrier space of the spin-$j$ representation by $\mathcal H_j$ and let $\mathcal H^*_j$ be its dual. Then, according to the Peter-Weyl theorem, $L_2[SU(2),\dd h]$ can be decomposed as
\begin{equation}
	L_2[SU(2),\dd h]  \simeq\bigoplus_j(\mathcal H_j\otimes\mathcal H^*_j)
\end{equation}
and any function $f$ on that space can be written as
\begin{equation}
	f(h) = \sum_j \hat{f}^{mn}_j\, D^j_{mn}(h),
\end{equation}
where $D^j_{mn}(h)$ are the $SU(2)$ Wigner matrices in the spin-$j$ representation, $m,n$ with $-j\leq m,n\leq j$ denote the magnetic numbers, and the expansion coefficients\footnote{These coefficients are the $SU(2)$ analogues of the Fourier expansion coefficients on $U(1)$.} are explicitly given by
\begin{equation}\label{eq:ExpansionCoefficient}
	\hat{f}^{mn}_j = d_j \innerp{D^{j}_{mn}(h)}{f}_\textsf{kin}= d_j \int_{SU(2)}\dd h\, \overline{D^{j}_{mn}(h)}\,f(h),
\end{equation}
where $d_j := 2j+1$ denotes the dimension of the representation space. Notice that the Wigner matrices provide a linear map from $\mathcal H_j$ onto itself and from basic linear algebra we know that any such endomorphism can naturally be identified with an element in $\text{End}(\mathcal H_j)\simeq\mathcal H_j\otimes\mathcal H^*_j$. Due to the orthogonality relation
\begin{equation}
	\int_{SU(2)}\dd h\, \overline{D^{j'}_{m'n'}(h)}\, D^{j}_{mn}(h) = \frac{1}{d_j}\delta^{jj'}\,\delta_{mm'}\,\delta_{nn'},
\end{equation}
which is provided by the Peter-Weyl theorem, we can think of the Wigner matrices as basis elements of $\mathcal H_j\otimes\mathcal H^*_j$. As can be check by a direct computation, this basis is overcomplete. This means there is a resolution of identity which can be written as
\begin{equation}
	\delta(g h^{-1}) = \sum_{j}\sum_{-j\leq m, n\leq j} d_j\, D^{j}_{mn}(g) D^j_{nm}(h^{-1}) =: \sum_j d_j\, \Tr_j[gh^{-1}],
\end{equation}
for any given $g,h\in SU(2)$. The right hand side is a convenient short hand notation which we will use very frequently later on.\\
It is straightforward to generalize these considerations to the space $L_2[SU(2)^L, \dd h]$ and one finds, not very surprisingly, the following vector space isomorphisms:
\begin{equation}
	L_2[SU(2)^L, \dd h] \simeq \bigotimes_{\ell\in\Gamma}\left[\bigoplus_{j}(\mathcal H_j\otimes\mathcal H^*_j)\right] \simeq \bigoplus_{j_\ell}\left[\bigotimes_{\ell\in\Gamma}(\mathcal H_j\otimes\mathcal H^*_j)\right].
\end{equation}
When exchanging the tensor product and the direct sum we introduced the label $j_\ell$ which is simply the spin number associated to the link $\ell$ of the graph $\Gamma$. A functional $\Psi_{\Gamma, f}$ on this space can now be expanded as
\begin{equation}
	\Psi_{\Gamma,f}[A] = \sum_{j_1,\dots,j_L} \hat{f}^{m_1,\dots,m_L, n_1,\dots, n_L}_{j_1,\dots, j_L} D^{j_1}_{m_1, n_1}(h_1)\cdots D^{j_L}_{m_L,n_L}(h_L),
\end{equation}
where the expansion coefficients are given by generalizing \eqref{eq:ExpansionCoefficient} to
\begin{equation}
	\hat{f}^{m_1,\dots,m_L, n_1,\dots, n_L}_{j_1,\dots, j_L} = \innerp{D^{j_1}_{m_1, n_1}(h_1)\cdots D^{j_L}_{m_L,n_L}}{\Psi_{\Gamma,f}}_\textsf{kin}.
\end{equation}
The structures discussed here also have a nice pictorial representation. Since we are working at the level of a fixed graph $\Gamma$, we can simply draw all its links and nodes and then assign an irreducible spin $j$ representation to every link. This is sometimes referred to as ``coloring'' the graph and it amounts to assigning every link a Wigner matrix $D^j_{mn}$.

Before concluding this subsection we remark that we succeeded in carrying out the first step in the Dirac quantization program. The Hilbert space $\mathcal H_\textsf{kin}$ is a well-defined and well-behaved space on which we can represent the holonomy-flux algebra. In fact, on the kinematical Hilbert space we can represent this algebra through the familiar Schr\"{o}dinger representation where the holonomies are taken to be the configuration space variables which act by multiplication


\subsection{The $SU(2)$ invariant Hilbert Space $\mathcal H_\textsf{G}$}
Wa accomplished to carry out the first step in the Dirac quantization program and identify a kinematical Hilbert space $\mathcal H_\textsf{kin}$. This space is  well-defined, admits a simple basis and, importantly, allows us to represent the holonomy-flux variables as operators in the Ashtekar-Lewandowski representation. Consequently, we can move to the next step which is the implementation of the constraints. As discussed in subsection~\ref{ssec:CanonicalAnalysis}, the smeared Gauss constraints generates $SU(2)$ transformations. Therefore, the condition that the action of the quantum Gauss constraint on a cylindrical function is zero is equivalent to demanding the cylindrical functions to be invariant under $SU(2)$ transformations:
\begin{align}\label{eq:SU(2)Invariance}
	\Psi[h_1,\dots,h_L] &= \Psi[U_{s(\gamma_1)} h_1 U^\dagger_{t(\gamma_1)},\dots, U_{s(\gamma_L)} h_L U^\dagger_{t(\gamma_L)}]
\end{align}
Recall from subsection~\ref{ssec:HolonomyFlux} how the holonomy transforms under $SU(2)$ transformation to see that we can write \eqref{eq:SU(2)Invariance} at the level of a fixed graph $\Gamma$ as
\begin{equation}\label{eq:TransformedStates}
	\Psi_{\Gamma, f}[A] = \sum_{j_1,\dots,j_L}\hat f^{m_1,\dots,m_L,n_1,\dots,n_L}_{j_1,\dots,j_L}D^{j_1}_{m_1,n_1}(U_{s(\gamma_1)} h_1 U^\dagger_{t(\gamma_1)})\cdots D^{j_L}_{m_L,n_L}(U_{s(\gamma_L)} h_L U^\dagger_{t(\gamma_L)}),
\end{equation}
where $\Lambda_{s(\gamma_i)}$ and $\Lambda_{t(\gamma_i)}$ denote the values of the $SU(2)$ element $\Lambda(x)$ at the source and target of the $i$-th path, respectively.\\
Notice that the $SU(2)$ transformations only act at the beginning and end points of the paths, i.e. at the so-called nodes. Hence, the states which solve the quantum Gauss constraint have to be elements of the Hilbert space $\mathcal H_\textsf{G}:=L_2[SU(2)^L/SU(2)^N]$. In order to determine the form of these states, we can exploit the group property $D^j_{ab}(hg) = D^j_{ac}(h)D^j_{cb}(g)$ of the Wigner matrices to rewrite the summand of~\eqref{eq:TransformedStates} as
\begin{equation}
	\prod_{k=1}^L D^{j_k}_{m_k, a_k}(U_{s(\gamma_k)}) D^{j_k}_{b_k n_k}(U^\dagger_{t(\gamma_k)})\hat f^{m_1,\dots,m_L,n_1,\dots,n_L}_{j_1,\dots,j_L} \prod_{k=1}^L D^{j_k}_{a_k b_k}(h_k).
\end{equation}
From this it can be seen that the requirement for $\Psi_{\Gamma, f}[A]$ to be invariant under $SU(2)$ transformations at the nodes of the graph $\Gamma$ translates into the requirement for $f^{m_1,\dots,m_L,n_1,\dots,n_L}_{j_1,\dots,j_L}$ to be invariant when acted upon by representation matrices. Tensorial objects with this kind of invariance property are called \textit{intertwiners}. More precisely, an intertwiner is a tensor $\iota^{n_1,n_2,\dots,n_L}$ with $L$ indices which lives in the tensor product space $\bigotimes_{k=1}^L\mathcal H_{j_k}$ which satisfies the covariance condition
\begin{equation}
	D^{j_1}_{m_1 n_1}(U)D^{j_2}_{m_2 n_2}(U)\cdots D^{j_L}_{m_L n_L}(U) \,\iota^{n_1,n_2,\dots,n_L} = \iota^{m_1,m_2,\dots,m_L}
\end{equation}
and which is normalized
\begin{equation}
	\overline{\iota^{n_1,n_2,\dots,n_L}}\iota_{n_1,n_2,\dots,n_L} = 1.
\end{equation}
In general, there is more than one intertwiner in the space $\bigotimes_{k=1}^L\mathcal H_{j_k}$. In fact, intertwiners span the invariant subspace $\text{Inv}[\bigotimes_{k=1}^L\mathcal H_{j_k}]$ of $\bigotimes_{k=1}^L\mathcal H_{j_k}$. If $L=3$, i.e. when the space in question has the form $\mathcal H_{j_1}\otimes\mathcal H_{j_2}\otimes\mathcal H_{j_3}$, then the invariant subspace is simply
\begin{equation}
	\text{Inv}[\mathcal H_{j_1}\otimes\mathcal H_{j_2}\otimes\mathcal H_{j_3}] = \mathbb C
\end{equation}
and it is therefore one dimensional. The only element in this space is then the so-called $3j$-symbol
\begin{equation}
	\iota^{m_1 m_2 m_3} = \begin{pmatrix}
		j_1 & j_2 & j_3 \\
		m_1 & m_2 & m_3
	\end{pmatrix},
\end{equation}
which vanish unless the magnetic indices satisfy $m_1+m_2+m_3 = 0$, the spins respect the triangular inequalities $|j_1-j_2|\leq j_3 \leq j_1+j_2$, and $j_1+j_2+j_3$ is an integer number. 
If the invariant tensor product space has four components, i.e.
\begin{equation}
	\mathcal K_{j_1,\dots,j_4} := \text{Inv}[\mathcal H_{j_1}\otimes\mathcal H_{j_2}\otimes\mathcal H_{j_3}\otimes\mathcal H_{j_4}],
\end{equation}
then one can show, using repeatedly the well-known Clebsch-Gordan decomposition formula $\mathcal H_{j_1}\otimes\mathcal H_{j_2} = \bigoplus_{j=|j_1-j_2|}^{j_1+j_2}\mathcal{H}_j$, that the space $\mathcal K_{j_1,\dots, j_4}$ contains several invariant components. That is to say, its dimension is in general larger than one and using the Clebsch-Gordan formula one finds
\begin{equation}
	\dim[\mathcal K_{j_1,\dots,j_4}] = \min[j_1+j_2, j_3+j_4] - \max[|j_1-j_2|, |j_3-j4|]+1.
\end{equation}
The so-called four-valent intertwiners which span this this space are given by
\begin{equation}
	\iota^{m_1 m_2 m_3 m_4}_k = (-1)^{j_1-j_2+\mu}\begin{pmatrix}
		j_1 & j_2 & k\\
		m_1 & m_2 & \mu
	\end{pmatrix}
	\begin{pmatrix}
		j_3 & j_4 & k\\
		m_3 & m_4 & -\mu
	\end{pmatrix},
\end{equation}
with $\mu=-m_1-m_2=m_3+m_4$ and $\max[|j_1-j_2|, |j_3-j4|] \leq k \leq \min[j_1+j_2, j_3+j_4]$.
Let us recall that we actually set out to find invariant tensors in order to write down states which solve the quantum Gauss constraint. The intertwiners we just discussed can indeed be used to achieve our goal since
\begin{equation}
	\hat f^{m_1,\dots,m_L,n_1,\dots,n_L}_{j_1,\dots,j_L} = \mathcal C_{j_1,\dots, j_L} \iota^{m_1\dots m_L}\iota^{n_1\dots n_L},
\end{equation}
where $\mathcal C_{j_1,\dots, j_L}$ is a mere normalization factor, has precisely the correct covariance properties to render the state $\Psi_{\Gamma, f}$ invariant under $SU(2)$ transformations at the nodes. Therefore, a generic state in the Hilbert space $\mathcal H_\textsf{G}$ which is generated by the kernel of the quantum Gauss constraint can be written as a linear combination
\begin{equation}
	\Psi[h_1,\dots, h_L] = \sum_{j_1,\dots, j_L}\mathcal C_{j_1,\dots, j_L}\Psi_\Gamma(h_1,\dots, h_L)
\end{equation}
of so-called \textit{gauge-invariant spin network states} or simply \textit{spin networks} $\Psi_\Gamma(h_1,\dots, h_L)$. These are explicitly given by
\begin{equation}
	\Psi_\Gamma(h_1,\dots, h_L) := \bigotimes_{\no\in\Gamma}\iota_\no \cdot_\Gamma \bigotimes_{\ell\in\Gamma} D^{j_\ell}(h_\ell)
\end{equation}
where the first product is over all nodes $\no$ of the graph $\Gamma$ and all intertwiners $\iota_\no$ associated to these nodes, while the second product is over all links $\ell$ and the Wigner matrices in the $j_\ell$-representation associated to these links. The symbol $\cdot_\Gamma$ means that the contraction between the intertwiners and the Wigner matrices is dictated by the topology of the graph $\Gamma$.  
\begin{center}
	\begin{figure}[htb]
		\centering
			\includegraphics[width=0.25\textwidth]{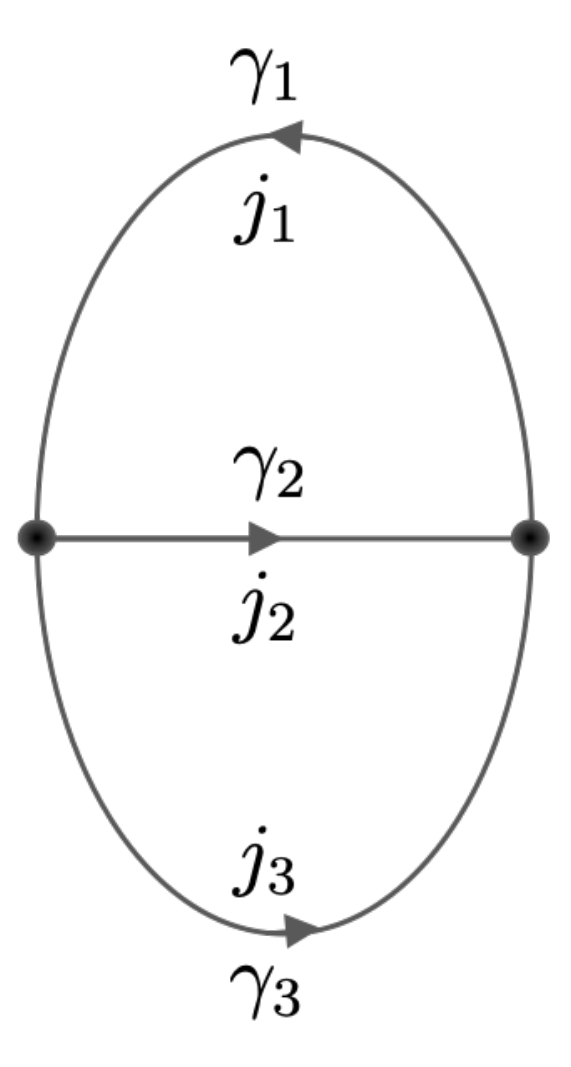}
			\caption{The theta graph}
			\label{fig:ThetaGraph}
	\end{figure}
\end{center}
Before concluding this subsection, let us give a brief example of a spin network function by considering the theta graph shown in Figure~\ref{fig:ThetaGraph}.
Since the two nodes are three valent, the intertwiner spaces associated to them are one-dimensional. There is therefore only one possible choice for the intertwiner -- the $3j$-symbol. These two symbols have to be contracted with the three Wigner matrices in the $j_\ell$-representation associated to the three links. Hence, we find that the spin network function for the theta graph is
\begin{equation}
	\Psi_{\theta}(h_1, h_2, h_3) = \begin{pmatrix}
		j_1 & j_2 & j_3\\
		m_1 & m_2 & m
	\end{pmatrix}
	\begin{pmatrix}
		j_1 & j_2 & j_3\\
		n_1 & n_2 & n
	\end{pmatrix}
	D^{j_1}_{m_1 n_1}(h_1) D^{j_2}_{m_2 n_2}(h_2) D^{j_3}_{m n}(h_3)
\end{equation}
with $m=-m_1-m_2$ and $n=-n_1 -n_2$ in order for the $3j$-symbols not to be zero. There is an implicit sum over all spins and magnetic numbers.

\section{Coherent and Semi-Classical Spin Network States}
The focus so far has been on presenting a background independent theory of quantum gravity. This lead to the description of $SU(2)$ and spatially diffeomorphism invariant states -- spin networks and $s$-knots. We have also seen that the spectra of the area and volume operator are discrete. Even though both spectra crowd very quickly and approach an apparent continuum behavior, the question remains if we can give a semi-classical description of geometry from LQG. The key to answer this question are coherent states.\\
In subsection~\ref{ssec:2_DefOfCoherentStates} we give a definition of coherent states and present a method due to Thiemann to generate candidate states which in some cases can be turned into proper coherent states. We illustrate this method for ordinary quantum mechanics and already mention at this point that these states will play an important role in the first part of chapter~\ref{chapter_5}.\\
Subsection~\ref{ssec:ThiemannStates} is devoted to a class of coherent semi-classical states for LQG introduced by Thiemann and Winkler. These so-called heat kernel states were the basis for further developments by Bianchi, Magliaro and Perini who established a connection between semi-classical coherent LQG states and twisted geometry. This will be reviewed in subsection~\ref{ssec:ExtrinsicStates} and will also be important in chapters~\ref{chap:CLQG} and~\ref{chapter_5}.\\
The last subsection contains unpublished work by the author where the resolution of identities for the heat kernel states in the twisted geometry parametrization will be proved.


\newpage
\subsection{Definition of Coherent States and the Complexifier Method}
\label{ssec:2_DefOfCoherentStates}
Let us jump right to the definition of coherent states given by Thiemann in~\cite{ThiemannBook}:
\begin{mydef}{ Coherent States}{}\label{def:CoherentState}
	Let $\mathcal P$ be a phase space and $\mathcal{\hat O}$ an algebra of linear operators on the Hilbert space $\mathcal H$. A collection of states $\{\Psi_{q,p}\}_{(q,p)\in\mathcal P}\in\mathcal H$ is then said to be coherent if
	\begin{itemize}
		\item[1)] There exist elementary operators $\hat g$ such that $\hat g\, \psi_{q,p} = g(q,p)\,\psi_{q,p}$;
		\item[2)] the Heisenberg uncertainty relation $\Delta q\,\Delta p = \frac12\vert\langle\psi_{q,p}\vert [\hat q,\hat p]\vert\psi_{q,p}\rangle\vert$ is saturated;
 		\item[3)] there is a resolution of identity $\id_\mathcal{H} = \int_\mathcal{P}\dd\nu(q,p)\,\psi_{q,p} \innerp{\psi_{q,p}}{\cdot}_\textsf{kin}$ for some measure $\nu$ on $\mathcal{P}$;
 		\item[4)] for any $(q,p)\in\mathcal{P}$ the overlap function $(q',p')\mapsto \vert\innerp{\psi_{q,p}}{\psi_{q',p'}}\vert^2$ is peaked on a phase space cell of Liouville volume $\frac12\vert\langle\psi_{q,p}\vert [\hat q,\hat p]\vert\psi_{q,p}\rangle\vert$.
	\end{itemize}
\end{mydef}
Informally, the first condition tells us that the coherent states form an overcomplete basis of the kinematical Hilbert space, the second condition is necessary (though not sufficient) to minimize the uncertainty of \textit{both} conjugate variables and the third condition tells us that the states are highly localized on a single phase space point $(q,p)$.
The key question is now how to construct kinematical states with these properties. This is a non-trivial problem as there is no known algorithm to construct coherent states, but there is at least a guideline developed by Thiemann~\cite{Thiemann:1995,Thiemann:1995b}. As explained in a very didactical manner in~\cite{ThiemannBook}, it is possible to use known methods to construct coherent states for the harmonic oscillator as a template, but to strip them off anything that belongs exclusively to the harmonic oscillator. This observation is at the core of Thiemann's method as it allows to construct candidate coherent states for other quantum systems. As it turns out, the main object needed for this is the so-called \textit{complexifier}~\cite{Thiemann:1995,Thiemann:1995b,ThiemannBook}.
\begin{mydef}{ Complexifier}{}\label{def:Complexifier}
	A complexifier is a positive definite function $C:\mathcal P\to\mathbb R$ on the phase space $\mathcal P$ which is smooth almost everywhere and whose Hamiltonian vector field $\rchi_C$ is nowhere vanishing on the configuration space $\mathcal C$. Moreover, for each point $q\in\mathcal C$ the function $p\mapsto C_q(p):=C(q,p)$ grows stronger than linearly with $\|p\|_q$ where $p$ is a local momentum coordinate and $\|\cdot\|_q$ is a suitable norm on $T^*_q\mathcal C$.	
\end{mydef}
The role of the complexifier is, as we will now see, two-fold. First of all, consider the delta distribution $\delta_{q}(x)$ with respect to some measure $\mu$ and with support on $x=q$, i.e. $\int f(x)\,\delta_{q}(x)\dd\mu = f(q)$. In quantum mechanics we routinely encounter these kind of distributions where, even though they are not part of the kinematical Hilbert space, they capture the idea of particles being ``localized'' in one point. Of course, by the Heisenberg uncertainty relation, this implies that the momentum of the particle in question is completely unknown or spread out.\\ That is where the complexifier comes into play. Its positive definitiveness suggests we can promote it to a positive-definite self-adjoint operator $\hat C$ on the kinematical Hilbert space $\mathcal H$ in question. According to definition~\ref{def:Complexifier} it grows stronger than linearly with $\|p\|_q$ and we can therefore define the one parameter family of states
\begin{equation}\label{eq:SmoothState}
	\psi^t_{q}(x) := \e^{-\frac{t}{\hbar}\hat C}\,\delta_{q}(x),
\end{equation}
with \textit{semi-classicality parameter} $t>0$ and where $\e^{-\frac{t}{\hbar}\hat C}$ acts as a smoothening operator. This means that the states $\psi^t_{q}(x)$ have a chance of being square integrable and may belong to $\mathcal H$. Moreover, the spread $\Delta p$ of the momentum operator on these states is finite.\\
There are now two possibilities to manipulate the spreads $\Delta x$ and $\Delta p$. Either by changing the value of the parameter $t$ or, in analogy with the harmonic oscillator, we may complexify $q\to z\in\mathbb C$. The second option is more promising as it works for the harmonic oscillator and it has the potential to satisfy the second and third condition in definition~\ref{def:CoherentState}. But how should we choose $z$?\\
That is where the complexifier comes into play for a second time: Let $(q,p)$ be local coordinates on the phase space $\mathcal P$. Then define~\cite{ThiemannBook}
\begin{equation}\label{eq:ComplexVariables}
	z(q,p) := \sum_{n=0}^\infty\frac{i^n}{n!} \{q,C(q,p)\}_n,
\end{equation}
with the inductively defined $n$-th order Poisson brackets,
\begin{equation}
	\{q,C\}_{n+1} = \{ \{q,C\}_{n}, C \} \quad\text{with}\quad \{q,C\}_0 := q,
\end{equation}
defines local complex coordinates\footnote{Hence the name \textit{complexifier} for the function $C$.} on $\mathcal P$, provided $z(q,p)$ and $\overline{z}(q,p)$ are invertible. Locally, the invertibility of these functions is guaranteed by the conditions imposed on $C$ by definition~\ref{def:Complexifier}. Next, let us define the (unnormalized) state
\begin{equation}\label{eq:DefCandidate}
	\psi^t_{q,p}(x) := \left[\psi^t_{q}(x)\right]_{q\rightarrow z(q,p)} = \left[\e^{-\frac{t}{\hbar}\hat{C}}\delta_{q}(x)\right]_{q\rightarrow z(q,p)}.
\end{equation} 
by promoting the complexifier to a self-adjoint operator. This has also the effect that the complex variable $z$ defined in~\eqref{eq:ComplexVariables} becomes an operator. More precisely, we transition to the quantum theory by promoting $q$ to the multiplicative position operator $\hat q$ and turning the Poisson bracket into a commutator. Then one can show~\cite{ThiemannBook} that $\hat z$ is given by
\begin{equation}
	\hat z =\e^{-\frac{t}{\hbar}\hat C}\, \hat q\, \e^{\frac{t}{\hbar}\hat C}.
\end{equation} 
With this it is straightforward to see that $\hat{z}$ is an eigenstate of~\eqref{eq:DefCandidate} with eigenvalue $z(q,p)$ since
\begin{align}
	\hat{z}\, \psi^t_{q,p}(x) &= \left[\e^{-\frac{t}{\hbar}\hat{C}}\,\hat{q}\, \underset{=\id}{\underbrace{\e^{\frac{t}{\hbar}\hat C} \e^{-\frac{t}{\hbar}\hat C}}}\delta_{q}(x)\right]_{q\rightarrow z(q,p)} = \left[q\e^{-\frac{t}{\hbar}\hat{C}}\delta_{q}(x)\right]_{q\rightarrow z(q,p)} \notag\\
	&= z(q,p)\, \psi^t_{q,p}(x).
\end{align}
Above we used that $\delta_q(x)$ is a generalized eigenstate of $\hat q$.
This eigenstate property is analogous to the property of harmonic oscillator coherent states which are eigenstates of the annihilation operator. We can therefore take this property to be a good sign and also check that $\psi^t_{q,p}$ has a chance of being peaked on $(q,p)$. Indeed, by defining
\begin{equation}\label{eq:XandYOps}
	\hat x := \frac12\left(\hat z+ \hat z^\dagger\right)\quad\text{and}\quad \hat y := \frac12\left(\hat z - \hat z^\dagger\right)
\end{equation}
and using similar techniques as before one can show
\begin{align}
	\langle \hat x\rangle &:= \frac{\langle\psi^t_{q,p}|\,\hat x\,|\psi^t_{q,p}\rangle}{\|\psi^t_{q,p}\|^2} = \frac12\left(z(q,p) + \overline{z}(q,p)\right)\notag\\
	\langle \hat y\rangle &:= \frac{\langle\psi^t_{q,p}|\,\hat y\,|\psi^t_{q,p}\rangle}{\|\psi^t_{q,p}\|^2} = \frac12\left(z(q,p) - \overline{z}(q,p)\right).
\end{align}
Finally, it is also easy to compute the uncertainties:
\begin{equation}\label{eq:MinimizedHeisenberg}
	\Delta x := \frac{\langle \psi^t_{q,p}|\,(\hat x - \langle \hat x\rangle)^2\,| \psi^t_{q,p}\rangle}{\|\psi^t_{q,p}\|^2} = \frac12 \left|\langle [\hat x, \hat y]\rangle\right| = \Delta y.
\end{equation} 
That is, the states $\psi^t_{q,p}$ possess unquenched uncertainties for the operators $\hat x$ and $\hat y$ defined above and saturate Heisenberg's uncertainty relations. With that, the states defined by~\eqref{eq:DefCandidate} satisfy two of the four conditions required by definition~\ref{def:CoherentState}. It should be emphasized that this is true irrespective of the quantum mechanical system in question as we have not chosen any Lagrangian or specified a system in any other way! These are therefore good candidates for coherent states, but the remaining conditions in~\ref{def:CoherentState} need to be checked on a case by case basis.

Before moving to the next subsection, where coherent semi-classical states for LQG will be introduced, we illustrate the complexifier method by constructing coherent states for ordinary quantum mechanical systems. These states will play an important role in chapter~\ref{chapter_5}.\\
First of all, we choose the following quadratic phase space function as complexifier: 
\begin{equation}
	C := \frac{p^2}{2 \sigma^2}.
\end{equation}
Here, $\sigma>0$ is an arbitrary parameter and $\sigma^2$ has units of momentum over length or, equivalently, action over length squared. With this choice it follows immediately that the first and second order Poisson brackets are given by
\begin{align}
	\{q,C\} = \frac{1}{\sigma^2}p\quad\text{and}\quad \{\{q,C\},C\} = 0
\end{align}
which implies that all higher order brackets vanish and the complex variable $z$ is therefore given by  
\begin{equation}
	z = \sum_{n=0}^\infty\frac{i^n}{n!}\{q,C\}_{(n)} = q + \frac{i}{\sigma^2}p.
\end{equation}
Next, we promote the complexifier to a self-adjoint hermitian operator $\hat{C}$ on the Hilbert space $L_2[\mathbb R^n, \dd^n x]$. Determining the state $\psi^\sigma_{q,p}(x)$ as defined in~\eqref{eq:DefCandidate} and normalizing it amounts to executing the following three steps:
\begin{itemize}
	\item[\textcolor{white}{\textit{ii}} (\textit{i})] Compute the action of $\e^{-\hat{C}/\hbar}$ on $\delta^{(n)}(x-q)$.
		\begin{align}
			\e^{-\hat{C}/\hbar}\delta^{(n)}(x-q) &= \e^{-\frac{\hat{p}^2}{2\sigma^2 \hbar}}\innerp{x}{q} = \e^{-\frac{\hat{p}^2}{2\sigma^2 \hbar}}\int_{\mathbb R^n} \innerp{x}{p}\innerp{p}{q}\,\dd^n p\notag\\
			&=\frac{1}{(2\pi\hbar)^n} \int_{\mathbb R^n}\e^{\frac{i}{\hbar}p x}\e^{-\frac{i}{\hbar}p q}\e^{-\frac{p^2}{2\sigma^2 \hbar}}\,\dd^n p \notag\\
			&=\left(\frac{\sigma}{\sqrt{2\pi\hbar}}\right)^n\e^{-\frac{\sigma^2}{2\hbar}(x-q)^2}
 		\end{align}	
 	\item[\textcolor{white}{\textit{i}} (\textit{ii})] Perform the complexification $q\rightarrow z=q+\frac{i}{\sigma^2}p$.
 		\begin{align}
 				\left[\e^{-\frac{\sigma^2}{2\hbar}(x-q)^2}\right]_{q\rightarrow q+\frac{i}{\sigma^2}p} = \e^{-\frac{\sigma^2}{2\hbar}(x-q)^2}\e^{\frac{i}{\hbar}p(x-q)}\e^{\frac{p^2}{2\sigma^2\hbar}}
 		\end{align}
	\item[(\textit{iii})] Normalize the state $\psi^\sigma_{q,p}(x) = \mathcal N \left(\frac{\sigma}{\sqrt{2\pi\hbar}}\right)^n \e^{-\frac{\sigma^2}{2\hbar}(x-q)^2}\e^{\frac{i}{\hbar}p(x-q)}\e^{\frac{p^2}{2\sigma^2\hbar}}$.
		\begin{align}
			|\mathcal N|^2 \left(\frac{\sigma^2}{2\pi\hbar}\right)^n \e^{\frac{p^2}{2\sigma^2\hbar}} \int_{\mathbb R^n}\e^{-\frac{\sigma^2}{\hbar}(x-q)^2} \,\dd^n x \overset{!}{=} 1\quad\Rightarrow\quad \mathcal N = \left(\frac{2\sqrt{\pi\hbar}}{\sigma}\right)^n \e^{-\frac{p^2}{\sigma^2\hbar}}
		\end{align}
\end{itemize}
Hence, we finally find that kinematical coherent states $\psi^\sigma_{q,p}(x)$ can be written as
\begin{equation}
	\psi^\sigma_{q,p}(x) = \left(\frac{\sigma^2}{\pi\hbar}\right)^{\frac{n}{4}} \e^{-\frac{\sigma^2}{2\hbar}(x-q)^2}\e^{\frac{i}{\hbar}p(x-q)}
\end{equation}
and it is not difficult to see that they are elements of the Hilbert space $L_2[\mathbb R^n, \dd^n x]$. It is also an easy exercise to show
\begin{align}
	\langle\psi^\sigma_{q,p}| \hat x | \psi^\sigma_{q,p}\rangle &= q\notag\\
	\langle\psi^\sigma_{q,p}| \hat p| \psi^\sigma_{q,p}\rangle &= p,
\end{align}
which means these states are peaked on the prescribed phase space point $(q,p)$. To be more precise, this means that we can freely choose some \textit{data} $(q,p)$, on which the states depend parametrically, and it is guaranteed that $q$ is the expectation value of the position operator while $p$ is the expectation value of the momentum operator.\\
Furthermore one can show
\begin{align}
	\Delta x^2 &:= \langle\psi^\sigma_{q,p}|\hat{x}^2|\psi^\sigma_{q,p}\rangle - \langle\psi^\sigma_{q,p}|\hat{x}|\psi^\sigma_{q,p}\rangle = \frac{\hbar}{2\sigma^2}\notag\\
	\Delta p^2 &:=\langle\psi^\sigma_{q,p}|\hat{p}^2|\psi^\sigma_{q,p}\rangle -\langle\psi^\sigma_{q,p}|\hat{p}|\psi^\sigma_{q,p}\rangle = \frac{\hbar \sigma^2}{2},
\end{align}
which implies $\Delta x\Delta p=\frac{\hbar}{2}$. Notice that the uncertainties are not unquenched\footnote{This is not in contradiction to what we said below equation~\eqref{eq:MinimizedHeisenberg} since this is only true for the operators $\hat x$ and $\hat y$ defined in~\eqref{eq:XandYOps}.} and that $\sigma^2$ can in principle be chosen such that either $\Delta x$ or $\Delta p$ is completely spread. Or it can be chosen such that the uncertainties in both variables are minimized and the state becomes semi-classical (a more precise definition will be provided below).\\
We conclude this subsection by remarking that a simple computation shows that these coherent states provide a resolution of identity with respect to the measure $\frac{\dd^n\, q\dd^n p}{(2\pi\hbar)^n}$:
\begin{equation}\label{eq:PQResolution}
	\int_{\mathbb R^{2n}}\frac{\dd^n q\,\dd^n p}{(2\pi\hbar)^n}\ket{\psi^\sigma_{q,p}}\bra{\psi^\sigma_{q,p}} = \id.
\end{equation}
Notice that the integration is over the data $(q,p)$. This is a point that is often overlooked and sometimes causes confusion. We will re-encounter this identity in chapter~\ref{chapter_5}.

\subsection{Thiemann's Heat Kernel States}
\label{ssec:ThiemannStates}
In~\cite{Thiemann:2001a}, Thiemann studied two different complexifier operators defined on the Hilbert space $L_2[SU(2)^L/SU(2)^N]_\Gamma$ and introduced candidate coherent states for LQG. Here, $\Gamma$ is a graph, $L$ denotes the number of links $\ell$ and $N$ the number of nodes $\no$. Of particular interest are the so-called heat kernel states, which use the $SU(2)$ heat kernel as complexifier and which are defined as
\begin{equation}\label{eq:ThiemannState}
	\Psi_{\Gamma, H_{\ell}}^{t}(h_\ell) := \int_{SU(2)^N}\left(\prod_{\no}\,\dd h_{\no(\ell)}\right)\, \prod_{\ell} K_{\ell}^{t} (h_\ell, h_{t(\ell)}\, H_{\ell}\, h_{s(\ell)}^{-1}).
\end{equation}
This $1$-parameter family of states is $SU(2)$ gauge-invariant due to the integrals over $h_{\no(\ell)}$. The labels $s(\ell)$ and $t(\ell)$ denote source and target node of the link $\ell$, $t$ is the semi-classicality parameter and $K^t(h, H)$ is the $SU(2)$ heat kernel with a complexified $SU(2)$ element as second argument.\\ 
Since $SU(2)^{\mathbb{C}}\simeq SL(2, \mathbb{C})$, $H$ is taken to be an element of $SL(2, \mathbb{C})$\footnote{$SL(2,\mathbb{C})$ is isomorphic to $SU(2) \times su(2) \simeq T^*SU(2)$ which corresponds to the (linkwise, not gauge invariant) classical phase space associated to the Hilbert space on a graph.} and this can be seen as the analogue of $q\rightarrow z(q,p)$ discussed in the foregoing subsection. The Wigner D-matrices of the $SU(2)$ heat kernel in \eqref{eq:ThiemannState} are defined by analytical extension to the group $SL(2,\mathbb{C})$\footnote{The explicit defining expression for the analytically extended matrix elements $D^j_{m n}$ can be found in \cite{RuhlBook} and \cite{Barrett:2009} and it provides in fact an analytic extension to $GL(2,\mathbb{C})$.}. Concretely, $K^{t}(h, H)$ is given in the spin-representation by
\begin{equation}
	K^{t}(h, H) = \sum_{j} d_j \e^{-j(j+1) t} \Tr\left[D^{(j)}(h H^{-1})\right].
\end{equation}
The states $\Psi_{\Gamma, H_\ell}^{t}(h_\ell)$ provide an overcomplete basis of the kinematical LQG Hilbert space. That is, there is a resolution of identity
\begin{equation}
	\delta_\Gamma(h_l, h'_l) = \int_{SL(2, \mathbb{C})^L}\left(\prod_l\Omega_{2 t}(H_l)\,\dd H_l\right) \Psi^{t}_{\Gamma, H_l}(h_l)\,\overline{\Psi^{t}_{\Gamma, H_l}(h'_l)},
\end{equation}
where the identity operator $\id_\Gamma$ on $\mathcal{H}_\Gamma$ is given in the holonomy representation by the delta distribution $\delta_\Gamma$ on $SU(2)^L / SU(2)^N$ (a precise definition will be provided in subsection~\ref{ssec:SemiclassicalResOfId}).\\
The measure $\Omega_{2t}(H)$ with respect to which the above resolution of identity holds is given by the heat kernel on the quotient space $\slc/SU(2)$, i.e.
\begin{equation}\label{eq:QuotientMeasure}
	\Omega_{2t}(H) := \int_{SU(2)}F_{2t}(H g)\, \dd g, 
\end{equation}
where $F_{2t}$ is the heat kernel on $\slc$. An explicit formula for the measure is known for the parametrization corresponding to the polar decomposition of $\slc$,
\begin{equation} \label{eq:polarDecomp}
	H =  h\, \e^{\vec{p}\cdot\frac{\vec{\sigma}}{2}}.
\end{equation}
Here, $h \in SU(2)$ and $\vec{p}$ is a vector in $\mathbb{R}^3$. It can then be shown that the measure in this parametrization is given by~\cite{Thiemann:2001a, Bianchi:2010}
\begin{equation} \label{eq:polarMeasure}
	\Omega_{2t}(h\, \e^{\vec{p}\cdot\frac{\vec{\sigma}}{2}}) = \frac{\e^{\frac{t}{2}}}{(2 \pi t)^{\frac{3}{2}}}\frac{\vert \vec{p} \vert}{\sinh \vert \vec{p} \vert}\e^{-\frac{\vert p \vert^2}{2 t}}\quad\quad\quad \dd H = \frac{\sinh^2\vert\vec{p}\vert}{\vert \vec{p}\vert^2}\,\dd h\,\dd^3\vec{p}.
\end{equation}
This shows that one of the conditions of definition~\ref{def:CoherentState} is satisfied and that the states belong to the kinematical LQG Hilbert space. In a series of articles, Thiemann and Winkler~\cite{Thiemann:2001a,Thiemann:2001b,Thiemann:2001c,Thiemann:2001d} proceeded in showing that all conditions of~\ref{def:CoherentState} are met. Moreover, they also showed that when the parameter $t$ is chosen appropriately, these states are semi-classical\footnote{Hence the name \textit{semi-classicality parameter} for $t$.} in the following sense:
\begin{mydef}{ Semi-Classical States}{}\label{def:SemiclassicalState}
		Let $\mathcal P$ be a phase space, $\mathcal H$ a kinematical Hilbert space and $\mathcal{\hat O}$ an algebra of linear operators on $\mathcal H$. A collection of states $\{\psi_{q,p}\}_{(q,p)\in\mathcal P}$ is said to be semi-classical provided that for any $\hat O, \hat O'\in\mathcal{\hat O}$ and any generic point $(q,p)\in\mathcal P$ the following properties hold
		\begin{itemize}
			\item[1)]  Expectation value property: $\left\vert\frac{\langle\psi_{q,p}\vert\, \hat O\, \vert\psi_{q,p}\rangle}{O(q,p)}-1\right\vert\ll 1$
			\item[2)] Ehrenfest property: $\left\vert\frac{\langle\psi_{q,p}\vert \,[\hat O,\hat O']\, \vert\psi_{q,p}\rangle}{i \hbar \{O(q,p),O'(q,p)\}} -1\right\vert\ll 1$
			\item[3)] Small fluctuation property: $\left\vert\frac{\langle\psi_{q,p}\vert\, \hat O^2\, \vert\psi_{q,p}\rangle}{\langle\psi_{q,p}\vert \hat O \vert\psi_{q,p}\rangle^2} -1\right\vert \ll 1$
		\end{itemize}
\end{mydef}
Let us emphasize what this means: The states~\eqref{eq:ThiemannState} provide us with an overcomplete basis for the kinematical LQG Hilbert space, hence allowing us to express any state in this basis. They also allow us to \textit{choose data} $H_\ell$ on every link of the graph $\Gamma$ and it is guaranteed that the states are peaked on these prescribed data. Moreover, the states satisfy well-defined coherence and semi-classicality conditions.\\
Of course, the word ``semi-classicality'' implies some sort of geometrical framework. It is however not clear what is the geometrical content of $\Psi^t_{\Gamma, H_\ell}$ and how the data $H_\ell$ factors in. This will become clearer in the next subsection.

\subsection{Extrinsic Coherent States}
\label{ssec:ExtrinsicStates}
\begin{center}
	\begin{figure}[htb]
		\centering
			\includegraphics[width=0.65\textwidth]{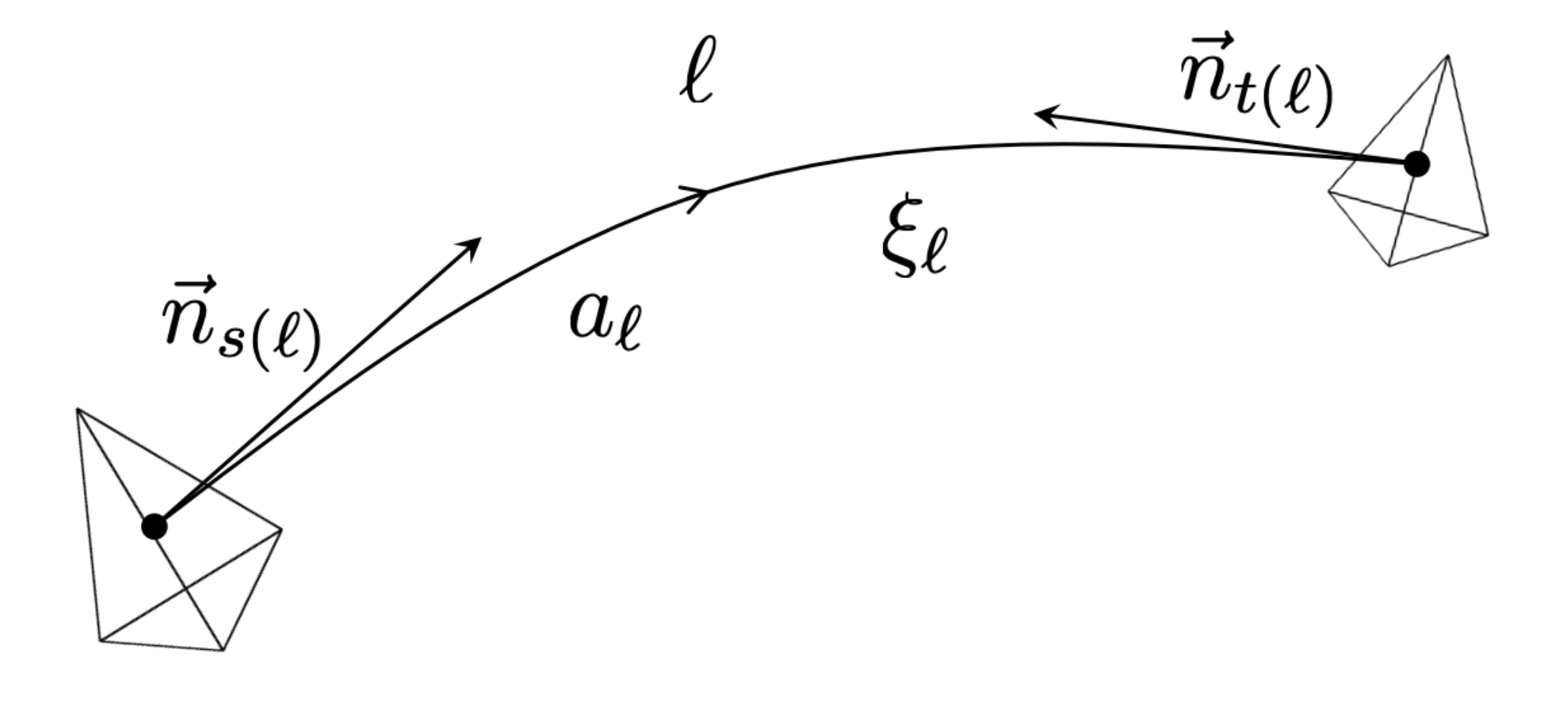}
			\caption{A geometrical visualization of the data $(a_\ell, \xi_\ell, \vec{n}_{s(\ell)}, \vec{n}_{t(\ell)})$.}
			\label{fig:BndData}
	\end{figure}
\end{center}

A key property of the heat kernel states is that they allow us to choose some data $H_\ell$ and the states are then automatically peaked on these data. Given that the heat kernel states also satisfy semi-classicality conditions, we would intuitively expect that $H_\ell$ encodes the geometry of $\Gamma$. This was indeed shown by the work of Bianchi, Magliaro and Perini~\cite{Bianchi:2010} who considered the Cartan decomposition of $H^{-1}_\ell$:
\begin{equation}\label{eq:HParam}
	H^{-1}_\ell = n_{s(\ell)}\,\e^{(\eta_\ell + i \gamma \xi_\ell)\frac{\sigma_3}{2}}\, n_{t(\ell)}^{-1},
\end{equation}
where
\begin{equation}\label{eq:DefNElement}
	n_{\no(\ell)} := \e^{-i \phi_{\no(\ell)} \frac{\sigma_3}{2}}\,\e^{-i \theta_{\no(\ell)} \frac{\sigma_2}{2}}
\end{equation}
is the $SU(2)$ group element associated to the unit vector
\begin{equation}
	\vec{n} = (\sin\theta_{\no(\ell)} \cos\phi_{\no(\ell)},\, \sin\theta_{\no(\ell)}\sin\phi_{\no(\ell)},\,\cos\theta_{\no(\ell)})^\transpose,
\end{equation}
while $\xi_\ell\in[0,4\pi)$ is an angle, $\gamma$ denotes the Barbero-Immirzi parameter and $\eta_\ell$ is a positive real number.  It follows that the six degrees of freedom of $H_\ell$ can now be represented by the six parameters $(\eta_\ell, \xi_\ell, \vec{n}_{s(\ell)}, \vec{n}_{t(\ell)})$.\\
These parameters possess indeed a geometrical interpretation as explained in~\cite{Bianchi:2010} and further emphasized in~\cite{Rovelli:2010}: If we assume $\Gamma$ to be a graph which consists of $4$-valent nodes, we can think of every half-link as representing a triangle of a tetrahedron. The parameter $\eta_\ell$ is then related to the area of the two triangles associated to the same link $\ell$. To be more precise, we assign a dimensionless parameter $a_\ell = \frac{\eta_\ell -t}{2t}$ to every link, for reasons that will become clear later on, and this parameter is related to the actual area $A_\ell$ of the triangles via $A_\ell \equiv \gamma l^2_\text{pl} a_\ell$. Furthermore, $\vec{n}_{s(\ell)}$ and $\vec{n}_{t(\ell)}$ are the unit normal vectors to the two triangles, cf. Figure~\ref{fig:BndData}. Lastly, there is the parameter $\xi_\ell$ which in~\cite{Rovelli:2010} has been shown to be a discrete measure for the extrinsic curvature. In the same paper it was shown that
\begin{equation}
	n_{s(\ell)}\,\e^{i\gamma\xi_\ell\frac{\sigma_3}{2}}\,n^{-1}_{t(\ell)}
\end{equation}
can be understood as the holonomy of the Ashtekar-Barbero connection. With this parametrization, it is furthermore possible to make contact with the phase space of \textit{closed twisted geometries} \cite{Freidel:2010}. This phase space, $P_\Gamma = \largetimes_\ell T^{*}SU(2)_\ell \simeq \largetimes_\ell \left(\mathbb{R}^+_\ell\times S^1_\ell \times S^2_\ell \times S^2_\ell \right)$, is spanned precisely by the six parameters $(\eta_\ell, \xi_\ell, \vec{n}_{s(\ell)}, \vec{n}_{t(\ell)})$ and it describes discretized geometries consisting of flat tetrahedra, characterized by their four normals and their four areas, which are glued together. 
The tetrahedra are glued together along their triangles and the areas of these connected triangles match. That is why we only need one variable for the area per link. However, the shapes of these triangles do in general not match. Since every tetrahedron can be used to define a local metric, we see that going from one tetrahedron to an other one induces a discontinuity in the metric due to the shape-mismatch. Twisted geometries therefore describe discrete, discontinous geometries which reduce to Regge geometries when the shapes of the triangles match \cite{Rovelli:2010,Freidel:2010}. We also remark that the phase space of closed twisted geometries is in one-to-one correspondence with the kinematical (gauge-invariant) phase space of LQG, $SU(2)^N/SU(2)^L$, for graphs with four-valent nodes \cite{Rovelli:2010,Freidel:2010}.\\
By introducing this parametrization we have a clearer understanding of the geometrical meaning of the heat kernel states. Showing all parameters explicitly, we can write these states in the twisted geometry parametrization as
\begin{equation}
	\Psi_{\Gamma, H_\ell}^t(h_\ell) = \prod_{\ell\in\Gamma}\sum_{j_\ell} d_{j_\ell} \,\e^{-j_\ell(j_\ell+1) t} \sum_{m, n, k} D^{j_\ell}_{mn}(h_\ell) D_{nk}^{j_\ell}(n_{t(\ell)}) D_{km}^{j_\ell}(\e^{i \gamma\xi_\ell \frac{\sigma_3}{2}}n_{s(\ell)}^\dagger)\,\e^{\eta_\ell k},
\end{equation} 
where we have momentarily dropped the gauge-averaging integrals. In later chapters it will be useful to assume $\eta_\ell$ to be a large number, which is tantamount to assuming the states to describe geometries with large areas. Then, we can use the so-called highest weight approximation
\begin{equation}\label{eq:HighestWeightApprox}
	D^{j_\ell}_{ab}(\e^{(\eta_\ell + i\gamma\xi_\ell)\frac{\sigma_3}{2}}) = \delta_{ab}\e^{\eta_\ell j_\ell}\left(\delta_{a j_\ell}\e^{i\xi_\ell j_\ell}+\mathcal{O}(\e^{-\eta_\ell})\right)\quad\text{for } \eta_\ell\to+\infty.
\end{equation}
This approximation allows us to rewrite the product of Wigner matrices as
\begin{equation} \label{eq:LS_States}
\Phi_{\Gamma, j_\ell, \vec{n}_{s(\ell)}, \vec{n}_{t(\ell)}}(h_\ell) := \sum_{m_s, m_t} D^{j_\ell}_{j_\ell m_t}(n_{t(\ell)}^\dagger)\; D^{j_\ell}_{m_t m_s}(h_\ell)\; D^{j_\ell}_{m_s j_\ell}(n_{s(\ell)}),
\end{equation}
which we recognize as a Livine-Speziale coherent states~\cite{Livine:2007}\footnote{To be more precise, these are the gauge-\textit{variant} Livine-Speziale states since we dropped the gauge-averaging integrals.}. These states, sometimes also referred to as intrinsic coherent states~\cite{RovelliVidottoBook}, describe a geometry composed of glued-together tetrahedra, but they are oblivious to the extrinsic geometry. The information about the extrinsic part of the geometry is coded in the exponentials of the heat kernel states which after applying the highest weight approximation read
\begin{equation}\label{eq:ExponentialAfterApprox}
	\exp(-j_\ell(j_\ell+1)t + \eta_\ell j_\ell + i\xi_\ell j_\ell).
\end{equation}
Notice that the heat kernel states are defined up to a normalization and we can therefore complete the square in~\eqref{eq:ExponentialAfterApprox}, i.e.
\begin{equation}\label{eq:CompleteSquare}
	-j_\ell(j_\ell+1)t + \eta_\ell j_\ell = -\left(j_\ell - \frac{\eta_\ell-t}{2t}\right)^2 t + \frac{(\eta_\ell-t)^2}{4t},
\end{equation} 
and throw away the last term in~\eqref{eq:CompleteSquare} which only depends on the data $\eta_\ell$ and the parameter~$t$. Hence, when putting everything back together, we find that the heat kernel states in the highest weight approximation can be written as
\begin{equation}\label{eq:ExtrinsicState}
	\Psi_{\Gamma, H_\ell}^{t}(h_\ell) = \sum_{\{j_\ell \}} \prod_\ell d_{j_\ell} \e^{-\left(j_\ell - a_\ell\right)^2 t \,+\, i \gamma \xi_\ell j_\ell}\, \Phi_{\Gamma, j_\ell, \vec{n}_{s(\ell)}, \vec{n}_{t(\ell)}}(h_\ell),
\end{equation}
where we defined
\begin{equation}
	a_\ell := \frac{\eta_\ell - t}{2t}.
\end{equation}
We will often refer to these states as \textit{extrinsic coherent states}. Notice that the Gaussian weights $\exp(-(j_\ell-a_\ell)^2t)$ suggest that the spins are all peaked on $a_\ell$ and indeed it was shown in~\cite{Bianchi:2010} that the expectation value of the area operator on these states is given by $A_\ell = \gamma l^2_\text{pl} a_\ell$. We will therefore regard $a_\ell$ as a ``dimensionless'' area variable and from now on, whenever we talk about extrinsic coherent states, we completely forget about $\eta_\ell$. Instead, we assume the prescribed data to be given by $(a_\ell, \xi_\ell, \vec{n}_{s(\ell)}, \vec{n}_{t(\ell)})$.\\
As discussed in the previous subsection, these states are peaked on the data $H_\ell$. Now  we know that these data describe an intrinsic and extrinsic geometry and, moreover, also describe a point in the phase space of closed twisted geometries. The situation is therefore analogous to the quantum mechanical case discussed in subsection~\ref{ssec:2_DefOfCoherentStates}. There, the coherent states were also peaked on a point $(q,p)$ in phase space. However, remember that the uncertainties $\Delta x$ and $\Delta p$ depended on the parameter $\sigma$, which plays the same role as $t$ does here.\\
By tuning $t$ it should therefore be possible to control the spread in the conjugate variables and therefore the extrinsic states can be made to be not only coherent, but also semi-classical~\cite{Bianchi:2010}. For this to be the case, $t$ has to be chosen such that the spread in the areas $\Delta A_\ell$ and the spread in the holonomies $\Delta h_\ell$ are much smaller than the expectation values of the corresponding operators. This requirement translates into
\begin{equation}
	\Delta A_\ell \sim \frac{l^2_\text{pl}}{\sqrt{t}}\ll A_\ell \quad \text{and}\quad \Delta h_\ell \sim \sqrt{t}\ll 1 \quad\forall\ell\in\Gamma.
\end{equation}
as shown in~\cite{Bianchi:2010}. When the two conditions are combined, one obtains what we call the semi-classicality condition:
\begin{align}\label{eq:SemiClassicality}
	l^2_\text{pl} \ll \sqrt{t}\, A_\ell \ll A_\ell\quad\Leftrightarrow\quad 1\ll \sqrt{t}\, a_\ell \ll a_\ell.
\end{align}
The second version of the condition, which is dimensionless and in terms of the parameter $a_\ell$, is for later reference. For $t$ to satisfy the semi-classicality condition, it may be assumed that this parameter is of the form
\begin{align}
	t = \left(\frac{l^2_\text{pl}}{A_\ell}\right)^{n}\quad\text{with } n\in (0, 2),
\end{align}
where $A_\ell$ is a typical macroscopic area in the triangulation. A natural choice would be $n=1$, which makes $t$ a rather small parameter. This parametrization of $t$ also coincides with the one originally considered by Thiemann in~\cite{Thiemann:2001a}.

In summary, we have seen in this subsection that Thiemann's heat kernel state can be parameterized in terms of (closed) twisted geometries. In the large area limit, the states take on a particularly simple form -- these are the so-called extrinsic coherent states~\eqref{eq:ExtrinsicState}. We can freely specify the data $(a_\ell, \xi_\ell, \vec{n}_{s(\ell)}, \vec{n}_{s(\ell)})$ and it is guaranteed that the states will be peaked on the intrinsic and extrinsic three-geometry described by these data. Furthermore, by tuning the semi-classicality parameter $t$ we can minimize the spreads $\Delta A_\ell$ and $\Delta h_\ell$ and thereby render the extrinsic states not only coherent, but also semi-classical.\\
Notice that the resolution of identity that holds for the heat kernel states no longer holds for the extrinsic states. The reason is the use of the highest weight approximation~\eqref{eq:HighestWeightApprox} involved in the transition from heat kernel states to extrinsic states. However, it is possible to derive a measure and prove a resolution of identity for Thiemann's heat kernel states in the twisted geometry parametrization. We will do so in the next subsection and thereby close a gap in the existing literature.

\subsection{Resolution of Identity for the Heat Kernel States in the Twisted Geometry Parametrization}
\label{ssec:SemiclassicalResOfId}
Here we consider Thiemann's heat kernel states and we provide the measure for the resolution of identity in the twisted geometry parametrization. This new result will be necessary in order to take into account the arguments in the recent work by R.~Oeckl \cite{Oecklpredictiveframeworkquantum2018}, which suggests that the measure must be considered in the definitions of the observables studied in \cite{RealisticObs,
TimeScale} for a black hole to white hole transition.

In order to find an adequate measure for the resolution of identity, we consider the function~\eqref{eq:QuotientMeasure}. Since it is a function on $\slc/SU(2)$, it has to be $SU(2)$ invariant and therefore it has to be true for $H_\ell$ in the Cartan decomposition~\eqref{eq:HParam} that
\begin{equation}
	\Omega_{2t}\left(n_{s(\ell)}\,\e^{\eta_\ell\frac{\sigma_3}{2}}\,\e^{i \gamma \zeta_\ell\frac{\sigma_3}{2}}\, n_{t(\ell)}^{-1}\right) = \Omega_{2t}\left(\e^{\eta_\ell\frac{\sigma_3}{2}}\right) = \frac{\e^{-\frac{t}{2}}}{(2 \pi t)^{\frac{3}{2}}}\frac{\eta_\ell}{\sinh \eta_\ell}\,\e^{-\frac{\eta^2_\ell}{2 t}},
\end{equation}
where the $SU(2)$ elements $n_{s(\ell)}$ and $n_{t(\ell)}$ drop out from the right hand side because of $SU(2)$ invariance and we also used~\eqref{eq:polarMeasure}. To proceed we notice that the measure $\dd H_\ell$ in the Cartan decomposition reads~\cite{RuhlBook}
\begin{equation}\label{eq:CartanMeasure}
	\dd H_\ell = \frac{\sinh^2 \eta_\ell}{4\pi}\,\dd\eta_\ell\,\dd u_\ell\,\dd v_\ell,
\end{equation}
where $\dd u_\ell$ and $\dd v_\ell$ are $SU(2)$ Haar measures. The resolution of identity in the twisted geometry parametrization does not immediately follow from these expressions because of the following subtlety regarding the $SU(2)$ Haar measures above:  The polar decomposition of  \eqref{eq:polarDecomp} for $H_\ell$ is unique 
and it is a parametrization by six real parameters. The twisted geometry parametrization \eqref{eq:HParam} for $H_\ell$ is not unique. There is a $U(1)$ gauge choice to be made, since there are seven real parameters to be integrated over in \eqref{eq:CartanMeasure}. As ansatz, we choose to drop the $\zeta_\ell$ integration in $\dd u_\ell$ such that the measure becomes proportional to the standard measure on the two-sphere $\mathcal S^2$. The measure $\dd v_\ell$ remains the standard $SU(2)$ Haar measure. Concretely, we define
\begin{align}
	\dd u_\ell &:= \mathcal N \sin\theta_{s(\ell)}\,\dd \phi_{s(\ell)}\,\dd \theta_{s(\ell)} = \mathcal N\,\dd^2 \vec{n}_{s(\ell)}\,\notag\\
	\dd v_\ell &:= \frac{1}{(4\pi)^2}\sin\theta_{t(\ell)}\,\dd \phi_{t(\ell)}\,\dd \theta_{t(\ell)} \,\dd \xi_{\ell} = \frac{1}{(4\pi)^2}\dd^2 \vec{n}_{t(\ell)} \,\dd\xi_{\ell},
\end{align}
where $\mathcal N$ is a possible normalization constant. The full ansatz for the resolution of identity measure in the twisted geometry parametrization then reads
\begin{align}\label{eq:MeasureAnsatz}
	\Omega_{2t}(H_\ell)\,\dd H_\ell = \frac{\mathcal{N}}{(4\pi)^2}\,\Omega_{2t}\left(\e^{\eta_\ell\frac{\sigma_3}{2}}\right) \,\sinh^2\eta_\ell\,\dd \eta_\ell\,\dd \xi_\ell\,\dd^2 \vec{n}_{s(\ell)}\,\dd^2\vec{n}_{t(\ell)}.
\end{align} 
The normalization $\mathcal{N}$ is determined by requiring that the ``volume" of this measure integrated over $\slc$ is the same in the polar decomposition and the Cartan decomposition:
\begin{equation}\label{eq:MeasureAnsatz}
	\int_{\slc}\mathcal{N}\,\Omega_{2t}\left(\e^{\eta\frac{\sigma_3}{2}}\right)\,\sinh^2\eta\,\dd \eta\,\dd \xi\,\dd^2 \vec{n}_{s}\,\dd^2\vec{n}_{t} \overset{!}{=} \int_{\slc}\Omega_{2t}(\e^{\vec{p}\cdot\frac{\vec{\sigma}}{2}})\,\frac{\sinh^2\vert\vec{p}\vert}{\vert \vec{p}\vert^2}\, \dd h\, \dd^3\vec{p}.
\end{equation}
On the left hand side the two integrations over $\mathcal{S}^2$ and the integration over $\xi$ give an overall factor of $(4\pi)^3$ and on the right hand side the $SU(2)$ integration over $h$ gives one. There remain only the $\eta$ and the $\vec{p}$ integrations. From \eqref{eq:polarMeasure} we know that the integrand on the right hand side only depends on the norm $\vert \vec{p}\vert$. We can therefore change to polar coordinates for the vector $\vec{p}$ which gives another factor of $4 \pi$ from the angular integration. The remaining integrals over $\eta$ and $p$ are of exactly the same form and they are both non zero since the integrand is positive definite. Hence, we can divide both sides by whatever value these integrals give. Collecting all  the remaining numerical factors and solving for $\mathcal{N}$ we find $\mathcal N = 1$.\\
The measure for the resolution of identity in the twisted geometry parametrization therefore reads
\begin{align}\label{eq:TheMeasure}
	&\Omega_{2t}(\e^{\eta_\ell\frac{\sigma_3}{2}})\,\dd H_\ell = \frac{\e^{-\frac{t}{2}}}{(4\pi)^2(2\pi t)^{\frac{3}{2}}} \eta_\ell \sinh \eta_\ell\,\e^{-\frac{\eta_\ell^2}{2t}}\,\dd\eta_\ell\,\dd\zeta_\ell\,\dd^2 \vec{n}_{s(\ell)}\,\dd^2 \vec{n}_{t(\ell)}\notag\\
	&\quad\quad\quad\quad\quad\eta_\ell\in \mathbb{R}^+,\quad \zeta_\ell\in [0, 4\pi), \quad\vec{n}_{s(\ell)}\in\mathcal{S}^2, \quad\vec{n}_{t(\ell)}\in\mathcal{S}^2.
\end{align}
We now proceed to show by direct computation that this measure does indeed give the resolution of identity for the heat kernel states \eqref{eq:ThiemannState} in the twisted geometry parametrization. To simplify the notation and render the computations more readable, we will drop the gauge-averaging integrations over $SU(2)$ and only consider a single link. This is sufficient to illustrate all the relevant steps in the proof of the resolution of identity. Under these simplifying assumptions, the identity to prove reads
\begin{equation}\label{eq:ResOfId}
\delta(hh'^\dagger) = \int_{\mathbb{R}^+}\dd\eta\,\nu_t(\eta)\int_{0}^{4\pi}\dd\xi\int_{\mathcal{S}^2}\dd^2 \vec{n}_s\int_{\mathcal{S}^2}\dd^2 \vec{n}_t\, \Psi_{H}^t(h)\,\overline{\Psi_H^t(h')},
\end{equation}
where 
\begin{align}
	\delta(hh'^\dagger) = \sum_j d_j \Tr_j[hh'^\dagger] = \sum_j d_j \sum_{|m|\leq j}\sum_{|n| \leq j}D^j_{mn}(h)D^j_{nm}(h'^\dagger)
\end{align}
is the Dirac distribution on $SU(2)$ and $\nu_t(\eta)$ is given by
\begin{align}\label{eq:TwistedMeasure}
	\nu_t(\eta) := \frac{\e^{-\frac{t}{2}}}{(4\pi)^2(2\pi t)^{\frac{3}{2}}} \eta \sinh \eta\,\e^{-\frac{\eta^2}{2t}}.
\end{align}
The states are explicitly given by 
\begin{align}
	\Psi_{\Gamma, H}^t(h) &= \sum_j d_j \,\e^{-j(j+1) t} \sum_{m, n, k} D^j_{mn}(h) D_{nk}^j(n_t) D_{km}^j(\e^{i \gamma\xi \frac{\sigma_3}{2}}n_s^\dagger)\,\e^{\eta k}     \notag\\
	\overline{\Psi_{\Gamma, H}^t(h')} &= \sum_{j'} d_{j'}\,\e^{-j'(j'+1) t}\sum_{m', n', k'} D_{n' m'}^{j'}(h'^\dagger)\,D_{k' n'}^{j'}(n_t^\dagger) D_{m' k'}^{j'}(n_s \e^{-i \gamma \xi \frac{\sigma_3}{2}})\, \e^{\eta k'}.
\end{align}
By noticing that $n_s = \e^{-i\phi_s\frac{\sigma_3}{2}}\e^{-i\theta_s\frac{\sigma_3}{2}}$ (see definition~\eqref{eq:DefNElement}) lives in a subspace of $SU(2)$ we can introduce the auxiliary variable $g:=n_s \e^{-i\gamma \xi \frac{\sigma_3}{2}}$, which is a genuine $SU(2)$ element. This allows us to perform the $n_s$ and the $\xi$ integration simultaneously by virtue of the Peter-Weyl theorem
\begin{align}
	A &:= \int_{\mathcal{S}^2}\dd^2\vec{n}_s \int_0^{4\pi}\dd\xi\, \Psi_{H}^t(h)\,\overline{\Psi_H^t(h')} = (4\pi)^2\int_{SU(2)}\dd g \, \Psi_{H}^t(h)\,\overline{\Psi_H^t(h')}\notag\\
	&=(4\pi)^2 \delta^{j j'} \delta_{m m'} \delta_{k k'} \sum_j d_j \e^{-2j(j+1) t} \sum_{m, n, k, n'} D_{mn}^j(h) D_{nk}^j(n_t) D_{n' m}^j(h'^\dagger) D_{k n'}^j(n_t^\dagger) \e^{-2\eta k}.
\end{align}
The quantity $A$ has been introduced for later convenience and should remind us that we performed the $\dd^2\vec{n}_s$ and $\dd\xi$ integrations in~\eqref{eq:ResOfId}.\\
To perform the next integration we notice that $n_t$ is also parametrized as $n_t = \e^{- i \phi_t\frac{\sigma_3}{2}}\,\e^{-i\theta_t\frac{\sigma_2}{2}}$ (see again definition~\eqref{eq:DefNElement}) and that therefore we have
\begin{align}
	D_{nk}^j(n_t) = \e^{-i\phi_t n} d_{nk}^j(\theta_t) \quad\text{and}\quad D_{k n'}^j(n_t^\dagger) = \e^{i\phi_t n'}\, d_{k n'}^j(-\theta_t),
\end{align}
which follows from elementary properties of Wigner matrices and which implies that the $n_t$ integration is zero unless the matrix indices satisfy $n=n'$. This allows us to do the following step:
\begin{align}
	\int_{\mathcal{S}^2}\dd^2 \vec{n}_t\, D_{nk}^j(n_t) D_{kn'}^j(n_t^\dagger) &=  \delta_{n n'} \int_{\mathcal{S}^2}\dd^2 \vec{n}_t\, D^j_{nk}(n_t) D^j_{kn}(n_t^\dagger)\,\overset{=1}{\overbrace{\frac{1}{4\pi}\int_0^{4\pi} \dd\alpha\, \e^{i\alpha k}\,\e^{-i\alpha k}}}\notag\\
	&= (4\pi) \delta_{n n'} \int_{SU(2)} \dd g' D_{nk}^j(g') D_{kn}^j(g'^\dagger) = \frac{4\pi \delta_{n n'}}{d_j}.
\end{align}
By using $n=n'$ and inserting an identity we were able to define the auxiliary variable $g' := n_t\,\e^{-i\alpha \frac{\sigma_3}{2}}$ and use again the Peter-Weyl theorem. Hence we find:
\begin{align}
	B:=\int_{\mathcal{S}^2}\dd^2\vec{n}_t \,A = (4\pi)^3 \sum_j \e^{-2j(j+1)t}\sum_{m, n} D^j_{mn}(h)D_{nm}(h'^\dagger)\sum_k\e^{-2\eta k}.
\end{align}
The last sum is easily performed by recognizing that it can be split into two geometric sums. The result of this simple computation is
\begin{align}
	\sum_{\vert k\vert\leq j} \e^{-2\eta k} = \frac{\sinh\left((2j+1)\eta\right)}{\sinh\eta},
\end{align}
which holds for both, integer and half-integer values of $j$. What is left to do is the integral over $\eta$ which gives
\begin{align}
	(4\pi)^3\int_{\mathbb R^+}\dd \eta\,\nu_t(\eta)\frac{\sinh\left((2j+1)\eta\right)}{\sinh\eta} = d_j\,\e^{2j(j+1)t}.
\end{align}
Putting everything together we obtain
\begin{align}
	\int_{\mathbb{R}^+}\dd\eta \,\nu_t(\eta) B = \sum_j d_j \sum_{m, n}D_{mn}^j(h) D_{nm}^j(h'^\dagger) = \delta(h h'^\dagger).
\end{align}
This completes the proof of the resolution of identity on a single link and for the gauge-variant states. With a little more effort and minor changes in the steps illustrated in this subsection one can also prove 
\begin{align}
	\delta_\Gamma(h, h'^\dagger) = \int_{\slc^L} \left(\prod_\ell\nu_{t_{\ell}}(\eta_\ell)\dd\eta_\ell\,\dd\zeta_\ell\,\dd\vec{n}_{s(\ell)}\,\dd\vec{n}_{t(\ell)}\right) \Psi_{\Gamma, H_\ell}^{t}(h)\overline{\Psi_{\Gamma, H_\ell}^{t}(h')}
\end{align}
for the gauge-invariant states \eqref{eq:ThiemannState} in the twisted geometry parametrization on a general graph $\Gamma$. The Dirac distribution $\delta_\Gamma$ on $SU(2)^L\setminus SU(2)^N$ is explicitly given by
\begin{align}
	\delta_\Gamma(h, h'^\dagger) = \int_{SU(2)^N}\left(\prod_\no\dd h_{\no(\ell)}\right) \int_{SU(2)^N}\left(\prod_\no\dd \tilde{h}_{\no(\ell)}\right)\prod_\ell \delta\left(h_{t(\ell)}^\dagger h h_{s(\ell)}\,\left(\tilde{h}_{t(\ell)}^\dagger h' \tilde{h}_{s(\ell)}\right)^\dagger\right).
\end{align}
We conclude that the integration measure giving the resolution of identity for the gauge-variant as well as the gauge-invariant heat kernel states in the twisted geometry parametrization on an arbitrary graph $\Gamma$ is given by~\eqref{eq:TheMeasure}.

%% file: Content/Chapter_3.tex
\chapter{Spin Network Histories}\label{chap:MathI}


An adequate choice of variables and a rigorous implementation of the Dirac quantization program has lead to an enormous progress in quantum gravity. In the framework of LQG it proved to be surprisingly simple to implement the Gauss and vector constraint and to characterize their solutions by spin network states and $s$-knots. The scalar constraint, however, is sensitive to quantization ambiguities and the dynamics of quantum gravity therefore remains largely unknown terrain.\\
In an attempt to shed more light on the dynamics of LQG and circumvent the complications of the scalar constraint, a path integral quantization of first oder gravity was attempted \cite{Reisenberger:1994,Reisenberger:1997, Reisenberger:1996}. This set in motion the development of several models \cite{Reisenberger:1997,Freidel:1998,Iwasaki:2000,Barrett:1997,Markopoulou:1997,Gambini:2001,Capovilla:1992,Engle:2007c,Freidel:2007} which can be summarized under the collective name \textit{spin foam model}. In this chapter, we will not go into the details of all these models (see \cite{Perez2012Review} though for a nice review) and instead content ourselves with presenting the main ideas and provide some of the necessary mathematical framework.\\
Section~\ref{sec:PathIntegralsAndSpinFoams} illuminates the connection between path integrals and spin foams and it is explained how LQG fits into the picture. In section~\ref{sec:MathBackground} we introduce the notion of simplicial triangulations and other terminology which will be important for understanding chapters~\ref{chap:CLQG} and~\ref{chapter_5}. This section also contains some unpublished results by the author about the triangulation of spacetime regions of topology $I\times\Sigma$ and $I^2\times\mathcal{S}^2$ in particular, which are relevant for the black hole to white hole tunneling model.

\section{On Path Integrals and Spin Foams}\label{sec:PathIntegralsAndSpinFoams}
Following a suggestion by Wheeler, Misner \cite{Misner:1957} considered defining quantum gravity through a Feynman path integral \cite{Feynman:1948}. The idea was to consider a four dimensional manifold $\mathcal M$ with boundaries $\Sigma$ and $\Sigma'$ and to endow these boundary manifolds with an intrinsic three dimensional geometry. That is to say, the boundaries carry two equivalence classes of metrics, $[q_{ab}]$ and $[q'_{ab}]$. In the quantum theory, these geometries are described by state vectors $\ket{[q_{ab}]}$, $\ket{[q'_{ab}]}$ and the path integral can be formally written as
\begin{equation}
	\innerp{[q_{ab}]}{[q'_{ab}]} = \int_{[g]\vert_{\Sigma}=[q_{ab}]}^{[g]\vert_{\Sigma'}=[q'_{ab}]}\mathcal D[g]\, \e^{i S([g])}.
\end{equation}
The integration is performed over all interpolating four dimensional geometries which have $[q_{ab}]$ and $[q'_{ab}]$ as their boundary. In other words, one needs to integrate over all metrics $g_{\mu\nu}$ and mod out the group $\textsf{Diff}(\mathcal M)$.\\
This is a beautiful idea, but there are various difficulties with this approach. First and foremost, it is not clear how to define the integration measure $\mathcal D[g]$ or how to characterize the diffeomorphism invariant information of the equivalence class $[g]$. Moreover, as we mentioned in chapter~\ref{chap:intro}, the canonical theory of quantum gravity in metric variable has never been properly defined and therefore there is no notion of boundary states $\ket{[q_{ab}]}$ of geometry.\\
The LQG framework offers a promising way out of some of these difficulties. First of all, there is a well defined notion of three dimensional (quantum) geometry in terms of spin network states. These states capture diffeomorphism invariant aspects of the boundary geometry as, for instance, area and volume eigenvalues. Since these states are defined on an abstract graph, they are intrinsically discrete and thereby suggest the use of discretization methods to solve the problem of defining the integration measure $\mathcal D[g]$ or the path integral as a whole. This is analogous to Feynman's original work on quantum mechanics.\\
As is well-known, the quantum mechanical path integral computes the matrix elements of the unitary evolution operator $\hat U(t)$. It thereby also provides a solution to the evolution equation. That is, given a state $\psi_0(x)$ from the kinematical Hilbert space, it can be shown that the ``evolved'' state $\psi(x,t) = \hat U(t)\, \psi_0(x)$ solves the Hamiltonian constraint\footnote{In quantum mechanics this constraint is of course simply the Schr\"{o}dinger equation.} $\hat H\,\psi(x,t) = 0$. It is therefore natural to try and define the path integral of quantum general relativity in the same way.

\subsection{From the canonical Theory to Spin Foams}
\label{ssec:FromCanonicalToSF}
\begin{center}
	\begin{figure}
		\centering
		\includegraphics[width=0.6\textwidth]{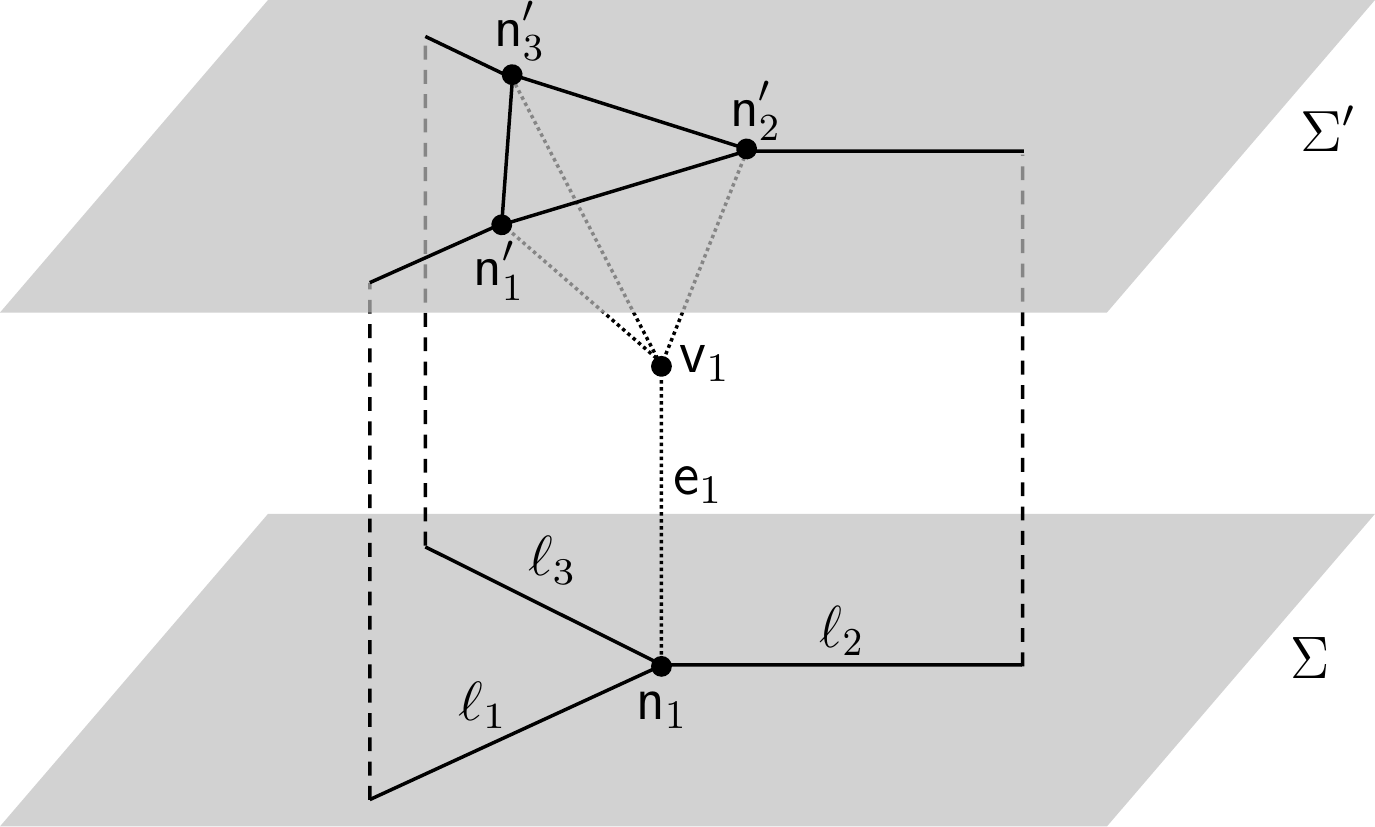}
		\caption{The action of the scalar constraint generates new links a nodes which gives rise to a foam like structure.}
		\label{fig:BasicSpinFoam}
	\end{figure}
\end{center}

Similarly to the vector constraint discussed in section \ref{ssec:2_DiffeosAndSknots}, the scalar constraint possesses a continuous spectrum containing zero. Its solutions can therefore not be found within the kinematical Hilbert space $\mathcal H_\textsf{kin}$. Rather, the solutions lie in the larger space $\textsf{Cyl}^*$ and the physical Hilbert space $\mathcal H_\textsf{phys}$ is therefore not a proper subspace of $\mathcal H_\textsf{kin}$. We may nevertheless introduce a formal ``projector'' $P$ which sends elements of the kinematical Hilbert space into solutions of the scalar constraint\footnote{The operator $P$ cannot be a true projector since it maps outside of its domain and $P^2$ is therefore ill-defined.}~\cite{Rovelli:1998}. Formally, we can write
\begin{equation}
	P = \prod_{x\in\mathcal M}\delta\left(\hat S(x)\right) = \int\mathcal D[N]\, \e^{i \hat S[N]},
\end{equation}
where $\hat S[N]$ is the smeared scalar constraint and $N$ denotes a laps function. We have not chosen a specific implementation of the scalar constraint and in the sequel we will only need general properties of $\hat S$ which are shared by several of the proposed variants. From the definition of $P$ it is also apparent that the integral needs to be regularized. Following \cite{Rovelli:1998} we restrict the domain of integration to
\begin{equation}
	\vert N(x)\vert \leq T
\end{equation}
for some $T>0$ and then we consider the matrix elements
\begin{equation}\label{eq:MatrixElements}
	\langle s\vert P \vert s'\rangle_\textsf{diff} = \int_{\vert N(x)\vert \leq T}\mathcal D[N]\langle s| e^{i \hat S[N]}|s'\rangle_\textsf{diff} = \int_{\vert N(x)\vert \leq T}\mathcal D[N]\langle s| \sum_{n=0}^{\infty} \frac{i^n}{n!}\hat S[N]^n |s'\rangle_\textsf{diff},
\end{equation}
where $\ket{s}$, $\ket{s'}$ are $s$-knot states. Knowing these matrix elements is tantamount to knowing the physical inner product: 
\begin{equation}
	\innerp{s}{s'}_\textsf{phys} := \langle s\vert P \vert s'\rangle_\textsf{diff}
\end{equation}
The scalar constraint in the expansion \eqref{eq:MatrixElements} acts on $s$-knot states and, quite independently from any quantization ambiguities \cite{Thiemann:1996}, acts on the nodes of the spin network. Generically, its action creates new links and nodes and therefore changes the spin network. An illustration of this process is given in Figure~\ref{fig:BasicSpinFoam}.\\ This also suggests that the matrix elements in the power series \eqref{eq:MatrixElements} represent transitions between two different spin networks and this gives rise to foam like combinatorial structures such as the one depicted in Figure~\ref{fig:BasicSpinFoam}. In~\cite{Rovelli:1998} it was shown that the integration over the laps in \eqref{eq:MatrixElements} can be performed and the resulting power series in the cut-off $T$ is of the form
\begin{equation}\label{eq:GeneralExpansion}
	\langle s\vert P \vert s'\rangle_\textsf{diff} = \sum_{n=0}^\infty \frac{(i T)^n}{n!}\sum_{\text{\text{nodes},\text{links}}}\prod_\text{vertices \ve} A(\ve).
\end{equation}
The functions $A(\ve)$ are so-called ``vertex functions'' which depend on the spin numbers of adjacent edges and faces. The precise mathematical definition is not relevant here. What is of interest to us is the picture that emerges: The original graph on which the $s$-knot is defined is extended into a foam like structure by virtue of the operator $\hat S$. The general form of the expansion \eqref{eq:GeneralExpansion} implies that one has to sum over all possible interpolating foam like structures between $s$ and $s'$. From very simple ones such as in Figure~\ref{fig:BasicSpinFoam} to structures with many links and nodes and complicated combinatorics. Finally, the extended combinatorial structure is colored by spin numbers through the vertex function. Hence, \textit{spin foams} are born.\\
This general picture does not depend on the details of the quantization of the scalar constraint. However, the values of the matrix elements depend on the precise definition of $\hat S$ and are therefore sensitive to quantization ambiguities. This attempt to define the dynamics of LQG therefore encounters the same difficulties as the canonical approach.

\subsection{Path Integral Quantization of BF Theory}\label{ssec:PathIntegralQuantOfBFTheory}
A possible way around the difficulties encountered in the previous section is to follow Feynman's lead more directly and try to quantize the gravitational field by discretizing it and define a path integral starting from the exponential of the classical action.\\
Before trying to quantize General Relativity on a four-dimensional Lorentzian spacetime, however, it is instructive to consider a simpler gauge theory which nevertheless shares some of GR's properties. This theory, known as BF theory, has no local degrees of freedom and it is background independent in the sense that it can be formulated on differential manifolds of any dimension without the use of a  metric. For concreteness we assume the differential manifold $\mathcal M$ to be compact, orientable, and four-dimensional and $G$ to be a compact Lie group (see \cite{Baez:1997} for the general case). BF's field content then consists of a connection $\omega$ on a principal $G$ bundle of $\mathcal M$ and a $2$-form $B$ which takes values in the Lie algebra $\mathfrak g$. Its action can be written as
\begin{equation}\label{eq:BFAction}
	S[B,\omega] := \int_\mathcal{M} \Tr[B\wedge F(\omega)],
\end{equation}
where $F(\omega)$ is the curvature $2$-form associated with $\omega$ and $\Tr[\cdot]$ denotes the Killing form in the adjoint representation. Our goal now is to discretize this action and define a path integral for BF theory. To that end, we first of all introduce a discretization $\mathcal T$ of spacetime in terms of $4$-simplices -- four-dimensional generalizations of tetrahedra. Since $B$ as well as $F$ are $2$-forms, it is natural to try to discretize them by integrating them over two-dimensional surfaces. In the present case, this would amount to integrating $B$ and $F$ over the triangles appearing in the $4$-simplices of the discretization. However, it seems somewhat unnatural to associate two distinct variables to the same discrete structure.\\
A more sophisticated way to decide how to discretize the fields is by invoking the branch of mathematics known as discretization theory. Even though it is a very interesting topic and highly relevant to our purposes, we will not discuss it in detail\footnote{A brief introduction to simplicial triangulation is given in the next section.} and instead just point the reader toward the pertinent literature~\cite{ThiemannBook,Adams:1996}. In particular, in~\cite{ThiemannBook} it is shown how discretization theory naturally leads us to integrate the $B$ field over triangles while the curvature $2$-form is naturally integrated over a two-dimensional object which is ``dual'' to the aforementioned triangle. This duality is akin of the duality provided by the Hodge star operator which, in four dimensions, sends $2$-forms to $2$-forms and which can also be visualized as two orthogonal planes. More concretely we can picture this triangle as being spanned by $\dd x_1\wedge \dd x_2$ while its dual is spanned by $\star(\dd x_1\wedge \dd x_2) = \dd x_3\wedge \dd x_4$.\\
Without attempting to make this more rigorous, we claim that the discretized BF action~\eqref{eq:BFAction} on the triangulation $\mathcal T$ can be written as~\cite{ThiemannBook}
\begin{align}
	S_\mathcal{T}[B,\omega] = \sum_{t\in\mathcal T}\Tr\left[\left(\int_{\star t}F\right)\left(\int_{t}B\right)\right],
\end{align}
where the sum is over all triangles in the triangulation. From now on, we will call $\star t$ a \textit{face} and denote it by $\fa$. Furthermore it is convenient to introduce the notation
\begin{equation}
	B_\fa := \int_{t(\fa)}B\,\in\mathfrak g,
\end{equation}
where we use the fact that there is a one-to-one relationship between triangles and their dual faces.\\ 
For the curvature $2$-form, which is an element of the group $G$, it is easier to approximate it by the holonomy around a closed loop instead of actually compute the integral over $\star t = \fa$. This closed loop has to lie in the plane $\star t$ and we call the segments which constitute these loops \textit{edges} and we denote them by $\ed$. More precisely, these edges belong to the boundary of $\fa$.\\
To proceed and construct a holonomy around a closed loop, we associate a $G$ group element to every edge $\ed$ by
\begin{equation}
	h_\ed := \textsf{P}\,\e^{-\int_\ed \omega}\,\in G.
\end{equation}
Multiplying all the holonomies which belong to the same face allows us to introduce the face holonomy
\begin{equation}
	h_\fa := h_{\ed_1}\cdots h_{\ed_n}\quad\text{with}\quad \ed_1,\dots,\ed_n\in\fa.
\end{equation}
Expanding the face holonomy in a power series up to the linear term results in
\begin{equation}
	h_\fa = \id_{G} + F(\fa)\quad\text{with}\quad F(\fa) := \int_\fa F\,\in G.
\end{equation}
Due to the skew-symmetry\footnote{The skew-symmetry implies that $B$ is trace-free and so we get $\Tr[h_\fa B_\fa] = \Tr[B_\fa + F(\fa)B_\fa] = Tr[F(\fa)B_\fa]$.} of the $B$ field, we can rewrite the discretized BF action, up to higher order correction terms, as
\begin{equation}
	\sum_{\fa}\Tr[F(\fa)B_\fa] = \sum_{\fa}\Tr[h_\fa\, B_\fa].
\end{equation}
To define the BF path integral we also need to discretize the functional measures $\mathcal D[B]$ and $\mathcal D[\omega]$. Given that the $B$-field possesses six independent components, we choose to discretize $\mathcal D[B]$ by the product $\prod_{\fa}\dd^6 B_\fa$ of Lebesgue measures. For $\mathcal D[\omega]$, on the other hand, we choose $\prod_{\ed}\dd h_\ed$, where $\dd h_\ed$ is the Haar measure on $G$.\\
Notice that we also silently passed from working on a triangulation to working exclusively on the dual quantities $\fa$ and $\ed$. These are the only structures required to define the discretized variables, the action, and the measures. With these choices we can finally write the partition function of BF theory as
\begin{equation}\label{eq:PartitionFuntionDef}
	Z_{BF} := \int_{(G)^{\# \ed}}\prod_{\ed}\dd h_\ed\int_{(\mathbb R^6)^{\#\fa}}\prod_{\fa}\dd^6 B_\fa \, \e^{i\sum_{\fa}\Tr[h_\fa\, B_\fa]}.
\end{equation}
The integration over the discrete $B$-field is easy to perform and results in a product of six delta distributions on the real line. It is however more common to replace this product by the delta distribution of the group $G$, even though this is not completely justified from a mathematical point of view. By doing so, one obtains the partition function
\begin{equation}
	Z_{BF} = \int_{(G)^{\# \ed}}\prod_{\ed}\dd h_\ed\,\prod_{\fa}\delta_{G}(h_{\ed_1}\dots h_{\ed_n}).
\end{equation}
By virtue of the Peter-Weyl theorem, we can rewrite the delta distribution as
\begin{equation}\label{eq:DeltaOnGroup}
	\delta_{G}(h) = \sum_{\rho}d_\rho \Tr[\rho(h)],
\end{equation}
where $\rho$ denotes the irreducible unitary representations of $G$ and $d_\rho$ is the dimension of these representations. Using~\eqref{eq:DeltaOnGroup}, the partition function assumes the form
\begin{equation}
	Z_{BF}= \sum_{\{\rho\}}\int_{(G)^{\#\ed}}\prod_{\ed}\dd h_\ed \prod_{\fa} d_{\rho_\fa}\, \Tr[\rho_\fa(h_{\ed_1})\cdots \rho_\fa(h_{\ed_n})].
\end{equation}
We can now group together all group elements $\rho_\fa(h_\ed)$ which belong to the same edge but to different faces. This allows us to define the projector
\begin{equation}
	P^\ed_\textsf{inv}(\rho_1, \rho_2,\dots,\rho_n) := \int_G\dd h_\ed\, \rho^{\fa_1}_{m_1 n_1}(h_\ed) \rho^{\fa_2}_{m_2 n_2}(h_\ed)\cdots \rho^{\fa_n}_{m_n n_n}(h_\ed)
\end{equation}
onto the space $\text{Inv}[\rho_1\otimes\rho_2\otimes\cdots\otimes\rho_n]$. Due to the group multiplication property of the representation matrices $\rho^\fa_{mn}$ and the $G$-invariance of the Haar measure it is easy to see that this group averaging procedure generates a tensor which is invariant under $G$ transformations. In other words, this procedure defines an intertwiner and it allows us to re-write the BF partition function as
\begin{equation}
	Z_{BF} = \sum_{\{\rho\}} \prod_{\fa}d_{\rho_\fa}\prod_{\ed} P^\ed_\textsf{inv}(\rho_1, \rho_2,\dots,\rho_n),
\end{equation}
where the intertwiners are contracted among each other according to the pattern provided by the dual structure defined by the edges and faces.\\
When everything is done more carefully, it can be seen that the edges and faces give rise to a foam like structure. A very simple example can be seen in Figure~\ref{fig:BasicSpinFoam}. This is the first similarity with the previous quantization approach. The second similarity is the appearance of intertwiners. In the first approach discussed in subsection~\ref{ssec:FromCanonicalToSF}, the intertwiners are ``hidden'' in the definition of the $s$-knot states. Here, they are much more explicit.\\
Notice that the foam like structures in the two approaches have very different origins. In the canonical approach, the foam is being generated by the scalar constraint while the foam of the covariant theory emerges through the hand chosen triangulation. Nevertheless, the fact that the same qualitative picture crystallizes from the covariant and the canonical theory and the fact that intertwiners play an important role hints at a connection between the two. This observation is important: The canonical theory has a very rigorous definition and it is based on a well-tested quantization method. The covariant theory, on the other hand, rests on a much more \textit{ad hoc} definition such as the one of the partition function given in~\eqref{eq:PartitionFuntionDef}. The fact that there is some sort of connection between the two approaches is encouraging enough to attempt a definition of quantum gravity through a discretized path integral while trying to make contact with the canonical theory. But to do that, we need a better understanding of spacetime triangulations.

\newpage
\section{Mathematical Background}\label{sec:MathBackground}
In the previous sections we were deliberately vague about what precisely is meant by ``discretization of spacetime'' and the meaning of the dual objects. In this section we close this gap and provide the necessary mathematical structures required to illuminate certain aspects of spin foam models. Some of the vocabulary introduced here will be crucial for understanding the definition of the CLQG amplitude given in chapter~\ref{chap:CLQG}.
Apart from standard definitions of simplicial complexes and related concepts (subsection~\ref{ssec:SimpComp}), we also discuss a simplicial triangulation algorithm for manifolds of topology $I\times\Sigma$ based on unpublished work by the author (subsections~\ref{ssec:AnExample} and~\ref{ssec:Algorithm}). We then conclude with a complement on how to define angles between Lorentzian vectors (subsection~\ref{ssec:LorentzianAngles}).

\subsection{Simplicial Complexes and their Dual}\label{ssec:SimpComp}
\begin{figure}[!htbp]
	\centering
	\includegraphics[width=0.45\textwidth]{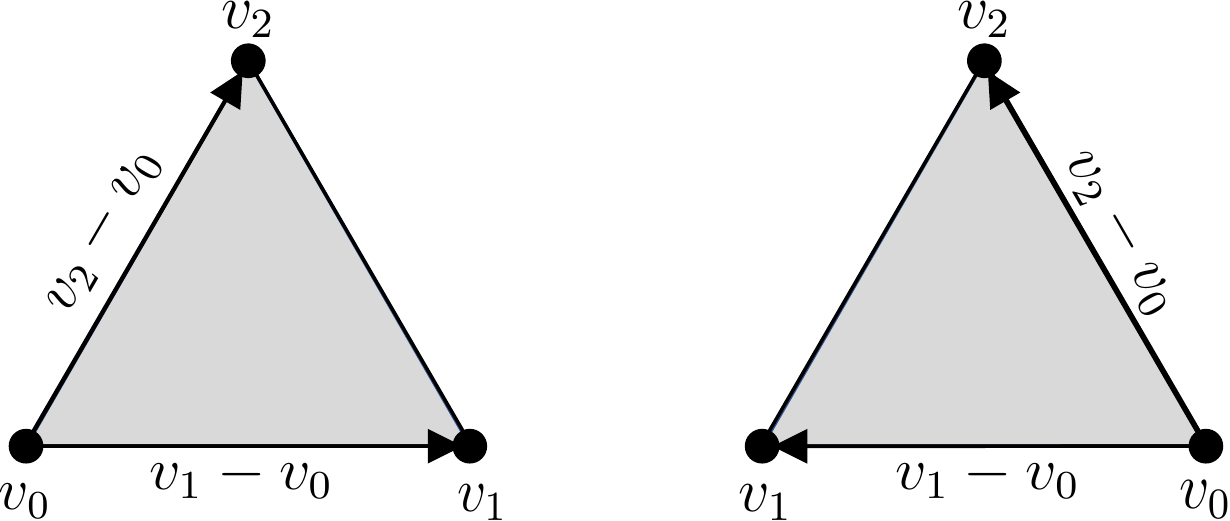}	
	\caption{The orientation of a $2$-simplex in terms of its basis vectors.}
	\label{fig:Orientations}
\end{figure}

\begin{center}
	\begin{figure}[h]
		\centering
		\includegraphics[width=1\textwidth]{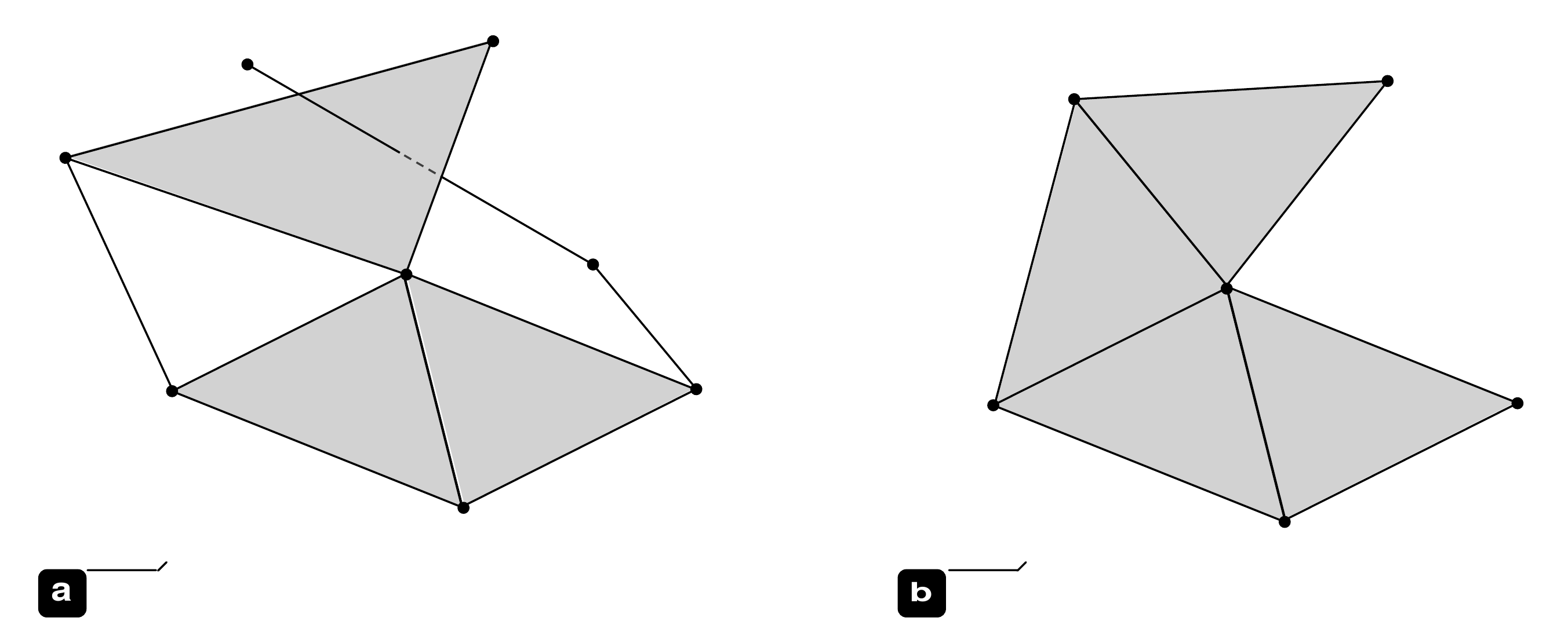}
		\caption{An example of a what is not a homogeneous $p$-simplex (a),  a homogeneous $p$-simplex (b).}
		\label{fig:NoIntersection}
	\end{figure}
\end{center}

\begin{center}
	\begin{figure}[h]
		\centering
		\includegraphics[width=1\textwidth]{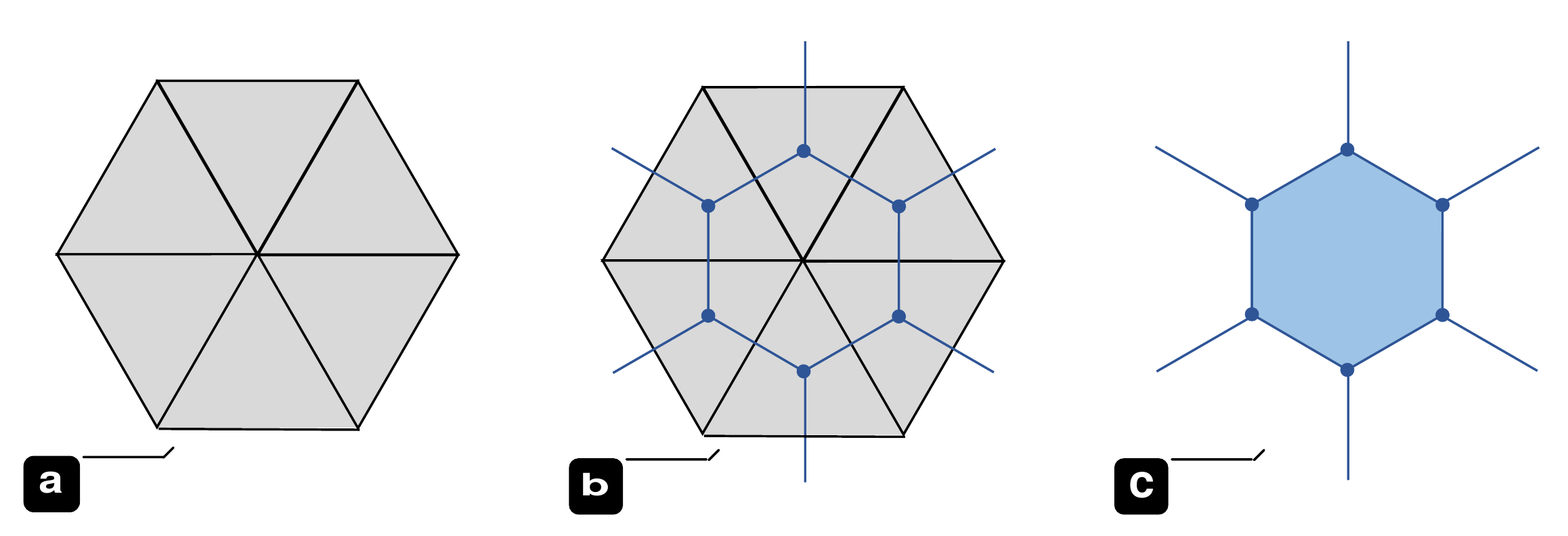}
		\caption{A two-dimensional simplicial triangulation (a), vertices and edges on top of the triangulation (b), the associated $2$-complex (c).}
		\label{fig:ExampleTwoComplex}
	\end{figure}
\end{center}

In order to make the notion of  ``discretization of spacetime'' more precise we start with the definition of simplices.
\begin{mydef}{ $p$-Simplex}{psimplex}
	A $p$-simplex $\sigma^{(p)}=[v_0,v_1,\dots,v_p]$ embedded in $\mathbb R^d$ is the convex hull of $p+1$ affinely independent vertices $v_0,v_1\dots, v_p\in\mathbb R^d$. That is, $\sigma^{(p)}$ is the set of points
	\begin{equation*}
		\sigma^{(p)} := \left.\left\{\sum_{i=0}^{p}t_i\, v_i\, \right\vert\, \sum_{i=0}^{p}t_i = 1\,\text{ with }\, t_i\geq 0\,\forall i\right\},
	\end{equation*}
	which spans a $p$-dimensional vector space.\label{def:Simplex}
\end{mydef}
From the definition it is clear that points, lines, triangles, and tetrahedra are $0$-, $1$-, $2$-, and $3$-simplices, respectively. Due to the affine independence of the vertices of $\sigma^{(p)}$ we can take the $p$ vectors $v_1-v_0,\dots,v_p-v_0$ as a basis of the $p$-dimensional vector space. It also immediately follows that there is a natural notion of orientation induced by the order in which the vertices $v_i$ appear in the list $[v_0,v_1,\dots, v_p]$. Interchanging any two of the vertices is the same as going from a right-handed to a left-handed basis. This is illustrated in Figure~\ref{fig:Orientations}. The permutation of vertices cannot be described by the action of an $SO(p)$ rotation but rather by an $O(p)$ transformation. Hence, a $p$-simplex can be assigned an orientation in the same sense the Euclidean space can be oriented. Thanks to the representation of a $p$-simplex as a list of vertices we can define orientation as follows:\\
\textcolor{white}{empty line}\\
\begin{mydef}{ Orientation of a $p$-Simplex}{Orientation}
	Let $\pi\in S_{p+1}$ be a permutation on the set of indices. We say that two $p$-simplices $[v_0,v_1,\dots,v_p]$ and $[v_{\pi(0)},v_{\pi(1)},\dots,v_{\pi(p)}]$ are equally oriented if $\pi$ is an even permutation. Otherwise they are oppositely oriented. \label{def:Orientation}
\end{mydef}
Consider a tetrahedron embedded in $\mathbb R^3$. It is clear from definition~\ref{def:Simplex} that removing one of the vertices from the list $[v_0,v_1,v_2,v_3]$ defines a triangle and this triangle belongs to the boundary of the tetrahedron. There are four different ways to remove a vertex from the list defining the tetrahedron and we obtain four different triangles. These are precisely the $2$-simplices which constitute the boundary of the tetrahedron. We therefore naturally arrive at the following definition:
\begin{mydef}{ The Boundary of a $p$-Simplex}{Boundary}
	The boundary $\partial\sigma^{(p)}$ of a $p$-simplex consists of $(p+1)$ different $(p-1)$-simplices which are obtained from $\sigma^{(p)}$ by omitting one vertex. That is, the boundary $(p-1)$-simplices are defined by $\sigma^{(p-1)}_i := [v_0,\dots,\arc{v}_i,\dots, v_p]$ for $0\leq i\leq p$, where the arc denotes omission. The boundary is then given by $\partial\sigma^{(p)}=\bigcup_{i=0}^p\sigma^{(p-1)}_i$.\label{def:Boundary}
\end{mydef}
Clearly we can remove more than one vertex from a given list and thereby obtain further subsimplices. In the example of the tetrahedron considered above, removing two vertices results in a list with only two remaining vertices and these define a line, i.e. a $1$-simplex. \\
In general, removing $k$ vertices from $[v_0,v_1,\dots, v_p]$ leaves us with a $p-k$-simplex contained in the original $p$-simplex. As there are $\binom{p+1}{k+1}$ possibilities to remove $k$ elements from a list of length $p+1$ we just proved
\begin{mylemma}{ Number of subsimplices in $\sigma^{(p)}$}{Subsimplices}
	The number of $(p-k)$-simplices contained in a $p$-simplex, which we denote as $N^p_k$, is given by
	\begin{equation*}
		N^p_k = \binom{p+1}{k+1}\quad\text{for } 0\leq k\leq p.
	\end{equation*}	\label{lem:Number}
\end{mylemma}
By integrating the volume form of $\mathbb R^p$ over the set of points defined in~\ref{def:Simplex}, i.e. over~$\sigma^{(p)}$, we obtain the volume of the $p$-simplex.
\begin{mythm}{ Volume of $\sigma^{(p)}$}{Volume}
		The volume of $\sigma^{(p)}$ is given by $\text{Vol}(\sigma^{(p)}) = \frac{1}{p!}\left|\det(v_1-v_0,\dots,v_p-v_0)\right|$, where $(v_1-v_0,\dots,v_p-v_0)$ is the determinant of the matrix which has the column vector $v_i-v_0$ as its $i$-th column. 
\end{mythm}
The proof of this theorem is not particularly difficult, but rather lengthy and we therefore do not include it here. Instead, we proceed and introduce yet another definition which finally formalizes what we mean by a ``discretization of spacetime''. 
\begin{mydef}{ Homogeneous simplicial $p$-Complex}{pComplex}
	A homogeneous simplicial $p$-complex $\mathcal C$ is a collection of $p$-simplices $\sigma^{(p)}_i$, $i=1,\dots, N$, which satisfy the following conditions:
	\begin{itemize}
		\item[1.] All the subsimplices of each $\sigma^{(p)}_i\in\mathcal C$ also belong to $\mathcal C$.
		\item[2.] The intersection of any two simplices $\sigma^{(k)}_i,\sigma^{(l)}_j$ with $0\leq k,l\leq p$ and $1\leq i,j\leq N$ is either empty or a subsimplex of both, $\sigma^{(k)}_i$ and $\sigma^{(l)}_j$.
		\item[3.] Every simplex of dimension less than $p$ is a subsimplex of some $\sigma^{(p)}_i\in\mathcal C$.
	\end{itemize}\label{def:pComplex}
\end{mydef}
The last two conditions are particularly important for our purposes. Condition 2. guarantees that two simplices in the complex cannot intersect, they can only be joined along common subsimplices (cf. Figure~\ref{fig:NoIntersection}).\\
Condition 3. implies that every subsimplex belongs to a $p$-simplex of the complex and therefore we can say, informally, that every homogeneous $p$-complex looks like a collection of $p$-simplices which are glued together along their $(p-1)$-subsimplices.  This construction is what formalizes ``discretization of spacetime''. 
In fact, Tullio Regge used a simplicial complex to triangulate spacetime and define a discrete version of General Relativity~\cite{Regge:1961}.\\
While this is a fascinating subject, we will not pursue it further and instead refer the reader to the original work of Regge and related work by Sorkin~\cite{Regge:1961, Sorkin:1975}. Here, we shall focus on the introduction of a different structure which will play an important role in the quantum theory: the dual $2$-complex. There is a precise mathematical definition of what a $2$-complex is, which can for example be found in~\cite{ThiemannBook}. However, the definition is overly complicated for something which is actually very simple and we shall therefore refrain from writing it here. Instead, we give a more colloquial description starting from a two-dimensional example:\\
Consider the two-dimensional triangulation shown in Figure~\ref{fig:ExampleTwoComplex} a. Now, mark a point, called a \textit{vertex} $\ve$, inside of every triangle (cf. Figure~\ref{fig:ExampleTwoComplex} b). Then, connect neighboring vertices by lines, the so-called \textit{edges} $\ed$. Notice how every edge intersects precisely one side of a triangle. Finally, remove the triangulation from the picture and what remains is a collection of vertices, edges and so-called faces $\fa$, which are bounded by the vertices and edges (cf. Figure~\ref{fig:ExampleTwoComplex} c). This collection of zero dimensional (vertices), one dimensional (edges) and two dimensional (faces) objects constitute the $2$-complex of the triangulation shown in Figure~\ref{fig:ExampleTwoComplex} a.\\
In the general case, one proceeds analogously to the above example. Assuming we are given a $p$-dimensional simplicial triangulation and we are asked to construct its $2$-complex, we would represent every $p$-simplex $\sigma^{(p)}_i$ by a vertex $\ve_i$.
$\phantom{.}$\\
 Then, we would determine the boundary of every $\sigma^{(p)}_i$ and represent every $(p-1)$-dimensional boundary simplex $\sigma^{(p-1)}_{i}$ by an edge $\ed$ which emanates from the vertex $\ve_i$. Of course, because the $p$-simplices are glued together, there will be $(p-1)$-simplices which belong to two different $p$-simplices. Whenever $\sigma^{(p)}_i$  and $\sigma^{(p)}_j$ share a $(p-1)$-simplex, we can connect the edge which emanates from $\ve_i$ and the edge which emanates from $\ve_j$. It is also convenient to call this edge $\ed_{ij}$ to indicate which two vertices it is connecting. Finally, what is left to do is to identify all the faces $\fa$ which are bounded by a number of vertices and edges.\\
We also say that $p$-simplices are dual to vertices, $(p-1)$-simplices are dual to edges and $(p-2)$-simplices are dual to faces. In four dimensions this leads to the following relations:
\begin{center}
\begin{tabular}{ll}
	\rowcolor{gray}
	\textcolor{white}{Object in the triangulation} & \textcolor{white}{Object in the $2$-complex}\\
	\rowcolor{myGray}
	$4$-simplex & vertex $\ve$\\
	\rowcolor{myGray}
	tetrahedron & edge $\ed$\\
	\rowcolor{myGray}
	triangle & face $\fa$
\end{tabular}
\end{center}
Notice that some of the $(p-1)$-simplices lie on the boundary of the triangulation and their dual edge is therefore not attached to anything. If we only consider these boundary $(p-1)$-simplices, we can construct what is called the boundary graph $\Gamma$. To do that, we simply represent each $(p-1)$ simplex by a node $\no$. These nodes are connected among each other in the same way the $(p-1)$-simplices are connected. That is to say, whenever two $(p-1)$-simplices share a $(p-2)$-simplex we connect them by a \text{link} $\ell$. We also often say that nodes are dual to $(p-1)$-simplices on the boundary of the triangulation and links are dual to $(p-2)$-simplices shared between these $(p-1)$-simplices. In four dimensions (i.e. on three dimensional boundaries) this leads to the following relations:
\begin{center}
	\begin{tabular}{ll}
	\rowcolor{gray}
		\textcolor{white}{Object on the boundary} & \textcolor{white}{Object in the boundary graph $\Gamma$}\\
	\rowcolor{myGray}
		tetrahedron & node $\no$\\
	\rowcolor{myGray}	
		triangle & link $\ell$
	\end{tabular}
\end{center}
This is all the structure we need for the quantum theory presented in chapter~\ref{chap:CLQG}. A more sophisticated examples involving triangulations and $2$-complexes will be discussed in the next subsection and it will culminate in the development of a simplicial triangulation algorithm.

\subsection{An Example: Triangulation and dual Complex of $I\times S^2$}
\label{ssec:AnExample}
In two or three dimensional spaces it is possible to visualize the process of triangulation. However, the amount of information one has to keep track of can quickly become overwhelming and it is difficult to imagine how to apply ``visual'' methods to four dimensional spaces. We therefore seek an algorithm which reliably generates a triangulation from some given initial data.\\
To be more precise, we assume we already know the triangulation of some initial data hypersurface $\Sigma$ and we look for the triangulation of $I\times\Sigma$, where $I:=[I_i,I_f]$ is a finite interval. In order to appreciate the difficulties associated with this problem and develop some intuition, let us consider the triangulation of $I\times \mathcal S^2$ by visual methods. This triangulation is relevant for current spin foam computations of the black hole to white hole transition\footnote{These more advanced computations have not been completed yet and are not part of this thesis.} where a spacetime region of topology $I^2\times\mathcal S^2$ is being considered. The manifold $I\times\mathcal S^2$ is the boundary of $I^2\times\mathcal S^2$.

\begin{center}
	\begin{figure}[h]
		\centering
		\includegraphics[width=0.4\textwidth]{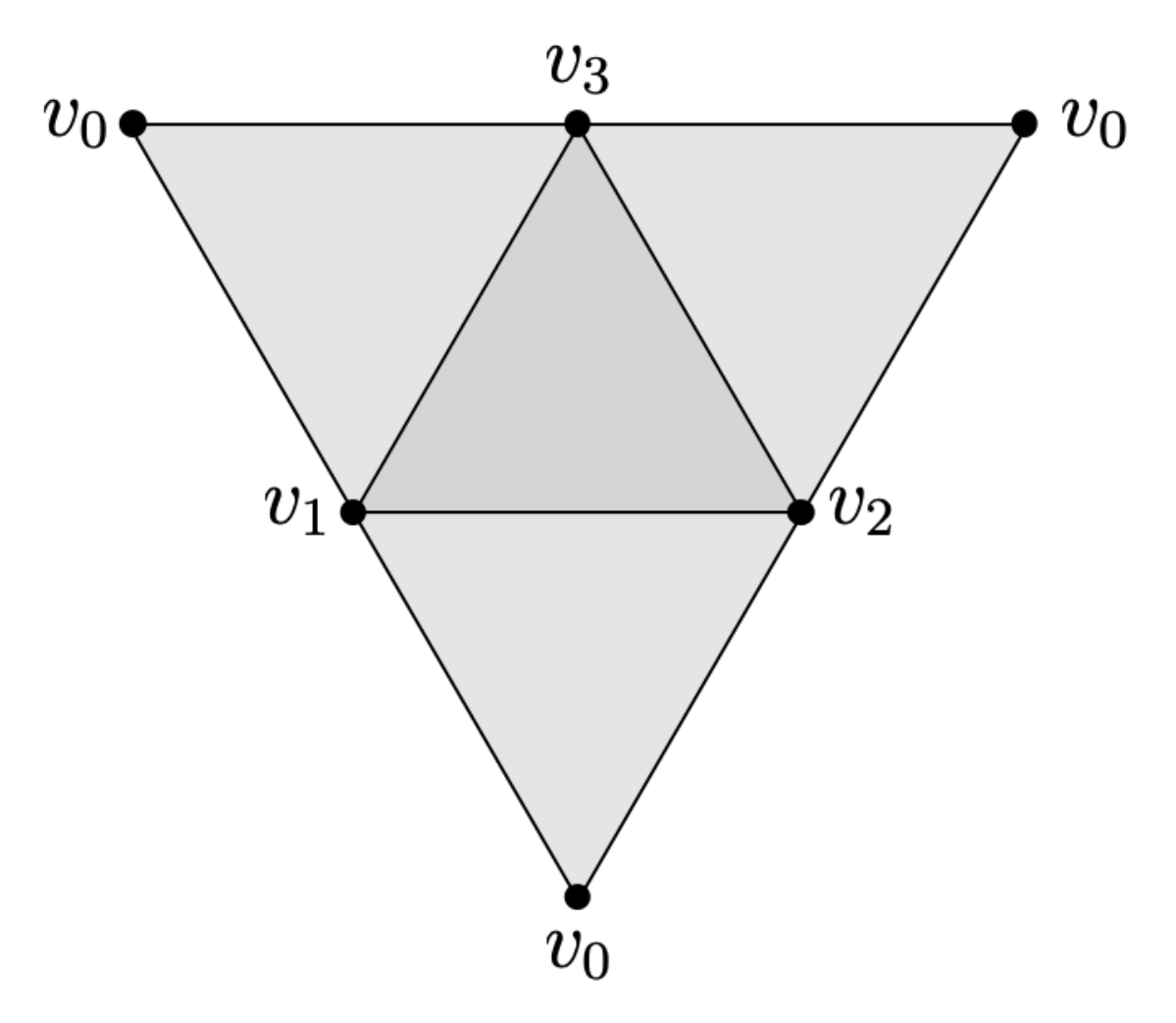}
		\caption{An unfolded tetrahedron}
		\label{fig:UnfoldedTetrahedron}
	\end{figure}
\end{center}

\begin{center}
	\begin{figure}[h]
		\centering
		\includegraphics[width=0.75\textwidth]{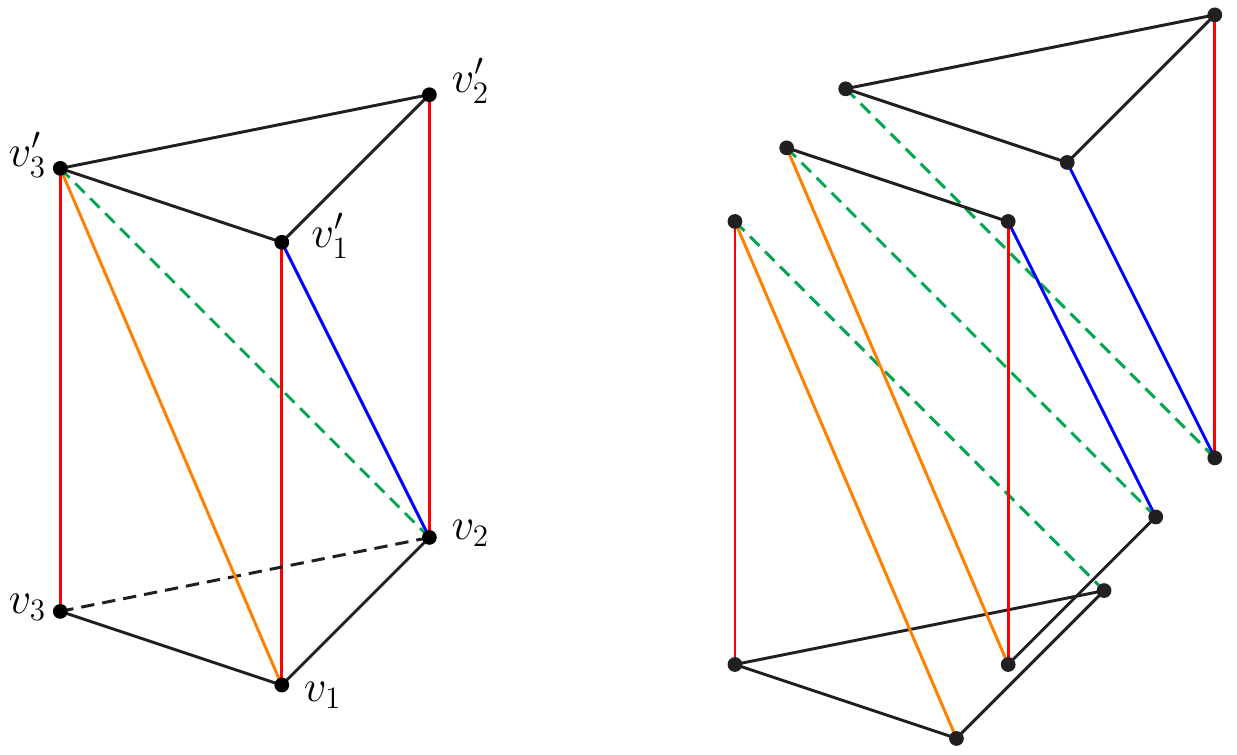}
		\caption{A simplicial triangulation of $I\times\Delta$.}
		\label{fig:FirstSteps}
	\end{figure}
\end{center}

\begin{center}
	\begin{figure}[h]
		\centering
		\includegraphics[width=0.7\textwidth]{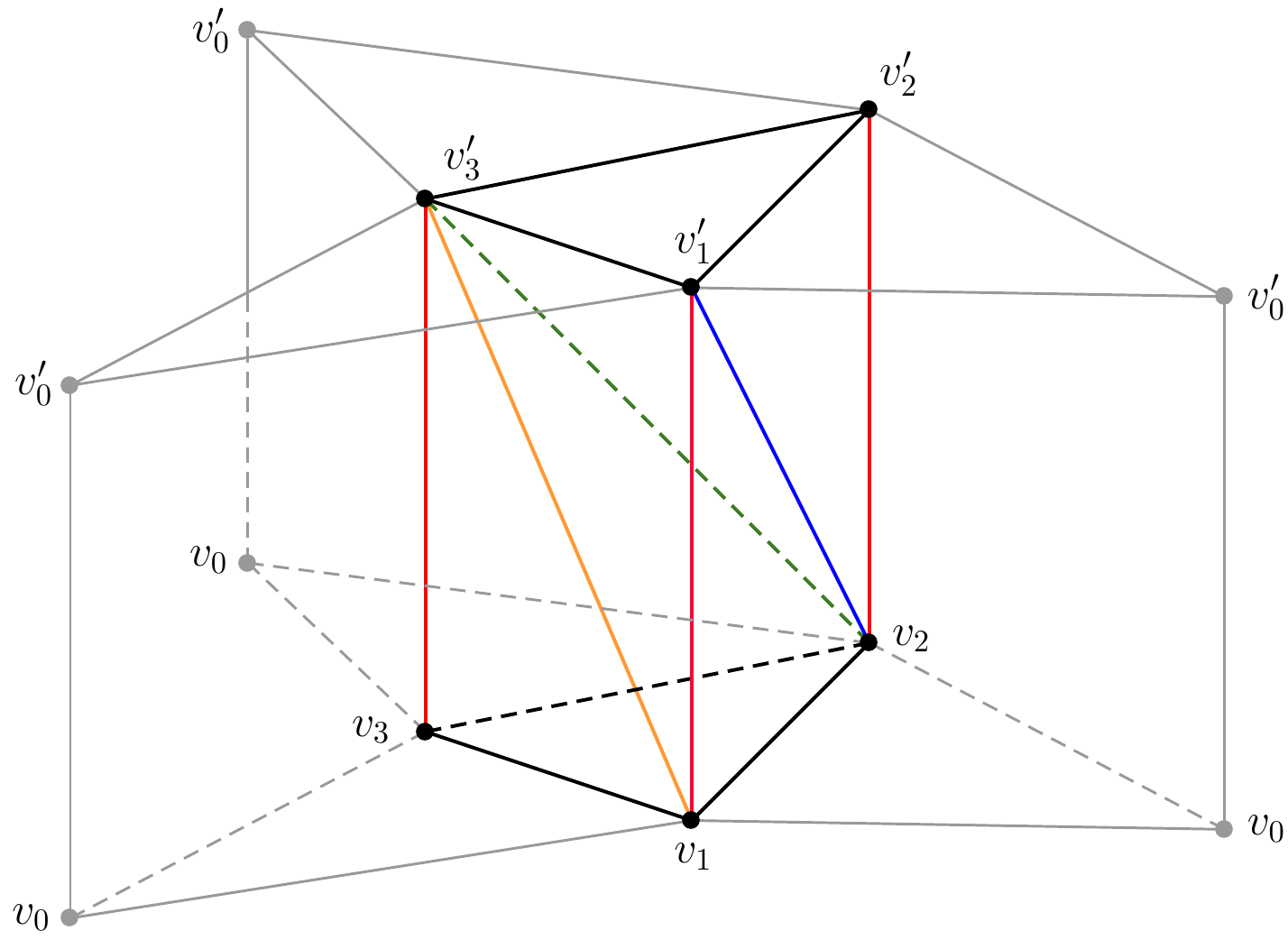}
		\caption{The next step toward a simplicial triangulation of $I\times\mathcal S^2$.}
		\label{fig:SecondStep}
	\end{figure}
\end{center}

\begin{center}
	\begin{figure}[h]
		\centering
		\includegraphics[width=0.75\textwidth]{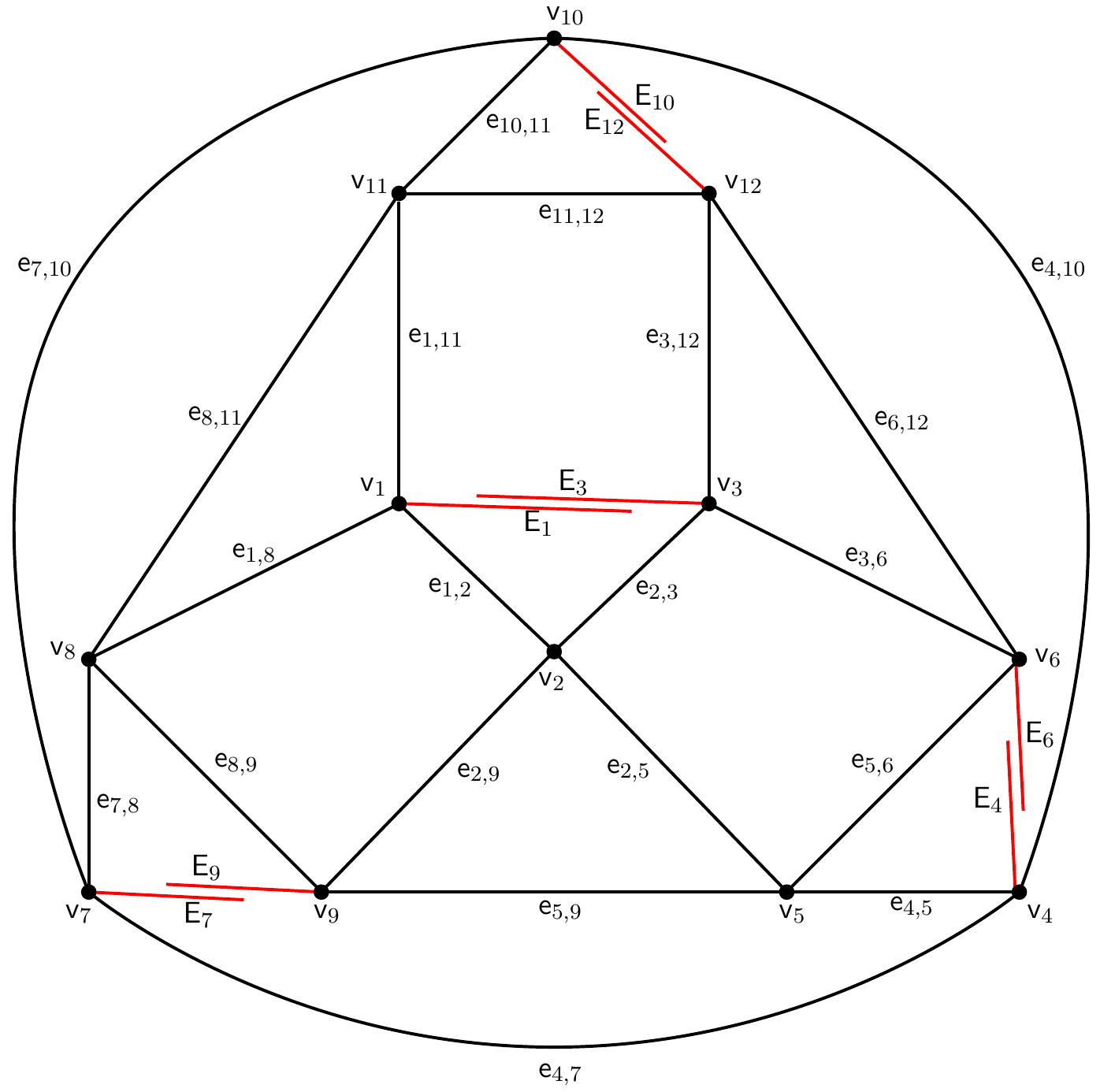}
		\caption{The $2$-complex of $I\times S^2$}
		\label{fig:2ComplexIS2}
	\end{figure}
\end{center}

\begin{center}
	\begin{figure}[h]
		\centering
		\includegraphics[width=0.75\textwidth]{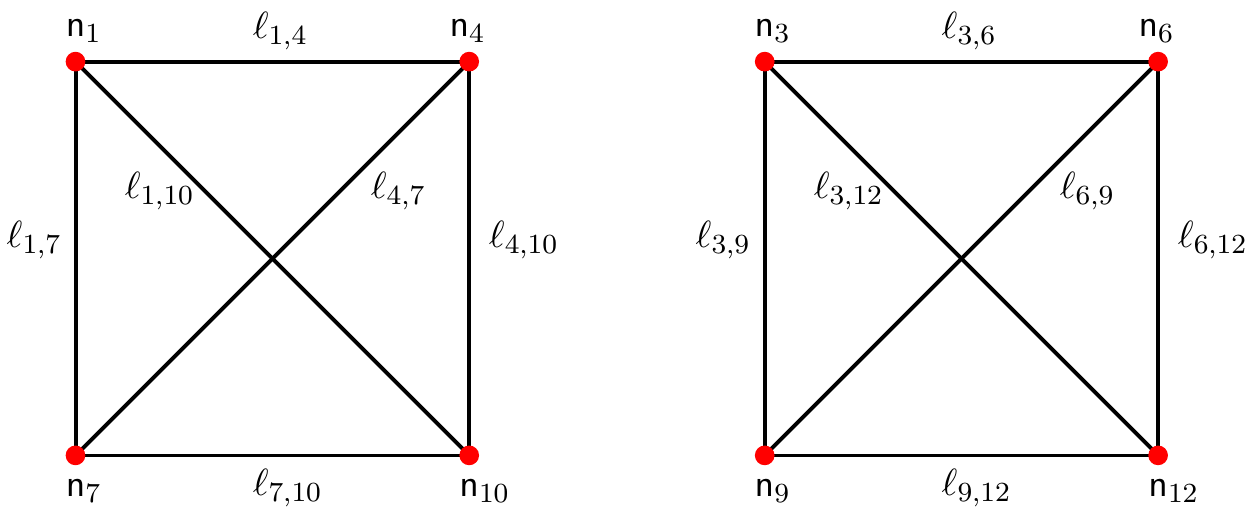}
		\caption{The boundary graph of $I\times S^2$}
		\label{fig:BndGraphIS2}
	\end{figure}
\end{center}

First of all, we note that $I\times\mathcal S^2$ can be pictured as a small sphere sitting inside a larger sphere and the space between them is filled out. So let us start with the triangulation of $\mathcal S^2$ before considering the effect of $I$. The simplest triangulation of a sphere consists of four triangles joined together such that they form the boundary of a tetrahedron. Our starting point will be the ``unfolded'' tetrahedron shown in Figure~\ref{fig:UnfoldedTetrahedron}. Furthermore we first consider an isolated triangle, say $\Delta:=[v_1,v_2,v_3]$, and we concentrate on constructing $I\times\Delta$. To that end, we produce a copy of $\Delta$ which we call $\Delta':=[v'_1,v'_2,v'_3]$ and we place it in a plane parallel to $\Delta$. 
By connecting the vertex $v_i$ to $v'_i$ by a straight line for $i=1,2,3$ (cf. Figure~\ref{fig:FirstSteps}), we obtain a triangulation of $I\times\Delta$, albeit a non-simplicial one. In order to make the triangulation simplicial we need to add further lines which cut the prism $I\times\Delta$ into simplicial building blocks, i.e. into tetrahedra. The result of the steps just described is shown in Figure~\ref{fig:FirstSteps}.
The fact that we could draw and visualize the triangulation was very helpful in cutting the prism into tetrahedra. A moment of thought will convince the reader that it is possible to draw lines on a prism such that the resulting triangulation becomes inconsistent because there will be intersecting tetrahedra.\\
Now that we understand how to construct the triangulation of $I\times\Delta$ we can simply repeat the process for the other three triangles in the triangulation of $S^2$. This is partially shown in Figure~\ref{fig:SecondStep}.
 Gluing these separate triangulations together according to the connectivity of the hollow tetrahedron we started with then guarantees that we obtain a consistent triangulation of $I\times S^2$\footnote{Notice that we drew $[v_0,v_1,v_2,v_3]$ and $[v'_0,v'_1,v'_2,v'_3]$ as if they had the same size. This is only for convenience. One of these tetrahedra, say $[v'_0,v'_1,v'_2,v'_3]$, should really be larger so that when all the $v'_3$'s and all the $v_3$'s are identified, the smaller tetrahedron is contained in the larger one.}. The end result is unfortunately too complicated to show here as one obtains, as had to be expected, a hollow tetrahedron (i.e. a triangulated sphere) located inside a larger tetrahedron (i.e. the second triangulated sphere) and a bunch of tetrahedra between this inner and outer boundary. This can be nicely visualized though using Mathematica.\\
Nevertheless, we can perform a consistency check and count the number of $p$-simplices. With the aid of Figure~\ref{fig:FirstSteps} and Figure~\ref{fig:SecondStep} we easily find that there are $8$ vertices, $22$ edges, $28$ faces and $12$ tetrahedra. As a consistency check for these numbers we can compute the Euler characteristic, for which we expect $\rchi(I\times S^2) = \rchi(I)\rchi(S^2) = 1\cdot 2$, and we find
\begin{equation}
	\rchi = 8 - 22 + 28 -12 = 2,
\end{equation}
 in agreement with our expectations. Finding the triangulation is however only the first step in spin foams. What we really need, as we will see later, is the $2$-complex of the triangulation. Constructing the $2$-complex for $I\times S^2$ by visual methods is almost impossible and so we have to resort to building lists. 
 By this we mean that we assign each of the $8$ vertices a label so that we can represent the $12$ tetrahedra by expressions of the form  $[v_0,\dots,v_3]$.  Applying the boundary operator to these $12$ lists gives us expressions of the form $\partial [v_0,v_1,v_2,v_3] = [v_1,v_2,v_3]\cup[v_0,v_2,v_3]\cup[v_0,v_1,v_3]\cup[v_0,v_1,v_2]$. Hence we find $4\cdot 12$ triangles. Some of these will appear twice because the tetrahedra are connected. We can then represent each tetrahedron by a vertex label $\ve$ and whenever a triangle appears twice, we know that we have to connect two vertices by an edge $\ed$. Let's say the triangle $[v_0,v_1,v_2]$ is part of the boundary of the tetrahedron $\ve_1$ and also of the boundary of the tetrahedron $\ve_2$, then there has to be a link $\ed_{12}$ connecting these tetrahedra (or rather their dual vertices $\ve_1$ and $\ve_2$). This is a tedious procedure which eventually results in the $2$-complex shown in Figure~\ref{fig:2ComplexIS2}.
Notice that the red edges are only connected to one vertex which means they are dual to boundary triangles. These are the only triangles which are not connected to any other triangles and therefore only appear once when we compute $\partial[v_0,\dots,v_3]$. Notice furthermore that there are eight red edges, exactly as we would expect since the boundary of $I\times\mathcal S^2$ consists of two disconnected spheres, as can easily be visualized or determined mathematically: $\partial(I\times S^2) = (\partial{I}\times S^2)\cup(I\times \partial S^2) = \{I_i,I_f\}\times S^2$.\\
The same procedure just explained for determining the $2$-complex of the triangulation of $I\times\mathcal S^2$ can also be applied to the collection of lists generated by $\partial[v_0,\dots,v_3]$. This even more tedious procedure gives us the $2$-complex of the boundary of $I\times\mathcal S^2$. The result consists of two disjoint graphs, as shown in Figure~\ref{fig:BndGraphIS2}. The red dots are nodes, which represent where the red edges intersect the boundary, and they are dual to triangles. The links $\ell$ are dual to the sides of the triangle and it can be seen that each graph represents the $2$-complex of the boundary of a single tetrahedron. 
This result is nicely consistent with $\partial(I\times S^2) = \{I_i,I_f\}\times S^2$. But what if we had to construct the triangulation of $I^2\times\mathcal S^2$ and determine its $2$-complex?\\
It is difficult to imagine how to achieve this if we have to rely on visual methods and the complexity involved in constructing the $2$-complex by inspecting large lists of vertices would soon become overwhelming. What we need is a simple algorithm which can be implemented on a computer and which generates triangulations and $2$-complexes.

\subsection{A Triangulation Algorithm for Manifolds of Topology $\mathbf{\mathcal M\simeq I\times\Sigma}$}\label{ssec:Algorithm}
The example of the previous section clearly shows that even in two or three dimensional spaces we cannot always rely on visual methods to construct a triangulation or its dual. Instead, we have to resort to large lists and tediously analyze its entries. Doing the same for a four-dimensional triangulation seems an almost impossible feat, especially because we cannot visualize what we are doing.\\
All this calls for an algorithmic approach which can be easily implemented on a computer. Assuming that we already know a homogeneous simplicial triangulation of the $d$ dimensional hypersurface $\Sigma$, the algorithm should accept a list of the form $[\sigma^{(d)}_1,\sigma^{(d)}_2,\dots,\sigma^{(d)}_N]$ as input, where each $\sigma$ represents a $d$-simplex, and create a similar list representing the $(d+1)$-dimensional triangulation of $I\times\Sigma$.\\
We can regard the triangulation of $\Sigma$ as being a $d$-dimensional hypersurface embedded in $\mathbb R^{d+1}$ and multiplying this hypersurface by a finite interval has the effect of dragging every vertex along a line in the $d+1$ direction to a new $d$-dimensional surface which is ``parallel'' to the first one. This operation doubles the number of vertices, which is in agreement with $\partial(I\times\Sigma) = (\{I_i,I_f\}\times\Sigma)\cup(I\times\partial\Sigma)$.\\
Now, let us focus on only one $d$-simplex of $\Sigma$,
\begin{equation}
	\sigma^{(d)} = [v_0,v_1,\dots,v_d].
\end{equation}
To embed this simplex into $\mathbb R^{d+1}$ we define $V_k:=(v_k,I_i)^\transpose$, where $I_i$ is the minimum of the interval $I=[I_i,I_f],$ and for the new set of vertices which lie in the second hypersurface we can write $U_k:=(v_k,I_f)^\transpose$, with $I_f$ being the maximum of $I$. We then proceed by defining the following $(d+1)$-simplices:
\begin{align}\label{eq:List}
	\sigma^{(d+1)}_0 &= [V_0,V_1,\dots,V_d,U_0]\notag\\
	\sigma^{(d+1)}_1 &= [V_1,V_2,\dots,V_d,U_0,U_1]\notag\\
	\sigma^{(d+1)}_2 &= [V_2,V_3,\dots,V_d,U_0,U_1,U_2]\notag\\
					 &\textcolor{white}{=}\vdots\notag\\
	\sigma^{(d+1)}_{d} &= [V_d,U_0,\dots,U_{d-1},U_d]
\end{align}
This collection of simplices provides us with a consistent triangulation of $I\times\sigma^{(d)}$. We show this as follows: The idea behind the construction~\eqref{eq:List} is to first extend $\sigma^{(d)}$ to a $(d+1)$-simplex $\sigma^{(d+1)}_0$ by adding the vertex $U_0$. Then, add a new $(d+1)$-simplex by choosing one of the newly created faces in $\sigma^{(d+1)}_0$ as ``base'' for $\sigma^{(d+1)}_1$. This ensures that $\sigma^{(d+1)}_0\cap\sigma^{(d+1)}_1$ is a $d$-simplex and that there are no other intersections. This process is then iterated and every newly created $(d+1)$-simplex can only have simplicial intersections with some of its neighbors, but there cannot be any non-simplicial intersections.\\
In order to use one of the newly created faces of $\sigma^{(d+1)}_0$ as ``base'' to create $\sigma^{(d+1)}_1$, we clearly need the vertex $U_0$ to be contained in $\sigma^{(d+1)}_1$. Moreover, we also need an other vertex, one which is not contained in $\sigma^{(d+1)}_0$. We choose $U_1$ and in turn we remove one of the $V$ vertices. We choose $V_1$. Hence, we have $\sigma^{(d+1)}_1 = [V_1,V_2,\dots,V_d,U_0,U_1]$ which has the $d$-simplex $[V_1,V_2,\dots,U_0]$ in common with $\sigma^{(d+1)}_0$.\\
This process of adding and removing vertices is iterated as shown in the list~\eqref{eq:List}, until we cannot add or remove any more vertices. This leaves us with $d+1$ simplices of dimension $d+1$ which, by construction, are joined together along $d$-simplices. There are not other intersections between these simplices.\\
To be more precise, the construction just described ensures that
\begin{itemize}
	\item[a)] $\sigma^{(d+1)}_0$ and $\sigma^{(d+1)}_d$ have each exactly one intersection: $\sigma^{(d+1)}_0$ intersects only with $\sigma^{(d+1)}_1$ while $\sigma^{(d+1)}_d$ intersects only with $\sigma^{(d+1)}_{d-1}$. The intersection occurs along one of the $d$-dimensional subsimplices. All the other $d$-dimensional subsimplices of $\sigma^{(d+1)}_0$ and $\sigma^{(d+1)}_d$ lie on the boundary of $I\times\sigma^{(d)}$.
	\item[b)] $\sigma^{(d+1)}_i$ for $0 < i < d$ intersects with two $(d+1)$-simplices. In fact, $\sigma^{(d+1)}_i$ intersects with $\sigma^{(d+1)}_{i-1}$ and $\sigma^{(d+1)}_{i+1}$ in a $d$-simplex.
	\item[c)] All $d$-subsimplices which are not involved in an intersection lie on the boundary of $I\times\sigma^{(d)}$.
\end{itemize}
This can all be verified by carefully looking at the boundary of $\sigma^{(d+1)}_i$,
\begin{align}
	\partial\sigma^{(d+1)}_i =& [V_{i+1},\dots,V_d,U_0,\dots, U_i]\cup[V_i,\dots,V_d,U_0,\dots,U_{i-1}]\cup\notag\\
		&\bigcup_{j=i+1}^d[V_i,\dots,\arc{V}_j,\dots,U_0,\dots, U_i]\bigcup_{k=0}^{i-1}[V_i,\dots,V_d,U_0,\dots,\arc{U}_k, \dots, U_i],
\end{align}
for $0<i<d$. From this general expression for the boundary, it is clear that the  two terms in the first line appear also for the boundary expansion for $\sigma^{(d+1)}_{i-1}$ and $\sigma^{(d+1)}_{i+1}$. Thus, claim b) is verified. Moreover, one sees that the second line only contains $d$-subsimplices of $\sigma^{(d+1)}$ which lie on the boundary of $I\times\sigma^{(d)}$. A moment of reflection will convince the reader that this is the case because both, $V_i$ and $U_i$ appear in the same list. Or put differently: The boundary of $I\times\sigma^{(d)}$ always needs to contain at least one $1$-simplex of the form $[V_i,U_i]$.\\
The two terms in the first line, on the other hand, cannot lie on the boundary of $I\times\sigma^{(d)}$ because when $U_i$ appears in the list, $V_i$ is missing and vice verse. Hence, these terms must belong to the interior of $I\times\sigma^{(d)}$.\\
What we have shown so far is that most $d$-simplices lie on the boundary of $I\times\sigma^{(d)}$ and that there are simplicial intersections in the interior. Non-simplicial intersections are not possible. It remains to show that the sum of the volumes of the $d+1$ simplices defined in~\eqref{eq:List} is equal to the volume of $I\times\sigma^{(d)}$.\\
By Fubini's theorem we easily find what we would intuitively expect:
\begin{align}
	\text{Vol}(I\times\sigma^{(d)}) &= \int_{I\times\sigma^{(d)}}\dd^d x\, \dd y = \int_{\sigma^{(d)}}\dd^d x\int_I\dd y\notag\\
	&= \frac{(I_f-I_i)}{d!}\left|\det(v_1-v_0,v_2-v_0,\dots,v_d-v_0)\right|\notag\\
	&= (I_f-I_i)\text{Vol}(\sigma^{(d)}),
\end{align}
where we have used theorem~\ref{thm:Volume}. To compute the volume of $\sigma^{(d+1)}_k$, for $0\leq k\leq d$, we notice that the matrix\footnote{Every entry should be read as a column vector and so the matrix is of dimension $(d+1)\times(d+1)$.}
\begin{align}
	M^{d+1}_k &:= (V_{k+1}-V_k,V_{k+2}-V_k,\dots, U_0-V_k,\dots, U_k-V_k)\notag\\
		&\textcolor{white}{:}= \begin{pmatrix}
		\begin{array}{cccccc}
			v_{k+1}-v_{k} & v_{k+2}-v_k & \cdots & v_0-v_k & \cdots & 0\\
			0 & 0 & \cdots & I_f-I_i & \cdots & I_f - I_i
		\end{array}
		\end{pmatrix}
\end{align}
always contains a $d\times d$ sub-matrix of edge vectors constructed from the vertices $(v_0,v_1,\dots,v_d)$ alone and that the last column contains only zeros except in the last row. Using the Laplace expansion along the last column we then easily find that the determinant of $M^{d+1}_k$ is given by the determinant of the $d\times d$ sub-matrix times the length of the interval, $(I_f-I_i)$. Therefore
\begin{align}
	\left|\det M^{d+1}_k\right| &= (I_f-I_i)\left|\det(v_1-v_0,v_2-v_0,\dots,v_d-v_0)\right|\notag\\
	&=d!\,(I_f-I_i)\,\text{Vol}(\sigma^{(d)}).
\end{align}
From this result we straightforwardly deduce
\begin{align}
	\sum_{k=0}^{d} \text{Vol}(\sigma^{(d+1)}_k) &= \frac{1}{(d+1)!}\sum_{k=0}^{d} \left|\det M^{d+1}_k\right| \notag\\
	&= \frac{d!(d+1)}{(d+1)!}\,(I_f-I_i)\,\text{Vol}(\sigma^{(p)}) = \text{Vol}(I\times\sigma^{(p)}),
\end{align}
which proves that the sum of volumes of the $(d+1)$-simplices defines in~\eqref{eq:List} is equal to the volume of $I\times\sigma^{(d)}$. This also concludes the proof that~\eqref{eq:List} provides us with a consistent simplicial triangulation of $I\times\sigma^{(d)}$.\\
Let us observe that we can also perform an easy consistency check by computing the Euler characteristic of $I\times\sigma^{(d)}$ from~\eqref{eq:List}, for which we should obtain $\rchi(I\times\sigma^{(d)}) = \rchi(I)\,\rchi(\sigma^{(d)}) = 1$. Since the characteristic of $I\times\sigma^{(d)}$ is really the characteristics of $d+1$ simplices of dimension $d+1$ glued together in a specific way, it must be of the form $\rchi(I\times\sigma^{(d)}) = (d+1)\rchi(\sigma^{(d+1)}) - (\textit{correction for gluing})$. From the algorithm it is evident that every $(d+1)$-simplex $\sigma^{(d+1)}_k$ is glued to its neighbor $\sigma^{(d+1)}_{k+1}$ along exactly one $d$-subsimplex. Thus, at every gluing we are over-counting vertices, edges, faces etc. of an amount corresponding to one $d$-subsimplex. Since in the construction~\eqref{eq:List} there are precisely $d$ ``gluings'', the correction factor has to be $d\,\rchi(\sigma^{(d)})$. This yields $\rchi(I\times\sigma^{(d)})=(d+1)\rchi(\sigma^{(d+1)})-d\,\rchi(\sigma^{(d)})$. Using lemma~\ref{lem:Number}, we can compute the characteristic of $\sigma^{(d)}$ as
\begin{align}
	\rchi(\sigma^{(d)}) &= \sum_{n=0}^d(-1)^n\,N^{d}_n = \sum_{n=0}^d(-1)^n\binom{d+1}{n+1}\notag\\
	&\overset{k=n+1}{=} \sum_{k=1}^{d+1}(-1)^{k+1}\binom{d+1}{k} = 1+(-1)\sum_{k=0}^{d+1} (-1)^{k}\binom{d+1}{k}\notag\\
	&= (-1)(1-1)^{d+1} + 1 = 1.
\end{align}
In the third line we completed the sum to a binomial sum so that we were able to apply the binomial theorem to deduce that the sum is equal to $(1-1)^{d+1}=0$. To compensate for the completion we needed to add a $1$ for the $k=0$ term and hence we find $\rchi(\sigma^{(d)})=1$ for all $d$\footnote{Of course we could simply have used the fact that every $d$-simplex is homeomorphic to a $d$-ball $B_d$ and that $\rchi(B_d)=1$ for all $d$.}. Hence, the expected result,
\begin{equation}
	\rchi(I\times\sigma^{(d)}) = (d+1)\rchi(\sigma^{(d+1)})-d\,\rchi(\sigma^{(d)}) = 1,
\end{equation}
follows and this completes the consistency check.\\
Finally, the triangulation of $I\times\Sigma$ can be constructed using~\eqref{eq:List} applied to every single $d$-simplex in $\Sigma$. The connectivity of the $\sigma^{(d)}$ contained in $\Sigma$ is inherited by the $\sigma^{(d+1)}$ simplices in $I\times\Sigma$, which allows to construct the full triangulation of $I\times\Sigma$.\\
This algorithm and the procedure just described have been implemented on a computer by the author in order to construct the triangulation of $I^2\times\mathcal{S}^2$, starting from the triangulation of $I\times\mathcal{S}^2$ discussed in the previous subsection. The results are too complex to be reported here in detail. However, we can say that the following number of $p$-simplices was found by the computer program:
\begin{center}
	\begin{tabular}{lr}
	\rowcolor{gray}
		\textcolor{white}{$p$-Simplex} & \textcolor{white}{Count}\\
	\rowcolor{myGray}
		$4$-Simplices & $48$\\
	\rowcolor{myGray}	
		$3$-Simplices & $144$\\
	\rowcolor{myGray}	
		$2$-Simplices & $156$\\	
	\rowcolor{myGray}	
		$1$-Simplices & $74$\\
	\rowcolor{myGray}	
		$0$-Simplices & $16$\\
	\end{tabular}
\end{center}
The Euler characteristic computed from these numbers yields precisely the expected result:
\begin{equation}
	\rchi(I^2\times\mathcal{S}^2) = 16-74+156-144+48 = 2.
\end{equation}

\subsection{On Euclidean and Lorentzian Angles}
\label{ssec:LorentzianAngles}
\begin{center}
	\begin{figure}[htb]
		\begin{minipage}[c]{.45\textwidth}
			\includegraphics[width=\textwidth]{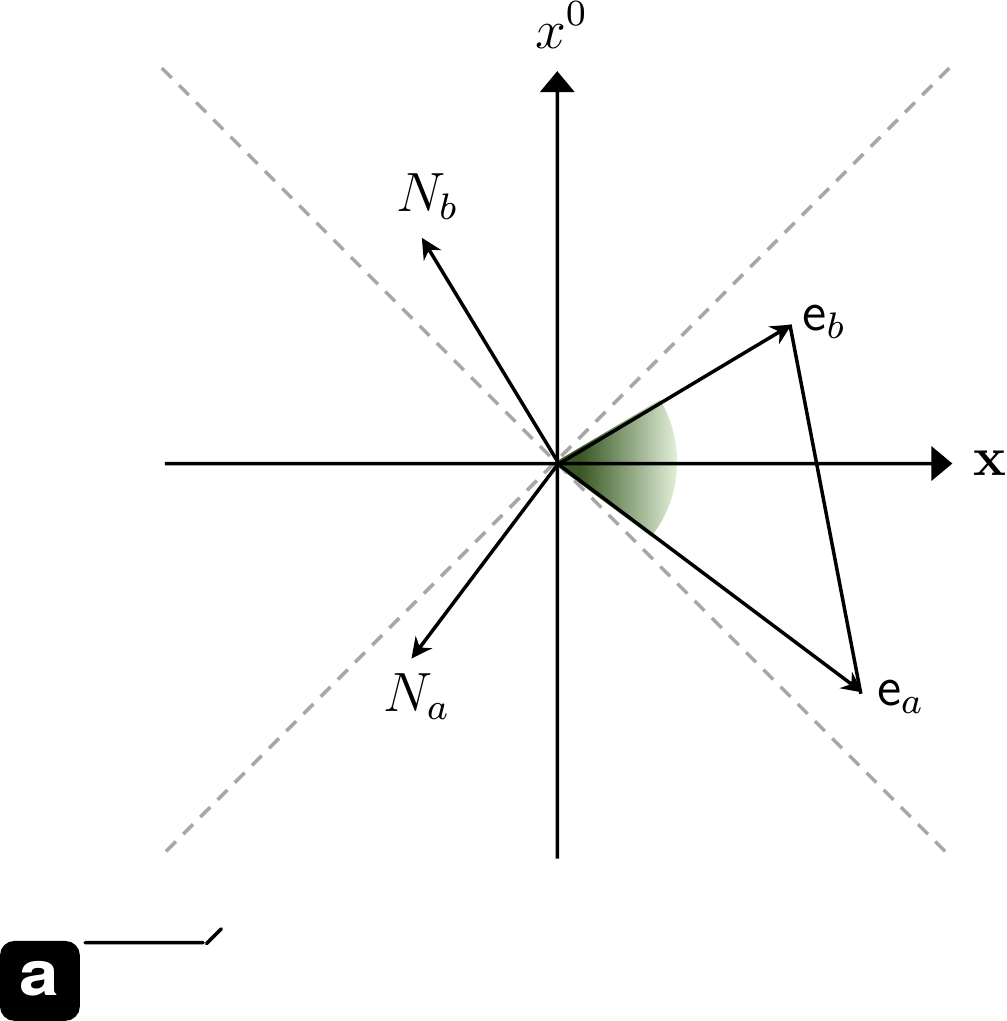}
		\end{minipage}
		\hfill
		\begin{minipage}[c]{.45\textwidth}
			\includegraphics[width=\textwidth]{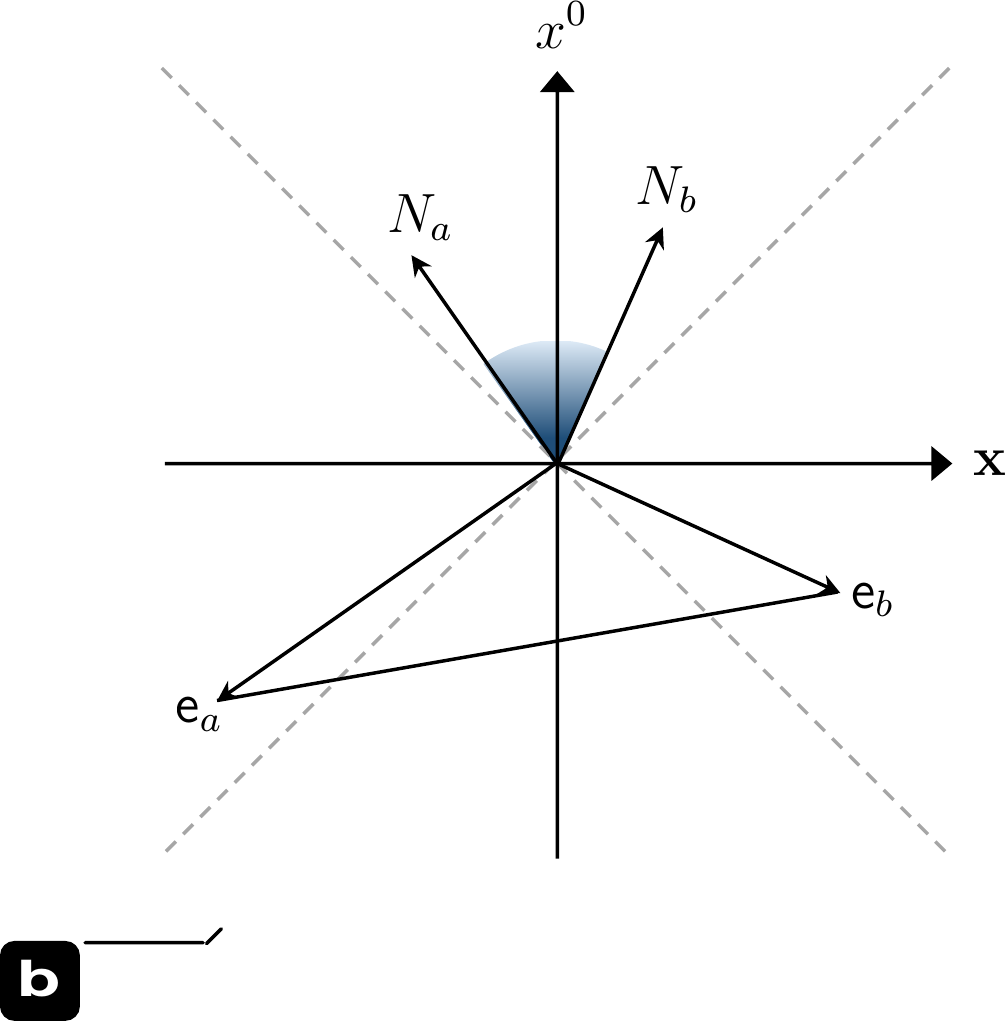}		
		\end{minipage}
		\caption{A thin wedge (a) and a thick wedge (b)}	
		\label{fig:ThinThickWedges}
	\end{figure}
\end{center}

\begin{center}
	\begin{figure}[htb]
		\centering
		\includegraphics[width=0.6\textwidth]{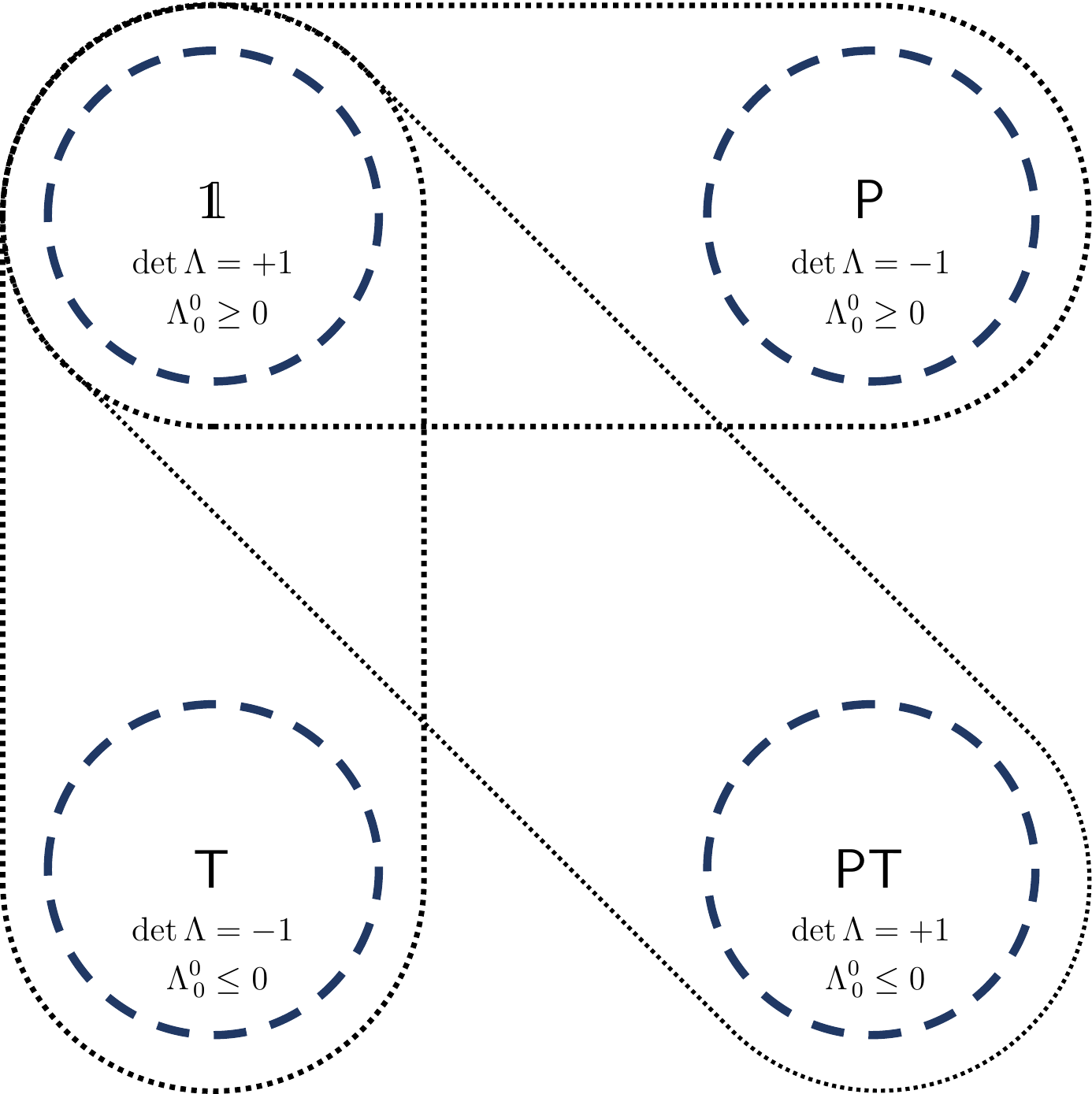}
		\caption{The four components of the Lorentz group}
		\label{fig:LOrentzGroup}
	\end{figure}
\end{center}

\begin{center}
	\begin{figure}[htb]
		\begin{minipage}[c]{.45\textwidth}
			\includegraphics[width=\textwidth]{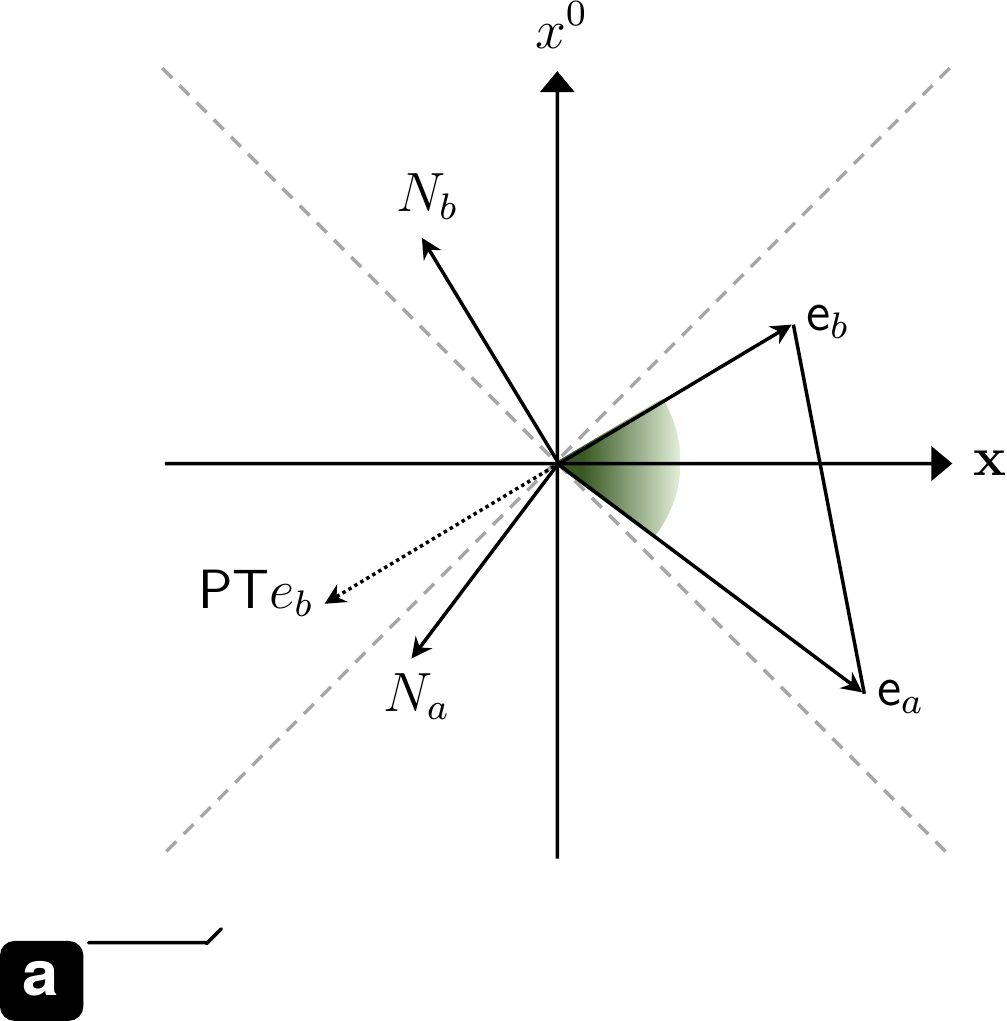}
		\end{minipage}
		\hfill
		\begin{minipage}[c]{.45\textwidth}
			\includegraphics[width=\textwidth]{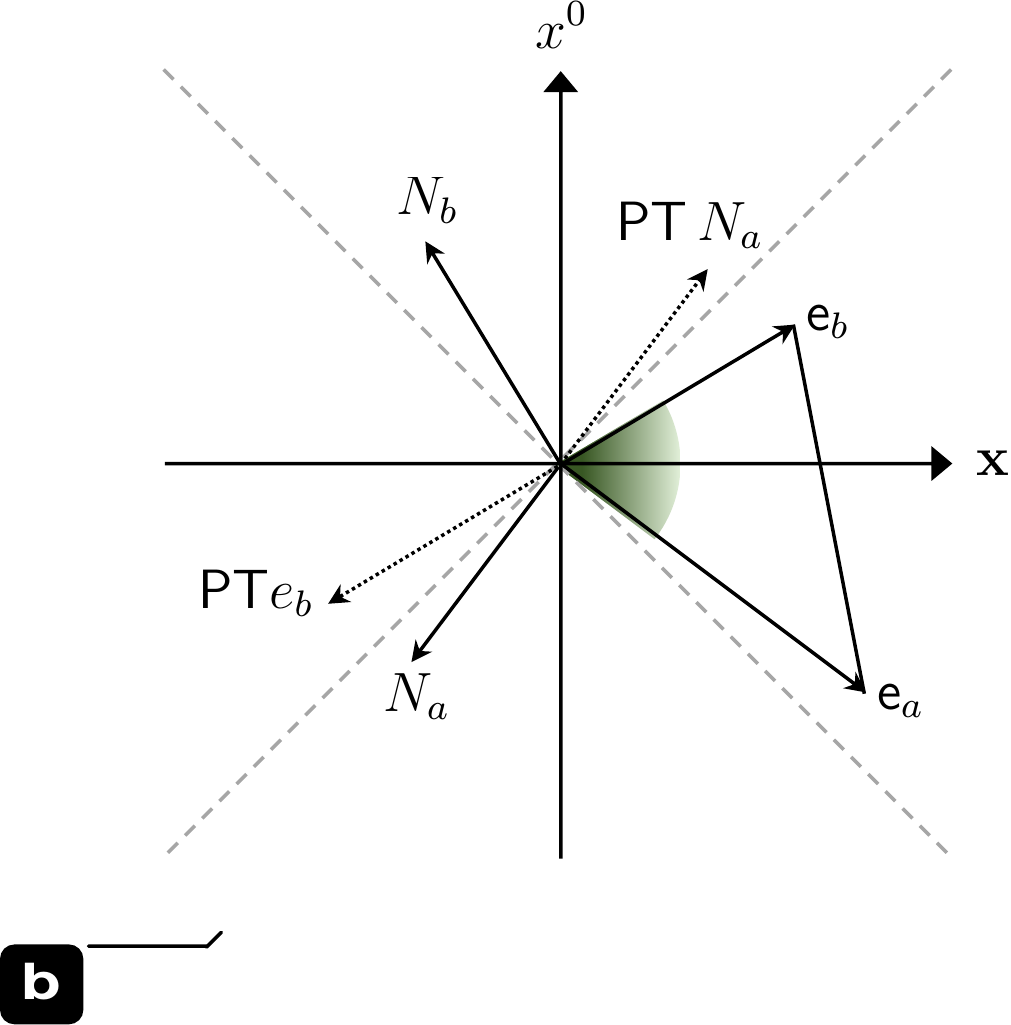}
		\end{minipage}
		\caption{First step in the construction: $\textsf{PT}$ transformation of $e_b$ (a); Second step: $\textsf{PT}$ transformation of $N_a$ (b)}	
		\label{fig:Construction}
	\end{figure}
\end{center}

This subsection, while still important for simplicial geometry, is mainly intended for later reference. We give a brief definition of angles in Euclidean and Lorentzian spaces of any dimension. For the Lorentzian case, we generalize a definition given by Barrett and Foxon~\cite{Barrett:1993} for $1+1$ dimensional  Minkowski space and we clarify certain aspects of the construction which are important for applications in spin foams and which are a common source of confusion.\\
Let us begin with the definition of angle in $n$-dimensional Euclidean space. 
\begin{mydef}{ Euclidean angle}{EuclideanAngle}
\label{EuclideanAngle}
	Let $u,v\in\mathbb R^n$ be two non-zero vectors and $n>1$. Then there is an $SO(n)$ element which takes $\frac{u}{\|u\|}$ into $\frac{v}{\|v\|}$. The Euclidean angle is defined as the rotation parameter $\theta\in[0,\pi]$ of this group element.  
\end{mydef}
At first this definition seems odd since $SO(n)$ is a Lie group with $\frac{n(n-1)}{2}$ parameters. To reduce it to only one parameter $\theta$, it seems we either have to work in $n=2$ dimensions or use a clever choice of vector space basis.\\
It turns out that it is a little bit of both, but the definition makes perfectly sense in $n$ dimensions and it is completely basis independent. Infact, the above definition implies that $\theta$ is given by
\begin{equation}\label{eq:EuclideanFormula}
	\langle u,v\rangle = \|u\| \, \|v\|\,\cos\theta\quad\forall u,v\in\mathbb R^n.
\end{equation} 
To see this, we first assume that $u$ and $v$ are linearly independent. Notice then that $R$ only needs to act in the plane $P:=\text{span}\{u,v\}$ spanned by $u$ and $v$. In terms of this plane, the original vector space can be decomposed as $\mathbb R^n \simeq P\oplus P^\bot$, where $P^\bot$ denotes the $(n-2)$-dimensional orthogonal complement to $P$. On $P\oplus P^\bot$, we can decompose the rotation matrix $R$ as $R=S\oplus\id_{(n-2)}$ with $S\in SO(2)$. In other words: $R$ acts on the two-dimensional space $P$ and leaves the orthogonal complement $P^\bot$ invariant.\\
With these definitions and decompositions it is then easy to derive an explicit expression for the angle between $u$ and $v$. To that end, we start from the condition
\begin{equation}
	v = \frac{\|v\|}{\|u\|} R\,u,
\end{equation}
stated in definition~\hyperref[EuclideanAngle]{3.2.5}. By multiplying this equation from both sides with $u$, using the canonical scalar product on $\mathbb R^n$, we obtain
\begin{align}
	\langle u, v\rangle &= \frac{\|v\|}{\|u\|} \langle u, R\,u\rangle\notag\\
				  &= \frac{\|v\|}{\|u\|} \langle u, S\,u\rangle\notag\\
				  &= \frac12 \|u\|\, \|v\|\, \Tr[S]\notag\\
				  &= \|u\|\, \|v\|\, \cos\theta.
\end{align}
From the first to the second line we used $R=S\oplus\id_{n-2}$. The second line can then be seen as a projection on the $uu$-component of $S$, which in turn can be expressed as $\langle u, S \, u\rangle = \frac{1}{2}\Tr[S]\,\|u\|^2$. This follows from the fact that $S$ is two-dimensional and satisfies $S^\transpose = S^{-1}$ as well as $\det S=1$, since it is in $SO(2)$. This immediately implies that the diagonal elements of $S$ have to be equal and hence $\Tr[S] = \frac{2}{\|u\|^2} \langle u, S\, u\rangle$\footnote{This can easily be checked by working in the basis $e_1$, $e_2$ defined by $u=\|u\|\, e_1$ and $v=\|v\| e_2$. The result is basis-independent, though.}. Given that the trace is basis-independent, we can evaluate it in the basis for which
\begin{equation}
	S = \begin{pmatrix}
		\cos\theta & -\sin\theta\\
		\sin\theta & \cos\theta
	\end{pmatrix}.
\end{equation}
This finally gives us the last line which is precisely what we are used to from two- and three-dimensional Euclidean geometry. If $u$ and $v$ do not span a plane, they are either parallel or anti-parallel. The above construction can be adapted in an obvious way from which one obtains $R=\id_n \Rightarrow \theta = 0$ when $u$ and $v$ are parallel and $R=-\id_n\Rightarrow \theta = \pi$ when $u$ and $v$ are anti-parallel. Hence, we confirm the well-known Euclidean formula~\eqref{eq:EuclideanFormula} for any dimension $n>1$.\\
The advantage of defining Euclidean angles through $SO(n)$ rotations rather than through formula~\eqref{eq:EuclideanFormula} is that this procedure can easily be generalized to Lorentzian angles.\\
To that end, let $\mathcal M$ stand for the manifold $(\mathbb R^{1+n}, \eta)$, where $\eta$ is a metric of signature $(1,n)$, and call it Minkowski space of dimension $d=1+n$. Furthermore, let $\mathcal I^+_p$ ($\mathcal I^-_p$) denote the future (past) light cone of the event $p$ in Minkowski space. We can then define two types of angles for timelike and spacelike vectors
\begin{mydef}{ Interior and Exterior Angles}{}
		Let $e_a,e_b\in\mathcal M$ be two spacelike vectors, normalized to $+1$, and let $N_a, N_b$ be two timelike vectors which satisfy $\eta(e_a, N_a)=0$ and $\eta(e_b, N_b)=0$ and which are normalized to $-1$.
		\begin{itemize}
			\item[1)] \textbf{Interior angle:} $\text{span}\{e_a,e_b\}$ does not contain any of the past (future) light cone of the origin. This is called a \textit{thin wedge}. Then there is an element of the proper orthochronous Lorentz group $SO^+(1,n)$ which takes $e_a$ into $e_b$. The \textit{interior angle} between $e_a$ and $e_b$ is defined as the (positive) boost parameter $\xi_\textsf{int}\in\mathbb R^+$ of this $SO^+(1,n)$ element.
			\item[2)] \textbf{Exterior angle:} $\text{span}\{e_a,e_b\}$ contains the past (future) light cone of the origin. This is a \textit{thick wedge}. Then there is an element of $SO^+(1,n)$ which takes the outward pointing normal $N_a$ to the outward pointing normal $N_b$. The \textit{exterior angle} between $N_a$ and $N_b$ is defined as minus the boost parameter of this $SO^+(1,n)$ element, i.e. $\xi_\textsf{ext}\in\mathbb R^-$.
		\end{itemize} 
		\label{def:LorentzianAngles}
\end{mydef}
Figure~\ref{fig:ThinThickWedges} a shows a thin wedge with the corresponding interior angle. Using the above definition, we can derive an explicit expression for computing $\xi_\textsf{int}$. To that end, we translate the above definition into
\begin{equation}\label{eq:LorentzianEq}
	e_b = \Lambda\, e_a,
\end{equation}
where $\Lambda\in SO^+(1,n)$ and the norm is taken with respect to the Minkowski metric~$\eta$. Just as in the Euclidean case, we notice that the boost which takes one spacelike vector into the other only needs to act in the plane $P:=\text{span}\{e_a,e_b\}$ while leaving the orthogonal complement invariant. That is, we can decompose $\Lambda$ as $\Lambda = K\oplus\id_{n-1}$, where $K\in SO^+(1,1)$ is a $2\times 2$ boost matrix.\\
If we now take the scalar product of equation~\eqref{eq:LorentzianEq} with $e_a$ we obtain
\begin{align}\label{eq:CompIntAngle}
	\eta(e_a,e_b) &= \eta(e_a,\Lambda\,e_a)\notag\\
				  &= \eta(e_a,K\,e_a)\notag\\
				  &= \frac12 \Tr[K]\notag\\
				  &= \cosh \xi_\textsf{int}.
\end{align}
$\phantom{.}$\\
From the first to the second line we used the decomposition $\Lambda = K\oplus\id_{n-1}$. From the second to the third line we used the fact that this decomposition is defined with respect to the basis $e_a$, $e_b$ of $P$ and hence we can view $\eta(e_a,K\,e_a)$ as a projection on one of the diagonal components of $K$. Since $K\in SO^+(1,1)$, it follows that it is a $2\times 2$ matrix which satisfies $\det K=1$ as well as $K^{-1}=\eta K^\transpose \eta$. This in turn implies that its diagonal elements are equal and therefore $\eta(e_a,K\,e_a) = \frac12 \Tr[K]$. Given that the trace is basis-independent, we can evaluate it in the basis in which 
\begin{equation}
	K = \begin{pmatrix}
		\cosh\xi & \sinh\xi\\
		\sinh\xi & \cosh\xi
	\end{pmatrix}.
\end{equation}
This finally implies the last line in~\eqref{eq:CompIntAngle} and we find that the interior angle can be explicitly written as
\begin{equation}\label{eq:XiInt}
	\xi_{\textsf{int}} = \text{arcosh}\left(\eta(e_a,e_b)\right).
\end{equation}
Notice that this holds for all spacelike vectors $e_a$, $e_b$ which live in the same wedge of Minkowski space. When the two vectors are parallel, then $\xi_\textsf{int} = 0$. They cannot be anti-parallel because that would define a thick wedge.\\
Through similar reasoning we can arrive at an explicit expression for exterior angles associated with thick wedges (cf. Figure~\ref{fig:ThinThickWedges} b):
\begin{equation}
	\xi_{\textsf{ext}}=-\text{arcosh}\left(-\eta(N_a, N_b)\right),
\end{equation}
where the minus sign is conventional since $\xi_\textsf{ext}$ is defined to be a negative number.\\
It is worth noticing that there is an unexpected relation between the interior and exterior angle. In fact, we can prove that $\xi_{\textsf{int}}=-\xi_{\textsf{ext}}$. To see this, we start from the orthogonality relation $\eta(e_a, N_a) = 0$ which we ``boost'' with $\Lambda(\xi_\textsf{int})$, i.e. 
\begin{equation}
	\eta(\Lambda(\xi_\textsf{int})\, e_a,\Lambda(\xi_\textsf{int})\, N_a) = \eta(e_b, \Lambda(\xi_\textsf{int})\, N_a) = 0.
\end{equation}
Here we used equation~\eqref{eq:LorentzianEq} and the fact that Lorentz transformations leave the Minkowski scalar product invariant. Comparing the above boosted equation with the second orthogonality relation,
\begin{equation}\label{eq:BoostedEquationForRelation}
	\eta(e_b, N_b) = \eta(e_b, \Lambda(-\xi_\textsf{ext})\, N_a) = 0,
\end{equation}
we conclude
\begin{equation}
	\xi_\textsf{int} = -\xi_\textsf{ext}.
\end{equation}
This relation allows us to compute interior angles using solely the normal vectors $N_a$ and $N_b$ and without ever referring to $e_a$ and $e_b$. There are two ideas which will lead us toward the desired result:
\begin{itemize}
	\item[1)] Intuitively, if we wanted to use directly $N_a$ and $N_b$, which live in two different light cones,  to define an internal angle pertaining to a thin wedge, we would have to find a Lorentz transformation which consists of two separate pieces. The first piece allows $N_a$ to cross two null hypersurfaces such that it comes to lie in the same light cone as $N_b$. The second piece then boosts $N_a$ such that it coincides with $N_b$.
	\item[2)] The relation $\xi_\textsf{int} = -\xi_\textsf{ext}$ can be interpreted as telling us that there exists an auxiliary thick wedge with auxiliary normals for which we can compute an external angle. This angle is then related to an actual internal angle.
\end{itemize}
Notice that the Lorentz transformation described above cannot be an element of the proper orthochronous group $SO^+(1,n)$. But recall that the Lorentz group possesses four topologically distinct components (cf. Figure~\ref{fig:LOrentzGroup}). We can use these additional transformations shown in Figure~\ref{fig:LOrentzGroup} to realize the Lorentz transformation described above and this in turn gives us a guide line for the construction of the auxiliary structures.\\
Clearly, we cannot use parity reversal ($\textsf{P}$) since its sole effect is to mirror $N_a$ along the $x^0$ axis. Time reversal ($\textsf{T}$) is also not an option since it would project $N_a$ into the same light cone as $N_b$, but reverse the spacetime orientation. Thus, there is no proper orthochronous Lorentz boost which takes $\textsf{T} N_a$ to $N_b$. What remains is the concatenation $\textsf{PT} = \textsf{TP} = -\id_{n+1}$ which leads us into the subgroup with $\det\Lambda = +1$ (i.e. the spacetime orientation is preserved) but reversed time direction. This transformation has the effect that $\textsf{PT} N_a$ and $N_b$ lie in the same light cone and all we need now is a boost which moves $\textsf{PT} N_a$ into $N_b$.
This $\textsf{PT}$ transformation can also be taken as the basis for constructing the auxiliary thick wedge. To that end, act with $\textsf{PT}$ on $e_b$. This has the effect of reversing its direction and it now lives in the left spacelike wedge of Minkowski space (see Figure~\ref{fig:Construction} a). The vectors $e_a$ and $\textsf{PT} e_b$ can now be seen as two sides of a thick wedge. However, while $N_b$ is outward pointing, $N_a$ points inside of the thick wedge. To remedy that, we act with $\textsf{PT}$ on $N_a$, which has the effect illustrated in Figure~\ref{fig:Construction}~b.\\
Now we have a proper thick wedge, spanned by $e_a$ and $\textsf{PT}  e_b$, which contains the past light cone and two future pointing normal vectors, $\textsf{PT}  N_a$ and $N_b$. \\
The exterior angle between these normals is given by
\begin{equation}
	\xi_\textsf{ext} = -\text{arcosh}(-\eta(N_a, \textsf{PT} N_b)).
\end{equation}
But we know that $\xi_\textsf{int} = -\xi_\textsf{ext}$ and so it follows that
\begin{equation}
	\xi_\textsf{int} = \text{arcosh}(-\eta(N_a, N_b)) = \text{arcosh}(\eta(N_a, N_b)).
\end{equation}
$\phantom{.}$\\
We now claim that this internal angle is independent of the auxiliary structure used to derive it and that it is equal to the internal angle~\eqref{eq:XiInt} derived from definition~\hyperref[def:LorentzianAngles]{3.2.6}, i.e. $\xi_{\textsf{int}}=\text{arcosh}(\eta(e_a, e_b))$.\\
To see this, we only need to notice that equation~\eqref{eq:BoostedEquationForRelation}, which was used to derive $\xi_{\textsf{int}}=-\xi_{\textsf{ext}}$, remains invariant if we $\textsf{PT}$-transform it. That is,
\begin{equation}\label{eq:FirstPiece}
	\eta(e_b, \Lambda(-\xi_\textsf{ext} N_a)) = \eta(\textsf{PT}e_b,\Lambda(-\xi_\textsf{ext}) \textsf{PT} N_a) = 0.
\end{equation}
In other words: This equation cannot distinguish between the original thin and thick wedges and the auxiliary thick wedge constructed using $\textsf{PT}e_b$ and $\textsf{PT} N_a$. Moreover, we recognize that $\Lambda(-\xi_\textsf{ext}) \textsf{PT}$ is precisely the Lorentz transformation described above: The first piece, $\textsf{PT}$, maps $N_a$ into the same light cone as $N_b$. The second piece then boosts $\textsf{PT} N_a$ into $N_b$. Hence, our intuition is confirmed. Equation~\eqref{eq:FirstPiece} together with $\xi_\textsf{int} = -\xi_\textsf{ext}$ also confirms what we claimed: $\xi_\textsf{int}= \text{arcosh}(\eta(N_a, N_b))$ is independent of the auxiliary structure used to derive it and it is equal to $\text{arcosh}(\eta(e_a, e_b))$, which we derived directly from definition~\hyperref[def:LorentzianAngles]{3.2.6}.\\
 It is important to notice that there is at least one other definition of Lorentzian angles in the literature. Sorkin~\cite{Sorkin_phd} and Neiman~\cite{Neiman:2013} use an analytical continuation of the Euclidean formula:
 \begin{equation}\label{eq:AlternativeDef}
 	\cos\phi := -\eta(N_a, N_b),
 \end{equation} 
where $N_a$ and $N_b$ are timelike vectors which are normalized to $-1$. The resulting angle $\phi$ is in general imaginary because $|\eta(N_a, N_b)|>1$. It can be shown, that $\text{Im}[\phi]$ is negative when $N_a$ and $N_b$ belong to different light cones and positive otherwise. This is precisely the opposite of the sign convention chosen in definition~\hyperref[def:LorentzianAngles]{3.2.6}. Furthermore, when $N_a$ and $N_b$ do not live in the same light cone, $\phi$ acquires a real part which is exactly~$\pi$. This allows us then to write the angle $\phi$ as
\begin{equation}
	\phi = m_{ab}\pi + i \text{Im}[\phi],
\end{equation}
where $m_{ab}$ is zero when $N_a$ and $N_b$ belong to the same light cone and $m_{ab}=1$ otherwise. If we insert this angle into the definition~\eqref{eq:AlternativeDef} we find
\begin{align}
	\eta(N_a, N_b) &= -\cos\phi = -\cos(m_{ab}\pi + i\text{Im}[\phi])\notag\\
	&= -(-1)^{m_{ab}}\cosh(\text{Im}[\phi])
\end{align}
We see that for $m_{ab}=1$ this definition of angle reduces to $\eta(N_a, N_b) = \cosh(\text{Im}[\phi])$, with $\text{Im}[\phi]$ negative, while for $m_{ab}=0$ we obtain $\eta(N_a, N_b) = -\cosh(\text{Im}[\phi])$ with $\text{Im}[\phi]$ positive. Hence, we obtain exactly the same expressions as for the internal and external angles, provided we identify $\xi\equiv - \text{Im}[\phi]$.

\begin{center}
	\begin{figure}[htb]
		\centering
		\begin{minipage}[c]{.45\textwidth}
			\includegraphics[width=\textwidth]{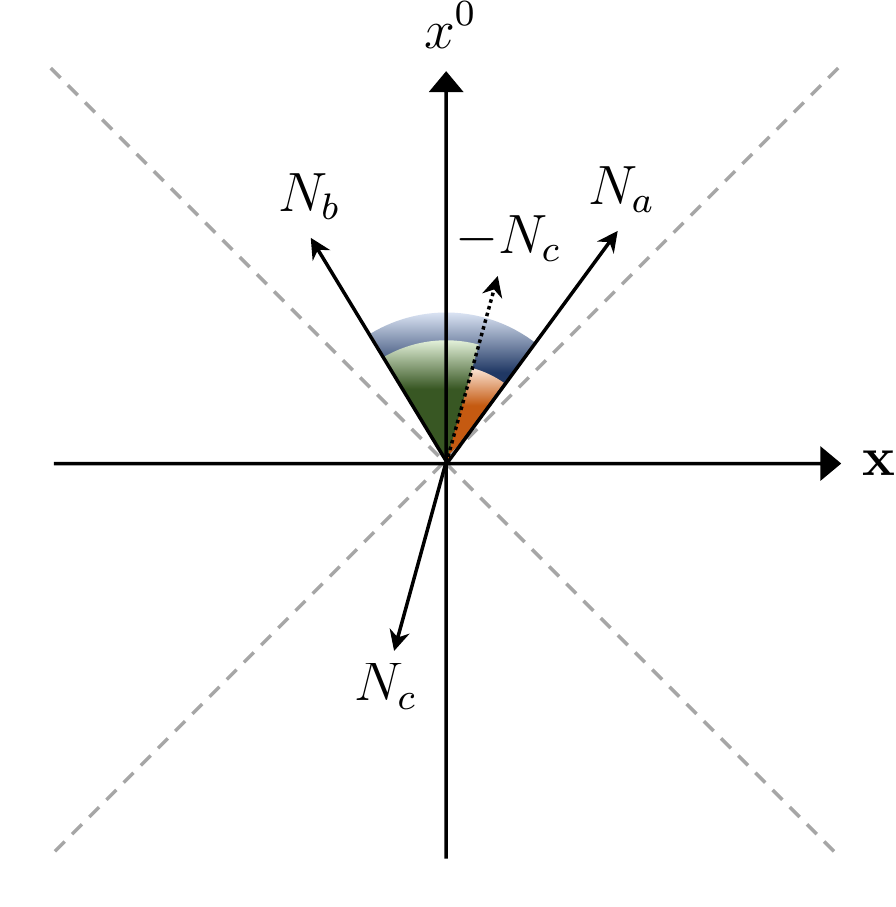}
		\end{minipage}
		\caption{Computation of a Lorentzian deficit angle}
		\label{fig:LorentzianDeficitAngle}
	\end{figure}
\end{center}

To conclude this subsection, let us remark that the definition of interior and exterior angle given here also allows us to talk about Lorentzian deficit angles. In the Euclidean case, we would define a deficit angle $\varepsilon^\text{E}$ as
\begin{equation}
	\varepsilon^\text{E} := 2\pi -\sum_{i=1}^N \theta_i.
\end{equation}
Hence, if we can move around a hinge without obstruction, the sum of the $\theta_i$ is equal to $2\pi$ and the deficit angle vanishes.\\
In the Lorentzian case, we can define the deficit angle as
\begin{equation}
	\varepsilon^{\text{L}} := \sum_{i=1}^N\xi_i,
\end{equation}
where $\xi_i$ are internal and external angles. This deficit angle has the same property as the Euclidean deficit angle: If we can move around a hinge without obstruction, the $\xi_i$ sum to zero. To see this, it suffices to picture three timelike vectors such as the ones shown in Figure~\ref{fig:LorentzianDeficitAngle}. To move around the origin, we need to compute three angles and to do so, we use the constructing described further above in the text. That is, we $\textsf{PT}$-transform the vector $N_c$. Then, all angles can be computed by boosting the vectors into each other. However, it should be remembered that the interior angle between $N_a$ and $N_c$ and also the interior angle between $N_b$ and $N_c$ is positive while the exterior angle between $N_a$ and $N_b$ is negative. One therefore finds that the sum of these angles vanishes and hence there is zero deficit. 
When verifying this statement quantitatively, one has to keep in mind that the vectors $N_a$, $N_b$ and $N_c$ are not truly independent. They are all attached to thin and thick wedges and changing one vector means changing the shape of a wedge. This in turn can induce changes in one of the other vectors. An example: $N_b$ and $N_c$ belong to the thin wedge lying in the right spacelike portion of Minkowski space, while $N_a$ and $N_c$ belong to the thin wedge in the left portion of Minkowski space. Changing $N_c$ so it lies on one line with $N_a$ really means making the thin wedge on the right larger and degenerating the thin wedge on the left to a single line. Bringing  $N_c$ even closer to the null hypersurface on the left is not possible because this would mean the degenerate thin wedge open up again, but now with $N_c$ and $N_a$ being inward point instead of outward pointing. Hence, the whole picture would have to be changed to describe this situation.

%% file: Content/Chapter_4.tex
\chapter{Covariant Loop Quantum Gravity}
\label{chap:CLQG}
Covariant Loop Quantum Gravity (CLQG) has  emerged from a number of different research directions \cite{Livine:2007,Engle:2007a, Engle:2007b, Engle:2007c,Pereira:2007,Freidel:2007,Kaminski:2009} all aimed at providing a consistent path integral quantization of General Relativity.\\
All approaches have in common that they pay special attention to how the simplicity constraint is imposed at the quantum level, hence avoiding the difficulties which brought down the Barrett-Crane model \cite{Barrett:1997, Barrett:1999}. While some of the details in the definition of CLQG may differ from approach to approach and have undergone changes over the years, the main results remain unaffected: The boundary Hilbert space of CLQG agrees with the kinematical Hilbert space of canonical LQG \cite{Engle:2007a, Kaminski:2009}, the graviton propagator and the three-point function possess the tensorial structure expected from perturbative quantum gravity \cite{Alesci:2007a,Alesci:2007b,Alesci:2008b,Alesci:2008, Bianchi:2012, Rovelli:2011}, the classical limit produces the action of area-Regge calculus \cite{Barrett:2009b, Barrett:2009,Conrady:2008,Han:2011b, Han:2011, HanKrajewski:2013}, the area and volume operators are well-defined and the spectra match the ones from canonical LQG \cite{Ding:2009, Ding:2010}, the theory is ultraviolet-finite, a positive cosmological constant can by included \cite{Fairbairn:2010, Han:2010, Han:2011c} (which incidentally renders the theory also infrared-finite), and the entropy for non-extremal black holes of the Kerr-Newman family can be computed \cite{Bianchi:2012b}. Moreover, the theory has been extended to matter couplings with fermions and Yang-Mills fields \cite{Bianchi:2010c,RovelliVidottoBook} and it has been applied to cosmology \cite{Bianchi:2010d, Bahr:2017b, Gozzini:2019}.\\
There are of course also open problems: there is no consensus on how to determine the continuum limit of the theory \cite{Rovelli:2010c, Dittrich:2013,Dittrich:2014, Delcamp:2016}, the renormalization group flow is currently under active investigation \cite{Bahr:2014,Bahr:2017,Bahr:2018}, the classical limit seems to produce only flat geometries (flatness problem, \cite{Bonzom:2009,Magliaro:2011,Hellmann:2013,Perini:2012,Oliveira:2017}) and there seems to be more than one limit (cosine problem, \cite{Mikovic:2011, Engle:2011a, Engle:2011b, Engle:2012, Vojinovic:2013}). An other problem, which will be the main focus of chapter~\ref{chapter_5}, is the current lack of a systematic method to compute transition amplitudes. A problem which impedes the application of CLQG to physical scenarios such as the black hole to white hole transition \cite{RealisticObs, BH2WH}.\\
In the present chapter, however, we will work under the premise that CLQG is a good candidate theory of quantum gravity. We do not attempt to ``derive'' CLQG from classical GR and instead just review the main concepts and ultimately define the theory.

\section{Transition Amplitudes in Covariant Quantum Gravity}
\label{sec:TransitionAmplitudesInCLQG}

\subsection{General Principles}
\label{ssec:GeneralPrinciples}
In covariant quantum theories, dynamics can be described by associating transition amplitudes to spacetime regions. This is known as the general boundary formalism \cite{Oeckl:2003,Oeckl:2005,Oeckl:2016} and the construction proceeds as follows:\\
Consider a compact spacetime region $\mathcal R$ with boundary $\Sigma:=\partial \mathcal R$. Associate to this boundary a Hilbert space $\mathcal H_\Sigma$ of states $\ket{\Psi}$ which represent (quantum) geometry. Moreover, define a linear functional $\bra{W}:\mathcal H_\Sigma\to\mathbb C$, called the \textit{propagator}, which associates to every boundary state a \textit{transition amplitude} $\innerp{W}{\Psi}$. These amplitudes entirely encode the dynamics of the system and, apart from subtleties related to the normalization of the propagator and the states, the modulus squared $|\innerp{W}{\Psi}|^2$ is the probability associated to the process described by the boundary state $\ket{\Psi}$. \\
Applying these ideas to define a path integral for quantum gravity requires some adaptations. Following Feynman \cite{Feynman:1948}, we define the path integral on a discrete structure. Hence, instead of working directly with a compact spacetime region we consider the $2$-complex $\mathcal C$ obtained from a simplicial triangulation of $\mathcal R$, which now is assumed to have a spacelike\footnote{For spin foam models with timelike components see for instance \cite{Conrady:2010,Conrady:2010b,Kaminski:2017,Liu:2018}.} boundary $\Sigma$. The boundary of $\mathcal C$ gives rise to the graph $\Gamma:=\partial \mathcal C$ on which we define the Hilbert space\footnote{For a derivation of the boundary Hilbert space see for instance \cite{Engle:2007c,Ding:2009,Ding:2010,Kaminski:2009}.} $\mathcal H_\Gamma = L_2[SU(2)^L/SU(2)^N]_\Gamma$. This is of course just the kinematical Hilbert space of canonical LQG and it contains states describing quantum geometry. Recall from chapter~\ref{chap:2} that working with a fixed graph is tantamount to truncating the degrees of freedom of continuum GR to finitely many excitations of geometry.\\
Defining the propagator $\bra{W}$ is a more challenging task and it will be the main focus of the next subsection.

\subsection{The Propagator}
\label{ssec:ThePropagator}
\begin{center}
	\begin{figure}[htb]
		\begin{minipage}[c]{.52\textwidth}
			\includegraphics[width=\textwidth]{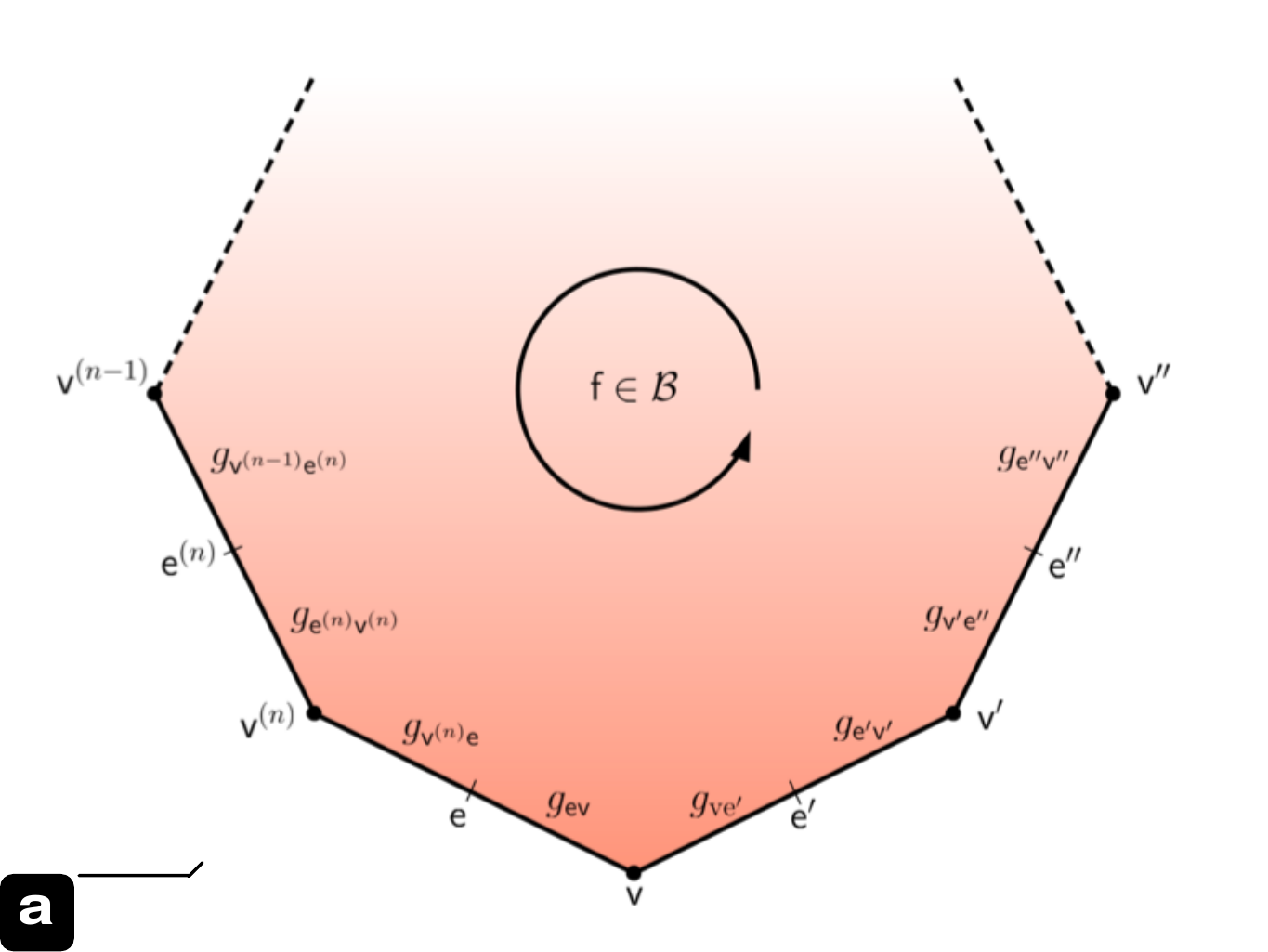}
		\end{minipage}
		\begin{minipage}[c]{.52\textwidth}
			\includegraphics[width=\textwidth]{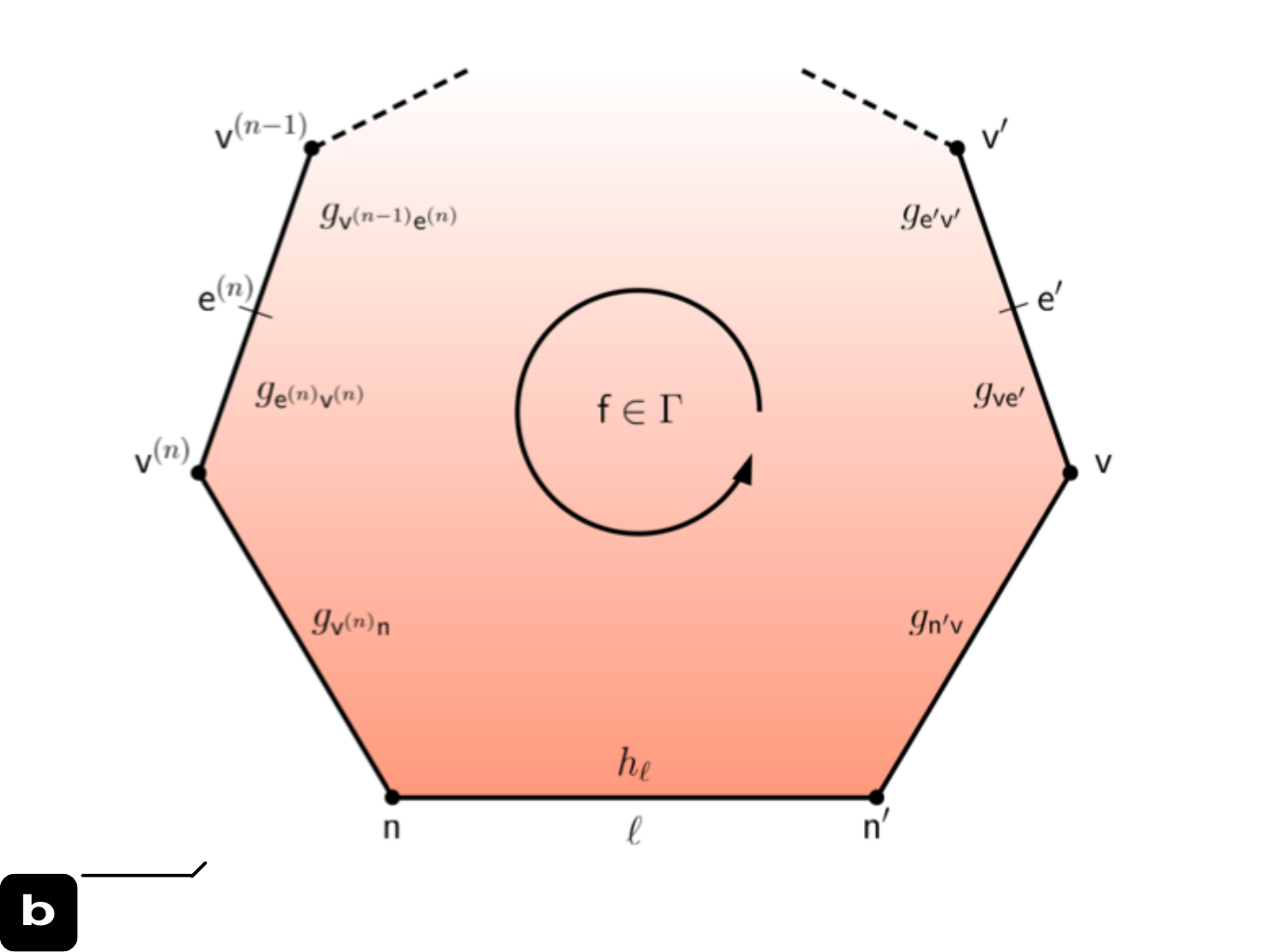}	
		\end{minipage}
		\caption{Conventions for the bulk (a) and boundary (b) notation.}	
		\label{fig:Notation}
	\end{figure}
\end{center}

It is convenient to distinguish between the boundary $\Gamma$ of the $2$-complex $\mathcal C$ and its bulk $\mathcal B := \mathcal C\setminus\Gamma$. Since the $2$-complex is assumed to be dual to a $4$-dimensional simplicial triangulation of a spacetime region, its interior consists of a collection of five-valent vertices $\ve$ which are connected by edges $\ed$ which in turn bound faces $\fa$. All vertices strictly belong to the interior of $\mathcal C$, but some of the edges which emanate from a vertex intersect the boundary and therefore end in a node $\no$. These nodes are four-valent and are connected by links $\ell$. The boundary graph $\Gamma$ is then simply the collection of all these nodes and links.\\
Faces either belong to the interior or to the boundary. We say a face belongs to the interior when it is bound by vertices and edges and we write $\fa\in\mathcal B$. If a face is bound by vertices, edges, nodes and links it is said to be a boundary face and we write $\fa\in\Gamma$.\\
In order to construct a map $\bra{W_\mathcal{C}}:\mathcal H_\Gamma\to\mathbb C$ we associate a $\slc$ group element $g_{\ve\ed}$ to every half edge of the interior of $\mathcal C$ (see Figure~\ref{fig:Notation}) and the convention $g_{\ve\ed} = g^{-1}_{\ed\ve}$ is assumed to hold. If an edge originating from $\ve$ terminates in a node $\no$ it is not split in two and a single group element $g_{\ve\no}\in\slc$ is associated to it. Links carry $SU(2)$ group elements $h_\ell$ and all faces, whether they are internal or boundary faces, are colored by a half-integer spin $j_\fa\neq 0$. Moreover, all faces are oriented and this in turn induces an orientation on the edges and links (see Figure~\ref{fig:Notation}).\\
Lastly, we need to introduce the so-called $Y$-map. It provides us with a unitary injection $Y:\mathcal{V}^j\rightarrow \mathcal{H}^{(\gamma j, j)}$ such that $\ket{jm}\mapsto\ket{\gamma j, j; jm}$.\\
With these definitions we can now associate an amplitude $A_\fa$ to every face of $\mathcal C$. For internal faces, the construction proceeds as follows: At every vertex $\ve$ build the product of ingoing group element $g_{\ed\ve}$ and outgoing group element $g_{\ve\ed'}$, i.e. $g_{\ed\ve}g_{\ve\ed'}$. Every such product is multiplied from the left by $Y^\dagger$ and from the right by $Y$, yielding a term of the form $Y^\dagger g_{\ed\ve}g_{\ve\ed'} Y$ at every vertex. Finally, all these terms are multiplied together with their nearest neighbors around the face, following the face's orientation (cf. Figure~\ref{fig:Notation} a): 
\begin{equation}
	Y^\dagger g_{\ed\ve}g_{\ve\ed'} Y Y^\dagger g_{\ed'\ve'}g_{\ve'\ed''} Y\dots Y^\dagger g_{\ed^{(n)}\ve^{(n)}}g_{\ve^{(n)}\ed} Y
\end{equation}
The face amplitude for internal faces is then defined as
\begin{align}\label{eq:FaceAmplitudes}
	A_\fa &:= \sum_{j_\fa}d_{j_\fa} \,  \Tr_{j_\fa}\left[\prod_{\ve\in\fa} Y^{\dagger} g^{-1}_{\ve\ed}\,g_{\ve\ed'} Y\right]\notag\\
		&:= \sum_{j_\fa} d_{j_\fa} \Tr_{j_\fa}\left[ Y^\dagger g_{\ed\ve}g_{\ve\ed'} Y Y^\dagger g_{\ed'\ve'}g_{\ve'\ed''} Y\cdots Y^\dagger g_{\ed^{(n)}\ve^{(n)}}g_{\ve^{(n)}\ed} Y \right] \quad\text{for }\fa\in\mathcal{B},
\end{align}
where the summation in $j_\fa$ runs over $\frac{1}{2}\mathbb{N}\backslash\{0\}$ in half-integer steps. The first line is a convenient short form that is useful for later computations while the trace of the second line can be explicitly written as
\begin{align}\label{eq:def_trace}
	\Tr_{j_\fa}\left[\prod_{\ve\in\fa} Y^{\dagger} g^{-1}_{\ve\ed}\,g_{\ve\ed'}  Y\right] = \sum_{\{m_\ed\}} D^{(\gamma j_\fa, j_\fa)}_{j_\fa m_\ed j_\fa m_\ed'}(g_{\ed\ve}g_{\ve\ed'})  D^{(\gamma j_\fa, j_\fa)}_{j_\fa m_{\ed'} j_\fa m_{\ed''}}(g_{\ed'\ve'}g_{\ve'\ed''}) \dots D^{(\gamma j_\fa, j_\fa)}_{j_\fa m_{\ed^{(n)}} j_\fa m_{\ed}}(g_{\ed^{(n)}\ve^{(n)}}g_{\ve^{(n)}\ed}),
\end{align} 
where $D^{(\gamma j, j)}_{j m j m'}(g)$ are $\slc$ representation matrices and $\sum_{\{m_\ed\}}$ is a short hand notation for multiple sums (in this case, over all magnetic indices $m_\ed$ appearing in \eqref{eq:def_trace} in the product of representation matrices).\\
The face amplitude $A_\fa(h_\ell)$ for boundary faces is defined analogously, the only differences being that edges terminating in nodes are not split into half edges and therefore carry only one $\slc$ group element and there is one $SU(2)$ group element $h_\ell$ on each link (see Figure~\ref{fig:Notation} b). 
\begin{align}\label{eq:BoundaryFaceAmplitude}
	A_\fa(h_\ell) &:= \sum_{j_\fa}d_{j_\fa} \Tr_{j_\fa}\left[Y^\dagger g_{\ve \no'}^{-1} g_{\ve\ed'} Y \left(\prod_{\ve\in\fa} Y^{\dagger} g^{-1}_{\ve\ed'}\,g_{\ve\ed} Y\right) Y^\dagger g_{\ve^{(n)}\ed^{(n)}}^{-1} g_{\ve^{(n)}\no} Y h_\ell^{-1}\right]  \notag\\
	&= \sum_{j_\fa} d_{j_\fa} \Tr_{j_\fa}\left[Y^\dagger g_{\no'\ve}g_{\ve\ed'} Y Y^\dagger g_{\ed'\ve'}g_{\ve'\ed''} Y \dots Y^\dagger g_{\ed^{(n)}\ve^{(n)}}g_{\ve^{(n)}\no} Y h_\ell^{-1} \right] & \text{for }\fa\in\Gamma.
\end{align}
Using the face amplitudes $A_\fa$ and $A_\fa(h_\ell)$ we can define the propagator $W_\mathcal{C}(h_\ell)$ associated to the $2$-complex $\mathcal C$ as
\begin{align}\label{eq:AmplitudeMap}
	W_\mathcal{C}(h_\ell) &:= \mathcal{N}_\mathcal{C}\int_{\slc}\left(\prod_{\ve}\dd \arc{g}_{\ve\ed}\right)\left(\prod_{\fa\in\mathcal{B}} A_\fa\right)\left(\prod_{\fa\in\Gamma} A_\fa(h_\ell)\right),
\end{align}
where $\mathcal{N}_\mathcal{C}$ is a normalization factor which depends on the combinatorial structure of $\mathcal C$ and $\dd\arc{g}_{\ve\ed}$ indicates that one $\slc$ integration per vertex has to be dropped. This regularization procedure is necessary, as one can convince oneself by looking at a single vertex, because there is one redundant integration for each vertex and since $\slc$ is not compact this leads to a divergent result \cite{Barrett:1999,Engle:2008}.\\
Hence, what remains is a product over four $\slc$ Haar measures per vertex, which are explicitly given by
\begin{align}
	\dd g = \frac{\dd\beta \,\dd\overline\beta \,\dd\gamma \,\dd\overline\gamma \, \dd\delta \,\dd\overline\delta }{|\delta|^2}\quad\text{for}\quad g=\begin{pmatrix}
		\alpha & \beta \\
		\gamma & \delta 
	\end{pmatrix}\in\slc.
\end{align}
The propagator~\eqref{eq:AmplitudeMap} should really be understood as a map which assigns complex numbers to states $\Psi\in\mathcal H_\Gamma$ through
\begin{equation}
	\innerp{W_\mathcal{C}}{\Psi} := \int_{SU(2)^L}\left(\prod_\ell \dd h_\ell\right) W_\mathcal{C}(h_\ell) \Psi(h_\ell).
\end{equation}
This completes the definition of the propagator.

\section{Path Integral Form of the Propagator}
\label{sec:PathIntegralForm}
Currently, the only systematic approach to analytically evaluate CLQG propagators is the stationary phase approximation first employed by \cite{Barrett:2009b, Barrett:2009} and then further developed by \cite{HanKrajewski:2013} and \cite{Han:2011b, Han:2011}. A crucial step required before the stationary phase approximation can be applied is the conversion of the propagator~\eqref{eq:AmplitudeMap} into a form more reminiscent of a standard path integral.\\
In this section we will perform this conversion in a novel way and we will break it down into several steps. First, we consider $2$-complexes without boundary in order to discuss the main steps without too many complications. The generalization to $2$-complexes with boundaries follows then very quickly. Finally, we introduce the so-called holomorphic transition amplitude in its path integral form. This amplitude will be of major importance in chapter~\ref{chapter_5}, where we develop a new analytical evaluation method for CLQG transition amplitudes in the semi-classical regime.

\subsection{The Propagator on a $2$-Complex without Boundary}
\label{ssec:PropagatorNoBoundary}
To recast the CLQG propagator \eqref{eq:AmplitudeMap} in a path integral form we need to change the representation. That is, we choose a representation of the principal series of $\slc$ and its subgroup $SU(2)$ on the space of homogeneous functions $\mathcal{H}^{(k, p)}$ in two complex variables $\textbf{z}=(z^0, z^1)^{\transpose}\in\mathbb{C}^2$. For a pedagogical introduction to the representation theory of $SU(2)$ and $\slc$ we refer the reader to \cite{Martin-Dussaud:2019}. Here, we shall only recall what is necessary for the calculations that follow:\\
The unitary, irreducible, infinite dimensional representations of the principal series of $\slc$ on $\mathcal{H}^{(k, p)}$ are labeled by two parameters, $(k, p)\in\mathbb{R}\times\frac{1}{2}\mathbb{Z}$. In terms of these parameters the homogeneity of a function $F\in\mathcal{H}^{(k, p)}$ can be expressed as
\begin{align}
	F(\lambda \textbf{z}) = \lambda^{i k + p -1}\,\overline{\lambda}^{i k - p -1}F(\textbf{z})\quad \forall\lambda\in \mathbb{C}\backslash\{0\}.
\end{align}
Furthermore, the space $\mathcal{H}^{(k, p)}$ decomposes as $\mathcal{H}^{(k, p)}\simeq\bigoplus_{j=\vert p\vert}^{\infty}\mathcal{V}^j$, where $\mathcal{V}^j$ is the space of homogeneous polynomials of degree $2j$ in two complex variables. The $Y$-map introduced in subsection~\ref{ssec:ThePropagator} provides us with a unitary injection
\begin{align}
	Y:\mathcal{V}^j\rightarrow \mathcal{H}^{(\gamma j, j)}\quad,\quad f(\textbf{z}) \mapsto F(\textbf{z})=\innerp{\textbf{z}}{\textbf{z}}^{i\gamma j - j-1}f(\textbf{z})\quad\forall f\in\mathcal V^j,
\end{align}
where $\innerp{\textbf x}{\textbf y} = \overline x^0 y^0 + \overline x^1 y^1$ is the $SU(2)$ (but not $\slc$) invariant inner product on~$\mathbb C^2$. Starting from the polynomial basis\footnote{Notice that the position of the spinorial index in $(z^0, z^1)^\transpose$ is irrelevant and we will often place it where there is enough space in order to improve readability.} 
\begin{align}
	P_m^j(\textbf{z}) = \left[\frac{(2j)!}{(j+m)!(j-m)!}\right]^{\frac{1}{2}}z_0^{j+m}\,z_1^{j-m},\quad m\in\{-j, ...,j\}
\end{align}
of $\mathcal V^j$, we can construct a basis of $\mathcal{H}^{(\gamma j, j)}$ by acting on $P^j_m(\textbf{z})$ with the $Y$-map:
\begin{align}\label{eq:SLPolynomials}
	\phi_{m}^{(\gamma j, j)}(\textbf{z}):= Y\rhd P_m^j(\textbf{z}) = \sqrt{\frac{d_j}{\pi}} \innerp{\textbf{z}}{\textbf{z}}^{i \gamma j -j-1} P_m^j(\textbf{z}),
\end{align}
This basis is orthonormal with respect to the inner product $\langle\cdot,\cdot\rangle:\mathcal H^{(\gamma j, j)}\times H^{(\gamma j, j)} \to \mathbb C$ defined by
\begin{align}
	\langle f, g \rangle := \int_{\CP}\dd \Omega \, \overline{f(\textbf{z})}\, g(\textbf{z}),\quad\forall f, g\in\mathcal H^{(\gamma j, j)},
\end{align}
where $\dd \Omega = \frac{i}{2}(z^0\dd z^1 - z^1\dd z^0)\wedge(\overline{z}^0\dd \overline{z}^1 - \overline{z}^1\dd \overline{z}^0)$ is a homogeneous and $\slc$ invariant measure on $\CP$. By virtue of this inner product, the $\slc$ representation matrices can be written as \cite{RuhlBook,Barrett:2009}
\begin{align}\label{eq:SL2CWigner}
	D_{j\,m\,j\,m'}^{(\gamma j, j)}(g) \equiv \bra{j m}Y^\dagger g Y \ket{j m'} = \int_{\CP}\dd \Omega\, \overline{\phi_{m}^{(\gamma j, j)}(\textbf{z})}\, \phi_{m'}^{(\gamma j, j)}(g^\transpose\textbf{z}).
\end{align}
If $g$ lies in the $SU(2)$ subgroup of $\slc$, then the usual Wigner $D$-matrices are recovered, see \cite{RuhlBook, Martin-Dussaud:2019}. Equation \eqref{eq:SL2CWigner} is crucial in what follows since it is the key to rewrite the trace of the face amplitude \eqref{eq:FaceAmplitudes} in the new representation. From the definition of the trace \eqref{eq:def_trace} together with \eqref{eq:SL2CWigner} it follows that
\begin{align}\label{eq:ExpandedTrace}
	\Tr_{j_\fa}\left[\prod_{\ve\in\fa} Y^{\dagger} g^{-1}_{\ve\ed}\,g_{\ve\ed'}  Y\right] &= \sum_{\{m_\ed\}}\prod_{\ve\in\fa}D_{j_\fa\,m_\ed\, j_\fa\, m_{\ed'}}^{(\gamma j_\fa, j_\fa)}(g_{\ve\ed}^{-1} g_{\ve\ed'})\notag\\
	&= \sum_{\{m_\ed\}} \prod_{\ve\in\fa}\int_{\CP}\dd\Omega_{\ve\fa}\,\overline{\phi_{m_\ed}^{(\gamma j_\fa, j_\fa)}(\textbf{z}_{\ve\fa})}\,\phi_{m_{\ed'}}^{(\gamma j_\fa, j_\fa)}(g_{\ve\ed'}^\transpose\, (g_{\ve\ed}^{-1})^\transpose\,\textbf{z}_{\ve\fa})\notag\\
	&= \sum_{\{m_\ed\}} \prod_{\ve\in\fa}\int_{\CP}\dd\Omega_{\ve\fa}\,\overline{\phi_{m_\ed}^{(\gamma j_\fa, j_\fa)}(g_{\ve\ed}^\transpose\,\textbf{z}_{\ve\fa})}\,\phi_{m_{\ed'}}^{(\gamma j_\fa, j_\fa)}(g_{\ve\ed'}^\transpose\,\textbf{z}_{\ve\fa}).
\end{align}
To get the last line we performed the change of integration variables $\textbf{z}_{\ve\fa}\rightarrow g_{\ve\ed}^\transpose\,\textbf{z}_{\ve\fa}$ and used the $\slc$ invariance of the measure $\dd\Omega_{\ve\fa}$. By exploiting the fact that the trace \eqref{eq:ExpandedTrace} appears under an integral with an $\slc$ Haar measure in \eqref{eq:AmplitudeMap}, it is possible to perform the replacement $g_{\ve\ed}\rightarrow \overline{g}_{\ve\ed}$ on all group variables\footnote{The bar indicates complex conjugation.} and define spinorial variables which are associated to vertices and half edges of a given face:
\begin{align}
	\textbf{Z}_{\ve\ed\fa} := g_{\ve\ed}^{\dagger}\,\textbf{z}_{\ve\fa}\quad\text{and}\quad \textbf{Z}_{\ve\ed'\fa} := g_{\ve\ed'}^{\dagger}\,\textbf{z}_{\ve\fa}.
\end{align}
Using the explicit expression \eqref{eq:SLPolynomials} for the basis functions $\phi_m^{(\gamma j, j)}$ brings the original trace into the form
\begin{align}\label{eq:TraceNewRep}
	&\Tr_{j_\fa}\left[\prod_{\ve\in\fa} Y^{\dagger} g^{-1}_{\ve\ed}\,g_{\ve\ed'}  Y\right] = \sum_{\{m_\ed\}} \prod_{\ve\in\fa}\int_{\CP}\dd\Omega_{\ve\fa}\,\overline{\phi_{m_\ed}^{(\gamma j_\fa, j_\fa)}(\textbf{Z}_{\ve\ed\fa})}\,\phi_{m_{\ed'}}^{(\gamma j_\fa, j_\fa)}(\textbf{Z}_{\ve\ed'\fa}) \notag\\
	&=\sum_{\{m_\ed\}}\prod_{\ve\in\fa}\frac{d_{j_\fa}}{\pi}\int_{\CP}\dd\Omega_{\ve\fa}\,\innerp{\textbf{Z}_{\ve\ed\fa}}{\textbf{Z}_{\ve\ed\fa}}^{-i\gamma j_\fa-j_\fa-1}\innerp{\textbf{Z}_{\ve\ed'\fa}}{\textbf{Z}_{\ve\ed'\fa}}^{i\gamma j_\fa-j_\fa-1}P_{m_{\ed'}}^{j_\fa}\left(\textbf{Z}_{\ve\ed'\fa}\right)\,P_{m_\ed}^{j_\fa}\left(\overline{\textbf{Z}}_{\ve\ed\fa}\right).
\end{align}
In the above expression, the spinorial inner products do not depend on any magnetic indices $m_\ed$. Hence, the sums only extend over the $SU(2)$ basis polynomials $P^j_m$. There are two such polynomials per edge $\ed$ which depend on the same magnetic index $m_\ed$ (as there are two $\textbf Z$ spinors per edge, but pertaining to different vertices). This follows from the contraction pattern in \eqref{eq:def_trace}. Consequently, the sum $\sum_{\{m_\ed\}}$ decomposes into a certain number\footnote{The number of sums is equal to the number of edges which constitute the face.} of single, independent sums of the form
\begin{align}\label{eq:SumOverMagInd}
	\sum_{\vert m_{\ed'}\vert \leq j_\fa}P_{m_{\ed'}}^{j_\fa} \left(\textbf{Z}_{\ve\ed'\fa}\right)\, P_{m_{\ed'}}^{j_\fa}\left(\overline{\textbf{Z}}_{\ve'\ed'\fa}\right) &= \sum_{\vert m_{\ed'}\vert \leq j_\fa}\frac{(2j_\fa)!}{(j_\fa+m_{\ed'})!(j_\fa-m_{\ed'})!}\left(\overline{\textbf{Z}}^0_{\ve'\ed'\fa}\,\textbf{Z}^0_{\ve\ed'\fa}\right)^{j_\fa+m_{\ed'}}\,\left(\overline{\textbf{Z}}^1_{\ve'\ed'\fa}\,\textbf{Z}^1_{\ve\ed'\fa}\right)^{j_\fa-m_{\ed'}} \notag\\
	&=\sum_{s=0}^{2j_\fa}\binom{2j_\fa}{s}\left(\overline{\textbf{Z}}^0_{\ve'\ed'\fa}\,\textbf{Z}^0_{\ve\ed'\fa}\right)^s \left(\overline{\textbf{Z}}^1_{\ve'\ed'\fa}\,\textbf{Z}^1_{\ve\ed'\fa}\right)^{2j_\fa-s}\notag\\
	&=\left(\overline{\textbf{Z}}^0_{\ve'\ed'\fa}\,\textbf{Z}^0_{\ve\ed'\fa} + \overline{\textbf{Z}}^1_{\ve'\ed'\fa}\,\textbf{Z}^1_{\ve\ed'\fa}\right)^{2j_\fa} =\innerp{\textbf{Z}_{\ve'\ed'\fa}}{\textbf{Z}_{\ve\ed'\fa}}^{2j_\fa}.
\end{align}
In the first line we use the definition \eqref{eq:SLPolynomials} of $P_{m_{\ed'}}^{j_\fa}$ and in the second line we performed the change of summation variable $s=j_\fa+m_{\ed'}$. The resulting binomial sum is trivial and yields the result on the third line. Plugging \eqref{eq:SumOverMagInd} into \eqref{eq:TraceNewRep} and changing from a product over vertices $\ve\in\fa$ to an equivalent product over edges $\ed\in\fa$ brings the bulk face amplitude into the form
\begin{align}\label{eq:RewrittenBulkFace}
	A_\fa &= \sum_{j_\fa}d_{j_\fa}\prod_{\ed\in\fa}\frac{d_{j_\fa}}{\pi}\int_{\CP}\dd\tilde{\Omega}_{\ve\ed\fa}\,\frac{\innerp{\textbf{Z}_{\ve'\ed'\fa}}{\textbf{Z}_{\ve\ed'\fa}}^{2j_\fa}}{\innerp{\textbf{Z}_{\ve'\ed'\fa}}{\textbf{Z}_{\ve'\ed'\fa}}^{i\gamma j_\fa+j_\fa} \innerp{\textbf{Z}_{\ve\ed'\fa}}{\textbf{Z}_{\ve'\ed'\fa}}^{-i\gamma j_\fa+j_\fa}}\notag\\
	&=\sum_{j_\fa}d_{j_\fa}\prod_{\ed\in\fa}\frac{d_{j_\fa}}{\pi}\int_{\CP}\dd\tilde{\Omega}_{\ve\ed\fa}\,\e^{i L_\fa\left[g_{\ve\ed}, \textbf{z}_{\ve\fa}\right]}&\forall\fa\in\mathcal B.
\end{align}
As in \cite{HanKrajewski:2013} and \cite{Han:2011} we introduced the rescaled measure 
\begin{align}\label{eq:RescaledMeasure}
	\dd\tilde{\Omega}_{\ve\ed\fa} & := \frac{\dd \Omega_{\ve\fa}}{\innerp{\textbf{Z}_{\ve'\ed'\fa}}{\textbf{Z}_{\ve'\ed'\fa}} \innerp{\textbf{Z}_{\ve\ed'\fa}}{\textbf{Z}_{\ve'\ed'\fa}}}
\end{align}
and a Lagrangian $L_\fa\left[g_{\ve\ed}, \textbf{z}_{\ve\fa}\right]$ associated to internal faces:
\begin{align}\label{eq:BulkLagrangian}
	L_\fa[ g_{\ve\ed}, \textbf{z}_{\ve\fa}] & :=	 \gamma j_\fa\sum_{\ed\in\fa} \left[\ln \frac{\innerp{\textbf{Z}_{\ve\ed'\fa}}{\textbf{Z}_{\ve\ed'\fa}}}{\innerp{\textbf{Z}_{\ve'\ed'\fa}}{\textbf{Z}_{\ve'\ed'\fa}}}  -\frac{i}{\gamma}  \ln\frac{\innerp{\textbf{Z}_{\ve'\ed'\fa}}{\textbf{Z}_{\ve\ed'\fa}}^{2}}{\innerp{\textbf{Z}_{\ve'\ed'\fa}}{\textbf{Z}_{\ve'\ed'\fa}} \innerp{\textbf{Z}_{\ve\ed'\fa}}{\textbf{Z}_{\ve'\ed'\fa}}}\right].
\end{align}
If the $2$-complex had no boundary, i.e. if $\Gamma = \emptyset$, then the CLQG propagator in its path integral form would read
\begin{align}
	W_\mathcal{C} &= \mathcal N_\mathcal{C}  \int_{\slc}\left(\prod_\ve \dd\arc{g}_{\ve\ed}\right)\prod_{\fa\in\mathcal B} \left(\sum_{j_\fa}d_{j_\fa}\prod_{\ed\in\fa}\frac{d_{j_\fa}}{\pi}\int_{\CP}\dd\tilde{\Omega}_{\ve\ed\fa}\,\e^{L_\fa\left[g_{\ve\ed}, \textbf{z}_{\ve\fa}\right]}\right) \notag\\
	&= \mathcal N_\mathcal{C}  \sum_{\{j_\fa\}} \int_{\slc}\left(\prod_\ve \dd\arc{g}_{\ve\ed}\right)\left(\prod_{\fa\in\mathcal B} d_{j_\fa}\prod_{\ed\in\fa}\frac{d_{j_\fa}}{\pi}\int_{\CP}\dd\tilde{\Omega}_{\ve\ed\fa}\right)\e^{i \sum_{\fa\in\mathcal B}L_\fa\left[g_{\ve\ed}, \textbf{z}_{\ve\fa}\right]},
\end{align}
which is precisely the result obtained in \cite{HanKrajewski:2013} by different group theoretical methods.

\subsection{A $2$-Complex with Boundary}
\label{ssec:WithBoundary}
We now proceed to generalize the above result to a $2$-complex with boundary. To that end, it is necessary to also rewrite the trace of the boundary face amplitude \eqref{eq:BoundaryFaceAmplitude} in terms of functions on $\mathcal H^{(\gamma j, j)}$. From the definition \eqref{eq:BoundaryFaceAmplitude} it follows that the product over vertices can be treated in the same way as for the bulk face amplitude as none of the group elements lives on an edge which terminates in a node. In fact, the only group elements we need to consider here are the first two (i.e. $g^{-1}_{\ve\no'}$ and $g_{\ve\ed'}$) and the last three (i.e. $g^{-1}_{\ve^{(n)}\ed^{(n)}}$, $g_{\ve^{(n)}\no}$ and $h^{-1}_\ell$) in the trace of \eqref{eq:BoundaryFaceAmplitude} (see also Figure~\ref{fig:Notation} b). Writing $\left(\star\right)$ as placeholder for the product over vertices which takes the same form as before we get
\begin{align}\label{eq:BoundaryTraceLong}
	\Tr_{j_\fa} & \left[Y^\dagger g_{\ve \no'}^{-1} g_{\ve\ed'} Y \left(\prod_{\ve\in\fa} Y^{\dagger} g^{-1}_{\ve\ed'}\,g_{\ve\ed} Y\right) Y^\dagger g_{\ve^{(n)}\ed^{(n)}}^{-1} g_{\ve^{(n)}\no} Y h_\ell^{-1}\right]\notag\\
	&= \sum_{\{m_\ed\}} D^{(\gamma j_\fa, j_\fa)}_{j_\fa m_{\no'} j_\fa m_{\ed'}}(g_{\ve\no'}^{-1}g_{\ve\ed'}) \left(\star\right) D^{(\gamma j_\fa, j_\fa)}_{j_\fa m_{\ed^{(n)}} j_\fa m_{\no}}(g_{\ve^{(n)}\ed^{(n)}}^{-1}g_{\ve^{(n)}\no}) D^{j_\fa}_{m_\no m_{\no'}}(h^{-1}_\ell)\notag\\
	&= \sum_{\{m_\ed\}}\int_{\CP}\dd\Omega_{\ve\fa}\,\overline{\phi^{(\gamma j_\fa, j_\fa)}_{m_{\no'}}(\textbf z_{\ve\fa})}\phi^{(\gamma j_\fa, j_\fa)}_{m_{\ed'}}(g_{\ve\ed'}^\transpose(g_{\ve\no'}^{-1})^\transpose\textbf z_{\ve\fa})      \left(\star\right)       \int_{\CP}\dd\Omega_\ell\, \overline{\phi^{(\gamma j_\fa, j_\fa)}_{m_{\no}}(\textbf z_\ell)}   \phi^{(\gamma j_\fa, j_\fa)}_{m_{\no'}}((h^{-1}_\ell)^\transpose\textbf z_\ell)\times\notag\\
	&\quad\times \int_{\CP}\dd\Omega_{\ve^{(n)}\fa}\,\overline{\phi^{(\gamma j_\fa, j_\fa)}_{m_{\ed^{(n)}}}(\textbf z_{\ve^{(n)}\fa})}\phi^{(\gamma j_\fa, j_\fa)}_{m_{\no}}(g_{\ve^{(n)}\no}^\transpose(g_{\ve^{(n)}\ed^{(n)}}^{-1})^\transpose\textbf z_{\ve^{(n)}\fa}) \notag\\
	&= \sum_{\{m_\ed\}}\int_{\CP}\dd\Omega_{\ve\fa}\,\overline{\phi^{(\gamma j_\fa, j_\fa)}_{m_{\no'}}(\textbf Z_{\ve\no'\fa})}\phi^{(\gamma j_\fa, j_\fa)}_{m_{\ed'}}(\textbf Z_{\ve\ed'\fa}) \left(\star\right) \int_{\CP}\dd\Omega_{\ve^{(n)}\fa}\,\overline{\phi^{(\gamma j_\fa, j_\fa)}_{m_{\ed^{(n)}}}(\textbf Z_{\ve^{(n)}\ed^{(n)}\fa})}\phi^{(\gamma j_\fa, j_\fa)}_{m_{\no}}(\textbf Z_{\ve^{(n)}\no\fa}) \times\notag\\
	&\quad\times \int_{\CP}\dd\Omega_\ell\, \overline{\phi^{(\gamma j_\fa, j_\fa)}_{m_{\no}}(h^\transpose_\ell \textbf z_\ell)}   \phi^{(\gamma j_\fa, j_\fa)}_{m_{\no'}}(\textbf z_\ell),
\end{align}
where $\textbf{z}_\ell$ is formally associated to the link connecting $\no$ to $\no'$. To get the last line we exploited again the $\slc$ invariance of the measures and performed the same change of integration variables as in subsection~\ref{ssec:PropagatorNoBoundary}, thereby also introducing the spinorial variables
\begin{align}
	\textbf{Z}_{\ve\no'\fa} := g^\dagger_{\ve\no'} \textbf z_{\ve\fa}\quad\text{and}\quad \textbf Z_{\ve^{(n)}\no\fa} := g^\dagger_{\ve^{(n)}\no}\textbf z_{\ve^{(n)}\fa},
\end{align}
which are associated to the two edges which terminate in the nodes $\no$ and $\no'$, respectively. To proceed, we collect only the relevant terms in \eqref{eq:BoundaryTraceLong} and compute
\begin{align}
	\sum_{m_{\no}, m_{\no'}} &\overline{\phi^{(\gamma j_\fa, j_\fa)}_{m_{\no'}}(\textbf Z_{\ve\no'\fa})}\,    \phi^{(\gamma j_\fa, j_\fa)}_{m_{\no}}(\textbf Z_{\ve^{(n)}\no\fa})\,    \overline{\phi^{(\gamma j_\fa, j_\fa)}_{m_{\no}}(h^\transpose_\ell \textbf z_\ell)}\,   \phi^{(\gamma j_\fa, j_\fa)}_{m_{\no'}}(\textbf z_\ell)\notag\\
	&=\innerp{\textbf Z_{\ve\no'\fa}}{\textbf Z_{\ve\no'\fa}}^{-i \gamma j_\fa - j_\fa - 1} \innerp{\textbf Z_{\ve^{(n)}\no\fa}}{\textbf Z_{\ve^{(n)}\no\fa}}^{i \gamma j_\fa - j_\fa - 1}\innerp{\textbf z_\ell}{\textbf z_\ell}^{-2(j_\fa+1)}\times \notag\\
	&\quad \times \sum_{m_{\no}} P^{j_\fa}_{m_\no}(h^\dagger_\ell \overline{\textbf{z}}_\ell) P^{j_\fa}_{m_\no}(\textbf Z_{\ve^{(n)}\no\fa}) \sum_{m_{\no'}} P^{j_\fa}_{m_\no'}(\overline{\textbf{Z}}_{\ve\no'\fa}) P^{j_\fa}_{m_\no'}(\textbf z_\ell)\notag\\
	&=\innerp{\textbf Z_{\ve\no'\fa}}{\textbf Z_{\ve\no'\fa}}^{-i \gamma j_\fa - j_\fa - 1} \innerp{\textbf Z_{\ve^{(n)}\no\fa}}{\textbf Z_{\ve^{(n)}\no\fa}}^{i \gamma j_\fa - j_\fa - 1}\innerp{\textbf z_\ell}{\textbf z_\ell}^{-2(j_\fa+1)} \innerp{h^\transpose_\ell \textbf z_\ell}{\textbf Z_{\ve^{(n)\no\fa}}}^{2 j_\fa} \innerp{\textbf Z_{\ve\no'\fa}}{\textbf z_\ell}^{2 j_\fa}.
\end{align}
Plugging the above result back into \eqref{eq:BoundaryTraceLong} turns the trace into
\begin{align}
	\Tr_{j_\fa} & \left[Y^\dagger g_{\ve \no'}^{-1} g_{\ve\ed'} Y \left(\prod_{\ve\in\fa} Y^{\dagger} g^{-1}_{\ve\ed'}\,g_{\ve\ed} Y\right) Y^\dagger g_{\ve^{(n)}\ed^{(n)}}^{-1} g_{\ve^{(n)}\no} Y h_\ell^{-1}\right] \notag\\
	&= \left(\prod_{\ed\in\fa}\frac{d_{j_\fa}}{\pi}\int_{\CP}\dd\tilde{\Omega}_{\ve\ed\fa}\right)\left(\frac{d^3_{j_\fa}}{\pi^3}\int_{(\CP)^3}\dd\tilde{\Omega}_{\no\ell\no'}\right)\,\e^{i L_\fa\left[g_{\ve\ed}, \textbf{z}_{\ve\fa}\right] + i B_\ell[g_{\ve\no}, h_\ell, \textbf z_\ell]},
\end{align}
where we have defined the rescaled $(\CP)^3$ measures 
\begin{align}\label{eq:BndMeasure}
	\dd\tilde{\Omega}_{\no\ell\no'} &:= \frac{\dd \Omega_{\ve^{(n)}\fa}}{\innerp{\textbf Z_{\ve^{(n)}\no\fa}}{\textbf Z_{\ve^{(n)}\no\fa}}}\frac{\dd\Omega_\ell}{\innerp{\textbf z_\ell}{\textbf z_\ell}^2}\,\frac{\dd \Omega_{\ve\fa}}{\innerp{\textbf Z_{\ve\no'\fa}}{\textbf Z_{\ve\no'\fa}}}
\end{align}
associated to the vertices attached to the nodes $\no$, $\no'$ and to the link $\ell$. Moreover, the Lagrangian $L_\fa[g_{\ve\ed}, \textbf z_{\ve\fa}]$ is defined as in \eqref{eq:BulkLagrangian}, except that the sum over edges excludes the two edges attached to the nodes. These edges are accounted for in the newly defined boundary face Lagrangian
\begin{align}\label{eq:BoundaryFaceAction}
	B_\ell[g_{\ve\no}, \textbf{z}_\ell, h_\ell] &:= -i j_\ell \ln \frac{\innerp{\textbf{Z}_{\ve\no'\fa}}{\textbf z_\ell}^2}{\innerp{\textbf Z_{\ve\no'\fa}}{\textbf Z_{\ve\no'\fa}} \innerp{\textbf z_\ell}{\textbf z_\ell}} -i j_\ell \ln \frac{\innerp{h^\transpose_\ell \textbf z_\ell}{\textbf Z_{\ve^{(n)}\no\fa}}^2}{\innerp{\textbf z_\ell}{\textbf z_\ell}\innerp{\textbf Z_{\ve^{(n)}\no\fa}}{\textbf Z_{\ve^{(n)}\no\fa}}}\notag\\
	&\phantom{:=} + \gamma j_\ell  \ln\frac{\innerp{\textbf Z_{\ve^{(n)}\no\fa}}{\textbf Z_{\ve^{(n)}\no\fa}}}{\innerp{\textbf Z_{\ve\no'\fa}}{\textbf Z_{\ve\no'\fa}}}.
\end{align}
With this definition, the full CLQG propagator on a $2$-complex with boundary in its path integral form can finally be written as
\begin{align}\label{eq:PathIntegral}
	W_\mathcal{C}(h_\ell) = \mathcal N_\mathcal{C} \sum_{\{j_\fa\}} \int_{\slc}&\left(\prod_\ve \dd\arc{g}_{\ve\ed}\right)  \left(\prod_{\fa\in\mathcal C} d_{j_\fa}\prod_{\ed\in\fa}\frac{d_{j_\fa}}{\pi}\int_{\CP}\dd\tilde{\Omega}_{\ve\ed\fa}\right)\times\notag\\
	\times &\left(\prod_{\ell\in\Gamma}\frac{d^3_{j_\fa}}{\pi^3}\int_{(\CP)^3}\dd\tilde{\Omega}_{\no\ell\no'}\right)\e^{i \sum_{\fa\in\mathcal B} L_\fa + i \sum_{\ell\in\Gamma} B_\ell}.
\end{align}
For later convenience we also define the \textit{spin foam action}
\begin{equation}\label{eq:SpinFoamAction}
	S[g_{\ve\ed}, g_{\ve\no}, \textbf{z}_{\ve\fa}, \textbf{z}_\ell, h_\ell] :=  \sum_{\fa\in\mathcal B} L_\fa[g_{\ve\ed},\textbf{z}_{\ve\fa}] +  \sum_{\ell\in\Gamma} B_\ell[g_{\ve\no}, \textbf{z}_\ell, h_\ell].
\end{equation}
This action seems oddly asymmetric due to the presence of the $h^\transpose_\ell$ element in $B_\ell$ (see equation~\eqref{eq:BoundaryFaceAction}). This can in principle be remedied by arbitrarily decomposing $h_\ell$ into a product of two $SU(2)$ elements. This amounts to splitting the link into half links and makes \eqref{eq:BoundaryFaceAction} appear more symmetric. While nothing is gained by doing this splitting in the general case, it happens quite naturally when the propagator \eqref{eq:PathIntegral} is contracted with coherent semi-classical states, as we will see in the next subsection.

\subsection{Holomorphic Transition Amplitudes}
\label{ssec:HolomorphicTransitionAmplitude}
So far we were solely concerned with the propagator. However, what is of actual physical interest is the transition amplitude obtained from contracting the propagator with a boundary state. These amplitudes supposedly encode the full dynamics of quantum gravity.\\
In this subsection we will consider not just the contraction of~\eqref{eq:PathIntegral} with any boundary state, but more specifically with the coherent semi-classical states discussed in subsection~\ref{ssec:ExtrinsicStates}. These states offer the advantage that they are peaked on a discrete $3$-dimensional extrinsic and intrinsic geometry and therefore allow us to model physical scenarios such as the black hole to white hole transition \cite{RealisticObs, BH2WH} or cosmological transitions \cite{Bianchi:2010d}.\\
Contracting the propagator~\eqref{eq:PathIntegral} with a semi-classical coherent state~\eqref{eq:ExtrinsicState} results in what is known as a \textit{holomorphic transition amplitude} \cite{Bianchi:2010b, Perini:2012}:
\begin{equation}
	W_\mathcal{C}^{t_\ell}(H_\ell):= \left<W_\mathcal{C}\,\vline\,\Psi^{t_\ell}_{\Gamma, H_\ell}\right>:= \int_{SU(2)^L}\left(\prod_{\ell\in\Gamma}\dd h_\ell\right)\, W_\mathcal{C}(h_\ell) \, \Psi^{t_\ell}_{\Gamma, H_\ell}(h_\ell).
\end{equation}
In the above definition we dropped the $SU(2)$ integrals implicitly present in $\Psi^{t_\ell}_{\Gamma, H_\ell}$ as the $\slc$ integrals of $W_\mathcal{C}$ already take care of gauge-invariance.\\
Explicitly computing the above contraction is particularly easy as only the boundary face amplitude $A_\fa(h_\ell)$ times the state $\Psi^{t_\ell}_{\Gamma, H_\ell}$ need to be considered and the Peter-Weyl theorem can be employed. Hence, the relevant part of the computation yields
\begin{align}
	\int_{SU(2)^L}&\left(\prod_{\ell\in\Gamma} \dd h_\ell\right) A_\fa(h_\ell)\Psi^{t_\ell}_{\Gamma, H_\ell}(h_\ell) \notag\\
	&= \sum_{\{j_\ell\}}\prod_{\ell\in\Gamma}d_{j_\ell} \e^{-j_\ell(j_\ell+1) t_\ell}\Tr_{j_\ell}\left[Y^\dagger g_{\ve \no'}^{-1} g_{\ve\ed'} Y \left(\prod_{\ve\in\fa} Y^{\dagger} g^{-1}_{\ve\ed'}\,g_{\ve\ed} Y\right) Y^\dagger g_{\ve^{(n)}\ed^{(n)}}^{-1} g_{\ve^{(n)}\no} Y H_\ell^{-1}\right].
\end{align}
We observe that the integration simply exchanged the arbitrary $h_\ell^{-1}\in SU(2)$ with $H_\ell^{-1}\in \slc$, which is given by~\eqref{eq:HParam} and is completely determined by the boundary data\footnote{There is a slight abuse of notation here. Even though $H_\ell$ is in $\slc$, we are really talking about $D^j_{ab}(H^{-1}_\ell)$, where $D^j_{ab}$ is an analytically continued Wigner D-matrix.} $(\eta_\ell, \xi_\ell, \vec{n}_{s(\ell)}, \vec{n}_{t(\ell)})$. Moreover, the integration gave rise to a $\delta^{j_\fa j_\ell}$, which forces the spins $j_\fa$ which color the boundary face to be the same as the spins $j_\ell$ appearing in the boundary states and which live on the links.\newline
Rewriting this trace in terms of functions on $\mathcal H^{(\gamma j_\ell, j_\ell)}$ essentially involves the same steps as in the previous subsections. However, at this stage we will make use of an approximation. As we are interested in boundary states which are peaked on geometries with large areas, states with $\eta_\ell\gg 1$ that is, we will apply the highest weight approximation \cite{Bianchi:2010}:
\begin{align}
	D^{j_\ell}_{ab}(H^{-1}_\ell) = D^{j_\ell}(n_{s(\ell)}\e^{(\eta_\ell + i \gamma\zeta_\ell)\frac{\sigma_3}{2}}n^{-1}_{t(\ell)}) = D^{j_\ell}_{a j_\ell}(n_{s(\ell)}) D^{j_\ell}_{j_\ell b}(n^{-1}_{t(\ell)})\,\e^{(\eta_\ell + i \gamma\zeta_\ell)j_\ell}\left(1+\mathcal O(\e^{-\eta_\ell})\right)
\end{align} 
Notice that this approximation essentially amounts to splitting the link $\ell$ into two parts and $D^{j_\ell}_{a j_\ell}(n_{s(\ell)})$ is associated to the half link attached to the source node while $D^{j_\ell}_{j_\ell b}(n_{t(\ell)})$ is associated to the half link attached to the target node. These two $SU(2)$ Wigner matrices can be rewritten by making use of \cite{Barrett:2009}
\begin{align}\label{eq:ImportantRelation}
	\phi_{\;j}^{(\gamma j, j)}(n^\transpose\textbf{z}) = \sqrt{\frac{d_j}{\pi}}\innerp{\textbf{z}}{\textbf{z}}^{i\gamma j - j-1}\innerp{\overline{\textbf{n}}}{\textbf{z}}^{2j},
\end{align}
where $\textbf n$ is the spinor corresponding to the $SU(2)$ element $n$ (see \cite{Martin-Dussaud:2019} for the mathematical details). This yields
\begin{align}
	D^{j_\ell}_{a j_\ell}(n_{s(\ell)}) &= \frac{d_{j_\ell}}{\pi} \int_{\CP}\dd\Omega_{\no\ell}\,\innerp{\textbf z_{\no\ell}}{\textbf z_{\no\ell}}^{-2(j_\ell+1)}\innerp{\overline{\textbf{n}}_{s(\ell)}}{\textbf z_{\no\ell}}^{2j_\ell} \overline{P^{j_\ell}_a(\textbf z_{\no\ell})}\notag\\
	D^{j_\ell}_{j_\ell b}(n^{-1}_{t(\ell)}) &= \frac{d_{j_\ell}}{\pi} \int_{\CP}\dd\Omega_{\no'\ell}\,\innerp{\textbf z_{\no'\ell}}{\textbf z_{\no'\ell}}^{-2(j_\ell+1)}\innerp{\textbf z_{\no'\ell}}{\overline{\textbf{n}}_{t(\ell)}}^{2j_\ell} P^{j_\ell}_b(\textbf z_{\no'\ell}).
\end{align}
Repeating the same steps as in the previous subsections one arrives without much effort at
\begin{align}\label{eq:HolomorphicFaceAmplitude}
	\e^{-j_\ell(j_\ell+1) t_\ell}&\Tr_{j_\ell}\left[Y^\dagger g_{\ve \no'}^{-1} g_{\ve\ed'} Y \left(\prod_{\ve\in\fa} Y^{\dagger} g^{-1}_{\ve\ed'}\,g_{\ve\ed} Y\right) Y^\dagger g_{\ve^{(n)}\ed^{(n)}}^{-1} g_{\ve^{(n)}\no} Y H_\ell^{-1}\right]\notag\\
	&= \e^{\frac{(\eta_\ell-t_\ell)^2}{4t_\ell}}\left(\prod_{\ed\in\fa}\frac{d_{j_\ell}}{\pi}\int_{\CP}\dd\tilde{\Omega}_{\ve\ed\fa}\right)\left(\frac{d^4_{j_\ell}}{\pi^4}\int_{(\CP)^4}\dd\tilde{\Omega}_{s\ell t}\right)\,\e^{i(L_\fa + \tilde{B}_\ell + D_\ell)}\left(1+\mathcal O(\e^{-\eta_\ell})\right),
\end{align}
where we defined
\begin{align}
	\dd\tilde\Omega_{s\ell t} := \frac{\dd\Omega_{\ve^{(n)}\fa}}{\innerp{\textbf{Z}_{\ve^{(n)}\no\fa}}{\textbf{Z}_{\ve^{(n)}\no\fa}}}\,\frac{\dd\Omega_{\no\ell}}{\innerp{\textbf z_{\no\ell}}{\textbf z_{\no\ell}}^2}\,\frac{\dd\Omega_{\no'\ell}}{\innerp{\textbf z_{\no'\ell}}{\textbf z_{\no'\ell}}^2}\,\frac{\dd\Omega_{\ve\fa}}{\innerp{\textbf Z_{\ve\no'\fa}}{\textbf Z_{\ve\no'\fa}}}
\end{align}
plus the new boundary Lagrangian $\tilde{B}_\ell$ and the boundary data contribution $D_\ell$
\begin{align}\label{eq:GWeight}
	\tilde{B}_\ell[g_{\ve\no},\textbf{z}_{\ve\fa}, \textbf z_{\no\ell};\textbf n_{\no(\ell)}] &:= \gamma j_\ell \ln \frac{\langle \textbf{Z}_{\ve^{(n)}\no\ell} \vert \textbf{Z}_{\ve^{(n)}\no\ell}\rangle}{\langle \textbf{Z}_{\ve\no'\ell} \vert \textbf{Z}_{\ve\no'\ell} \rangle} - i j_\ell \ln \frac{\langle \textbf{Z}_{\ve\no'\ell} \vert\textbf{z}_{\no'\ell} \rangle^2 \langle \textbf{z}_{\no\ell}\vert \textbf{Z}_{\ve^{(n)}\no\ell} \rangle^2}{\langle \textbf{Z}_{\ve\no'\ell} \vert \textbf{Z}_{\ve\no'\ell} \rangle \langle \textbf{Z}_{\ve^{(n)}\no\ell} \vert \textbf{Z}_{\ve^{(n)}\no\ell} \rangle}
	\notag\\
	D_\ell[\textbf{z}_{\no\ell}; a_\ell, \xi_\ell, \textbf{n}_{s(\ell)}, \textbf{n}_{t(\ell)}]&:= -\gamma j_\ell \zeta_\ell  - i j_\ell \ln \frac{\langle \overline{\textbf{n}}_{s(\ell)}\vert \textbf{z}_{\no\ell} \rangle^2 \langle \textbf{z}_{\no'\ell} \vert \overline{\textbf{n}}_{t(\ell)} \rangle^2}{\innerp{\textbf{z}_{\no\ell}}{\textbf{z}_{\no\ell}}^2 \innerp{\textbf{z}_{\no'\ell}}{\textbf{z}_{\no'\ell}}^2} + i \left(j_\ell-a_\ell\right)^2 t_\ell.
\end{align}
The definitions of $\dd\tilde\Omega_{\ve\ed\fa}$ and $L_\ell[g_{\ve\ed}, \textbf z_{\ve\fa}]$ remain the same as in the previous subsection and the exponential pre-factor $\exp((\eta_\ell-t_\ell)^2/4t_\ell)$, which only depends on the data $\eta_\ell$ and the parameter $t_\ell$, arises from completing the square such that the Gaussian weight $\exp(-(j_\ell - a_\ell)^2 t_\ell)$ appears in \eqref{eq:GWeight}, with $a_\ell := \frac{\eta_\ell -t}{2t}$ (see subsection~\ref{ssec:ExtrinsicStates}). Absorbing the pre-factor into the normalization $\mathcal N_\mathcal{C}$ we finally arrive at the holomorphic transition amplitude in its path integral form:
\begin{align}\label{eq:HolomorphicAmplitude}
	W^{t_\ell}_\mathcal{C}(H_\ell) = \mathcal N_\mathcal{C} \sum_{\{j_\fa, j_\ell\}} \int_{\slc} & \left(\prod_\ve \dd\arc{g}_{\ve\ed}\right)\left(\prod_{\fa\in\mathcal C} d_{j_\fa}\prod_{\ed\in\fa}\frac{d_{j_\fa}}{\pi}\int_{\CP}\dd\tilde{\Omega}_{\ve\ed\fa}\right)\times\notag\\
	& \times\left(\prod_{\ell\in\Gamma}\frac{d^4_{j_\ell}}{\pi^4}\int_{(\CP)^4}\dd\tilde{\Omega}_{s\ell t}\right)\e^{i\sum_{\fa\in\mathcal B}L_\fa + i\sum_{\ell\in\Gamma}(\tilde{B}_\ell+D_\ell)}.
\end{align}
where for later convenience we also defined the \textit{holomorphic spin foam action}
\begin{equation}\label{eq:HolomorphicAction}
	A[g_{\ve\ed}, g_{\ve\no}, \textbf{z}_{\ve\fa}, \textbf{z}_{\no\ell} ;a_\ell, \xi_\ell, \textbf{n}_{s(\ell)}, \textbf{n}_{t(\ell)}] := \sum_{\fa\in\mathcal B} L_\fa+ \sum_{\ell\in\Gamma}\left(\tilde{B}_\ell + D_\ell\right) 
\end{equation}
The above amplitude and its dependence on the boundary data $(a_\ell, \xi_\ell, \textbf{n}_{s(\ell)}, \textbf{n}_{t(\ell)})$ will be the main focus of chapter~\ref{chapter_5}. There we will develop an approximation method which allows us to evaluate the holomorphic transition amplitude in its semi-classical regime. However, before we can proceed to this task we need to understand the propagators's classical limit.

\subsection{The Classical Limit and the Emergence of Geometry}
\label{ssec:ClassicalLimitGeometry}
When first confronted with the propagator~\eqref{eq:PathIntegral} one may wonder what it has to do with GR or even just pure geometry. It is completely expressed in terms of group variables and spinors and, even worse, all these variables are complex! Determining the classical limit of~\eqref{eq:PathIntegral} and establishing a connection with GR on a simplicial manifold is indeed a non-trivial task, albeit one that is very rewarding. After all, simplicial geometry miraculously emerges from~\eqref{eq:PathIntegral}!\\
In this subsection we will briefly review the main steps involved in the determination of the classical limit and simply state the main results. We refer the interested reader to the extensive literature \cite{Barrett:2009b, Barrett:2009,Conrady:2008,Han:2011b, Han:2011, HanKrajewski:2013} for the details. The results discussed here play an important role for the discussion that follows in subsection~\ref{ssec:ComputationBounceTime} and which will be continued in chapter~\ref{chapter_5}.

When trying to determine the classical limit of the quantum mechanical propagator, such as the one originally defined by Feynman \cite{Feynman:1948}, one compares the action $S[x]$ with $\hbar$ in the phase factor $\e^{\frac{i}{\hbar}S[x]}$. Assuming quantum effects are negligible when $S[x]$ is large in units of $\hbar$, we can effectively treat $\frac{1}{\hbar}$ as a large parameter and therefore apply a stationary phase approximation to the path integral.\\
In CLQG the idea is a similar one, but a small modification of the action \eqref{eq:SpinFoamAction} is necessary. Notice that every term in the Lagrangian \eqref{eq:BulkLagrangian} is multiplied by $\gamma j$. Using the LQG area spectrum in the large $j$ limit, we can rewrite this factor as
\begin{equation}
	A = \frac{8\pi\gamma G}{c^3} \hbar \sqrt{j(j +1)} \approx  \frac{8\pi\gamma G}{c^3} \hbar j \quad\Rightarrow\quad \gamma j = \frac{1}{\kappa} \frac{A}{\hbar},
\end{equation}
where $A$ denotes the area of a face or a link in the $2$-complex $\mathcal C$ and $\kappa = \frac{8\pi G}{c^3}$\footnote{Notice that this $\kappa$ in front of the action would be the wrong one for continuum GR, where its value should be $\kappa = \frac{16 \pi G}{c^4}$. But it has precisely the correct units and also the correct numerical factors for the action of (area) Regge calculus because $\sum_\fa A_\fa\theta_\fa \longrightarrow \frac{1}{2}\int \sqrt{-g} R \, \dd^4 x$ (see eg. \cite{Regge:1961, Sorkin:1975}).}. This trick allows us to introduce an area in the action \eqref{eq:SpinFoamAction} and $\hbar$. Hence, we can compare the action with $\hbar$ and, assuming quantum effects are negligible when the action is much larger than Planck's constant, we can attempt a stationary phase approximation. This approximation is applied to the integral over the $\slc$ elements and the spinorial variables, but not to the spin sums.\\
Instead of devising an approximation scheme for the spin sums, one usual considers the so-called \textit{partial amplitude} (in our terminology that would rather be the \textit{partial propagator}) which is simply the propagator~\eqref{eq:PathIntegral} without the spin sums. In order to apply the stationary phase approximation it is then assumed that all the spins are uniformly large which is the same as saying that the action \eqref{eq:SpinFoamAction} is much larger than $\hbar$. One then finds the following critical point equations \cite{HanKrajewski:2013}
\begin{align}\label{eq:CritEq}
	\delta_{g_{\ve\ed}} S \overset{!}{=} 0 & & \Leftrightarrow & & \sum_\fa j_\fa\frac{\langle \textbf{Z}_{\ve\ed\fa}| \vec{\sigma}|\textbf{Z}_{\ve\ed\fa}\rangle}{\|\textbf{Z}_{\ve\ed\fa}\|^2} & = 0\notag\\
	\text{Im}[S] \overset{!}{=} 0 & & \Leftrightarrow & & \frac{\textbf{Z}_{\ve\ed\fa}}{\|\textbf{Z}_{\ve\ed\fa}\|} &= \e^{i \alpha^\fa_{\ve\ve'}}\frac{\textbf{Z}_{\ve'\ed\fa}}{\|\textbf{Z}_{\ve\ed\fa}\|} \notag\\
	\delta_{\textbf{z}_{\ve\fa}}S \overset{!}{=} 0 & & \Leftrightarrow & & g_{\ve\ed}\frac{\textbf{Z}_{\ve\ed\fa}}{\|\textbf{Z}_{\ve\ed\fa}\|^2} &= g_{\ve\ed'}\frac{\textbf{Z}_{\ve\ed'\fa}}{\|\textbf{Z}_{\ve\ed'\fa}\|^2}
\end{align}
which have to be supplemented by boundary conditions. That is, we have to specify the boundary spins $j_\ell$ and the boundary spinorial variables $\textbf{z}_{\no\ell}$. This is equivalent to specifying the areas and unit normal vectors of all the boundary tetrahedra (which in turn is the same as fixing the intrinsic geometry of the boundary). The critical point equations~\eqref{eq:CritEq} supplemented by the boundary data $(j_\ell, \textbf{z}_{\no\ell})$ then provide us with a boundary value problem.\\
At first sight these equations may look obscure, but it turns out that they have a clear geometrical interpretation. The first equation is the easiest: Any spinor $\hat{\textbf{z}}$ which is normalized to unity with respect to the canonical scalar product on $\mathbb{C}^2$ is mapped to a unit $3$d vector through $\hat{\textbf{z}}^\dagger \vec{\sigma} \hat{\textbf{z}}$, where $\vec{\sigma}$ is the Pauli vector. Therefore, the first equation is simply the closure constraint which means that the critical point consists of a collection of tetrahedra. These tetrahedra have triangles with areas~$j_\fa$ and unit normal vectors $\frac{\langle \textbf{Z}_{\ve\ed\fa}| \vec{\sigma}|\textbf{Z}_{\ve\ed\fa}\rangle}{\|\textbf{Z}_{\ve\ed\fa}\|^2}$ attached to them. All tetrahedra are uniquely determined through the boundary data $(j_\ell, \textbf{z}_{\no\ell})$ up to inversion. We can also say that every edge of the $2$-complex corresponds to a tetrahedron when the critical point equations are satisfied. This is the first emergence of geometry from spin foams, but the question remains what type of geometry we are dealing with.\\
The second equation tells us that the spinors $\frac{\textbf{Z}_{\ve\ed\fa}}{\|\textbf{Z}_{\ve\ed\fa}\|}$ and $\frac{\textbf{Z}_{\ve'\ed\fa}}{\|\textbf{Z}_{\ve'\ed\fa}\|}$, which are both normalized to unity, agree up to a phase $\alpha^\fa_{\ve\ve'}$. Hence, they reproduce the same unit normal vector. Observe furthermore that these spinors are associated to different vertices, $\ve$ and $\ve'$, but they share a common face $\fa$ and an edge $\ed$. Since we have just established that edges are dual to tetrahedra and normalized spinors describe normal vectors to the triangles of these tetrahedra, we gain a notion of common frame and the second equation tells us (qualitatively, see \cite{Han:2011,HanKrajewski:2013} for details) how the tetrahedra are glued together.\\
The remaining critical point equation finally gives us a notion of parallel transport. By bringing $g_{\ve\ed}$ to the other side, we can interpret $g^{-1}_{\ve\ed}g_{\ve\ed'}$ as the $\slc$ transformation that takes us from the frame of $\ed$ to the frame of $\ed'$ within the same vertex $\ve$. This transformation can be used to define a (Lorentzian) dihedral angle between the normal vector to~$\ed$ and the normal vector to~$\ed'$, provided~$\ed$ and~$\ed'$ emanate from the same vertex. It can be shown~\cite{Barrett:2009,Han:2011} that this angle is given by
\begin{equation}
	\Theta^\fa_{\ed\ed'} + i\Pi^\fa_{\ed\ed'}\quad\text{with}\quad \Pi^\fa_{\ed\ed'}:=\begin{cases}
		\pi & \text{if } \ed \text{ and } \ed'\text{ span a thin wedge}\\
		0 & \text{if } \ed \text{ and } \ed'\text{ span a thick wedge}
		\end{cases},
\end{equation}
where $\Theta^\fa_{\ed\ed'}$ is a real number and the indices indicated that it is an angle between the normals to the tetrahedron $\ed$ and $\ed'$. This dihedral angle coincides precisely with the definition of Lorentzian angle given in subsection~\ref{ssec:LorentzianAngles}, including the thin wedge contribution $\Pi^\fa_{\ed\ed'}$ (see also Figure~\ref{fig:ThinThickWedges} a and b).\\
Using the third critical point equation in conjunction with the second one allows us to define a holonomy simply by multiplying all pairs  $g^{-1}_{\ve\ed}g_{\ve\ed'}$ around the face $\fa$. In \cite{Han:2011,HanKrajewski:2013} it is shown that this holonomy corresponds to a pure boost which leaves the face $\fa$ invariant and which amounts to the following sum of dihedral angles, which we call a \textit{Palatini deficit angle}:
\begin{equation}\label{eq:PalatiniDeficitAngle}
	\phi_\fa := \sum_{\ed\in\fa} \textsf{s}(\ed\cap\ed')\left(\Theta^\fa_{\ed\ed'} + i\Pi^\fa_{\ed\ed'}\right),
\end{equation}
where $\textsf{s}(\ed\cap\ed')\in\{-1,+1\}$. When $\textsf{s}(\ed\cap\ed')=1$ for all vertices of a face, we recover the definition of dihedral angle given at the end of subsection~\ref{ssec:LorentzianAngles} and we call it a~\textit{Regge deficit angle}.\\
The sign factor $\textsf{s}(\ve)$ can be traced back to the signed $4$-volume of the vertex $\ve:=\ed\cap\ed'$. By signed $4$-volume we mean
\begin{equation}\label{eq:FourVolume}
	\mathcal{V}^4(\ve) := e^{I}_{s_1}(\ve) e^{J}_{s_2}(\ve) e^{K}_{s_3}(\ve) e^{L}_{s_4}(\ve) \varepsilon_{IJKL},
\end{equation}
where $e^{I}_{s}$ are Minkowski vectors corresponding to the segment $s$ of the tetrahedron reconstructed from the boundary data $(j_\ell, \textbf{z}_{\no\ell})$ and the critical point equations \cite{Han:2011,HanKrajewski:2013}. The resulting $4$-volume~\eqref{eq:FourVolume} associated to the vertex $\ve$ can either be positive or negative, depending on how the Minkowski vectors are oriented. For notational convenience we can then define  $\textsf{s}(\ve) := \text{sign}(\mathcal V^4(\ve))$. These Minkowski vectors are of course just a discrete analogue of a co-tetrad field and the sign of the $4$-volume can therefore be understood as the sign of the determinant $\det e$ in the tetradic Palatini action~\eqref{eq:EH_vs_Palatini}. See \cite{Han:2011} and in particular \cite{Marios_phd} for a didactical discussion of this sign factor and its relation to the Palatini action.

For a given set of boundary data $(j_\ell, \textbf{z}_{\no\ell})$ there either exists a set of critical points $(\textbf{z}_{\ve\fa}, g_{\ve\ed})$ or there is no critical point. In the latter case, the propagator is exponentially suppressed. However, if there are critical points, there are several different geometries that can be realized. Following \cite{Barrett:2009, Dona:2019}, we distinguish between $4$-dimensional Lorentzian Regge geometries, $4$-dimensional Euclidean Regge geometries and vector geometries (which contain the geometries that in \cite{Barrett:2009} were called $3$-dimensional Regge geometries).\\
Notice also that we said there is a \textit{set} of critical points. This is indeed inevitable and one finds that whenever the boundary data $(j_\ell, \textbf{z}_{\no\ell})$ leads to a non-suppressed propagator, there are exactly $2^V$ critical points, where $V$ denotes the number of vertices of the spin foam. This is again related to the fact that at every vertex there are two differently oriented co-tetrad bases which are reconstructed by the boundary data and the critical point equations. See again \cite{Marios_phd} for a detailed and didactical explanation. Hence, when evaluating the spin foam action~\eqref{eq:SpinFoamAction} on a critical point, we need to distinguish between the different geometries\footnote{Here, we disregard vector geometries since they need a more careful discussion and they are not important for what follows.} and we need to keep track of the sign factor $\textsf{s}(\ve)$ since eventually we need to sum over all possible critical points. 
\begin{itemize}
	\item[1)] \textbf{$\mathbf{4}$d Lorentzian geometry:} The Hamilton-Regge function (i.e. the discrete action evaluated on a critical point) has the form
	\begin{equation}\label{eq:HamiltonRegge}
		S_\textsf{c} = \gamma\sum_{\fa\in\mathcal{B}}j_\fa \Theta_\fa + \sum_{\fa\in\mathcal{B}} j_\fa \Pi_\fa + \gamma \sum_{\ell\in\Gamma} j_\ell\, \textsf{s}\, \theta_\ell +\sum_{\ell\in\Gamma} j_\ell \Pi_\ell, 
	\end{equation}
	where we defined
	\begin{equation}
	\Theta_\fa := \sum_{\ed\in\fa} s(\ed\cap\ed') \Theta^\fa_{\ed\ed'}\quad\text{and}\quad \Pi_\fa := \sum_{\ed\in\fa} s(\ed\cap\ed')\Pi^\fa_{\ed\ed'}.
\end{equation}
The variable $\theta_\ell$ appearing in the boundary term is the angle between the $3$d vectors which are normal to the two triangles dual to the link $\ell$. Since these triangles belong to different tetrahedra which, in particular, can live in different light cones, there is also a term $j_\ell \Pi_\ell$ which accounts for thin wedges. That is, $\Pi_\ell$ is zero when the $3$d normal vectors span a thick wedge and $\Pi_\ell = \pi$ when the normal vectors span a thin wedge. 
\item[2)] \textbf{$\mathbf{4}$d Euclidean geometry:} The Hamilton-Regge function has the form
	\begin{equation}
		S_\textsf{c} = \gamma \sum_{\fa\in\mathcal{B}} j_\fa \Theta^\textsf{E}_\fa + \gamma \sum_{\ell\in\Gamma}j_\ell\,\textsf{s}\, \theta^\textsf{E}_\ell, 
	\end{equation}
	where $\Theta^\textsf{E}$ is a Euclidean deficit angle and $\theta^\textsf{E}_\ell$ is a Euclidean dihedral angle. 
\end{itemize}
In both discussed cases, the propagator takes the approximate form
\begin{equation}
	K \sim \sum_{\{\textsf{s}(\ve)\}}\frac{\e^{\frac{i}{\hbar}S_\textsf{c}}}{\sqrt{\det\textsf{Hess}}},
\end{equation}
where $\textsf{Hess}$ is the Hessian of the spin foam action and we deliberately neglected numerical factors, normalizations, phases and scalings. Interestingly, it has recently been shown that the Hessian for the Lorentzian CLQG propagator defined on a single vertex is non-degenerate~\cite{Kaminski:2019}.

\section{The Black Hole to White Hole Transition}
\label{sec:BH2WHTransition}
Black holes hide a singularity behind their trapping surface -- at least in theory. Singularities are hardly properties of black holes in Nature and it is generally believed that quantum gravity leads to a singularity-free description of collapsed stars. Such a singularity resolution has been observed in Loop Quantum Cosmology (LQC) and, inspired by this result, Rovelli and Vidotto \cite{Rovelli:2014} suggested that quantum gravitational effects not only prevent collapsing stellar matter from being compressed to a point, hence avoiding the black hole singularity from forming, but might even develop an effective repulsive force. Ultimately, this force would disrupt the collapsing star and lead to an explosion.\\
Remarkably, the idea that black holes might explode had already been put forward by Hawking in 1974 \cite{Hawking:1974} and 30 years later Ambrus and H\'{a}j\'{i}\v{c}ek \cite{Ambrus:2005}, who studied the quantization of spherical thin shells of null dust, discovered that the quantum theory not only contains collapsing solutions, but also expanding shells.\\ 
It was however not until 2014 that the idea of exploding black holes started to enjoy more interest from a wider range of researchers. In \cite{Haggard:2015}, Haggard and Rovelli studied the black to white hole transition from a semi-classical perspective and they found that a spacetime which describes this process and which satisfies the classical Einstein field equations everywhere except in a small region around the trapping horizon can be constructed. De Lorenzo and Perez \cite{DeLorenzo:2016} also considered a variation of the spacetime proposed by Haggard and Rovelli which accounts for the Eardley \cite{Eardley:1974} instability of white holes and both research groups defined an observable which, as particularly emphasized in \cite{TimeScale}, is well-defined and observer-independent.\\
The construction of the fireworks spacetime from the perspective of \cite{TimeScale} and the definition of the aforementioned observable will be the main focus of the next two subsections.

\subsection{Definition of the Model}
\label{ssec:DefOfModelAndBounceTime}
\begin{center}
	\begin{figure}[h] 
	\centering
	\includegraphics[scale=.90]{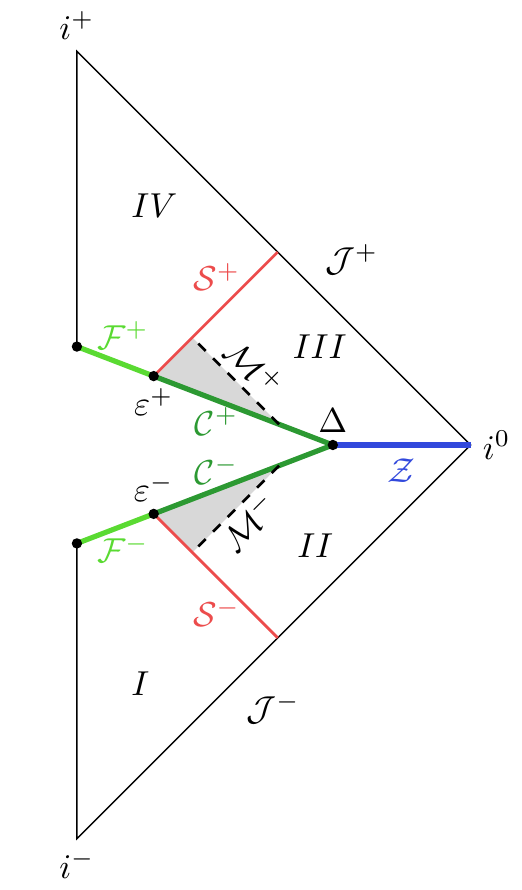} 
	\caption{The fireworks spacetime.}
	\label{fig:ansatz}
\end{figure}
\end{center}

We wish to construct a spacetime where there is a transition of a trapped region, formed by collapsing matter, to an anti-trapped region from which matter is being released. The transition is induced by quantum gravitational effects which are non-negligible only in a finite spatio-temporal region.\\
This quantum region is excised from the spacetime by introducing a spacelike compact interior boundary. Outside this region the metric solves Einstein's field equations exactly everywhere, including on the interior boundary. This is the so-called fireworks spacetime.\\
The construction of this spacetime presented in \cite{TimeScale} is based on the following simplifying assumptions:
\begin{itemize}
	\item Collapse and expansion of matter are modeled by thin shells of null dust of constant mass $m$. 
	\item Spacetime is spherically symmetric.
	\item Hawking radiation is being neglected\footnote{See \cite{Martin-Dussaud:2019b} for a recent model where this limitation has been removed.}. 
\end{itemize}
These assumptions determine the local form of the metric by virtue of Birkhoff's theorem \cite{HawkingEllisBook}, which in particular implies that the fireworks spacetime is locally but not globally isomorphic to portions of the Kruskal spacetime. Moreover, it follows that the geometry inside the null shells is Minkowski, while the geometry outside the shells is locally Kruskal with $m$ being the mass of the shells and spacetime is asymptotically flat.\\
An ansatz for a Carter-Penrose diagram of a fireworks spacetime is given in Figure~\ref{fig:ansatz}. In the construction that follows, we express the metric, the energy--momentum tensor and the expansions of null geodesics in Eddington-Finkelstein coordinates. For an equivalent reformulation in terms of Kruskal coordinates and the relation to the original construction of \cite{Haggard:2015} see \cite{TimeScale}.\\
The task is now to construct a metric such that the surfaces and regions in Figure~\ref{fig:ansatz} have the following properties:
\begin{itemize}
	\item $\mathcal{S}^-$ and $\mathcal{S}^+$ are null hypersurfaces and the junction condition \cite{Vaidya:1951,PoissonBook} on the intrinsic metric holds. In particular this means that there is an allowed discontinuity in their extrinsic curvature which results in a distributional contribution to $T_{\mu\nu}$ on $\mathcal{S}^-$ and $\mathcal{S}^+$ (see bellow). The energy-momentum tensor $T_{\mu\nu}$ vanishes everywhere else in the spacetime.
	\item The surfaces $\mathcal{F^+}$, $\mathcal{F^-}$, $\mathcal{C^+}$, $\mathcal{C^-}$ depicted in Figure~\ref{fig:ansatz} are spacelike. Their union $\mathcal{B} \equiv \mathcal{F}^-\cup\mathcal{C}^-\cup\mathcal{C}^+\cup\mathcal{F}^+$ constitutes the interior boundary $\mathcal{B}$. The intrinsic metric is matched on the spheres $\Delta$ and $\varepsilon^{\pm}$. Since the vertical lines from $i_+$ and $i_-$ are zero radius lines, the topology of the boundary $\mathcal{B}$ is that of a three-dimensional ball. The extrinsic curvature is discontinuous on $\varepsilon^{\pm}$, as discussed in the previous point, and is also discontinuous on $\Delta$ because of the requirement that $C^\pm$ are spacelike: A ball in Lorentzian space whose boundary is spacelike necessarily has corners, where the normal to the surface jumps from being future oriented to being past oriented.  
	\item The hypersurface $\mathcal{Z}$ is spacelike. The junction conditions for both the intrinsic metric and extrinsic curvature hold, including on the sphere $\Delta$. As we will see below, $\mathcal{Z}$ plays only an auxiliary role and need not be further specified. 
	\item $\mathcal{M^-}$ and $\mathcal{M^+}$ are marginally trapped (anti-trapped) surfaces and the shaded regions are trapped (anti-trapped). That is, the expansion of outgoing (ingoing) null geodesics vanishes on $\mathcal{M^-}$ ($\mathcal{M^+}$), is negative inside the shaded regions and positive everywhere else in the spacetime. 
\end{itemize}
Before explicitly constructing the metric, let us comment on the necessity of extending the interior boundary outside the (anti-)trapped regions. By Birkhoff's theorem, and as noted above, the marginally trapped and anti-trapped surfaces $\mathcal{M^-}$ and $\mathcal{M^+}$ can only be realized as being portions of the $r=2m$ surfaces of the Kruskal spacetime. If these do not meet the interior boundary, they must run all the way to null infinity. Thus, in order to have a consistent physical picture of the spacetime far away from the transition region, we must allow for non negligible quantum gravitational effects taking place in the vicinity, and crucially, outside, the (anti-)trapped surfaces.
Now we turn to the explicit construction of the fireworks metric in Eddington-Finkelstein (EF) coordinates. The union of the regions I and II of Figure~\ref{fig:ansatz} is coordinatized by ingoing EF coordinates $(v, r)$ while the union of the regions III and IV is described by outgoing EF coordinates $(u, r)$. The only junction condition to be considered is the one on $\mathcal{Z}$, which we describe bellow. It is easy to see that the radial coordinate $r$ will be trivially identified in the two coordinate systems. For the regions $\text{I}\cup \text{II}$ and $\text{III}\cup \text{IV}$ the metric reads
\begin{align} \label{eq:downMetricEF}
	\text{I}\cup \text{II}:& &\dd s^2 &= -\left(1-\frac{2m}{r} \Theta(v-v_{\mathcal{S}^-}) \right)\dd v^2 + 2 \dd v \, \dd r+ r^2 \dd \Omega^2\notag\\
	\text{III}\cup \text{IV}:& &\dd s^2 &= -\left(1-\frac{2m}{r} \Theta(u-u_{\mathcal{S}^+})\right)\dd u^2 -2\dd u\, \dd r + r^2 \dd \Omega^2,
\end{align}
where $\Theta$ is the Heaviside step function. The ingoing and outgoing EF times $v_{\mathcal{S}^-}$ and $u_{\mathcal{S}^+}$ denote the position of the shells $\mathcal{S^-}$ and $\mathcal{S^+}$ on $\mathcal I^-$ and $\mathcal I^+$ in these coordinates. As mentioned above, the two junction conditions on $\mathcal{Z}$ are satisfied by the identification of the radial coordinate along $\mathcal{Z}$ and the condition
\begin{equation}\label{eq:junctT}
	v - u \stackrel{\mathcal{Z}}{=} 2r^\star(r),
\end{equation}
where $r^\star(r)=r + 2m \ln{\left|\frac{r}{2m}-1 \right|}$. Notice that this relation is the usual coordinate transformation between $(v,r)$ and $(u,r)$. We also recall that the EF times are defined as $v=t+r^\star(r)$ and $u=t-r^\star(r)$, where $t$ is the Schwarzschild time.\\
Next we need to specify the range of the coordinates. Assume an explicit choice of boundary surfaces $\mathcal{B}$ has been given. Having covered every region of the spacetime by a coordinate chart, we can describe embedded surfaces. Since all surfaces $\Sigma$ appearing in Figure~\ref{fig:ansatz} are spherically symmetric, it suffices to represent the surfaces as curves in the $v-r$ and $u-r$ planes. Using a slight abuse of notation we write $v=\Sigma(r)$ or, in parametric form, $(\Sigma(r), r)$. The range of coordinates in $\text{I}\cup \text{II}$ and $\text{III}\cup \text{IV}$ is then given by the following conditions:
\begin{align}
	\text{I}\cup \text{II}:& & v&\in\mathbb R &  r&\in\mathbb R^+  & v&\leq \mathcal F^-(r) &  v&\leq \mathcal C^-(r) &  v&\leq \mathcal Z(r) \notag\\
	\text{III}\cup \text{IV}:& &  u&\in\mathbb R  & r&\in\mathbb R^+  & u&\geq \mathcal F^+(r) &  u&\geq \mathcal C^+(r) &  u&\geq\mathcal Z(r).
\end{align}
What remains to be checked is the presence of trapped and anti-trapped regions, as required by the ansatz Carter-Penrose diagram of Figure~\ref{fig:ansatz}. This is equivalent to the geometrical requirement that the spheres $\varepsilon^\pm$ have proper area less than $4 \pi (2m)^2$ while the sphere $\Delta$ has proper area larger than $4 \pi (2m)^2$. We may write this in terms of the radial coordinate as
\begin{equation}\label{eq:radiusRequirements}
	r_{\varepsilon^\pm} < 2m\quad\text{and}\quad r_{\Delta} > 2m.
\end{equation} 
Apart from this requirement, the areas of the spheres $\epsilon^\pm$ and $\Delta$ are left arbitrary. Since $\varepsilon^\pm$ and $\Delta$ are specified once the boundary is explicitly chosen, this is a condition on the allowed boundary surfaces that can be used as an interior boundary of a fireworks spacetime: $\mathcal{C^\pm}$ can be any spacelike surfaces that have their endpoints at a radius less and greater than $2m$, respectively, and which intersect in the latter endpoint. Since $\mathcal{C}^\pm$ are spacelike, it follows that we necessarily have a portion of the (lightlike) $r=2m$ surfaces in the spacetime along with trapped and anti--trapped regions. The conditions
\begin{equation}\label{eq:posDeltas}
	v_\Delta  \geq  v_{\mathcal{S}^-}\quad\text{and}\quad u_\Delta\leq u_{\mathcal{S}^+}
\end{equation}
for the coordinates of the sphere $\Delta$ follow from equation \eqref{eq:radiusRequirements} and the fact that $\mathcal{C}^\pm$ are taken spacelike. This completes the construction of the fireworks spacetime metric.\\
Let us point out that this spacetime is really a two--parameter family of spacetimes in the following sense. The geometry of the spacetime, up to the choice of the interior boundary $\mathcal{B}$, is determined once two dimension-full but coordinate independent quantities are specified. One parameter is the mass $m$ of the null shells $\mathcal{S}^\pm$ while the second parameter is the \emph{bounce time} $T$. We can express $T$ in terms of $u_{\mathcal{S}^+}$ and $v_{\mathcal{S}^-}$ simply by 
\begin{equation} \label{eq:BounceTimeEF}
T = u_{\mathcal{S}^+}-v_{\mathcal{S}^-}.
\end{equation}
As with the mass $m$, the bounce time $T$ is taken to be positive. Then, the fireworks geometry has \emph{two} characteristic physical scales: a length scale $G m/c^2 $ and a time scale $G T/c^3$, where we momentarily reinstated the gravitational constant $G$ and the speed of light $c$.\\ 
The role of the bounce time $T$ as the second spacetime parameter is obscure in the line elements \eqref{eq:downMetricEF}, but it can be made more explicit by performing the shift 
\begin{equation} \label{eq:timeShift}
	v \rightarrow v- \frac{v_{\mathcal{S}^-}+u_{\mathcal{S}^+}}{2}\quad\text{and}\quad u \rightarrow u- \frac{v_{\mathcal{S}^-}+u_{\mathcal{S}^+}}{2}
\end{equation}
This is an isometry, since $(\partial_v)^\alpha$ and $(\partial_u)^\alpha$ are the timelike Killing fields in each patch. It simply amounts to shifting simultaneously the origin of the two coordinates systems. The line elements \eqref{eq:downMetricEF} now read
\begin{align}\label{eq:downMetricNew}
	\text{I}\cup \text{II}:& &\dd s^2 &=  - \left(1- \frac{2m}{r}\,\Theta\left(v+\frac{T}{2}\right) \, \right)\dd v^2 + 2 \dd v \, \dd r+ r^2 \dd \Omega^2 \notag\\
	\text{III}\cup \text{IV}:& &\dd s^2 &= -\left(1-\frac{2 m}{r}\, \Theta\left(u-\frac{T}{2}\right) \, \right)\dd u^2 - 2 \dd v \, \dd r+ r^2 \dd \Omega^2,
\end{align}
which makes the role of $T$ as a spacetime parameter manifest. It is impossible to make both $v_{\mathcal{S}^-}$ and $u_{\mathcal{S}^+}$ disappear from the line elements by shifting the origins of the coordinate charts, the best we can do is remove one of the two or, as we did above, a combination of them. This observation emphasizes that the bounce time $T$ is a free parameter of the spacetime. Notice that the junction condition \eqref{eq:junctT} is unaffected by a simultaneous shifting of the form \eqref{eq:timeShift}.\\
Due to the junction conditions, one finds \cite{PoissonBook, blau} that the field equations are solved for the distributional energy momentum tensor 
\begin{align}
	& & \text{I}\cup \text{II}:& &T_{\mu\nu} &= \phantom{+} \frac{1}{4 \pi r^2}  \, \delta^v_\mu \delta^v_\nu\, \delta\left(v+\frac{T}{2}\right)& & \notag\\
	& & \text{III}\cup \text{IV}:& &T_{\mu\nu} &= - \frac{1}{4 \pi r^2}\, \delta^u_\mu \delta^u_\nu\,  \delta\left(u-\frac{T}{2}\right) & &
\end{align}
The expansion $\theta^-$ of outgoing null geodesics in the patch $\text{I}\cup\text{II}$ and the expansion $\theta^+$ of ingoing null geodesics in the patch $\text{III}\cup\text{IV}$ read 
\begin{align}
	\text{I}\cup \text{II}:& &\theta^- \equiv \nabla_\mu k^{-\mu} &= \Gamma^- \left(1- \frac{2 m}{r} \Theta\left(v+\frac{T}{2}\right)\right) \notag\\
	\text{III}\cup \text{IV}:& &\theta^+ \equiv \nabla_\mu k^{+\mu} &= -\Gamma^+ \left(1- \frac{2 m}{r} \Theta\left(u-\frac{T}{2}\right)\right)
\end{align}
where $k^{-\mu}$ and $k^{+\mu}$ are affinely parametrized tangent vectors of the null geodesics and $\Gamma^\pm$  are positive scalar functions which we will not need here, see \cite{PoissonBook, blau} for details. From these expressions, it follows that the spacetime possesses a trapped and an anti--trapped surface, defined as the locus where the expansions $\theta^-$ and $\theta^+$ vanish respectively, and which we identify with $\mathcal{M^-}$ and $\mathcal{M^+}$ in Figure \ref{fig:ansatz}. Thus, in EF coordinates, $\mathcal{M^\pm}$ are given by 
\begin{align}
	& &\mathcal{M^-}:& &  r=2m &\quad\text{and}\quad v\in \left(-\frac{T}{2}, \,\mathcal{C^-}(2m)\right) & & \notag\\
	& &\mathcal{M^+}:& &  r=2m &\quad\text{and}\quad u\in \left(\mathcal{C^+}(2m), \,\frac{T}{2}\right) & & 
\end{align}
As explained above, by the requirement $r_{\varepsilon^\pm} < 2m$ and $r_{\Delta} > 2m$, it will always be the case that the surfaces $\mathcal{M^\pm}$ are present in the spacetime, along with trapped and anti-trapped regions where $\theta^\pm$ are negative. We may explicitly describe the  trapped region as the intersection of the conditions $r<2m$, $v\in (-T/2,\mathcal{C^-}(2m))$, and $v \leq \mathcal{C^-}(r)$. Similarly, the anti-trapped region is given by  $r<2m$, $u\in \left(\mathcal{C^+}(2m),T/2\right)$ and $u \geq \mathcal{C^+}(r)$. The expansions $\theta^\pm$ are positive in the remaining spacetime.

\subsection{The Bounce Time $T$} 
\label{sec:BounceTime}
As we have seen, the fireworks spacetime is characterized by two parameters, one of which is the bounce time $T$. Intuitively, $T$ controls the time separation between the two shells. More precisely, as explained in \cite{Haggard:2015}, the bounce time has an operational meaning in terms of the proper time along the worldline of a stationary observer. That is, of an observer at a constant radius $r=r_0$, measuring the proper time $\tau_{r_0}$ between the events at which the worldline intersects the collapsing and expanding shells $S^\pm$. A straightforward calculation yields 
\begin{equation} 
\tau_{r_0}= \sqrt{1-\frac{2m}{r_0}} \left(u_{\mathcal{S}^+}-v_{\mathcal{S}^-} + 2 r^\star(r_0)\right).
\end{equation}
Note that to derive this expression we must add the contributions from the two line elements \eqref{eq:downMetricNew} and use the junction condition  \eqref{eq:junctT}. Using equation \eqref{eq:BounceTimeEF}, we have
\begin{equation} \label{eq:bouncetimeR}
T = \frac{\tau_{r_0}}{\sqrt{1-\frac{2m}{r_0}}}-2 r^\star(r_0). 
\end{equation}
Thus, the bounce time $T$ can be measured by an observer, provided the mass $m$ and the (coordinate) distance $r_0$ from the hole are known.\\ 
Let us now rephrase equation \eqref{eq:bouncetimeR} to emphasize the role of $T$ as a spacetime parameter, a coordinate and observer independent quantity, and its relation with the symmetries of the spacetime. The exterior spacetime described by the fireworks metric has the three Killing vector fields of a static spherically symmetric spacetime, a timelike Killing vector field generating time translation and two spacelike Killing vector fields that together generate spheres. The orbits $\Upsilon$ of the timelike Killing field are labelled by an area $A_\Upsilon$: The proper area of a sphere generated by the two spacelike Killing vector fields on any point on $\Upsilon$. This is of course the geometrical meaning of the coordinate $r$.\\ 
We can thus avoid mentioning any coordinates or observers and specify $T$ through the following geometrical construction. Consider any orbit $\Upsilon$ that does not intersect with the interior boundary surfaces $\mathcal{B}$. The proper time $\tau_\Upsilon$ is an invariant integral evaluated on the portion of $\Upsilon$ that lies between its intersections with the null hypersurfaces $S^\pm$. For any such $\Upsilon$, we have
\begin{equation} \label{eq:defBounceTime}
T = \frac{\tau_\Upsilon}{\sqrt{1-\frac{2m}{r_\Upsilon(A_\Upsilon)}}}-r^\star(A_\Upsilon). 
\end{equation}   
The bounce time $T$ is independent of the chosen orbit $\Upsilon$ and it is expressed only in terms of invariant quantities -- a proper area and a proper time. This expression can be taken to be the \emph{definition} of $T$.\\
In the next subsection we present a preliminary computation of $T$ from CLQG.

\subsection{CLQG Computation of the Black Hole to White Hole Transition}
\label{ssec:ComputationBounceTime}
\begin{center}
	\begin{figure}[htb]
		\centering
		\includegraphics[width=0.9\textwidth]{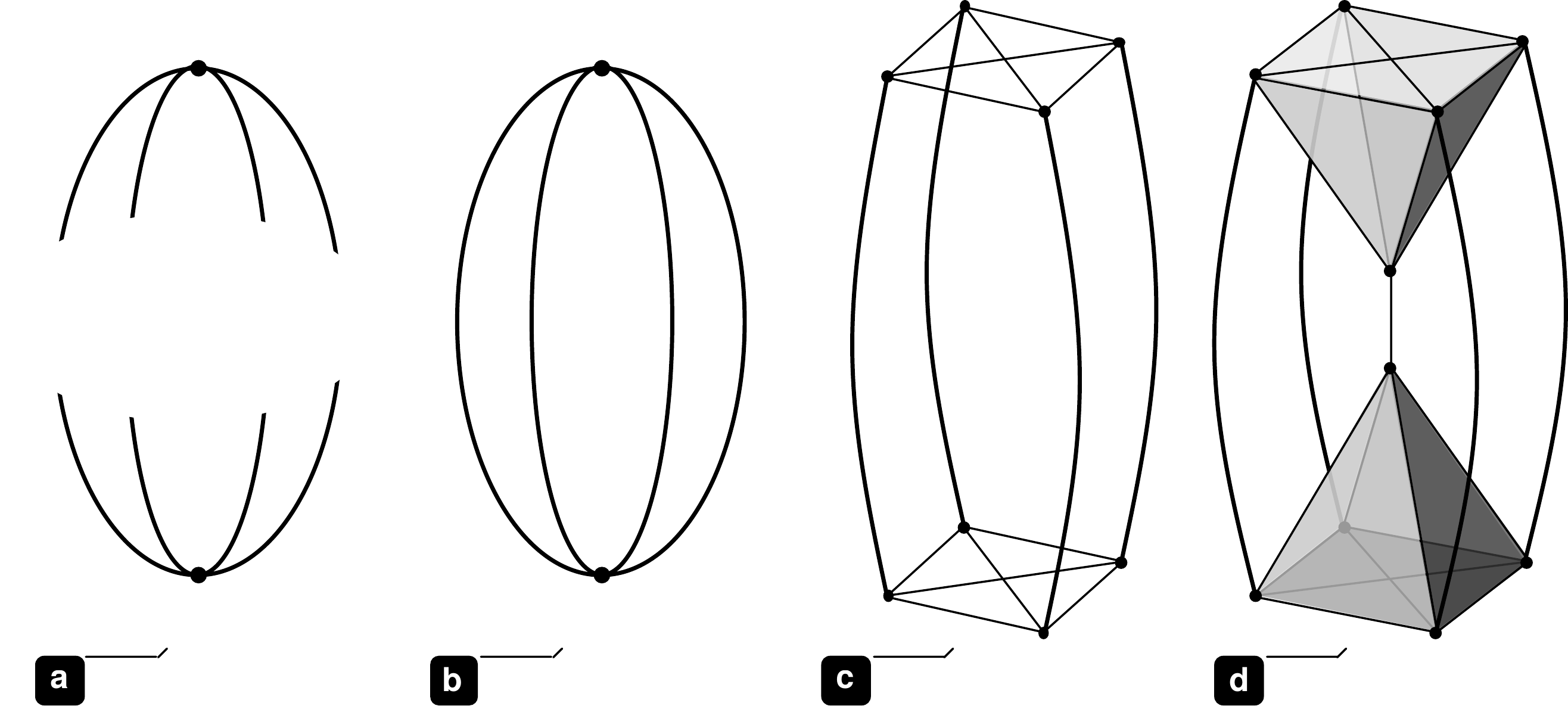}
		\caption{Step by step construction of the spin foam used in~\cite{RealisticObs} to model the black hole to white hole transition in CLQG.}
		\label{fig:SpinFoam}
	\end{figure}
\end{center}

Loosely speaking, the research efforts concerning the black hole to white hole transition model can be segmented into two categories: phenomenology and quantum gravity computations.\\
Soon after the inception of the model it was conjectured that the existence of primordial black holes might lead to observable signals in the skies and these could even be related to the as yet unexplained phenomenon of fast radio bursts \cite{Barrau:2014,Barrau:2014b,Barrau:2016,Barrau:2016b,Vidotto:2016}. More recent phenomenological models also attempt to account for dark matter through long-lived remnants resulting from the black hole to white hole transition~\cite{Remnants, Carballo-Rubio:2018,Rovelli:2018a,Rovelli:2018b,Rovelli:2018c}. Incidentally, these models also provide an answer to the information puzzle.\\
From the pure quantum gravity perspective, various research groups have tried to model the transition using different methods such as Euclidean path integrals \cite{Barcelo:2014a,Barcelo:2014b,Barcelo:2015}, symmetry reduced LQG models \cite{Corichi:2015, Ashtekar:2018}, Quantum Reduced Loop Gravity (QRLG) \cite{Alesci:2018,Alesci:2019}, and CLQG transition amplitudes \cite{RealisticObs}.\\
The strategy employed in \cite{RealisticObs} was to choose Lema\^{i}tre coordinates to describe the boundary $\mathcal B$ by two intersecting constant Lema\^{i}tre-time hypersurfaces. As it turns out, these hypersurfaces are intrinsically flat and topological $3$-balls, which greatly simplifies the triangulation of these surfaces. The simplest triangulation of a $3$-dimensional ball is of course in terms of a single (equilateral) tetrahedron. There are two tetrahedra, one for the $\mathcal F^-\cup\mathcal C^-$ part of the boundary and an other one for the $\mathcal F^+\cup\mathcal C^+$ part. Figure~\ref{fig:SpinFoam} a illustrates the two tetrahedra as nodes (i.e. we are working in the dual framework) with four emanating links which represent the triangles which constitute the tetrahedra.\\
The two hypersurfaces intersect each other in a $2$-sphere. Since the boundary of a tetrahedron is topologically a $2$-sphere, one can identify the four triangles of the ``past'' tetrahedron with the four triangles of the ``future'' tetrahedron. This is shown in Figure~\ref{fig:SpinFoam} b.\\ 
At this point it is convenient to split the two tetrahedra into four equal isosceles tetrahedra each. In the dual picture, this amounts to splitting the two nodes into four new nodes each which are interconnected in a specific way. Moreover, every one of the new nodes has a boundary triangle which is dual to a link which connects past and future tetrahedra. It is easy to see that this gives rise to the graph shown in Figure~\ref{fig:SpinFoam} c. This is the graph of the boundary triangulation.\\ 
Generalizing the boundary triangulation to a $4$-dimensional triangulation is easily achieved by adding two vertices to the graph shown in Figure~\ref{fig:SpinFoam} c and connecting them to the nodes and to each other by edges. This results in the spin foam represented in Figure~\ref{fig:SpinFoam} d.  
It consists of two vertices, dual to two four-simplices, nine edges (since one edge is used to connect the two vertices), eight nodes which are interconnected by $16$ links and $20$ faces. Notice that due to the simplicity of this spin foam, there are no internal faces. That is, all faces are adjacent to the boundary.\\
The discrete geometry described by this triangulation was studied in detail in \cite{RealisticObs}, which was a necessary step toward determining the boundary data for the semi-classical coherent states~\eqref{eq:ExtrinsicState}. It was found that the discrete extrinsic curvature on the top and bottom links of the boundary graph is given by purely numerical factors while all areas on all links only depend on the mass. For the discrete extrinsic curvature on the links connecting the top nodes to the bottom nodes the authors of \cite{RealisticObs} found that it depends on the spacetime parameters $m$ and~$T$. More precisely,
\begin{equation}
	\xi_0 = \frac{T}{2m}.
\end{equation} 
In the same paper, the amplitude corresponding to this spin foam was computed, albeit in a rather formal way. That is to say that it proved to be extremely difficult to extract physical predictions, such as the scaling of the bounce time with the black hole mass, from the formal expression found in \cite{RealisticObs}, despite the simplifications brought about by the chosen spin foam.\\
These difficulties initiated two independent research directions at the CPT Marseille. There is now considerable effort toward developing fast and efficient numerical methods to compute CLQG transition amplitudes and many exciting results have already been obtained \cite{Gozzini:2019,Dona:2019,Dona:2017,Dona:2018,Sarno:2018}.\\
The second direction is concerned with developing approximation methods to compute CLQG amplitudes analytically and the first results were obtained while trying to evaluate the black hole to white hole transition amplitude. We shall review these results here before generalizing them in chapter~\ref{chapter_5}.\\
 The strategy employed in~\cite{BH2WH,Marios_phd} is to use the holomorphic transition amplitude~\eqref{eq:HolomorphicAmplitude} in its general form rather than in its very specialized form derived in~\cite{RealisticObs}. Furthermore, it was assumed that the amplitude is defined on a $2$-complex without internal faces, just as for the spin foam in Figure~\ref{fig:SpinFoam} d, which is dual to a simplicial triangulation. In a strongly simplified notation we may therefore write
 \begin{equation}\label{eq:ExtremelySimpleAmplitude}
 	W^t_{\mathcal C}(H_\ell) = \sum_{j_\ell} \nu(j_\ell) \int_{\slc}\dd\mu(g) \int_{\CP} \dd\rho(\textbf{z}) \e^{i A[g, \textbf{z}; a,\xi, \textbf{n}_s, \textbf{n}_t]},
 \end{equation} 
 where $\nu(j)$ is a certain polynomial of boundary spins, $\dd\mu(g)$ is an appropriately regularized product of $\slc$ measures, $\dd\rho(\textbf{z})$ refers to a product of certain homogeneous and $\slc$ invariant measures on $\CP$, $g\in\slc$, $(a,\xi,\textbf{n}_s, \textbf{n}_t)$ is the boundary data, and $\textbf{z}\in\mathbb C^2$ denotes auxiliary spinorial variables. (see \eqref{eq:HolomorphicAmplitude} for the details). In the sequel we will assume $t$ to be a small parameter, so that the boundary states are semi-classical, and the dimensionless area parameters $a_\ell$ to be all of the same order of magnitude and much larger than unity. \\
 Notice that assuming there are no internal faces effectively removes all spin sums in the bulk and only leaves us the spin sums stemming from the boundary states which are regulated by the weight factors $\exp(-(j_\ell-a_\ell)^2 t)$ (these factors are now hidden inside the action $A$). Due to the exponential suppression behavior of the weight factors, we can truncate\footnote{That this is a good approximation of the actual sum follows from the fact that the partial amplitude is an oscillating and finite function of the spins. The Gaussian weights therefore strongly dominate. Further justification is provided by the procedure performed in \cite{Han_2016}, where the author introduced a regulator $\sim\e^{-j}$ to study phase transitions in large spin foams. Here, the coherent states naturally provide us with the even stronger regulator $\sim\e^{-j^2}$.} the sums and only consider a small summation domain around the peak values $j_\ell = a_\ell$. More precisely, we restrict the spin sums to
\begin{equation}
	D^k_{a} := \underset{\ell}{\largetimes}\left\{\floor*{a_\ell - \frac{k}{\sqrt{2t}}}, \floor*{a_\ell + \frac{k}{\sqrt{2t}}}\right\} \quad \text{with }\quad 0<k\in \mathbb{N}.
\end{equation}
The symbol $\floor*{x}$ denotes the floor function which here is defined to be the largest \textit{half integer} number equal to or less than $x$. The parameter $k$ acts as cut-off and it measures how many standard deviations $\sigma=1/\sqrt{2t}$ the summation moves away from the peak $a_\ell$. An issue that needs to be addressed is the fact that the spin sums cannot immediately be treated as independent. In fact, the summand vanishes when the triangle inequalities among the spins joining at a node are not satisfied. Put differently, the summand in~\eqref{eq:ExtremelySimpleAmplitude} vanishes whenever any one of the intertwiner spaces associated to the nodes of the $2$-complex is of dimension zero. Therefore, in order to treat the sums as independent, they need to be restricted to spin configurations for which the intertwiner space is always non-trivial. To implement this requirement we use the fact that the nodes of the $2$-complex are four-valent, since we assumed the $2$-complex to be dual to a simplicial triangulation, and we define the set
\begin{align}\label{eq:OmegaGamma}
	D_{\Gamma} &:= \left\{\{j_\ell\} \bigg\vert \dim\text{Inv}_{SU(2)}\left[\overset{4}{\underset{\ell=1}{\bigotimes}} \mathcal{H}_{j_\ell} \right] > 0 \quad\forall \no \in \Gamma\right\} \notag\\
	&\text{ }=\left\{\{j_\ell\} \bigg\vert \min\left(j_1+j_2, j_3+j_4\right) - \max\left(\vert j_1 - j_2\vert, \vert j_3 - j_4\vert\right) + 1 > 0 \quad\forall \no \in \Gamma\right\}.
\end{align}
This is the set of all spin configurations $\{j_\ell\}$ for which the intertwiner spaces over the whole boundary graph $\Gamma$ are non-trivial. To adequately truncate the spin sums we must now choose the cut-off parameter $k$ such that 
\begin{align}\label{eq:Implication}
	\{j_\ell\}\in D^k_a\subseteq D_\Gamma.
\end{align}
To rewrite this condition, it is convenient to split the boundary spins  $j_\ell$ into fixed \textit{background contributions} $a_\ell$ and \textit{fluctuations} $s_\ell$, i.e.
\begin{equation}\label{eq:JDecomposition}
	j_\ell =  a_\ell + s_\ell \quad\text{with}\quad s_\ell \in \CBr*{-\floor*{\frac{k}{\sqrt{2t}}}, \floor*{\frac{k}{\sqrt{2t}}}}\quad\forall\ell\in\Gamma.
\end{equation}
Combining \eqref{eq:JDecomposition} with \eqref{eq:Implication} allows us to write
\begin{align}
	&a_{\text{sum}} - \frac{2k}{\sqrt{2t}} + 1 > a_{\text{diff}} + \frac{2k}{\sqrt{2t}}\notag\\
	 a_\text{sum} := \min&\left(a_1 + a_2, a_3 + a_4\right) \quad\quad a_\text{diff} := \max\left(\vert a_1 - a_2\vert, \vert a_3 - a_4 \vert\right)
\end{align}
which can be rearranged to
\begin{equation}\label{eq:SemiclassicalityCondition}
	\left(a_\text{sum}-a_\text{diff}\right) \sqrt{t} > \frac{4k}{\sqrt{2}}-\sqrt{t} \approx \frac{4k}{\sqrt{2}}
\end{equation}
since $t$ was assumed to be much smaller than one. By the assumptions made on the area data $a_\ell$, the differences $a_\text{diff}$ are negligible\footnote{One can show $a_\textsf{sum} \geq 0.8\, a_\textsf{diff}$.} compared to the sums $a_{\text{sum}}$. Moreover, we assumed the coherent states to be semi-classical which means the semi-classicality condition $a_\ell\sqrt{t} \gg 1$ is satisfied (see subsection~\ref{ssec:ExtrinsicStates}). Hence, we can infer values of $k$ for which~\eqref{eq:SemiclassicalityCondition} holds. At this point, the semi-classicality condition can also be read as a geometricity condition on the coherent states. It imposes that the intrinsic states~\eqref{eq:LS_States} have spins which are well within the triangle inequalities. This in turn means that the coherent states are composed of a superposition of~\eqref{eq:LS_States} that describe a triangulation of a spacelike hypersurface.\\
The next step in the approximation is to use the stationary phase analysis discussed in~\ref{ssec:ClassicalLimitGeometry}. To that end, we assume for concreteness that the boundary data $(j_\ell, \xi_\ell, \textbf{n}_s$, $\textbf{n}_t)$ determines a Lorentzian geometry and that the normal vectors $\textbf{n}_s$, $\textbf{n}_t$ play no further role in the analysis. This amounts to getting rid of the $\slc$ and the spinorial integrals and the holomorphic action~\eqref{eq:HolomorphicAction} evaluated on a critical point becomes
\begin{equation}
	A(j_\ell; a_\ell, \xi_\ell) = \gamma\sum_{\ell\in\Gamma} j_\ell\left(\textsf{s}\, \theta_\ell+\frac{1}{\gamma}\Pi_\ell -\xi_\ell\right) + i\sum_{\ell\in\Gamma}\left(j_\ell-a_\ell\right)^2 t.
\end{equation}
Notice that the assumption that the vectors $\textbf{n}_s$, $\textbf{n}_t$ do not play a role effectively removed the last term from the data contribution~\eqref{eq:GWeight}. Moreover, we can use the split of $j_\ell$ into background contribution and fluctuation introduced in~\eqref{eq:JDecomposition} to rewrite the approximated holomorphic amplitude as
\begin{equation}
	W^t_\mathcal{C}(H_\ell) \sim \sum_{\{\textsf{s}(\ve)\}}\sum_{\{s_\ell\}} \prod_{\ell\in\Gamma}\e^{i\gamma a_\ell \left(\textsf{s}\, \theta_\ell+\frac{1}{\gamma}\Pi_\ell -\xi_\ell\right)}\e^{-s^2_\ell t + i\gamma s_\ell\left(\textsf{s}\, \theta_\ell+\frac{1}{\gamma}\Pi_\ell -\xi_\ell\right)},
\end{equation}
where for the time being we ignore the Hessian and the sum over~$\textsf{s}(\ve)$ is a sum over the $2^V = 4$ distinct critical points.\\
Due to the presence of the exponential damping factor $\exp(-s^2_\ell t)$ it can reasonably be expected that the sums converge very fast and that it is therefore a good approximation to remove the cut-off $k$ by formally sending it to infinity. After performing the change of summation variables $s_\ell\rightarrow n_\ell = 2 s_\ell$ we find a closed analytical expression for the spin sums:
\begin{equation}\label{eq:DoneSpinSum}
	\sum_{n_\ell = -\infty}^{\infty} \e^{-\frac{n^2_\ell}{4}t + \frac{i\gamma}{2} n_\ell\left(\textsf{s}\, \theta_\ell+\frac{1}{\gamma}\Pi_\ell -\xi_\ell\right)} = 2\sqrt{\frac{\pi}{t}} \e^{-\frac{\gamma^2 \Delta^2_\ell}{4 t}}\vartheta_3\left(-\frac{i \pi \gamma \Delta_\ell}{t}, \e^{-\frac{4 \pi^2}{t}}\right),
\end{equation} 
where we have introduced the abbreviation $\Delta_\ell:=\textsf{s}\, \theta_\ell+\frac{1}{\gamma}\Pi_\ell -\xi_\ell$ and
\begin{equation}
	\vartheta_3(u,q) := 1+ 2\sum_{n=1}^\infty q^{n^2}\cos(2 n u)
\end{equation}
is the third Jacobi theta function. This result looks complicated at first, but it is easy to see that the magnitude of the holomorphic transition amplitude is given by
\begin{equation}
	2\sqrt{\frac{\pi}{t}} \e^{-\frac{\gamma^2 \Delta^2_\ell}{4 t}},
\end{equation}
while the Jacobi theta function accounts for the $\frac{4\pi}{\gamma}$-periodicity in $\xi_\ell$ due to the $\e^{i\gamma s_\ell \Delta_\ell}$ term in the original spin sum. The magnitude could also have been obtained in a different way. We could have applied the Euler-Maclaurin theorem and approximate the spin sum by an integral which would have resulted in
\begin{equation}
	\int_{-\infty}^\infty \e^{-\frac{n^2_\ell}{4}t + \frac{i\gamma}{2} n_\ell\left(\textsf{s}\, \theta_\ell+\frac{1}{\gamma}\Pi_\ell -\xi_\ell\right)} \dd n_\ell = 2\sqrt{\frac{\pi}{t}} \e^{-\frac{\gamma^2 \Delta^2_\ell}{4 t}}.
\end{equation}
The reason why we loose the periodicity when using this approximation is because we ignored the higher order correction terms of the Euler-Maclaurin series.\\
Finally we find that on a $2$-complex with no internal faces, when the area data parameters $a_\ell$ are much larger than unity, $t$ is such that the boundary states are semi-classical and the normal vectors $\textbf{n}_s$, $\textbf{n}_t$ are ignored, the critical point corresponds to a non-degenerate Lorentzian geometry, the holomorphic transition amplitude can be approximated by
\begin{equation}\label{eq:OldApprox}
	W^t_\mathcal{C}(H_\ell) \sim \sum_{\{\textsf{s}(\ve)\}} \prod_{\ell\in\Gamma} \e^{i\gamma a_\ell \Delta_\ell}\e^{-\frac{\gamma^2\Delta^2_\ell}{4t}} \vartheta_3\left(-\frac{i \pi \gamma \Delta_\ell}{t}, \e^{-\frac{4 \pi^2}{t}}\right).
\end{equation}
We ignore contributions from numerical factors, normalizations and the Hessian. Although the results of~\cite{Kaminski:2019} guarantee that the Hessian is non-degenerate for Lorentzian critical points and the above expression is therefore well-defined.\\
There are two interesting features in this approximation: The amplitude is exponentially suppressed whenever $\Delta_\ell\neq 0$. That is, the amplitude is suppressed\footnote{The suppression is enhanced by the fact that $t$ is a small parameter by assumption while $\Delta$ is an angle of order one and $\gamma$ is also thought to be of order one.} whenever the boundary data $\xi_\ell$ does not match the angles $\textsf{s}\, \theta_\ell+\frac{1}{\gamma}\Pi_\ell$ computed from the stationary phase analysis. We may interpret this as an indication, that when the boundary data does not describe a classically allowed transition, the amplitude decays exponentially. In chapter~\ref{chapter_5} we will make this idea much more precise.\\
The second interesting feature concerns the sum over critical points. It can be shown~\cite{BH2WH,TimeScale} that because of the exponential damping factor, there is always one critical point which dominates over the others. The data provided by the semi-classical coherent boundary states seems to select a preferred critical point. We will return to this observation in chapter~\ref{chapter_5}.\\
Despite the many restrictions under which the above approximation is valid, we apply it to the black hole to white hole transition. The main motivation is the fact that the amplitude is suppressed when $\Delta_\ell\neq 0$ and this is precisely the case in the black hole to white hole transition because of the flip in extrinsic curvature. In the next subsection we use~\eqref{eq:OldApprox} to estimate the bounce time.

\subsection{Estimation of the Bounce Time from CLQG}
\label{ssec:EstimationBounceTime}
There exists a probabilistic interpretation of the transition amplitude due to Oeckl who developed a general formalism \cite{Oeckl:2003,Oeckl:2005,Oeckl:2016} and also applied it to the black hole to white hole transition \cite{Oeckl:2018}.\\
Thanks to Oeckls explanations we were able to develop an adaptation of his methods and estimate the bounce time as follows:\\
We take the point of view that $m$ and $T$ are partial observables in the sense of~\cite{Rovelli:2002} and that $T(m)$ is a complete observable. That is, $m$ and $T$ are both measurable quantities, but they cannot be predicted. The correlation $T(m)$ on the other hand side represents a complete observable which has to be predictable from a theory. Since our theory is quantum mechanical in nature, the correlation has to be a probabilistic one. Hence, we need to introduce a probability distribution related to our measuring/observing procedure.\\
To that end, consider the semi-classical coherent states $\Psi^t_{a_\ell, \xi_\ell}$ defined in~\ref{ssec:ExtrinsicStates} which satisfy the resolution of identity~\eqref{eq:ResOfId} with respect to the integration measure defined in~\eqref{eq:TwistedMeasure}. For convenience, we repeat the relevant expression here:
\begin{equation}
	\mu(\eta) = \frac{\e^{-\frac{t}{2}}}{(4\pi)^2(2\pi t)^{\frac{3}{2}}}\eta\,\sinh \eta\, \e^{-\frac{\eta^2}{2t}}.
\end{equation}
 Since $a_\ell$ and $\xi_\ell$ are both functions of $m$ and $T$, the states $\Psi^t_{a_\ell, \xi_\ell}$ and the holomorphic transition amplitude $W^t(m,T):=\innerp{W_\mathcal{C}}{\Psi^t_{a_\ell, \xi_\ell}}$ also become functions of $m$ and $T$. Following Oeckl, we now define a conditional probability distribution by
\begin{equation}
	P(T|m) := \frac{Q(m, T)}{Q_T(m)},
\end{equation}
where $Q(m, T)$ and $Q_T(m)$ are defined as
\begin{align}
	Q(m, T) &:= \mu(m,T)|W^t(m,T)|^2\notag\\
	Q_T(m) &:= \int_0^\infty \mu(m, T)|W^t(m,T)|^2\dd T.
\end{align}
Notice that the conditional probability $P(T|m)$ is automatically normalized by its definition. We can furthermore interpret this quantity as the conditional probability distribution for measuring $T$ for a given fixed mass $m$.\\
Observe furthermore that $P(T|m)$ does not depend on the normalization $\mathcal N_\mathcal{C}$ nor on the Hessian from the stationary phase approximation. The reason is that these quantities only depend on the combinatorial structure of the $2$-complex but not not on any boundary data. They can therefore be pulled out of the integral and they cancel in the fraction $\frac{Q(m,T)}{Q_T(m)}$. \\
Next we introduce the expectation value of $T$: 
\begin{equation}
	\tau(m) := \int T\, P(T|m)\, \dd T.
\end{equation}
This quantity provides us with the probabilistic correlation between $m$ and $T$ we were seeking. If we use the boundary data specified in~\cite{RealisticObs}, then we only need to consider the four links which connect the top nodes to the bottom nodes in Figure~\ref{fig:SpinFoam}. The reason is that only here does $T$ appear in the boundary data, all the other contributions drop out. By using the approximation~\eqref{eq:OldApprox} and $\xi_0=\frac{T}{2m}$, $\eta_0 = \frac{2m(1+\frac{T}{2m})}{\sqrt{2\gamma}}$ we can then write the expectation value $\tau(m)$ as
\begin{equation}
	\tau(m) = \int T\,  \frac{\mu(m,T)\, \e^{-\frac{\gamma^2 \Delta^2}{2t}}}{\int \mu(m,T)\,\e^{-\frac{\gamma^2 \Delta^2}{2t}}\,\dd T}\dd T.
\end{equation}
This integral is of course not easy to solve, but it has been studied numerically~\cite{TimeScale} and it has been found that
\begin{equation}
	\tau(m) \approx m f(\gamma),
\end{equation}
where $f$ is an unknown function that depends on $\gamma$ but otherwise nothing else. Recall that at the beginning of this subsection we set out to determine a correlation between the black hole mass $m$ and the bounce time $T$, where $T$ is supposed to be a long time scale which measures the lifetime of a black hole before it turns white. If we momentarily insert units, we find that
\begin{equation}
	\tau(m) \sim m\, \frac{t_p}{m_p} = m\frac{G}{c^3}
\end{equation}
seems to be wrong on two counts: It is observationally wrong because for a solar mass black hole this time scale would be of the order of hours which is in contradiction with the black holes we observe in the sky. It is also theoretically wrong because there is no $\hbar$ in this time scale. We would expect that taking a formal classical limit, i.e. $\hbar\to 0$, would send $\tau$ to infinity because the transition is supposedly a pure quantum phenomenon.\\
Interestingly, this result has also been obtained by two other research groups with quite different methods. Ambrus and H\'{a}j\'{i}\v{c}ek \cite{Ambrus:2005} computed what they called the scattering time for a quantized shell of null matter. They found that this shell can collapse and re-expand in time $m$, where the classical observable for this phenomenon was defined as in this thesis or as in~\cite{TimeScale}. Barcel\'{o}, Carballo-Rubio and Garay \cite{Barcelo:2016} used a Wick-rotated path integral with an interpolating Euclidean geometry for the quantum region and they found the same linear scaling of the bounce time.\\
Where did all these different approaches go wrong? In~\cite{TimeScale}, the following answer was proposed to this question: What these three approaches computed is not the bounce time, but rather what one may call the \textit{crossing time}. This is the characteristic time scale for the transition \textit{when it happens}. An analogy with quantum mechanics and nuclear decay can be very illuminating at this point:\\
Consider an electron gun that shoots an electrons toward a square potential of height $V_0$ in the region $a< x < b$. Let's assume that the electron is described by a Gaussian wave packet and that its momentum $p_0$ is too low to overcome the potential barrier. At some point, the electron hits the barrier and it will be reflected with a probability $1-p$. There is however a small probability, $p$, that the electron will tunnel through the potential and be detected on the other side. Assume the tunneling happens, what is the most probable time for the detector to click?\\
This time is what we might call the \textit{crossing time}, which mathematically we would define through the process
\begin{equation}
	\langle x_b, p_0| \e^{\frac{i}{\hbar}\hat{H} T}| x_a, p_0\rangle \sim \e^{-\frac{1}{\hbar}(2E T-p_0(x_b-x_a))}.
\end{equation}
In the above matrix element it is assumed that the electron tunnels through the potential and this process is dominated by transitions which take a time $T=\frac{p_0}{2E}(x_b-x_a)\sim\frac{m}{p_0}(x_b-x_a)$. Notice the absence of $\hbar$ in this  time scale. \\
In the case of the black hole to white hole transition we also \textit{assumed} the process would happen through the data we provided the boundary states and we found a time scale which does not depend on $\hbar$. Hence, we may say that CLQG predicts a crossing time of oder $m$ when the transition happens, which is exactly what one would expect since $\pi m$ is the time it takes timelike geodesics crossing the Schwarzschild horizon to reach the singularity.\\
Turning back to simple quantum mechanics, we can also consider a toy model of radioactive decay where a particle of mass $m$ is trapped inside a square potential of finite height and bounces back and forth between the walls. We can think of the particle as moving with a mean momentum $p_0$ within a box of size $L$. The bounce period is then easily estimated as $\Delta T = L\frac{m}{p_0}$. This gives us again a classical time scale and at each bounce there is a small probability $p$ for the particle to tunnel through the barrier. Hence, the probability to exit the barrier per unit time is given by $P\sim p/\Delta T$ and the probability for the particle to exit at time $T$ is determined through $\frac{\dd P}{\dd T} = -pP(t)$. This equation is easily solved by
\begin{equation}
	P(T) = \frac{1}{\tau}\e^{-\frac{T}{\tau}},
\end{equation}
where the integration constant $\tau$ is interpreted as the lifetime of the radioactive nucleus and we can estimate this time scale as
\begin{equation}
	\tau\sim \frac{1}{P}\sim\frac{\Delta T}{p}.
\end{equation}
We can therefore distinguish between three different time scales: the crossing time, the characteristic time $\Delta T$ and the lifetime $\tau$. As explained above, the CLQG computation and the other methods used to compute the black hole to white hole transition all seem to compute the crossing time. To estimate the bounce time, we can adapt the ideas used to estimate the lifetime of a nucleus. That is,
\begin{equation}
	\tau_\text{Bounce} \propto \frac{1}{p} \sim \frac{1}{|W^t(m,T)|^2}\sim \e^{\frac{m^2}{m^2_p} \Xi},
\end{equation}  
to leading order and where $\Xi$ is a numerical factor (an angle) \cite{TimeScale}. We deliberately skipped showing the various steps that lead to this result because it is a very preliminary computation. It is certainly consistent with observational data, since this is an unimaginably long time scale, and it is also theoretically more consistent since $\tau_\text{Bounce}\overset{\hbar\to 0}{\longrightarrow}+\infty$. It is also what one would naively expect for a tunneling phenomenon where the transmission probability is generally given by $p\sim\e^{-S_\text{E}}$, where $S_\text{E}$ is the Euclidean action. For Regge calculus this would become $p\sim\e^{m^2/m^2_p \theta}$, assuming the areas involved in the problem are close to the apparent black/white hole horizon.\\
However, this estimate is largely inconclusive. At best, it gives us a hint that the bounce time is much larger than the originally expected $m^2$ \cite{Haggard:2014}. This in itself poses a problem because in the construction of the fireworks spacetime we explicitly assumed Hawking evaporation not to play a role.\\
This is consistent with a bounce time scale of order $m^2$, but certainly not with $\e^{m^2/m^2_p \Xi}$. Hawking evaporation changes the geometry\footnote{A modified model which takes into account Hawking evaporation has recently been developed by Martin-Dussaud and Rovelli \cite{Martin-Dussaud:2019b}.} and therefore also affects the boundary data of the semi-classical coherent states and the details of the CLQG computation outlined in the previous subsection.\\
Other problems involve the fact that this computation has been carried out on a rather corse triangulation and it is not clear how interior faces would change this result. Also, we have seen in~\eqref{eq:DoneSpinSum} that there is a Jacobi theta function in the amplitude due to its periodicity in $\xi$. This function has been neglected in the above estimate. It is important to note that the periodicity in $\xi$ effectively puts constraints on how much extrinsic curvature we can encode through the coherent states. The formal relation between the holonomy $h_\text{AB}$ of the Ashtekar-Barbero connection and the discrete extrinsic $\xi$ curvature is \cite{Rovelli:2010}
\begin{equation}
	h_\text{AB}(\xi) \sim \e^{i\gamma \xi \frac{\sigma_3}{2}},
\end{equation}
which is $\frac{4\pi}{\gamma}$-periodic and therefore puts an upper bound on $\xi$ on each link of the boundary graph. This problem could be overcome by refining the triangulation, but then again the computation of the CLQG amplitude breaks down because of the introduction of internal faces.

%% file: Content/Chapter_5.tex
\chapter{On the Evaluation of holomorphic Transition Amplitudes}
\label{chapter_5}
Computing the black hole to white hole transition amplitude and extracting physical predictions from it has proven to be a very challenging task. Several obstacles have impeded progress in the investigation of this physical scenario. These obstacles involve conceptual issues such as how to define observables in a background independent quantum theory and the different time scales that appear in the transition model. Both issues were discussed in the previous chapter.\\
The other obstacle is of computational nature and is due to a lack of systematic methods to evaluate CLQG transition amplitudes. So far, the only computational tool available is the stationary phase approximation~\cite{Barrett:2009b, Barrett:2009,Conrady:2008,Han:2011b, Han:2011, HanKrajewski:2013} which has been developed by different research groups and it has been shown that it gives the classical limit of CLQG in terms of (area) Regge calculus.\\
However, this approximation method also played an important role in subsection~\ref{ssec:ComputationBounceTime} for estimating the black hole to white hole transition amplitude. This raises immediately a question: How can a method which has been shown to be related to the classical limit be used to describe a phenomenon which is clearly not classical? After all, there is no critical point for the chosen boundary data and this method should not be applicable!\\
This is an issue that has been repeatedly raised with the computation presented in~\ref{ssec:ComputationBounceTime} and it is indeed a justified criticism which needs to be addressed. To do so, we will first take two steps back in this chapter. In section~\ref{sec:QM_Continuum} we consider holomorphic transition amplitudes for (continuum) quantum mechanics. Already at this stage we will gain some insights which can teach us something about the much more complicated CLQG amplitudes -- or at least we gain a new perspective. More importantly, we succeed in developing an approximation method which allows us to evaluate holomorphic amplitudes away from critical points.\\
Before moving from simple quantum mechanics to CLQG, we add one layer of complexity to the quantum mechanical transition amplitude and study its discrete version. This cautious procedure reveals how many results from the continuum actually survive the transition to the discrete theory and it uncovers issues which are strictly tied to working with discrete structures. This will give us again some important pointers for spin foams.\\
In section~\ref{sec:SemiClassicalCLQG} we then extend the newly developed approximation method to CLQG, where it can be understood as a semi-classical expansion of the transition amplitude around a classical background spacetime.

\section{Holomorphic Amplitudes in Quantum Mechanics}\label{sec:QM_Continuum}

\subsection{Framing the Question}
\label{ssec:FramingTheQuestion}
Consider a point particle of mass $m$ moving in a potential $V(x)$. The motion of the point particle is then quantum mechanically described by the propagator
\begin{align}\label{eq:QMPropagator}
	K(x_i, t_i; x_f, t_f) := \int_{x_i}^{x_f}\mathcal{D}[x] \e^{\frac{i}{\hbar}S[x]},
\end{align}
where $x_i$ and $x_f$ are the fixed initial and final points of the path $x\equiv x(t)$ and
\begin{align}\label{eq:GeneralActionFunctional}
	S[x] = \int_{t_i}^{t_f}\left(\frac12 m \dot x^2-V(x)\right)\dd t
\end{align}
is the particle's action. Throughout this section it is assumed that $x$ lives in an $n$-dimensional configuration space $\mathcal C$. Let us further assume that at the initial time $t_i$ we measured the particle's position to be $q_i$ and its momentum to be $p_i$. These conditions can conveniently be modeled by saying that the particle's initial state is given by the coherent state
\begin{align}\label{eq:InitialState}
	\Psi^\sigma_{z_i}(x) =\left(\frac{\sigma^2}{\pi\hbar}\right)^{\frac{n}{4}}\e^{-\frac{\sigma^2}{2\hbar}(x-q_i)^2}\e^{\frac{i}{\hbar}p_i(x-q_i)}.
\end{align}
Recall from subsection~\ref{ssec:2_DefOfCoherentStates} that the complex variable  $z_i=q_i+i \sigma^{-2} p_i$ encodes the boundary data in the sense that the expectation values of the operators $\hat x$ and $\hat p$ are $q_i$ and $p_i$, respectively, while the parameter $\sigma>0$ is closely tied to the uncertainties $\Delta x$ and $\Delta p$ associated with the measuring process.\\
Given that the particle's initial state is~\eqref{eq:InitialState} and that its evolution is described by~\eqref{eq:QMPropagator}, we may now ask the question: What is the probability to find the particle in the final state $\Psi^\sigma_{z_f}(x)$, with $z_f = q_f + i \sigma^{-2} p_f$, after a time period $T:=t_f-t_i$ and how does this probability depend on the boundary data $(q_i,p_i;q_f,p_f)$?\\
In search for an answer, we are naturally led to consider the holomorphic transition amplitude
\begin{align}\label{eq:HolomorphicAmplitudeQM}
	W^\sigma(z_i, z_f, T) &:= \int_{\mathbb{R}^{2n}}\dd^n x_f\dd^n x_i\,\overline{\Psi^\sigma_{z_f}(x_f)}\,K(x_i, t_i; x_f, t_f)\,\Psi^\sigma_{z_i}(x_i)\notag\\
	&= \left(\frac{\sigma}{\sqrt{\pi\hbar}}\right)^n\int_{\mathbb{R}^{2n}}\dd^n x_f\dd^n x_i\int_{x_i}^{x_f}\mathcal{D}[x]\e^{-\frac{\sigma^2}{2\hbar}\left((x_i-q_i)^2+(x_f-q_f)^2\right)}\,\e^{\frac{i}{\hbar}\left(S[x]+p_i(x_i-q_i)-p_f(x_f-q_f)\right)}\notag\\
	&= \left(\frac{\sigma}{\sqrt{\pi\hbar}}\right)^n\int_{\mathbb{R}^{2n}}\dd^n x_f\dd^n x_i\int_{x_i}^{x_f}\mathcal{D}[x]\e^{\frac{i}{\hbar}A}.
\end{align}
In the last line we introduced the complex ``action''
\begin{align}\label{eq:ComplexAction}
	A[x, T;x_i,x_f;z_i,z_f] := S[x]+p_i(x_i-q_i)-p_f(x_f-q_f) +i \frac{\sigma^2}{2}\left((x_i-q_i)^2+(x_f-q_f)^2\right),
\end{align}
which depends on the path $x$ and the time interval $T$ through the action functional $S[x]$, but also on the boundary data $(q_i,p_i;q_f,p_f)$ as well as the end points $x_i, x_f$ of the path. The time interval $t_i-t_f$ makes its appearance only through the action functional. Surprisingly, the transition amplitude \eqref{eq:HolomorphicAmplitudeQM} can be evaluated exactly for the free particle and the harmonic oscillator (see Appendix~\ref{App:HolomorphicAmplitudes}). In general, however, we need to resort to approximations in order to evaluate $W^\sigma$. One possibility is to determine the asymptotic behavior of the transition amplitude for the case when $A$ is large compared to $\hbar$. How this can be achieved will be reviewed in the following section and the observations made there will serve as the starting point for a more refined approximation discussed later on.

\subsection{Asymptotic Behavior of the Transition Amplitude}\label{ssec:AsympBehavior}
Determining the asymptotic behavior of path integrals is of course nothing new and it is a topic discussed in every textbook on the subject of path integrals. See for example \cite{FeynmanBook, KleinertBook}. The key mathematical tool needed is the so-called stationary phase approximation and it tells us that path integrals are mostly supported on classical trajectories on which they take the form
\begin{equation}
	\int_{x_i}^{x_f}\mathcal{D}[x] \e^{\frac{i}{\hbar}S[x]} \approx \sqrt{\frac{i}{2\pi \hbar}\frac{\partial^2 S}{\partial x_i\partial x_f}} \e^{\frac{i}{\hbar}S(x_i,t_i; x_f, t_f)},
\end{equation}
up to corrections of order $\hbar$. Here, $S(x_i,t_i; x_f, t_f)$ denotes the Hamilton function, i.e. the action evaluated on a solution of the equations of motion. Of course we assumed that a classical solution to the equations of motion with boundary conditions $x(t_i)=x_i$ and $x(t_f) = x_f$ exists and that it is unique. If no such solution exists, the amplitude is suppressed, as it is for example the case in quantum mechanical tunneling. If the solution is not unique, the above asymptotic form is slightly modified to
\begin{equation}\label{eq:NonUniqueSolution}
	\int_{x_i}^{x_f}\mathcal{D}[x] \e^{\frac{i}{\hbar}S[x]} \approx \sqrt{\frac{i}{2\pi \hbar}}\, \sum_k\sqrt{\frac{\partial^2 S^{(k)}}{\partial x_i\partial x_f}} \e^{\frac{i}{\hbar}S^{(k)}(x_i,t_i; x_f, t_f)},
\end{equation}
where $k$ labels distinct solutions to the equations of motion. Remarkably, holomorphic amplitudes select only one term in the sum~\eqref{eq:NonUniqueSolution} and they allow us, under certain conditions, to precisely determine \textit{how} the amplitude is suppressed when there is no classical solution satisfying all boundary conditions. Both of these properties are directly relevant for Covariant LQG and it is therefore well-worth the time to discuss the asymptotic behavior of holomorphic transition amplitudes in a bit more detail. To begin, we need the following theorem~\cite{HormanderBook}:
\begin{mythm}{ Stationary Phase Approximation}{StatPhaseApprox}
		Let $K\subset\mathbb R^n$ be a compact set, $U$ an open neighborhood of $K$ and $k$ a positive integer. Furthermore, let $f\in C^{2k}_0(K)$ and $A\in C^{3k+1}(U)$. If $A$ satisfies $\text{Im}[A]\geq 0$ in $U$ as well as the so-called critical point equations $\text{Im}[A(x_0)]=0$, $A'(x_0) = 0$ with $A'(x) \neq 0$ in $K\setminus\{x_0\}$ and also the condition $\det A''(x_0) \neq 0$, then we have the following approximation
		\begin{equation*}
			\int_K f(x)\, \e^{i \lambda A(x)}\, \dd^n x = \left(\frac{2\pi}{\lambda}\right)^{\frac{n}{2}}\e^{\frac{\pi i}{4}\text{Ind}(A''(x_0))}\frac{f(x_0)\e^{i\lambda A(x_0)}}{\sqrt{|\det A''(x_0)}|} + \begin{smallmatrix}\!\mathcal{O}\!\end{smallmatrix}(\lambda^{-\frac{n}{2}}),
		\end{equation*}
		where $\lambda>0$ and $\text{Ind}(A''(x_0))$, the index of the Hessian matrix, is the number of negative eigenvalues of $A''(x_0)$.
\end{mythm}
Above we made use of Landau's little-O notation, $\begin{smallmatrix}\!\mathcal{O}\!\end{smallmatrix}(\lambda^{-\frac{n}{2}})$, to indicate that the higher order terms grow slower than $\lambda^{-\frac{n}{2}}$.\\
In the case of the holomorphic amplitude, $f$ is just a constant function and $A$ can be assumed to be smooth on all of $\mathbb R^n$. Moreover, the imaginary part of $A$ satisfies
\begin{equation}
	\text{Im}[A] = \frac{\sigma^2}{2}\left((x_i-q_i)^2+(x_f-q_f)^2\right) \geq 0
\end{equation}
and so we are left with the task of checking the critical point equations. These equations come in two types\footnote{We are strictly speaking cheating because we apply the stationary phase theorem to a functional integral. The steps that follow are nevertheless justified as can be seen by working with a discretized path integral \cite{FeynmanBook,KleinertBook}.}: For the functional integral over all paths $x(t)$ connecting $x_i$ to $x_f$ we need to take the variation of $A$ with respect to $x$ while for the integrals over the end points $x(t_i)\equiv x_i$ and $x(t_f)\equiv x_f$ it suffices to take ordinary derivatives. Obviously, the variation of $A$ with respect to the path $x(t)$ simply yields, as is well-known, the Euler-Lagrange equations:
\begin{align}\label{eq:ELeq}
	\delta_x A &= 0 \quad\Rightarrow\quad \PD{L}{x}-\frac{\dd}{\dd t}\PD{L}{\dot x} = 0\quad\text{ with }\quad x(t_i) = x_i \text{ and } x(t_f) = x_f.
\end{align}
These equations cannot yet be solved since the end points $x_i$ and $x_f$ have not been determined (remember that we are integrating over the end points). To remedy that, we compute the derivative of $A$ with respect to $x_i$ and $x_f$ which results in the following complex equations:
\begin{align}\label{eq:OtherEqs}
	\partial_{x_i} A &= 0\quad\Rightarrow\quad \left[\PD{S}{x_i} + p_i\right]+i\sigma^2(x_i-q_i) = 0 \notag\\
	\partial_{x_f} A &= 0\quad\Rightarrow\quad \left[\PD{S}{x_f}-p_f \right]+i\sigma^2(x_f-q_f) = 0.
\end{align}
Since the terms in the square brackets as well as the terms in the round brackets are real, it is straightforward to separate these equations into real and imaginary parts:
\begin{align}\label{eq:InterestingEq}
	\PD{S}{x_i} + p_i &=0 & x_i-q_i &=0 \notag\\
	\PD{S}{x_f}-p_f &=0 &  x_f-q_f &= 0.
\end{align}
The two equations on the right hand side obviously impose the boundary conditions that the trajectory $x(t)$ starts in $q_i$ and ends in $q_f$. Hence, the coherent states directly determine, not very surprisingly, the boundary conditions for the partial differential equation~\eqref{eq:ELeq}. Let us for the time being assume that there is a unique solution to the boundary value problem~\eqref{eq:ELeq} with initial and final point $q_i$ and $q_f$ and let us call this solution $x_\textsf{c}$. Plugging $x_\textsf{c}$ into the action functional defines the Hamilton function $S_\textsf{c}(x_i, t_i; x_f, t_f):=S[x_\textsf{c}]$. As is well-known, the derivatives of this function with respect to the initial and final positions of a physical trajectory (i.e. one that solves the equations of motion) are given by
\begin{equation}\label{eq:momenta}
	\PD{S_\textsf{c}}{x_i} = -p_i(x_i, t_i; x_f, t_f)\quad\text{and}\quad \PD{S_\textsf{c}}{x_f}=p_f(x_i, t_i; x_f, t_f).
\end{equation}
The functions $p_i(x_i, t_i, x_f, t_f)$ and $p_f(x_i, t_i, x_f, t_f)$ are the initial and final momenta the particle acquires on the physical trajectory. Using this relation between the derivatives of the Hamilton function and the momenta, the equations on the left hand side of~\eqref{eq:InterestingEq} now read
\begin{equation}\label{eq:MomentaEq}
	p_i - p_i(q_i,t_i; q_f, t_f) = 0\quad\text{and}\quad p_f(q_i,t_i; q_f, t_f)-p_f = 0,
\end{equation}
where we have also used $x_i = q_i$ and $x_f = q_f$. In other words, we find that the critical point equations are only satisfied if the boundary data $(q_i,p_i; q_f, p_f)$ has been chosen such that there exists a solution of the equations of motion with initial and final positions $q_i$ and $q_f$, respectively, and whose initial and final momenta are given by $p_i$ and $p_f$, respectively. If these conditions are met, and assuming that the non-degeneracy condition $\det A''(x_\textsf{c})$ holds, the amplitude is approximately given by
\begin{equation}\label{eq:ApproxHolAmplitude}
	W^\sigma(z_i, z_f, T) = \left(2\sigma^2 \frac{\det\nabla_{q_i}\nabla_{q_f}S_\textsf{c}}{\det\nabla^2_x A(x_\textsf{c})}\right)^{\frac12}\,\e^{\frac{i}{\hbar}S_\textsf{c}(q_i,t_i;q_f,t_f)} + \begin{smallmatrix}\!\mathcal{O}\!\end{smallmatrix}(\hbar^{\frac{1}{2}}),
\end{equation}
where we omitted an overall phase factor and $\det\nabla_{q_i}\nabla_{q_f} S_\textsf{c}$ is the Van Vleck-Pauli-Morrette determinant \cite{VanVleck:1928,Morrette:1951,PauliBook}. It emerges after applying the stationary phase approximation to the functional integral, while $\det\nabla^2_x A(x_\textsf{c})$ stems from approximating the two integrals over $x_i$ and $x_f$.\\
Two observations are in order: The first one concerns $\hbar$. Notice that we kept $\hbar$ explicit in~\eqref{eq:ApproxHolAmplitude} and recall from subsection~\ref{ssec:2_DefOfCoherentStates} that $\sigma^2$ has units of action over length squared. It follows that the prefactor is dimensionless and independent of $\hbar$. The modulus squared of $W^\sigma(z_i, z_f; T)$, which directly gives the probability of finding a particle which started in $q_i$ with momentum $p_i$ at the position $q_f$ with momentum $p_f$ after a time $T$ has elapsed, is completely independent of $\hbar$. In this sense, we can justifiably claim that the stationary phase approximation of the holomorphic amplitude yields the classical limit of the theory.\\
The second observation concerns the case where there is more than one solution to the Euler-Lagrange equations. Mathematically speaking, the Euler-Lagrange equations together with $x(t_i)=q_i$ and $x(t_f)=q_f$ constitute a boundary value problem (BVP). If a solution exists, there is no guarantee that it is unique. A simple example to illustrate this fact is the equation of motion of the harmonic oscillator, $\ddot x(t)+x(t) = 0$, with boundary conditions $x(0)=0$ and $x(\pi)=0$, which is solved by the one-parameter family of solutions $x(t) = C \sin(t)$. However, there is a different way to look at this problem as we also have to take into consideration the critical point equations~\eqref{eq:MomentaEq}. If it is possible to solve the momenta for the velocities $\dot x$, we can read~\eqref{eq:MomentaEq} as boundary conditions on the velocities or, more conveniently, re-express the BVP as an initial value problem (IVP). That is to say, we seek solutions of the Euler-Lagrange equations subjected to the initial value conditions $x(t_i) = q_i$ and $\dot x(t_i) = f(p_i)$, where $f$ is the function resulting from solving $p$ for $\dot x$. The advantage of this point of view is that IVP's for second order differential equations are guaranteed to have solutions and these solutions are unique!\\
At this point, it should be remembered that we limited ourselves to action functionals of the form~\eqref{eq:GeneralActionFunctional}, which ensure that the Euler-Lagrange equations are of second order and that the momenta can always be solved for the velocities. It follows that in the case of the holomorphic amplitude for this particular class of action functionals there is no need to sum over distinct classical solutions connecting the end points $q_i$, $q_f$ of the trajectory. The IVP possesses a unique solution and if the boundary data $(q_f,p_f)$ at the end point has been chosen appropriately, equation~\eqref{eq:ApproxHolAmplitude} is the unique classical approximation of the holomorphic amplitude.\\
We can think of the initial state $\Psi^\sigma_{z_i}(x)$ as representing a measurement which provides us with the initial value data $(q_i,p_i)$, the propagator~\eqref{eq:QMPropagator} encodes the dynamics and evolves this initial data along a unique classical trajectory while the final state  $\Psi^\sigma_{z_f}(x)$ represents a part of the question ``how likely is it to find the particle at $q_f$ with momentum $p_f$?''\footnote{Of course one can always interchange the role of the initial and final state without affecting the general picture employed here.}. 
Of course, one does not need to choose the data $(q_f,p_f)$ to lie on a classical trajectory which originates from $(q_i,p_i)$ and there are essentially three types of boundary data to be distinguished:
\begin{itemize}
	\item[i)] \textbf{Classical data:}\\ 
				The data $(q_i,p_i;q_f,p_f)$ has been chosen such that $(q_i, p_i)$ is connected to $(q_f, p_f)$ by a classical phase space trajectory. In this case we follow \cite{Bianchi:2010b} and call the data \textit{classical} and the amplitude $W^\sigma$ is approximately given by its classical limit~\eqref{eq:ApproxHolAmplitude}.
		
	\item[ii)] \textbf{Semi-classical data:}\\ 
				The trajectory which results from evolving the initial data $(q_i, p_i)$ passes through a small neighborhood of $(q_f,p_f)$, cf. Figure~\ref{fig:PSP}. In this case, there is still a classical trajectory but one or several of the critical point equations involving the final boundary data cannot be satisfied and the amplitude is therefore suppressed. This type of boundary data is called \textit{semi-classical} and in the next section we will see how the suppression factor can be explicitly determined. 				
	
	\item[iii)] \textbf{Non-classical data:}\\
				The trajectory obtained from evolving the initial data $(q_i, p_i)$ is not connected to the phase space point $(q_f,p_f)$ and it does not pass through a small neighborhood of $(q_f,p_f)$. In this case, there might be a complex trajectory connecting $(q_i, p_i)$ to $(q_f,p_f)$. This is for example the case in quantum mechanical tunneling and the complex trajectories can be found with the help of Picard-Lefschetz theory \cite{Tanizaki:2014,Behtash:2015}. We leave this important case open as subject for future investigations.    
\end{itemize}

\subsection{Semi-Classical Boundary Data}\label{ssec:BndData}
\begin{center}
\begin{figure}[h!]
	\centering
	\includegraphics[width=.5\textwidth]{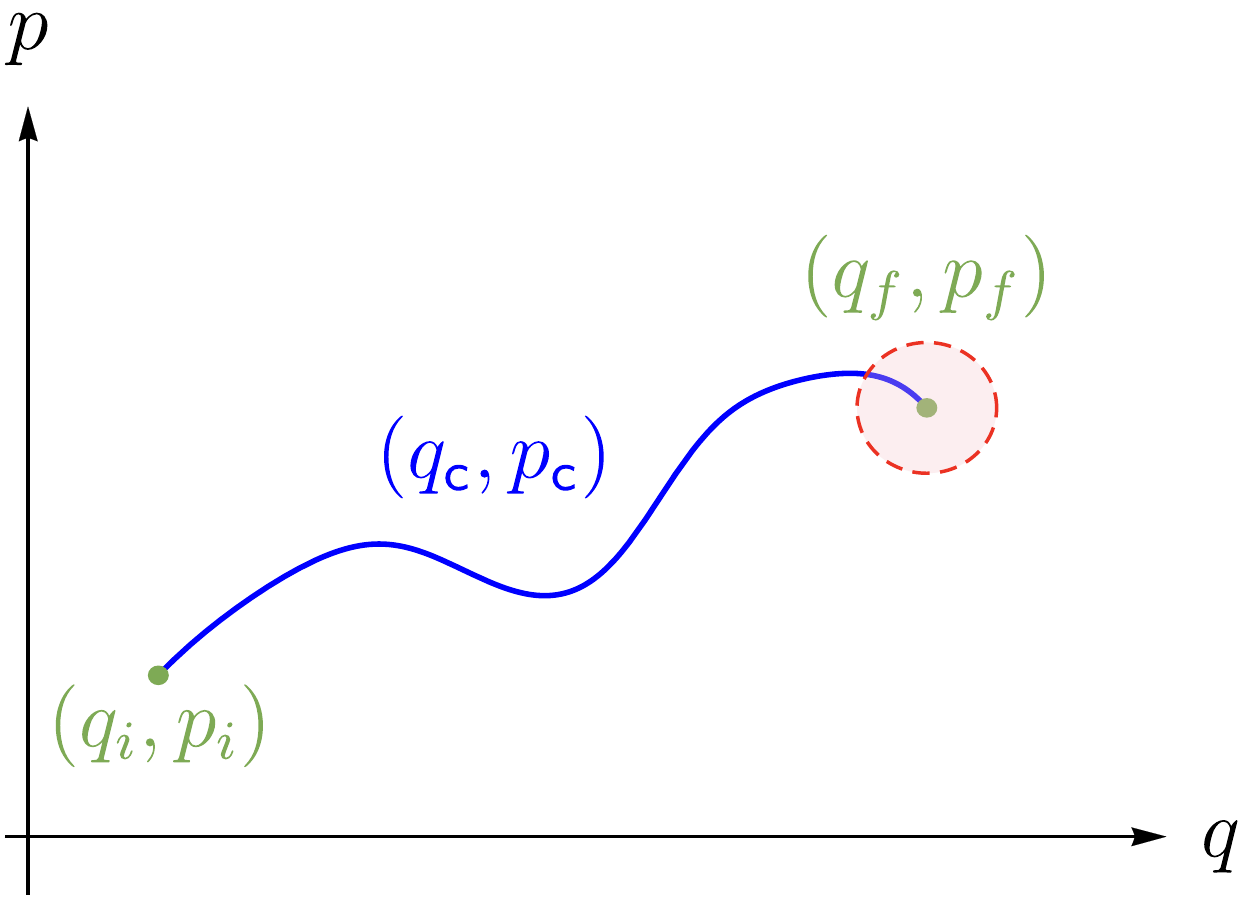}
	\caption{The modulus squared $|W^\sigma(q_i, p_i, t_i; q_f, p_f, t_f)|^2$ with $r_i=s_i = 0$ computes the probability to find a particle within a neighborhood of the classically expected end point of a phase space trajectory.}
	\label{fig:PSP}
\end{figure}
\end{center}

Let the path $x_\textsf{c}(t)$ satisfy the Euler-Lagrange equations with boundary value conditions $x_\textsf{c}(t_i)=x_i$ and $x_\textsf{c}(t_f) = x_f$. If the boundary data is semi-classical, $x_\textsf{c}(t)$ may fail to satisfy all of the critical point equations~\eqref{eq:InterestingEq} and the transition amplitude is therefore suppressed. The following technical result (theorem 7.7.12 in \cite{HormanderBook}) allows us to quantify what ``suppressed'' means, provided the boundary data $(q_i, p_i; q_f, p_f)$ lies in a neighborhood of the end points of the actual phase space trajectory $(q(t), p(t))$.
\begin{mythm}{ Generalized Stationary Phase Approximation}{GenStatPhaseApprox}
Let $A(y,v)$ be a complex valued $C^{\infty}$ function in a neighborhood $U$ of $(0,0)\in\mathbb R^{n+m}$, such that Im$[A]\geq 0$, Im$[A(0,0)]=0$, $\nabla_y A(0,0) = 0$ and $\det\nabla^2_y A(0,0)\neq 0$. Let $u$ be a $C^\infty$ function with compact support in $U$ and $0<\lambda\in\mathbb R$. Then
\begin{align*}
	\int_{\mathbb R^n} u(y,v) \e^{i\lambda A(y,v)}\dd^n y = \left(\frac{2\pi i}{\lambda}\right)^{\frac{n}{2}}\frac{u^{(0)}(0,v) \e^{i\lambda A^{(0)}(v)}}{\sqrt{(\det\nabla^2_y A(0,v))^{(0)}}}\, + \begin{smallmatrix}\!\mathcal{O}\!\end{smallmatrix}(\lambda^{-\frac{n}{2}}),	
\end{align*}
where the superscript $0$ signals that the corresponding function is the $0$-th order term in a Malgrange expansion.
\end{mythm}
Equipped with this new tool, the strategy is now to apply the usual stationary phase approximation to the propagator~\eqref{eq:QMPropagator} under the assumption that the BVP has a solution $x_\textsf{c}$ for the boundary values $\mathring{q}_i := x_\textsf{c}(t_i)$ and $\mathring{q}_f := x_\textsf{c}(t_f)$, while the generalized stationary phase approximation~\cref{thm:GenStatPhaseApprox} is reserved for the integrals needed to contract the propagator with the boundary states. Since $x_\textsf{c}(t)$ solves the Euler-Lagrange equations, it also generates the momenta $\mathring{p}_i := -\PD{S_\textsf{c}}{\mathring{q}_i}$ and $\mathring{p}_f:=\PD{S_\textsf{c}}{\mathring{q}_f}$ and therefore $(\mathring{q}_i,\mathring{p}_i; \mathring{q}_f, \mathring{p}_f)$ constitutes,  according to the classification introduced in the previous section, classical data. \\
The actual boundary data $(q_i,p_i; q_f, p_f)$ encoded into the coherent states may of course not coincide with the classical data and it is therefore convenient to parametrize the boundary data in terms of $(\mathring{q}_i,\mathring{p}_i; \mathring{q}_f, \mathring{p}_f)$ and variables $r_i,r_f, s_i, s_f,$ which measure the discrepancy between the two data sets:
\begin{align}
	q_i &= \mathring{q}_i + r_i &  p_i &= \mathring{p}_i + s_i\notag\\
	q_f &= \mathring{q}_f + r_f &  p_f &= \mathring{p}_f + s_f.
\end{align}
To bring the holomorphic amplitude into a form suitable for the application of theorem~\cref{thm:GenStatPhaseApprox}, we perform the change of integration variables $y:=(y_i,y_f)^\transpose:= (x_i-\mathring{q}_i, x_f-\mathring{q}_f)^\transpose$, which leaves the measure unaffected. However, in terms of the newly introduced variables, the complex action~\eqref{eq:ComplexAction} now reads
\begin{align}\label{eq:ActionInNewVariables}
	A(y,r,s) &= S_\textsf{c}(y_i+\mathring{q}_i, t_i; y_f+\mathring{q}_f, t_f) + (\mathring{p}_i + s_i)(y_i-r_i) - (\mathring{p}_f + s_f)(y_f-r_f)\notag\\
		 &\textcolor{white}{=} + i\frac{\sigma^2}{2}\left((y_i-r_i)^2 + (y_f-r_f)^2\right),
\end{align}
where the Hamilton function $S_\textsf{c}(x_i, t_i; x_f, t_f)$ appears because we assumed $x_\textsf{c}(t)$ to solve the Euler-Lagrange equations which arose from approximating the propagator and we used $x_i=y_i+\mathring{q}_i$ as well as a similar equation for $x_f$. It is now straightforward to check that all the conditions of the above theorem are met:
\begin{align}\label{eq:Conditions}
	\text{Im}[A(y,v)]&=\frac{\sigma^2}{2}\left((y_i-r_i)^2 + (y_f-r_f)^2\right)\geq 0\quad\Rightarrow\quad \text{Im}[\left.A\right|_\textsf{c}] = 0\notag\\
		\left.\nabla_y A\right|_\textsf{c} &= 	\left.\begin{pmatrix}
		\PD{S_\textsf{c}}{y_i}+\mathring{p}_i + s_i + i\sigma^2 (y_i-r_i)^2\\
		\PD{S_\textsf{c}}{y_f}-\mathring{p}_f - s_f + i\sigma^2 (y_f-r_f)^2
	\end{pmatrix}\right\vert_\textsf{c}
	= \begin{pmatrix}
		0 \\
		0
	\end{pmatrix},
\end{align}
where the notation $\left.A\right|_\textsf{c}$ signalizes that $A$ is being evaluated at the critical point $y=r=s=0$. The non-degeneracy of the Hessian  $\left.\nabla^2_y A\right|_\textsf{c} = \nabla^2_y S_\textsf{c} + i\sigma^2 \id_{n\times n}$ follows from the fact that $\nabla^2_y S_\textsf{c}$ is a real symmetric matrix and as such only has real eigenvalues. Hence, if $\det(\nabla^2_y S_\textsf{c} + i\sigma^2 \id_{n\times n}) = 0$ were true, $\nabla^2_y S_\textsf{c}$ would possess the purely imaginary eigenvalue $-i\sigma^2$, which is a contradiction.\\
To proceed we need to compute the $0$-th order Malgrange function $A^{(0)}$. According to Malgrange's preparation theorem (see for instance~\cite{HormanderBook}) any complex function $A(y,v)$ which is smooth in a neighborhood $U$ of $(0,0)\in\mathbb R^{n+m}$ can be written as
\begin{align}\label{eq:MalgrangeExpansion}
	A(y, v) = \sum_{|\alpha|<N} \frac{A^{(\alpha)}(v)}{\alpha!}\left(y-X(v)\right)^\alpha\, \mod I^N\quad\forall N\in\mathbb N,
\end{align}
where $A^{(\alpha)}, X\in C^\infty$ with $X(0)=0$, $\alpha$ is a multi-index, and $I$ denotes the ideal of functions generated by $\nabla_y A(y, v)$. This is a  technical and highly non-trivial piece of mathematics, but fortunately we do not need to delve too deep into the theory of ideals and it is easy to understand how to use this result to our advantage. First of all, according to~\cite{HormanderBook}, we can exploit the fact that the expansion is only defined modulo $I^N$ to set the vector $A^{(1)}=0$. After that, we are left with an expansion for $A(y,v)$ which is polynomial in $y$ and potentially non-polynomial but still smooth in $v$. This should be contrasted with a Taylor expansion which is polynomial in all variables and which satisfies the condition that the $k$-th derivative of the function at the expansion point $x_0$ equals the $k$-th derivative of the Taylor polynomial around that same point. However, unlike Taylor's theorem, Malgrange's preparation theorem does not provide a prescription to construct this expansion for any given function $A(y,v)$, it merely states that the right hand side of~\eqref{eq:MalgrangeExpansion} exists.\\
Hence, our task is now to determine the vector field $X$, and, assuming we chose $N=3$, the scalar  $A^{(0)}$ and the matrix $A^{(2)}$. As mentioned above, we can set $A^{(1)}=0$ and to determine the other unknown functions we compare the Taylor expansion
\begin{align}
	A(y,v) =& \left.A\right|_\textsf{c}	+ y^\transpose \underset{=0}{\underbrace{\left.\nabla_y A\right|_\textsf{c}}} + v^\transpose \underset{=0}{\underbrace{\left.\nabla_v A\right|_\textsf{c}}} + y^\transpose\underset{=:K}{\underbrace{\left.\nabla_v\nabla_y A\right|_\textsf{c}}} v + \frac{1}{2} y^\transpose \underset{=:H}{\underbrace{\left.\nabla^2_y A\right|_\textsf{c}}} y\notag\\
	 &+ \frac{1}{2} v^\transpose \underset{=:M}{\underbrace{\left.\nabla^2_v A\right|_\textsf{c}}} v + R_3(y,v)
\end{align}
with rest term $R_3(y,v)$ around the critical point $\textsf{c}$ order by order with the Malgrange expansion
\begin{align}
	A(y, v)= A^{(0)}(v) + \frac12 \left(y-X(v)\right)^\transpose \, A^{(2)}(v)\, \left(y-X(v)\right).
\end{align}
This comparison can be naturally structured as follows:
\begin{align}\label{eq:MalgrangeTaylor}
	\text{A)} &&  \left.A\right|_\textsf{c} + \frac12 v^\transpose M v  &= A^{(0)}(v) + \frac12 X(v)^\transpose\, A^{(2)}(v) \, X(v)   && \text{(independent of $y$)}\notag\\
	\text{B)} && y^\transpose K v &= -\frac12 \left[y^\transpose\, A^{(2)}(v)\, X(v) + X(v)^\transpose\, A^{(2)}(v)\, y \right] && \text{(linear in $y$)}\notag\\
	\text{C)} && \frac12 y^\transpose H y &= \frac12 y^\transpose \, A^{(2)}(v)\, y && \text{(quadratic in $y$)}
\end{align}
Equation~\eqref{eq:MalgrangeTaylor}~C) implies that the matrix $A^{(2)}(v)$ is equal to the non-degenerate Hessian matrix $H$. Then~\eqref{eq:MalgrangeTaylor}~B) implies
\begin{equation}
	y^\transpose K v = -y^\transpose H X(v)\quad\Rightarrow\quad X(v) = -H^{-1}Kv,
\end{equation}
where we used the symmetry of the Hessian and its non-degeneracy. Plugging the results of C) and B) into A) finally yields
\begin{align}
	A^{(0)}(v) = S_\textsf{c}(\mathring{q}_i, t_i; \mathring{q}_f, t_f) + \frac12 v^\transpose\left(M - K^\transpose H^{-1} K\right) v,
\end{align}
where we have used $\left.A\right|_\textsf{c} = S_\textsf{c}(\mathring{q}_i, t_i; \mathring{q}_f, t_f)$ and the $2n\times 4n$ matrix $K$ is explicitly given by
\begin{equation}
	K:= \left.\nabla_v \nabla_y A\right|_\textsf{c} = \begin{pmatrix}
		- i \sigma^2 \id_{n\times n} & 0 & \id_{n\times n} & 0 \\
		0 &  - i\sigma^2 \id_{n\times n} & 0 & -\id_{n\times n} 
	\end{pmatrix}
\end{equation}
while the non-degenerate $4n\times 4n$-matrix $M$ can be written as
\begin{equation}
	M:= \left.\nabla^2_v A\right|_\textsf{c} = \begin{pmatrix}
		 i\sigma^2 \id_{n\times n} & 0 & -\id_{n\times n} & 0\\
		0 & -i\sigma^2 \id_{n\times n} & 0 & \id_{n\times n} \\
		-\id_{n\times n} & 0 & 0 & 0\\
		0 & \id_{n\times n} & 0 & 0
	\end{pmatrix},
\end{equation}
from which one easily deduces $\det M = 1$, and the $2n\times 2n$ non-degenerate Hessian matrix reads
\begin{equation}
	H:= \left.\nabla^2_y A\right|_\textsf{c} = \left.\nabla^2_y S\right|_\textsf{c} + i \sigma^2 \id_{2n\times 2n}.
\end{equation}
Notice that for $v=0$, i.e. when there is neither discrepancy between the position data and the actual end points of the trajectory $x_\textsf{c}(t)$ nor discrepancy between the momentum data and the momenta computed from Hamilton's function, $A^{(0)}$ simply reduces to $S_\textsf{c}(q_i,t_i;q_f,t_f)$, exactly as we would have expected.\\
When $v$ is not the zero vector, $A^{(0)}(v)$ can acquire an imaginary part due to the presence of $i\sigma^2$ factors in the matrices $K,M,H$ and this imaginary part leads to a suppression of the transition amplitude. One may of course worry that the imaginary part Im$[A^{(0)}(v)]$ is smaller than zero, which would result in an exponentially enhanced amplitude which in turn is physically untenable. However, H\"{o}rmander~\cite{HormanderBook} puts our mind at ease:
\begin{mythm}{ Imaginary Part of $A^{(0)}$}{ImaginaryPart}
Let $A(y,v)$ be a complex valued $C^{\infty}$ function in a neighborhood $U$ of $(0,0)\in\mathbb R^{n+m}$, such that Im$[A]\geq 0$, Im$[A(0,0)]=0$, $\nabla_y A(0,0) = 0$ and $\det\nabla^2_y A(0,0)\neq 0$. Then
 \begin{equation*}
 	\text{Im}[A^{(0)}(v)] \geq C \left\|\text{Im}[X(v)]	\right\|^2
 \end{equation*}
 in a neighborhood of $v=0$ and for some constant $C>0$. If $\text{Im}[A]=0$ can only be attained at $(0,0)$ and nowhere else, then
 \begin{equation*}
 	\text{Im}[A^{(0)}(v)]\geq 0
 \end{equation*}
 and equality holds only at $v=0$.
\end{mythm}
Putting all the pieces together, we conclude that the holomorphic transition amplitude for semi-classical boundary data can be approximated by
\begin{equation}\label{eq:ApproxAmp}
	W^\sigma(z_i,z_f,T) = \left(2 \sigma^2\frac{ \det\nabla_{\mathring{q}_i}\nabla_{\mathring{q}_f} S_\textsf{c}}{\det H}\right)^\frac{1}{2}\exp\left(\frac{i}{\hbar} S_\textsf{c} + \frac{i}{2 \hbar} v^\transpose\left(M - K^\transpose H^{-1} K \right)v\right) + \begin{smallmatrix}\!\mathcal{O}\!\end{smallmatrix}(\hbar^{\frac{1}{2}}),
\end{equation}
where we dropped again an over-all phase factor. Because $\left.\nabla^2_y A\right|_{y=0}$ is independent of $v$ and, moreover, equivalent to $H=\left.\nabla^2_y A\right|_\textsf{c}$, the pre-factor turns out to be again exactly the same as in~\eqref{eq:ApproxHolAmplitude}. In particular, this means that it is again independent of $\hbar$. What is different, however, is the exponential which is not a pure phase any more. Instead, we find that the amplitude is suppressed by the factor
\begin{equation}\label{eq:SuppressionFactor}
	\exp\left(-\frac{1}{2\hbar}\text{Im}\left[v^\transpose\left(M - K^\transpose H^{-1} K \right) v\right]\right),
\end{equation}
which explicitly depends on $\hbar$ and where $\text{Im}\left[v^\transpose\left(M - K^\transpose H^{-1} K \right) v\right]$ is guaranteed to be smaller or equal to zero in a neighborhood of $v=0$ due to theorem~\cref{thm:ImaginaryPart}.\\
Just as~\eqref{eq:ApproxHolAmplitude} can be seen as the classical limit of the theory, we can think of~\eqref{eq:ApproxAmp} as a semi-classical approximation. In fact, the modulus squared of the transition amplitude~\eqref{eq:ApproxAmp} can be interpreted, for $r_i=s_i=0$, as the probability to find a particle, which started from the phase space point $(q_i, p_i)$, in a neigborhood of the classically expected final phase space point $(q_f, p_f)$. This situation is also illustrated in Figure~\ref{fig:PSP}.
As mentioned in the previous subsection, in the case of the free particle and the harmonic oscillator it is actually possible to compute the holomorphic transition amplitude analytically. In Appendix~\ref{App:HolomorphicAmplitudes} we make use of this fact and compare the exact analytical expressions with the approximation~\eqref{eq:ApproxAmp}. We find perfect agreement between $|W^\sigma_\textsf{exact}|^2$ and $|W^\sigma_\textsf{approx.}|^2$ in all examined scenarios.\\
While this is very reassuring, the question remains what all of this has to do with CLQG transition amplitudes. As we will discuss in more detail later on, the newly introduced methods of this section can be applied to holomorphic CLQG transition amplitudes, where they shed some new light on the classical limit as well as on the black hole to white hole computation of chapter~\ref{chap:CLQG}. However, so far we have been working with a continuum theory where we could use the calculus of variations and resort to well-known theorems about BVP's and IVP's. These tools are not available in a discrete theory and it is not clear if, or to what extent, the results reported thus far hold for a discretized holomorphic amplitude. It is important to address these concerns in the context of quantum mechanics, where we can always use the continuum theory to guide our way, before advancing to CLQG, where the continuum theory is unknown.

\subsection{Discretized Holomorphic Amplitude}\label{ssec:DiscreteQM}
Let us discretize the time interval $[t_i,t_f]$ by an equidistant partition of step length $\varepsilon:=\frac{t_f-t_i}{N}$ for some $N\in\mathbb N$. Let $\{t_k\}$ with $t_k:=t_i+\varepsilon k$ for $k\in\{0,1,\dots, N\}$ denote the set of instants of time and identify $t_0\equiv t_i$ and $t_N\equiv t_f$. The discretized path $x(t)$ with $x(t_i)=x_i$ and $x(t_f)=x_f$ is then given by the set $\{x_k\}$ with $x_k:=x(t_k)$, $k\in\{0,1,\dots, N\}$, and where we naturally identify $x_0\equiv x_i$ and $x_N\equiv x_f$. On occasion, we will visualize the discretized path as a collection of $N+1$ vertices, labeled by $x_k$, which are joined by $N$ line segments. We can then approximate the unitary time evolution operator $U(t_f,t_i) = \exp(-\frac{i}{\hbar}\int_{t_i}^{t_f}\hat H\,\dd t)$ in the usual way~\cite{FeynmanBook,KleinertBook} by the time-ordered product $U(t_f,t_i)\approx\prod_{k=1}^N U(t_k,t_{k-1})$. It follows that the propagator on discretized paths assumes the approximate form
 \begin{align}\label{eq:DiscretizedPropagator}
	K(x_i, t_i; x_f,t_f)&=\left<x_f|U(T)|x_i\right>\approx\big<x_f|\prod_{k=1}^N U(t_k,t_{k-1})|x_i\big>.
\end{align}
While not strictly necessary, it will prove to be very advantageous to rewrite the above propagator in terms of coherent heat kernel states. To that end, we need to use the easy to verify resolution of identity\footnote{Notice that in the resolution of identity we are integrating over the \textit{boundary data} $q$ and $p$. This is a point that sometimes causes confusion.}
\begin{align}\label{eq:HeatKernelResOfId}
	\int_{\mathbb{R}^{2n}} \frac{\dd^n q\, \dd^n p}{(2\pi \hbar)^n}\ket{\Psi^\rho_{q,p}}\bra{\Psi^\rho_{q,p}} = \id,
\end{align}
which we already encountered in chapter~\ref{chap:2}. We insert one such resolution of identity between each $U(\varepsilon)$ in the above product. That is a total of $N-1$ insertions. Moreover, we insert one resolution of identity between $\bra{x_f}$ and the first $U(\varepsilon)$ and an other one between the last $U(\varepsilon)$ and $\ket{x_i}$. This increases the total number of insertions to $N+1$ and the propagator assumes the following form:
\begin{align}\label{eq:PropagatorWithInsertions}
	\langle x_f|U(T)|x_i\rangle \approx & \int_{\mathbb R^{2n(N+1)}}\left(\prod^N_{k=0}\frac{\dd^n q_k\, \dd^n p_k}{(2\pi\hbar)^n}\right)\innerp{x_f}{\Psi^\rho_{q_N,p_N}}\notag\\
	&\langle\Psi^\rho_{q_N,p_N}|\bigg(\prod^{N-1}_{k=1}U(\varepsilon)|\Psi^\rho_{q_k,p_k}\rangle \langle \Psi^\rho_{q_k,p_k} |\bigg) U(\varepsilon)|\Psi^\rho_{q_0,p_0}\rangle	\innerp{\Psi^\rho_{q_0,p_0}}{x_i}.
\end{align}
A closer inspection of the product in~\eqref{eq:PropagatorWithInsertions} reveals that there are $N$ matrix elements of the form
\begin{equation}
	\langle\Psi^\rho_{q_k,p_k}|U(\varepsilon)|\Psi^\rho_{q_{k-1},p_{k-1}}\rangle\quad\text{for}\,\, k\in\{1,2,\dots, N\}.
\end{equation}
To evaluate these matrix elements, we need to insert resolutions of identity of the form $\int_{\mathbb R^n}\dd^n x_k \ket{x_k}\bra{x_k} =~\id$. We insert two resolutions of identity per matrix element: One between $\langle\Psi^\rho_{q_k,p_k}|$ and $U(\varepsilon)$ and a second one between $U(\varepsilon)$ and $|\Psi^\rho_{q_{k-1},p_{k-1}}\rangle$. This results in
\begin{equation}
	\int_{\mathbb R^n}\dd^n x_k\int_{\mathbb R^n}\dd^n x'_{k-1} \innerp{\Psi^\rho_{q_k,p_k}}{x_k}\langle x_k |U(\varepsilon)| x'_{k-1}\rangle\langle x'_{k-1}|\Psi^\rho_{q_{k-1},p_{k-1}}\rangle.
\end{equation}
As a notational convention we use that resolutions of identity inserted to the right of $U(\varepsilon)$ carry a prime, i.e. $\int_{\mathbb R^n}\dd^n x'_{k-1}| x'_{k-1}\rangle\langle x'_{k-1}|$. From the standard treatment of path integrals~\cite{FeynmanBook,KleinertBook} we know that
\begin{equation}
	\langle x_k|U(\varepsilon)| x'_{k-1}\rangle = \e^{\frac{i}{\hbar}\left(\frac{m}{2}\frac{(x_k-x'_{k-1})^2}{\varepsilon}-\varepsilon V(x'_{k-1})\right)},
\end{equation}
while $\langle x'_{k-1}|\Psi^\rho_{q_{k-1}, p_{k-1}}\rangle$ is given by (see subsection~\ref{ssec:2_DefOfCoherentStates})
\begin{equation}\label{eq:BraKetCoherentStates}
	\langle x'_{k-1}|\Psi^\rho_{q_{k-1}, p_{k-1}}\rangle \equiv \Psi^\rho_{q_{k-1},p_{k-1}}(x'_{k-1}) = \left(\frac{\rho^2}{\pi\hbar}\right)^{\frac{n}{4}}\e^{-\frac{\rho^2}{2\hbar}(x'_{k-1}-q_{k-1})^2}\e^{\frac{i}{\hbar}p_{k-1}(x_{k-1}-q_{k-1})}.
\end{equation}
The element $\langle\Psi^\rho_{q_k,p_k}|x_k\rangle$ has the same form, but is complex conjugate. With these results, we can express the matrix element $\langle\Psi^\rho_{q_k,p_k}|U(\varepsilon)|\Psi^\rho_{q_{k-1},p_{k-1}}\rangle$ as
\begin{equation}\label{eq:MatrixElementForSubstitution}
	\langle\Psi^\rho_{q_k,p_k}|U(\varepsilon)|\Psi^\rho_{q_{k-1},p_{k-1}}\rangle =\left(\frac{\rho^2}{\pi\hbar}\right)^{\frac{n}{2}}\int_{\mathbb R^n}\dd^n x_k\int_{\mathbb R^n}\dd^n x'_{k-1}\,\e^{\frac{i}{\hbar}L_{k,k-1}},
\end{equation}
where we introduced a bulk Lagrangian $L_{k,k-1}$ associated to the $k$-th line segment, i.e. the segment connecting the $k-1$ vertex to the $k$ vertex, and which is defined as
\begin{align}
	L_{k,k-1}:=& \frac{m}{2}\frac{(x_k-x'_{k-1})^2}{\varepsilon}-\varepsilon V(x'_{k-1}) + p_{k-1}(x'_{k-1}-q_{k-1}) - p_k(x_k-q_k)\notag\\
	 &+ i\frac{\rho^2}{2\hbar}\left((x_k-q_k)^2 + (x'_{k-1}-q_{k-1})^2\right).
\end{align}
We now have all the necessary parts to re-write the propagator~\eqref{eq:PropagatorWithInsertions}. However, we really want to discuss the discretized holomorphic amplitude
\begin{equation}\label{eq:HolAmpForDisc}
	W^\sigma(\mathring{z}_i,\mathring{z}_f,T) := \int_{\mathbb R^n}\dd^n x_i\int_{\mathbb R^n}\dd^n x_f\, \overline{\Psi^\sigma_{\mathring{q}_f,\mathring{p}_f}(x_f)}\,\langle x_f|U(T)|x_i\rangle\,\Psi^\sigma_{\mathring{q}_i, \mathring{p}_i}(x_i).
\end{equation} 
Before we proceed, a word about notation: In order to distinguish the boundary data $(\mathring{q}_i, \mathring{p}_i;\mathring{q}_f, \mathring{p}_f)$ from the auxiliary variables $(\{q_k\},\{p_k\})$ introduced by the $N+1$ resolutions of identity, we place a ring on top of it. Also, the parameter $\rho$ is part of the definition of the coherent states used in the resolution of identity, but it is completely devoid of physical meaning. Unlike the parameter $\sigma$ which appears through the boundary coherent states and is of physical significance.\\
 We can discretize the holomorphic amplitude~\eqref{eq:HolAmpForDisc} by replacing the propagator $\langle x_f|U(T)|x_i\rangle$ by the right hand side of~\eqref{eq:PropagatorWithInsertions} where the $N$ matrix elements involving the $U(\varepsilon)$ have been substituted by~\eqref{eq:MatrixElementForSubstitution}. Moreover, we need to use~\eqref{eq:BraKetCoherentStates}. The result of these operations reads
\begin{align}
	W^\sigma(\mathring{z}_i, \mathring{z}_f, T) &= \left(\frac{\sigma^2}{\pi\hbar}\right)^\frac{n}{2}\left(\frac{\rho^2}{\pi\hbar}\right)^{\frac{n(N+1)}{2}}\int_{\mathbb R^n}\dd^n x_i \int_{\mathbb R^n}\dd^n x_f \int_{\mathbb R^{nN}}\left(\prod^N_{k=1}\dd^n x_k\right)\times\notag\\
	&\phantom{==}\times \int_{\mathbb R^{n N}}\left(\prod^N_{k=1}\dd^n x'_{k-1}\right)\int_{\mathbb R^{2 n(N+1)}}\left(\prod^N_{k=0}\frac{\dd^n q_k\, \dd^n p_k}{(2\pi\hbar)^n}\right)\,\e^{\frac{i}{\hbar} A}.
\end{align}
The function $A$ denotes the discretized holomorphic action and it is explicitly given~by
\begin{align}\label{eq:DiscHolAction}
	A &= \sum_{k=1}^N\left(\frac{m}{2}\frac{(x_k-x'_{k-1})^2}{\varepsilon}-\varepsilon V(x'_{k-1})\right) + \sum_{k=1}^N\left(p_{k-1}(x'_{k-1}-q_{k-1}) - p_k(x_k-q_k)\right)\notag\\
	&\phantom{==}+i\frac{\rho^2}{2\hbar}\sum_{k=0}^N\left((x_k-q_k)^2 + (x'_k-q_k)^2\right) + p_N(x_f-q_N) - p_0(x_i-q_0)\notag\\
	&\phantom{==} + i\frac{\rho^2}{2\hbar}\left((x_i-q_0)^2 + (x_f-q_N)^2\right) + \mathring{p}_i(x_i-\mathring{q}_i) - \mathring{p}_f(x_f-\mathring{q}_f)\notag\\
	&\phantom{==} + i \frac{\sigma^2}{2\hbar}\left((x_i-\mathring{q}_i)^2 + (x_f-\mathring{q}_f)^2\right)
\end{align}
We recognize the first sum as being the discretized action functional~\eqref{eq:GeneralActionFunctional}, while the second and third sums arise from the coherent states used in the bulk (i.e. not at the end points) of the discretized path. All the terms that appear after the third sum are associated to the end points of the path.\\
Our goal is now to understand the classical limit of the discretized holomorphic amplitude. To that end, one can check without any difficulties that all conditions of theorem~\cref{thm:StatPhaseApprox} are satisfied. Hence, we can determine the critical point equations which, in particular, means that we need to take derivatives of $A$ with respect to $x_i$, $x_f$, $\{x_k\}$ and $\{x'_{k-1}\}$ for $k\in\{1,2,\dots,N\}$. This amounts to the following equations:
\begin{align}\label{eq:DiscreteCritPt}
	\PD{A}{x_i} \overset{!}{=}&\, 0 & & \Leftrightarrow & \mathring{p}_i-p_0 + i\frac{\rho^2}{\hbar}(x_i-q_0) + i\frac{\sigma^2}{\hbar}(x_i-\mathring{q}_i) =&\, 0\notag\\
	\PD{A}{x_f}\overset{!}{=}&\, 0 & & \Leftrightarrow & p_N -\mathring{p}_f + i\frac{\rho^2}{\hbar}(x_f-q_N) + i\frac{\sigma^2}{\hbar} (x_f-\mathring{q}_f) =&\, 0\notag\\
	\PD{A}{x_k} \overset{!}{=}&\, 0 & & \Leftrightarrow & \frac{m}{\varepsilon}(x_k-x'_{k-1})-p_k+ i\frac{\rho^2}{\hbar}(x_k-q_k) =&\, 0\notag\\
	\PD{A}{x'_{k-1}}\overset{!}{=}&\, 0 & & \Leftrightarrow & -\frac{m}{\varepsilon}(x_k-x'_{k-1}) - \varepsilon V'(x'_{k-1}) + p_{k-1}+i\frac{\rho^2}{\hbar}(x'_{k-1}-q_{k-1}) =&\, 0
\end{align}
Let us first focus on the first two equations and split them into real and imaginary parts. From the real parts we straightforwardly deduce 
\begin{equation}\label{eq:InitialP}
	p_0 = \mathring{p}_i\quad\text{and}\quad p_N = \mathring{p}_f.
\end{equation}
The imaginary part can be solved if we keep in mind that on the critical point, the imaginary part of the discretized holomorphic action~\eqref{eq:DiscHolAction} has to vanish. This then leads to the unique solutions
\begin{equation}\label{eq:InitialQ}
	x_i = q_0 = \mathring{q}_i\quad\text{and}\quad x_f = q_N = \mathring{q}_f.
\end{equation}
These solutions imply that the boundary data $(\mathring{q}_i, \mathring{p}_i;\mathring{q}_f, \mathring{p}_f)$ fixes the values of the auxiliary variables $(q_0,p_0; q_N, p_N)$ introduced by the resolutions of identity and they also fix the end points $x_i$ and $x_f$ of the propagator.\\
Next we consider the imaginary parts of the last two equations in~\eqref{eq:DiscreteCritPt}. These are unambiguously solved by
\begin{equation}
	x_k = q_k\quad\text{and}\quad x'_{k-1} = q_{k-1}\quad\text{for all} k\in\{1,2,\dots, N\},
\end{equation}
and these solutions also ensure $\text{Im}[A] = 0$ at the critical point. We can now use these solutions to simplify the real parts of the last two equations in~\eqref{eq:DiscreteCritPt}. These equations then become
\begin{align}
	\frac{m}{\varepsilon} (q_k-q_{k-1})-p_k &= 0\notag\\
	-\frac{m}{\varepsilon}(q_k-q_{k-1})-\varepsilon V'(q_{k-1}) + p_{k-1} &= 0.
\end{align}
By adding and subtracting the above equations from each other, we obtain the following solutions for the auxiliary variables $(\{q_k\}, \{p_k\})$
\begin{align}\label{eq:SimpleSolutions}
	p_k &= p_{k-1} - \varepsilon\PD{V(q_{k-1})}{q_{k-1}}\notag\\
	q_k &= q_{k-1} + \frac{\varepsilon}{m} p_k,	
\end{align}
which hold for all $k\in\{1,2,\dots, N\}$. At this point it is worth-while to pause and remark on the simplicity of these solutions, their transparent interpretation and the advantage they provide in numerical computations. The right hand side of the first equation in~\eqref{eq:SimpleSolutions} only depends on variables living on the $(k-1)$-th node of the discretized path while the second equation involves the position $q_{k-1}$ and the newly computed momentum $p_k$. Moreover, we know from equations~\eqref{eq:InitialP} and~\eqref{eq:InitialQ} that $p_0=\mathring{p}_i$ and $q_0=\mathring{q}_i$. Hence, it follows that the initial data $(\mathring{q}_i, \mathring{p}_i)$ is sufficient to recursively compute all $q_k$, $p_k$ with $k\in\{1,2,\dots, N\}$. Moreover, $\{q_k\}$ and $\{p_k\}$ are \textit{uniquely} determined by~\eqref{eq:SimpleSolutions} and the initial data. In other words: The equations~\eqref{eq:SimpleSolutions} together with the initial data guarantee the \textit{existence} and the \textit{uniqueness} of the discretized solution $\{q_k\}$ and can therefore be interpreted as the discrete analogue of an initial value problem (IVP).\\
This should be contrasted with the critical point equations obtained from the discretized propagator~\eqref{eq:DiscretizedPropagator} alone. That is, assume we only consider the propagator and we do not use the resolution of identity in terms of the coherent states $\Psi^\rho_{q_k,p_k}$. Then we would approximate the propagator in the usual way~\cite{FeynmanBook, KleinertBook}, which would give rise to the discrete action
\begin{equation}\label{eq:DiscAction22}
	S=\sum_{k=1}^N\left(\frac{m}{2}\frac{(x_k-x_{k-1})^2}{\varepsilon}-\varepsilon V(x_{k-1})\right)
\end{equation} 
and the critical point equations that result from this action are
\begin{equation}\label{eq:SlowEq}
\frac{m}{\varepsilon}(x_{k+1}-2x_k + x_{k-1})+\varepsilon\PD{V(x_k)}{x_k} \overset{!}{=} 0\quad\text{for all } k\in\{1,2,\dots, N-1\}.
\end{equation}
Notice that these equations are in general not linear because of the potential $V(x_k)$. Moreover, we do not have two initial conditions, but rather we have fixed end points,  $x_0\equiv x_i$ and $x_N\equiv x_f$. Hence, we can regard this equation as the discrete analogue of a boundary value problem (BVP), which we already encountered in subsection~\ref{ssec:AsympBehavior} when discussing the continuum theory.\\
In fact, these equations display some of the same problems as their continuous counterparts:  Solving~\eqref{eq:SlowEq} for $x_k$ in general requires inverting the potential~$V(x_k)$. This is a highly non-trivial problem and it necessitates the use of further approximation techniques. However, even when the potential is simple enough that it can be inverted it might result in a multivalued solution. Putting even this issue aside, finding a unique solution for $x_k$ is still not enough to solve the system~\eqref{eq:SlowEq} because of the presence of $x_{k+1}$. To find a solution to the whole system it is necessary to first solve these potentially non-linear equations formally (i.e. finding $x_k$ as function of other $x_k$'s), until one can use $x_0$ and $x_N$ to compute actual numerical values for the variables $\{x_k\}$. Evidently, this is a tedious and difficult exercise and there is no guarantee that a solution can be found. Plus, there might be several solutions and it is not clear how to discriminate between them. The IVP~\eqref{eq:SimpleSolutions} overcomes all these difficulties and it allows us to compute the discrete solution quickly end efficiently. An example will be provided further below.\\
However, these differences also raise some questions: What would have happened if we had not used the resolution of identity~\eqref{eq:HeatKernelResOfId} to derived the holomorphic action~\eqref{eq:DiscHolAction}? Would we run into trouble if we discretized the propagator in the traditional way and contracted it with coherent states? This procedure would give rise to a different discretized holomorphic action,
\begin{align}\label{eq:DiscHolActionDirect}
	\tilde{A} :=& \sum_{k=1}^N\left(\frac{m}{2}\frac{(x_k-x_{k-1})^2}{\varepsilon}-\varepsilon V(x_{k-1})\right) + \mathring{p}_i(x_i-\mathring{q}_i)-\mathring{p}_f(x_f-\mathring{q}_f)\notag\\
	&+i\frac{\sigma^2}{2\hbar}\left((x_i-\mathring{q}_i)^2 + (x_f-\mathring{q}_f)^2\right),
\end{align}
which is clearly the discrete analogue of the holomorphic action encountered in the foregoing subsection. The critical point equations for $x_k$ would be given by~\eqref{eq:SlowEq} plus
\begin{equation}
	x_0 \equiv x_i =\mathring{q}_i\quad\text{and}\quad x_N\equiv x_f=\mathring{q}_f.
\end{equation}
It seems that one runs into the same problems as with~\eqref{eq:SlowEq}. However, there are now modifications induced by the coherent states. In fact, now we also need to take derivatives with respect to $x_i\equiv x_0$ and $x_f\equiv x_N$. Let us have a closer look at the equation for $x_0$:
\begin{align}
	\PD{\tilde{A}}{x_0} \overset{!}{=}&\, 0 & & \Leftrightarrow & \mathring{p}_i-\frac{m}{\varepsilon}(x_1-x_0)-\varepsilon V'(x_0)  =&\, 0
\end{align}
Using $x_0 = \mathring{q}_i$, this equation can easily be solved for $x_1$ yielding
\begin{equation}\label{eq:SecondCondition}
	x_1 = \mathring{q}_i + \underset{=: p_1}{\underbrace{(\mathring{p}_i - \varepsilon V'(\mathring{q}_i))}}\frac{\varepsilon}{m}.
\end{equation}
This solution has precisely the same form as the second equation in~\eqref{eq:SimpleSolutions} and notice that it is crucial that we know the momentum data $\mathring{p}_i$. This ensures that we can actually compute $x_1$ and with this we know two initial points, $x_0$ and $x_1$. Hence, we find again that we are dealing with an IVP, not a BVP.\\
In fact, knowing $x_0$ and $x_1$ allows us to iteratively solve~\eqref{eq:SlowEq}. To see how this works, let us have a look at the equation for $k=1$:
\begin{equation}
	\frac{m}{\varepsilon}\left(x_2-2 x_1 + x_0\right) + \varepsilon V'(x_0) = 0.
\end{equation}
Using $x_0=\mathring{q}_i$ and $x_1 = \mathring{q}_i + p_1\frac{\varepsilon}{m}$, where $p_1$ is the auxiliary variable defined in~\eqref{eq:SecondCondition}, we can re-write this equation as
\begin{align}
	\frac{m}{\varepsilon}\left(x_2-x_1\right) + \underset{=-p_1}{\underbrace{\frac{m}{\varepsilon}\left(\mathring{q}_i-x_1\right)}}+\varepsilon V'(x_1) &= 0\notag\\
	\Rightarrow\quad\frac{m}{\varepsilon}\left(x_2-x_1\right)-\underset{=:p_2}{\underbrace{\left(p_1-\varepsilon V'(x_1)\right)}} &= 0
\end{align}
where we have introduced a new auxiliary variable, $p_2:=p_1-\varepsilon V'(x_1)$, which is completely determined by quantities we already know. We therefore find for $x_2$
\begin{equation}
	x_2 = x_1+ \frac{\varepsilon}{m}p_2.
\end{equation} 
The auxiliary variable $p_2$ and the value of $x_2$ are determined by precisely the same equations as~\eqref{eq:SimpleSolutions}! We can iterate the process just illustrated and determine all $x_k$ by splitting the equations in a smart way and introducing auxiliary variables $p_k$. Doing so amounts to proving that the solutions~\eqref{eq:SimpleSolutions} also solve the critical point equations derived from $\tilde{A}$!\\
This is a nice consistency check and we find some close analogies to the continuum case. Let us summarize our findings:
\begin{itemize}
	\item If we only consider the propagator, not the holomorphic amplitude, and we discretize it using standard techniques, we simply find the discretized action~\eqref{eq:DiscAction22}. The resulting critical point equations~\eqref{eq:SlowEq} are in general difficult to solve because we only have the initial and final point of the path as data and the equations are non-linear. This non-linear behavior can also give rise to multivalued solutions. Hence, we are confronted with a BVP, just as in the continuum case discussed in the previous subsection.
	\item If we discretize the holomorphic amplitude directly, i.e. if we discretize the propagator using standard techniques but also take into account the contraction with the coherent states, we find the discretized holomorphic action~\eqref{eq:DiscHolActionDirect}. This is precisely the discrete analogue of~\eqref{eq:ComplexAction} one would expect. Since we know the initial data $(\mathring{q}_i,\mathring{p}_i)$, we can solve the critical point equations iteratively be splitting them in a smart way. Hence, we can ``evolve'' the data $(\mathring{q}_i,\mathring{p}_i)$ in a unique way and we find a discrete analogue of an IVP. 
	\item Finally, we can also discretize the holomorphic amplitude by inserting resolutions of identity in terms of coherent states. The resulting discretized holomorphic action looks very complicated, but the critical point equations are very simple. In fact, the effect of the coherent states is that they perform the aforementioned splitting of equations for us and we can recognize more easily that there is a unique solution to the critical point equations. Moreover, this solution solves also the critical point equations of the discretized holomorphic amplitude described in the foregoing bullet point. This had to be expected since the two discretizations only differ by a number of insertions of identities! 
\end{itemize}
The idea that the critical point equations derived from the discrete holomorphic action~\eqref{eq:DiscHolAction} constitutes and IVP can be made even more precise. In fact, we recognize the equations~\eqref{eq:SimpleSolutions} to represent what in numerical mathematics are known as the equations of the \textit{forward Euler method} for numerical integration of ordinary differential equations (ODE's). This method can be summarized as follows~\cite{ButcherBook}:\\
\textcolor{white}{empty line}\\
Let $y'(t) =f(t, y(t))$ with $y(t_0)=y_0$ be the ODE to be solve. Furthermore, assume $f:[t_i,t_f]\times\mathbb{R}^N\to\mathbb{R}^N$ to be continuous and to satisfy the Lipschitz condition $\|f(t,y)-f(t,z)\|\leq L\|y-z\|$. Choose a step size $\varepsilon$ for every step and set $t_k=t_0 + \varepsilon k$. Then, a step from $t_{k-1}$ to $t_k$ is given by
\begin{equation}
	y_{k} = y_{k-1}+\varepsilon f(t_{k-1},y_{k-1}).
\end{equation}
The value $y_k$ is an approximate solution to the ODE at time $t_k$: $y_k\approx y(t_k)$.\\
\textcolor{white}{empty line}\\
It can be shown~\cite{ButcherBook} that the discrete solution $\{y_k\}$ provided by the Euler method converges to the smooth solution $y(t)$ of the original IVP. This follows from the following error estimates~\cite{ButcherBook}:
\begin{itemize}
	\item[i)] The error caused by one iteration, the so-called local truncation error (\textsf{LTE}), is given by 
	\begin{equation}
		\textsf{LTE} := y(t_0+\varepsilon)-y_0 = \frac{1}{2}\varepsilon^2 y''(t_0)+\mathcal{O}(\varepsilon^3),
	\end{equation}
	i.e. the error grows locally like $\sim \varepsilon^2$.
	
	\item[ii)] The cumulative error caused by many iterations, the so-called global truncation error (\textsf{GTE}), is bound from above by
	\begin{equation}
		\textsf{GTE}:=\|y(t)-\tilde y(t)\|\leq \frac{\varepsilon M}{2 L}\left(\e^{L(t-t_0)}-1\right),
	\end{equation}
	where $M$ is the upper bound on the second derivative of $y$, $|y''(t)|\leq M$, $L$ is a Lipschitz constant, and $\tilde y$ is the solution computed from the Euler method. In general the \textsf{GTE} overestimates the actual error.
\end{itemize}
Returning to the critical point equations~\eqref{eq:SimpleSolutions}, we see that  the Euler method implies that these equations approximate the classical phase space trajectory $(q(t), p(t))$ which starts in $(\mathring{q}_i, \mathring{p}_i)$.

\begin{center}
\begin{figure}[h!]
	\centering
	\includegraphics[width=.5\textwidth]{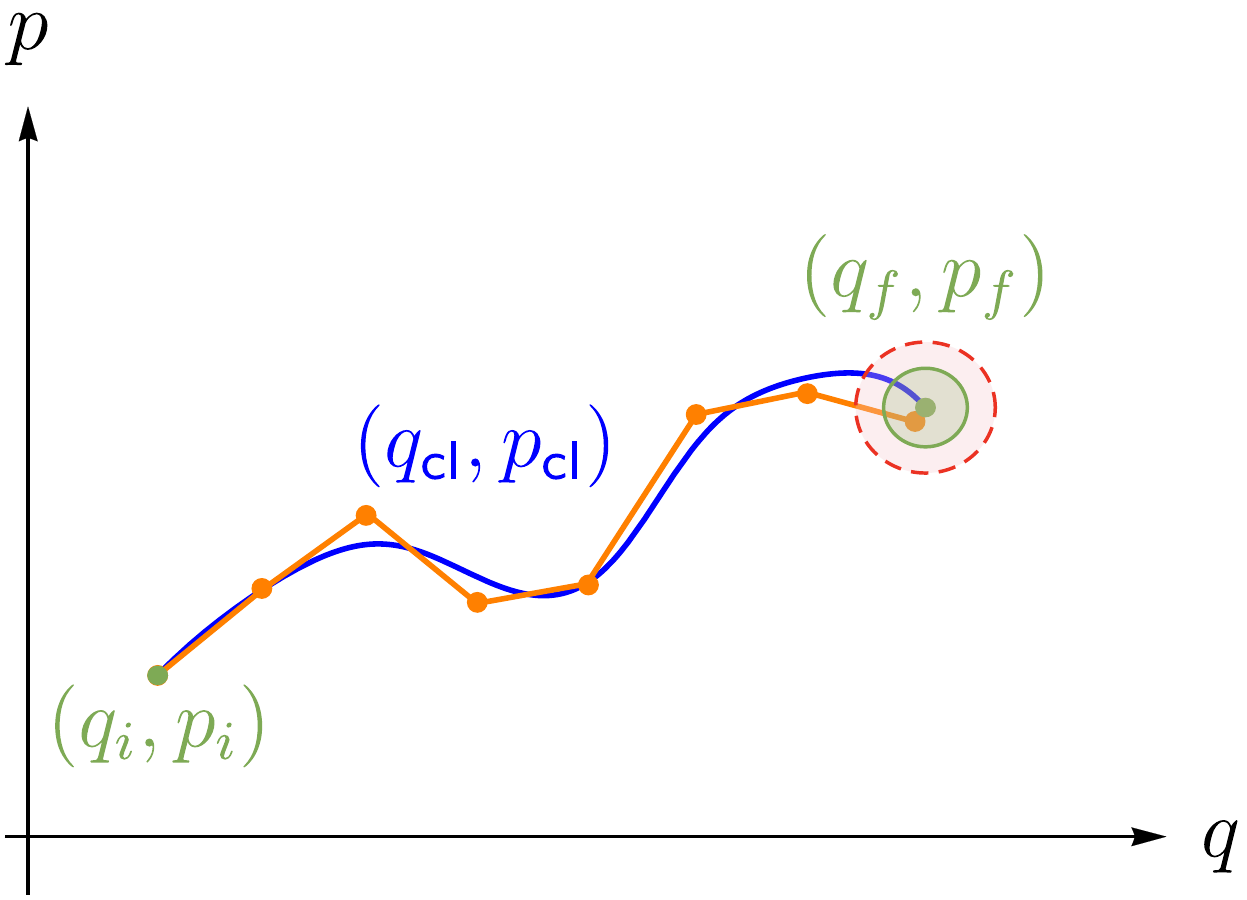}
	\caption{The qualitative picture of the discrete theory. It reproduces most aspects of the continuum theory.}
	\label{fig:DPSP}
\end{figure}
\end{center}

So far we assumed $\mathring{q}_i$, $\mathring{p}_i$ to represent the initial data and then, as long as $|\PD{V}{q_{k-1}}|<\infty$, the Euler method generates a discrete solution $(\{q_k\},\{p_k\})$. However, there is also the data $(\mathring{q}_f, \mathring{p}_f)$ on the end point of the discrete trajectory. Since we are working with a discretization, we introduce numerical errors by hand and it is therefore unlikely that the last critical point equation, $q_N=\mathring{q}_f$ and $p_N=\mathring{p}_f$, can be satisfied. It follows that the transition amplitude is suppressed and, moreover, the suppression factor is given by~\eqref{eq:SuppressionFactor}. Hence, the transition amplitude is not only suppressed for semi-classical boundary data, it has to be expected in general that it is suppressed simply because of numerical errors.\\
This is clearly an unsatisfactory situation and that is precisely where the above error estimates come to our help. Whether or not $q_N\neq \mathring{q}_f$ and $p_N\neq \mathring{p}_f$ is due to the non-existence of a classical trajectory or to numerical errors can be decided by
\begin{equation}
	\|q_N-\mathring{q}_f\| < \varepsilon\, K_q \lesssim\textsf{GTE}\quad\text{and}\quad \|p_N-\mathring{p}_f\|< \varepsilon\, K_p\lesssim\textsf{GTE}
\end{equation}
for some appropriately chosen constants $K_q$, $K_p$. In other words, we can introduce fiducial regions $\|q_N-\mathring{q}_f\|\lesssim \varepsilon\, K_q$ and $\|p_N-\mathring{p}_f\|\lesssim \varepsilon\, K_p$, for which we \textit{declare} the numerical solution to be accurate enough. The $\textsf{GTE}$ may serve as a guide line in establishing such regions. Hence, we can visualize what is happening in the discrete theory as in Figure~\ref{fig:DPSP}.
The initial data $(\mathring{q}_i,\mathring{p}_i)$ is evolved by the discrete propagator and the discrete phase space trajectory (orange line) can be made to converge to the classical phase space trajectory (blue curve). However, due to numerical errors, we may not be able to satisfy the last equation. Hence, we define a fiducial region (green disc) within which we can confidently ascribe any deviation between boundary data and computed solution to numerical errors. Within this region, we also disregard the suppression factor~\eqref{eq:SuppressionFactor}.\\
Alternatively, we can use the Euler method to evolve the data $(\mathring{q}_i, \mathring{p}_i)$ and use it to \textit{define} what $(\mathring{q}_f, \mathring{p}_f)$ should be. Then, we can study semi-classical deviations around the final point $(\mathring{q}_f, \mathring{p}_f)$ which has been determined by the discretized classical equations. This is precisely the strategy which we will use in the CLQG case in the next section.

Before proceeding to the next section, where we apply the approximation methods discussed here to the holomorphic CLQG transition amplitude, we make a final observation. We may regard the critical point equations~\eqref{eq:SimpleSolutions} also as discretized Hamiltonian equations of motion:
\begin{align}\label{eq:HamiltonianEOM}
		\PD{H}{q} &= -\dot p & & \longrightarrow & & & \left.\PD{V}{q}	\right|_{q=q_{k-1}} &=-\frac{p_k-p_{k-1}}{\varepsilon} & & \Leftrightarrow & &  p_k=p_{k-1}-\PD{V(q_{k-1})}{q_{k-1}} \notag\\
		\PD{H}{p} &= \dot q & & \longrightarrow & & & \left.\frac{p}{m}\right|_{p=p_k} &= \frac{q_k-q_{k-1}}{\varepsilon} & & \Leftrightarrow & & q_k = q_{k-1}+\frac{\varepsilon}{m} p_k
\end{align}
This fits very well into the phase space picture reported here, but at the time of writing this connection has not been fully explored. We keep it in mind though, as it would be interesting to see if in CLQG it is possible to establish a connection to the (discrete) canonical theory.

\section{Semi-Classical Transition Amplitudes in Spin Foams}
\label{sec:SemiClassicalCLQG}
Consider a compact spacetime region $\mathcal R$ enclosed by a three-dimensional spacelike hypersurface $\Sigma$. Triangulate both manifolds and assign an extrinsic coherent state~$\Psi^t_{\Gamma, H}$ to the boundary graph $\Gamma$. Choose the boundary data $(a_\ell, \xi_\ell, \vec{n}_{s(\ell)},\vec{n}_{t(\ell)})$ and the semi-classicality parameter $t$ such that the boundary state is peak on an intrinsic and extrinsic three-geometry with large areas $a_\ell$ and small spread in the conjugate variables $A_\ell$ and $h_\ell$. Then we can assign the holomorphic transition amplitude~\eqref{eq:HolomorphicAmplitude} to the triangulated spacetime region.\\
In a physical situation such as the one described in~\cite{Haggard:2014,RealisticObs}, we may further wish to divide the boundary into two pieces\footnote{Notice that we \textit{can} introduce this subdivision, but we \textit{do not have to}. The advantage is of course that we can talk about the transition from one geometry to an other geometry and relate more easily to processes such as the one described in \cite{RealisticObs}.}. Call them $\Sigma_\textsf{in}$ and $\Sigma_\textsf{out}$, and choose them such that $\Sigma=\Sigma_\textsf{in}\cup\Sigma_\textsf{out}$ and $\Psi^t_{\Gamma, H} = \Psi^t_{\Gamma_\textsf{in}, H_\textsf{in}}\otimes\Psi^t_{\Gamma_\textsf{out}, H_\textsf{out}}$. The state $\Psi^t_{\Gamma_\textsf{in}, H_\textsf{in}}$ associated to the triangulation of $\Sigma_\textsf{in}$ can then be taken to represent our knowledge of the spacetime geometry at a certain instant of time while the state $\Psi^t_{\Gamma_\textsf{out}, H_\textsf{out}}$ describes the spacetime geometry at a different instant of time. The question we are now interested in is: Given the initial state $\Psi^t_{\Gamma_\textsf{in},H_\textsf{in}}$ on the spacetime slice $\Sigma_\textsf{in}$, what is the amplitude to find the spacetime slice $\Sigma_\textsf{out}$ to be described by $\Psi^t_{\Gamma_\textsf{out},H_\textsf{out}}$, where $H_\textsf{out}$ deviates from the classically expected boundary data?\\
This setup is the quantum gravity analogue of the situation described in subsection~\ref{ssec:FramingTheQuestion} and we will emulate the strategy used for the quantum mechanical case to answer the above question. In the next subsection we will briefly discuss the classical limit of holomorphic CLQG amplitudes and contrast the results with the classical limit of the propagator discussed in~\ref{ssec:ClassicalLimitGeometry}.\\
We then follow up with developing an approximation for CLQG transition amplitudes which depend on semi-classical boundary data.

\subsection{Classical Limit of Holomorphic CLQG Amplitudes}
In chapter~\ref{chap:CLQG} we gave a detailed derivation of the holomorphic CLQG transition amplitude on a $2$-complex $\mathcal{C}$ with boundary graph $\Gamma$. In a strongly simplified notation, this amplitude can be re-written as
\begin{equation}\label{eq:SimplifiedTransAmp}
	W_\mathcal{C}^{t}(H_\ell):= \left<W_\mathcal{C}\,\vline\,\Psi^t_{\Gamma, H_\ell}\right> = \sum_{j_\fa, j_\ell}\nu(j) \int_{\slc}\dd\mu(g)\int_{\CP}\dd\rho(\textbf{z})\e^{i A[g, \textbf{z}; a, \xi, \textbf{n}_s, \textbf{n}_t]},
\end{equation}
where $\nu(j)$ is a certain polynomial of bulk and boundary spins, $\dd\mu(g)$ is an appropriately regularized product of $\slc$ Haar measures, $\dd\rho(\textbf{z})$ refers to a product of certain homogeneous and $\slc$ invariant measures on $\CP$ and $\textbf z\in\mathbb C^2$ denotes auxiliary spinorial variables. For detailed definitions of all measures and variables see chapter~\ref{chap:CLQG}~section~\ref{sec:TransitionAmplitudesInCLQG}. Recall also from that section that the holomorphic spin foam action is defined as
\begin{equation}\label{eq:CLQGComplexAction}
	A[g,\textbf z; a, \xi, \textbf{n}_s, \textbf{n}_t] := \sum_{\fa\in\mathcal B}L_\fa[g, \textbf z] + \sum_{\ell\in\Gamma}\left(\tilde{B}_\ell[g, \textbf z] + D_\ell[a, \xi, \textbf{n}_s, \textbf{n}_t]\right),
\end{equation}
where the bulk Lagrangian $L_\fa$, the boundary term $\tilde{B}_\ell$ and the contribution $D_\ell$ from the boundary data are given by
\begin{align}\label{eq:Lagrangians}
	L_\fa[g,\textbf z] &:= \gamma j_\fa \sum_{\ed\in\fa}\left( \ln\frac{\innerp{\textbf{Z}_{\ve\ed'\fa}}{\textbf{Z}_{\ve\ed'\fa}}}{\innerp{\textbf{Z}_{\ve'\ed'\fa}}{\textbf{Z}_{\ve'\ed'\fa}}} - \frac{i}{\gamma}\ln\frac{\innerp{\textbf{Z}_{\ve'\ed'\fa}}{\textbf{Z}_{\ve\ed'\fa}}^2}{\innerp{\textbf{Z}_{\ve'\ed'\fa}}{\textbf{Z}_{\ve'\ed'\fa}}\innerp{\textbf{Z}_{\ve\ed'\fa}}{\textbf{Z}_{\ve\ed'\fa}}}\right)\notag\\
	\tilde{B}_\ell[g, \textbf z] &:= \gamma j_\ell \left(\ln\frac{\innerp{\textbf{Z}_{\ve^{(n)}\no\ell}}{\textbf{Z}_{\ve^{(n)}\no\ell}}}{\innerp{\textbf{Z}_{\ve\no'\ell}}{\textbf{Z}_{\ve\no'\ell}}} - \frac{i}{\gamma}\ln\frac{\innerp{\textbf{Z}_{\ve\no'\ell}}{\textbf{z}_{\no'\ell}}^2\innerp{\textbf{z}_{\no\ell}}{\textbf{Z}_{\ve^{(n)}\no\ell}}^2}{\innerp{\textbf{Z}_{\ve\no'\ell}}{\textbf{Z}_{\ve\no'\ell}}\innerp{\textbf{Z}_{\ve^{(n)}\no\ell}}{\textbf{Z}_{\ve^{(n)}\no\ell}}}\right) \notag\\
	D_\ell[a,\xi,\textbf{n}_s, \textbf{n}_t] &:= i(j_\ell-a_\ell)^2 t -\gamma j_\ell \xi_\ell - i j_\ell\ln\frac{\innerp{\overline{\textbf n}_{s(\ell)}}{\textbf{z}_{\no\ell}}^2\innerp{\textbf{z}_{\no'\ell}}{\overline{\textbf{n}}_{t(\ell)}}^2}{\innerp{\textbf{z}_{\no\ell}}{\textbf{z}_{\no\ell}}\innerp{\textbf{z}_{\no'\ell}}{\textbf{z}_{\no'\ell}}}.
\end{align}
Considering that there is a large amount of indices, variables, and parameters, it is advisable recalling some of the definitions provided in chapter~\ref{chap:CLQG} to avoid confusion:
The indices $\ve,\ed,\fa$ stand for vertex, edge and face, respectively, and they are associated with the bulk $\mathcal B$ of the $2$-complex. Links and nodes, on the other hand, live on the boundary of the $2$-complex and they are designated by $\ell$ and~$\no$, respectively. Consequently, we think of the spinorial variables $\textbf{Z}_{\ve\ed\fa}$ as being associated with the bulk while $\textbf{z}_{\no\ell}$ belong to the boundary. There is one type of variable, $\textbf{Z}_{\ve\no\ell}:=g^\dagger_{\ve\no} \textbf{z}_{\no\ell}$, which carries mixed indices. These hybrid variables appear in the boundary term $\tilde{B}_\ell$ and they connect the boundary to the bulk.\\
Then, there is the boundary data which associates an area\footnote{$a_\ell$ is really dimensionless and related to an actual area $A_\ell$ through $a_\ell:=A_\ell /(\gamma l^2_\text{pl})$.} $a_\ell\in\mathbb R^+$ to every link of the boundary, an angle $\xi_\ell\in [0,4\pi)$ which encodes the extrinsic curvature, and two $3$d unit normal vectors, $\vec{n}_{s(\ell)}$ and $\vec{n}_{t(\ell)}$, associated to source and target of every link. The normal vectors do not enter directly into the complex action~\eqref{eq:CLQGComplexAction} but rather through their spinorial\footnote{The bar over $\textbf{n}$ in equation~\eqref{eq:Lagrangians} simply indicates complex conjugation} counter part defined by the mapping
\begin{equation}\label{eq:ParamNormalVec}
	\vec{n} = \begin{pmatrix}
		\sin\theta\cos\phi\\
		\sin\theta \sin\phi\\
		\cos\theta
	\end{pmatrix}\,\, \mapsto\,\, \textbf{n} = \begin{pmatrix}
		\cos\frac{\theta	}{2}\\
		\e^{i\phi}\sin\frac{\theta}{2}
	\end{pmatrix}
\end{equation}
with $\theta\in[0,\pi]$ and $\phi\in [0,2\pi)$. The holomorphic amplitude is a function of the boundary data $(a_\ell, \xi_\ell, \vec{n}_{s(\ell)}, \vec{n}_{t(\ell)})$.\\
Determining its classical limit is straightforward since we already know the classical limit of the propagator. If we momentarily forget about the variables associated to the boundary, i.e. $g_{\ve\no}$ and $\textbf{z}_{\no\ell}$, we know that the conditions of theorem~\cref{thm:StatPhaseApprox} hold and we can perform a stationary phase approximation of the bulk variables. Doing so yields of course exactly the same critical point equations as for the propagator. Extending the stationary phase analysis to the boundary variables requires us to check the conditions of theorem~\cref{thm:StatPhaseApprox} and we find that $\text{Im}[\tilde{B}_\ell]\geq 0$ is still trivially true (for the same reason that $\text{Im}[L_\fa]\geq 0$ is true, see eg. \cite{Barrett:2009,Han:2011, HanKrajewski:2013}) and $\text{Im}[\tilde{B}_\ell]\overset{!}{=} 0$ constitutes a critical point equation which is formally similar to the critical point equation $\text{Im}[L_\fa]\overset{!}{=} 0$ (cf. equation~\eqref{eq:CritEq}).\\
What remains to be checked is that the boundary data contribution $D_\ell$ satisfies these conditions. For the imaginary part of $D_\ell$ we find
\begin{equation}
	\text{Im}[D_\ell] = (j_\ell-a_\ell)^2 t - j_\ell \ln\frac{|\innerp{\overline{\textbf{n}}_{s(\ell)}}{\textbf{z}_{\no\ell}}|^2  \,  |\innerp{\textbf{z}_{\no'\ell}}{\overline{\textbf{n}}_{t(\ell)}}|^2}{|\innerp{\textbf{z}_{\no\ell}}{\textbf{z}_{\no\ell}}| \,  |\innerp{\textbf{z}_{\no'\ell}}{\textbf{z}_{\no'\ell}}|}\geq 0.
\end{equation}
Clearly, the first term is always larger or equal to zero. For the logarithmic term we can employ the Cauchy-Schwarz inequality, $|\innerp{\textbf{x}}{\textbf{y}}|^2 \leq \innerp{\textbf{x}}{\textbf{x}} \innerp{\textbf{y}}{\textbf{y}}$, to conclude that the logarithm is always smaller or equal to zero. Therefore, due to the minus sign in front of the logarithm, the whole sum is larger or equal to zero.\\
The next requirement is that the imaginary part vanishes on a critical point:
\begin{equation}\label{eq:SameReggeGeometry}
	\text{Im}[\left.D_\ell\right|_\textsf{c}] \overset{!}{=} 0 \quad\Leftrightarrow\quad \left\{ \begin{matrix}
		j_\ell & = & a_\ell\textcolor{white}{s(\ell)}\\
		\textbf{z}_{\no\ell} & = & \e^{-\frac{i}{2}\varphi_{s(\ell)}}\overline{\textbf{n}}_{s(\ell)}\\
		\textbf{z}_{\no'\ell} & = & \e^{\frac{i}{2}\varphi_{t(\ell)}}\overline{\textbf{n}}_{t(\ell)}
	\end{matrix}\right.
\end{equation}
Observe that the spinors $\overline{\textbf{n}}_{\no(\ell)}$ are normalized to one, as they correspond to unit normal vectors, and they force the spinors $\textbf{z}_{\no\ell}$ to be unimodular as well. This is potentially problematic since these spinors are in general not normalized and they not only appear in $D_\ell$, but also in $\tilde{B}_\ell$ together with the spinors $\textbf{Z}_{\ve\no\fa}$. However, from the requirement $\text{Im}[\tilde{B}_\ell]\overset{!}{=}0$ we find 
\begin{equation}
	\textbf{z}_{\no\ell} = \e^{i \alpha} \frac{\textbf{Z}_{\ve^{(n)}\no\ell}}{\|\textbf{Z}_{\ve^{(n)}\no\ell}\|}\quad \text{and}\quad \textbf{z}_{\no'\ell} = \e^{i \beta} \frac{\textbf{Z}_{\ve\no'\ell}}{\|\textbf{Z}_{\ve\no'\ell}\|},
\end{equation}
which also implies that $\textbf{z}_{\no\ell}$ is unimodular and therefore everything is nicely consistent. Since all the requirements of the stationary phase theorem are met, we can also impose the critical point equations $\delta_{g_{\ve\no}}A\overset{!}{=}0$ and $\delta_{\textbf{z}_{\no\ell}}A\overset{!}{=}0$. They are similar to the other two critical point equations in~\eqref{eq:CritEq} and do not add anything new. \\
Recall that when we considered just the plain propagator, without any boundary states, we had to choose boundary data $(j_\ell, \textbf{z}_{\no\ell}, \textbf{z}_{\no'\ell})$ in order to have a well posed boundary value problem. In the case of the holomorphic amplitude, the presence of the  boundary data contribution $D_\ell$ has the effect that $j_\ell$, $\textbf{z}_{\no\ell}$ and $\textbf{z}_{\no'\ell}$ are specified (up to the phases $\e^{\frac{i}{2}\varphi}$) by the boundary data $(a_\ell, \vec{n}_{s(\ell)},\vec{n}_{t(\ell)})$ provided through the semi-classical coherent states. Notice furthermore that the phase difference between the $\textbf{n}$ and $\textbf{z}$ spinors is not determined by any equation and it is irrelevant since both spinors reconstruct the same $3$d unit vector $\overline{\textbf{n}}^\dagger\,\vec{\sigma}\,\overline{\textbf{n}} = \textbf{z}^\dagger\,\vec{\sigma}\,\textbf{z}$. Hence we can say that $(j_\ell, \textbf{z}_{\no\ell}, \textbf{z}_{\no'\ell})$ and $(a_\ell, \vec{n}_{s(\ell)},\vec{n}_{t(\ell)})$ reconstruct the same $3$d discrete boundary geometry and the phases $\e^{\frac{i}{2}\varphi_{\no(\ell)}}$ are just gauge.\\
What is happening here is analogous to what happens in quantum mechanics, where the coherent states impose the boundary conditions $x_i=q_i$ and $x_f=q_f$. Here, the coherent states describe a whole $3$d geometry which serves as boundary condition for the critical point equations.\\
This situation can be slightly improved and brought even closer to the quantum mechanical case by introducing a new approximation. In subsection~\ref{ssec:ComputationBounceTime}, when applying the holomorphic amplitude to the black hole to white hole transition, we saw that when the semi-classicality condition~\eqref{eq:SemiClassicality} is satisfied, the spin sums over the boundary spins can be considered as being independent. This allows us then to apply the Euler-Maclaurin theorem to approximate the sums by integrals.
Applying these ideas to the current situation, we first of all promote all boundary spin sums in~\eqref{eq:SimplifiedTransAmp} to integrals: $\sum_{j_\ell}\rightarrow \int\dd j_\ell$. Next we extend the stationary phase approximation to these integrals. All conditions of theorem~\cref{thm:StatPhaseApprox} continue to hold since we are still talking about the action~\eqref{eq:CLQGComplexAction}. However, now we have an additional critical point equation which results from taking derivatives with respect to~$j_\ell$. It is thereby important to notice that both, $\tilde{B}_\ell$ and $D_\ell$ contain a logarithmic term with $\textbf{z}_{\no\ell}$ and $\textbf{z}_{\no'\ell}$ which is multiplied by $j_\ell$. On a critical point, which, in particular, is a point on which equation~\eqref{eq:SameReggeGeometry} is true, the logarithm of $\tilde{B}_\ell$ is equal to $\varphi_{s(\ell)}+\varphi_{t(\ell)}$, while the logarithm of $D_\ell$ is equal to $-\varphi_{s(\ell)}-\varphi_{t(\ell)}$. Hence, these two contributions cancel. The other logarithmic term in $\tilde{B}_\ell$ vanishes identically when evaluated on a critical point because the spinorial variables are unimodular. With these observations in mind we find
\begin{equation}\label{eq:ConditionOnExtrGeom}
	\left.\PD{A}{j_\ell}\right|_{\mathsf{c}} \overset{!}{=} 0\quad \Leftrightarrow\quad \textsf{s}\,\theta_\ell - \xi_\ell = 0,
\end{equation}
where the angle $\textsf{s}\,\theta_\ell$ is the angle that appears in the boundary term of the Hamilton-Regge function (cf. equation~\eqref{eq:HamiltonRegge}) and it can be computed as\footnote{Notice the curious similarity of equations~\eqref{eq:ConditionOnExtrGeom} and~\eqref{eq:SimilarEq} to $\left.\PD{S}{q_i}\right|_\textsf{c} + p_i = 0$ in mechanics and $\Pi^{ab} = \left.\frac{\delta}{\delta q_{ab}}\left(S_\text{EH} + S_\text{GHY}\right)\right|_{\text{EOM}}$ in General Relativity.}
\begin{equation}\label{eq:SimilarEq}
	\textsf{s}\,\theta_\ell =\frac{\partial}{\partial j_\ell}\left.\left(\sum_{\fa\in\mathcal{B}}L_\fa + \sum_{\ell\in\Gamma}\tilde{B}_\ell\right)\right|_{\textsf{c}}.
\end{equation}
We conclude that the holomorphic transition amplitude is not only suppressed when the BVP, which consists of the critical point equation subjected to the (intrinsic) Regge-like three-geometries $(j_\ell, \vec{n}_{s(\ell)}, \vec{n}_{t(\ell)})$ as boundary conditions, cannot be solved, but also when the extrinsic curvatures $\textsf{s}\,\theta_\ell$ (computed from the critical point equations) and $\xi_\ell$ (provided by the boundary data) do not match. This is analogous to what happens in quantum mechanics and it suggests there might be a formulation of the problem spin foam critical point equations as IVP which would remove superfluous critical points.

Finally, if we evaluate the holomorphic action on a critical point, as defined by the critical point equations~\eqref{eq:CritEq} for the propagator supplemented by the equations~\eqref{eq:SameReggeGeometry} and~\eqref{eq:ConditionOnExtrGeom}, we find that on the boundary the data $\xi_\ell$ selects only one critical point though $\textsf{s}\,\theta_\ell-\xi_\ell=0$. Whether or not the boundary data propagates into the bulk and also there selects only one of the two possible critical points at each vertex is an open question and under current investigation.

\subsection{Semi-Classical Holomorphic CLQG Amplitudes}
\label{ssec:SemiClassicalHolomorphicCLQGAmplitudes}
In order to understand the dependence of holomorphic CLQG amplitudes on their boundary data and describe their behavior away from critical points, we emulate the strategy developed for the quantum mechanical case in section~\ref{sec:QM_Continuum}. The main assumptions of this subsection are as follows:\\
We work with holomorphic transition amplitudes in the large area limit, $a_\ell\gg 1$, and with small semi-classicality parameter, $t\ll 1$, such that the semi-classicality condition $1\ll \sqrt{t} a_\ell \ll a_\ell$ holds.  In turn, this implies that the boundary spin sums can be treated as independent, as explained in subsection~\ref{ssec:ComputationBounceTime}, and we can approximate them by integrals.\\
Our goal is then to apply the usual stationary phase approximation, theorem~\cref{thm:StatPhaseApprox}, to the bulk variables of the holomorphic amplitude
\begin{equation}\label{eq:ModifiedTransAmp}
	W_\mathcal{C}^{t}(H_\ell) =  \sum_{j_\fa} \int_{\mathbb R^+}\nu(j)\,\dd j_\ell \int_{\slc}\dd\mu(g)\int_{\CP}\dd\rho(\textbf{z})\e^{i A[g, \textbf{z}; a, \xi, \textbf{n}_s, \textbf{n}_t]},
\end{equation}
while the generalized stationary phase approximation, theorem~\cref{thm:GenStatPhaseApprox}, is reserved for the boundary integrals over $j_\ell$.

We already know that all conditions for the application of theorem~\cref{thm:StatPhaseApprox} are met. For the application of theorem~\cref{thm:GenStatPhaseApprox} we need to keep in mind that there are two sets of variables, $y$ and $v$, which are related to the integration variable $j_\ell$ and the boundary data. Moreover, theorem~\cref{thm:GenStatPhaseApprox} assumes the critical point equations as well as the non-degeneracy condition to hold only if the critical point is located at $(y;v)=(0,0)$. In order to meet these requirements, we introduce the \textit{background data} $(\mathring{a}_\ell, \mathring{\xi}_\ell, \mathring{\textbf{n}}_{s(\ell)},\mathring{\textbf{n}}_{t(\ell)})$ which satisfies the critical point equations:
\begin{align}
\left\{ \begin{matrix}
		j_\ell & = & \mathring{a}_\ell\textcolor{white}{s(\ell)}\\
		\textsf{s}\,\theta_\ell & = & \mathring{\xi}_\ell\textcolor{white}{s(\ell)}\\
		\textbf{z}_{\no\ell} & = & \e^{-\frac{i}{2}\varphi_{s(\ell)}}\mathring{\overline{\textbf{n}}}_{s(\ell)}\\
		\textbf{z}_{\no'\ell} & = & \e^{\frac{i}{2}\varphi_{t(\ell)}}\mathring{\overline{\textbf{n}}}_{t(\ell)}
	\end{matrix}\right.
\end{align}
These data define a classical discrete background geometry, which is assumed to be Lorentzian, and around which we will later study semi-classical deviations. Based on this data we can also introduce the simple change of integration variables
\begin{equation}
	j_\ell\quad\longrightarrow\quad y_\ell:=j_\ell-\mathring{a}_\ell.
\end{equation}
This ensures that $y_\ell=0$ corresponds to the critical point. To define the variable~$v_\ell$ and to understand what happens when the actually chosen boundary data, $(a_\ell, \xi_\ell, \textbf{n}_{s(\ell)},\textbf{n}_{t(\ell)})$, does not coincide with the background data we parametrize $a_\ell$ and $\xi_\ell$ as follows:
\begin{equation}
	\xi_\ell = \mathring{\xi}_\ell + r_\ell \quad\text{and}\quad a_\ell = \mathring{a}_\ell + s_\ell.
\end{equation}
Since the angles $\xi_\ell$ take values in the interval $[0,\frac{4\pi}{\gamma}]$ it seems reasonable to restrict the deviation variables $r_\ell$ to the range $(-\frac{2\pi}{\gamma},\frac{2\pi}{\gamma})$, while for the macroscopic areas we can assume $s_\ell\in\mathbb R$ to belong to some open neighborhood of $a_\ell$. The parameters $r_\ell$ and $s_\ell$ can be understood as measures of deviation between the background data $(\mathring{a}_\ell, \mathring{\xi}_\ell)$ and the actually chosen boundary data $(a_\ell, \xi_\ell)$.\\
Measuring the discrepancy between the boundary normals $\textbf{n}$ and the background normals $\mathring{\textbf{n}}$ requires a little bit more care. Since these are unimodular spinors, a natural ansatz would be
\begin{equation}
	\textbf{n}_\no = U_\no \mathring{\textbf{n}}_\no,
\end{equation} 
where $U_\no$ is an $SU(2)$ matrix and the node label $\no$ stands representatively for either $s(\ell)$ or $t(\ell)$. However, we have to bear in mind that spinors which only differ by a phase factor have to be regarded as equivalent because they reconstruct the same unit vector. In other words, we have an equivalence relation $\textbf{n}_\no\sim \tilde{\textbf{n}}_\no \Leftrightarrow \textbf{n}_\no = \e^{i\varphi}\tilde{\textbf{n}}_\no$ and we have to exclude all $SU(2)$ transformations which satisfy the eigenvalue equation
\begin{equation}
	U_\no \mathring{\textbf{n}}_\no = \e^{i \varphi} \mathring{\textbf{n}}_\no
\end{equation}
for some $\varphi\in(0,2\pi)$. Zero is excluded from the interval because we want   a notion of ``$\textbf{n}_\no$ and $\mathring{\textbf{n}}_\no$ are the same, hence there is no deviation''. This amounts to describing the space $SU(2)/U(1)\simeq \mathcal{S}^2$ up to a point using spinors. It turns out that it is not actually difficult to achieve this description and explicitly construct matrices $U_\no$ which transform spinors by more than just a phase. But it takes up a lot of place and we therefore postpone the construction of $U_\no$ to Appendix~\ref{App:NEW} and just report the result here\footnote{It is not possible to describe all scenarios with just a single parametrization of $U_\no$. The result reported here in the main text only holds under the assumption that $\vec{n}:=\mathring{\textbf{n}}^\dagger\vec{\sigma}\mathring{\textbf{n}}$ is not parallel to the $z$-axis. A second parametrization, which also holds when $\vec{n}$ and the $z$-axis are parallel, is given in Appendix~\ref{App:NEW}.}:
\begin{equation}\label{eq:ExplicitParametrizationU}
	U_\no(\alpha_\no,\beta_\no) = \e^{i \alpha_\no \vec{w}(\beta_\no)\cdot\vec{\sigma}}\quad\text{with}\quad 	\vec{w}(\beta) = \begin{pmatrix}
		\sin\beta_\no\cos\theta\cos\phi + \cos\beta_\no\sin\phi\\
		-\cos\beta_\no\cos\phi +\cos\theta\sin\beta_\no\sin\phi\\
		-\sin\beta_\no\sin\theta
	\end{pmatrix}
\end{equation}
The angles $\theta\in(0,\pi)$ and $\phi\in[0,2\pi)$ parametrize the background unit vector $\vec{n}:=\mathring{\textbf{n}}^\dagger\vec{\sigma}\mathring{\textbf{n}}$ (cf. equation~\eqref{eq:ParamNormalVec}) while the angles $\alpha_\no\in[0,2\pi)$ and $\beta_\no\in[0,2\pi)$ are measures for the deviation of the background vector and the boundary data vector~$\textbf{n}^\dagger\vec{\sigma}\textbf{n}$.\\
Notice that $U_\no$ only depends on two parameters and that it can represent the identity. This is easily understood geometrically: When $\alpha_\no=\beta_\no=0$, it follows that $U_\no=\id_{2\times 2}$ and therefore there is no discrepancy between $\textbf{n}$ and $\mathring{\textbf{n}}$. However, when $\alpha$ and $\beta$ differ from zero, $U_\no$ can be seen to move around $\mathring{\textbf{n}}$ so that the vector $\textbf{n}^\dagger\vec{\sigma}\textbf{n}$ and $\mathring{\textbf{n}}^\dagger\, U^\dagger_\no\, \vec{\sigma} U_\no\, \mathring{\textbf{n}}$ point in the same direction on the two-sphere $\mathcal S^2$. It is in this sense that $\alpha_\no$ and $\beta_\no$ can be seen as measures of discrepancy between background and boundary data. Moreover, we can now define the vector $v_\ell$ as
\begin{equation}
	v_\ell := (r_\ell, s_\ell, \alpha_{s(\ell)}, \beta_{s(\ell)}, \alpha_{t(\ell)}, \beta_{t(\ell)})^\transpose.
\end{equation}
This is our second set of variables and we finally have that $(y_\ell, v_\ell)=(0;0)$, for all $\ell\in\Gamma$, describes the critical point on the boundary.\\
The next step in our approximation strategy is to perform the stationary phase approximation of the bulk and the boundary. From section~\ref{ssec:ClassicalLimitGeometry} we know that for the CLQG propagator we get the Lorentzian Hamilton-Regge function~\eqref{eq:HamiltonRegge}. Here we write it as
\begin{equation}\label{eq:CLQGHamiltonFunction}
	 S_\textsf{c}(j_\ell, \textbf{z}) := \left.\left(\sum_{\fa\in\mathcal{B}}L_\fa + \sum_{\ell\in\Gamma}\tilde{B}_\ell\right)\right|_\textsf{c} = \gamma \sum_{\fa\in\mathcal{B}} j_\fa \Theta_\fa + \gamma \sum_{\ell\in\Gamma} j_\ell\, \textsf{s}\, \theta_\ell.
\end{equation}
The difference to~\eqref{eq:HamiltonRegge} is that here we re-defined the deficit and dihedral angles as $\Theta_\fa\rightarrow\Theta_\fa - \frac{1}{\gamma}\Pi_\fa$ and $\theta_\ell \rightarrow \theta_\ell - \frac{\textsf{s}}{\gamma}\Pi_\ell$, respectively, so that we do not need to explicitly keep track of thin wedge contributions. The above Hamilton-Regge function can be used to re-write the holomorphic spin foam action~\eqref{eq:CLQGComplexAction} in the simpler form
\begin{equation}\label{eq:ActionForAnalysis}
	A[g,\textbf z; a, \xi, \textbf{n}_s, \textbf{n}_t] = S_\textsf{c}(j_\ell, \textbf{z}) + \sum_{\ell\in\Gamma}D_\ell[a,\xi,\textbf{n}_s, \textbf{n}_t].
\end{equation}
To proceed and use theorem~\cref{thm:GenStatPhaseApprox} it is of course necessary to re-express the holomorphic action~\eqref{eq:ActionForAnalysis} in terms of the newly introduced variables $(y_\ell;v_\ell)$:
\begin{align}\label{eq:ComplexActionInNewVariables}
	A = \gamma\sum_{\fa\in\mathcal{B}} j_\fa \Theta_\fa &- \sum_{\ell\in\Gamma}\left(\gamma(y_\ell-s_\ell +a_\ell)\,\textsf{s}\, r_\ell + i(y_\ell - s_\ell)^2 t\right)\notag\\
	&-i\sum_{\ell\in\Gamma}(y_\ell-s_\ell+a_\ell)\ln\frac{\langle \mathring{\overline{\textbf{n}}}_{s(\ell)}|U^\transpose_{s(\ell)}|\textbf{z}_{\no\ell}\rangle^2   \,   \langle \textbf{z}_{\no'\ell}|\overline{U}_{t(\ell)}|\mathring{\overline{\textbf{n}}}_{t(\ell)} \rangle^2}{\innerp{\textbf{z}_{\no\ell}}{\textbf{z}_{\no\ell}}\innerp{\textbf{z}_{\no'\ell}}{\textbf{z}_{\no'\ell}}}.
\end{align} 
It is also convenient to introduce the $L$-dimensional vector $y:=(y_1,y_2,\dots,y_L)^\transpose$ and the $6L$-dimensional vector $v:=(v_1,v_2,\dots,v_L)^\transpose$ where every ``component'' is itself a vector: $v_\ell := (r_\ell, s_\ell, \alpha_{s(\ell)}, \beta_{s(\ell)}, \alpha_{t(\ell)}, \beta_{t(\ell)})^\transpose$. With the help of the vectors $y$ and $v$ we can write down the second order Taylor expansion of $A$ in a compact form:
\begin{align}\label{eq:TaylorCLQG}
	A(y,v) =& \left.A\right|_\textsf{c}	+ y^\transpose \underset{=0}{\underbrace{\left.\nabla_y A\right|_\textsf{c}}} + v^\transpose \underset{=:V}{\underbrace{\left.\nabla_v A\right|_\textsf{c}}} + y^\transpose\underset{=:K}{\underbrace{\left.\nabla_v\nabla_y A\right|_\textsf{c}}} v + \frac{1}{2} y^\transpose \underset{=:H}{\underbrace{\left.\nabla^2_y A\right|_\textsf{c}}} y\notag\\
	 &+ \frac{1}{2} v^\transpose \underset{=:M}{\underbrace{\left.\nabla^2_v A\right|_\textsf{c}}} v + R_3(y,v).
\end{align}
Our goal is again to determine the $0$-th order Malgrange function $A^{(0)}$, for which we compare the above Taylor expansion order by order with
\begin{equation}
	A(y,v) = A^{(0)}(v) + \frac12 (y-X(v))^\transpose A^{(2)}(v) (y-X(v)),
\end{equation}
where we have again chosen the linear term to be zero. Since the Taylor expansion of the holomorphic spin foam action has essentially the same structure\footnote{The only difference is the term $v^\transpose V$ which is absent in the quantum mechanical case. However, as this term is independent of $y$, one easily infers from equation~\eqref{eq:MalgrangeTaylor} how it modifies the final result.} as the Taylor expansion in section~\ref{ssec:BndData}, we can immediately write down the function we seek:
\begin{equation}\label{eq:MainResult}
	A^{(0)}(v) = S_\textsf{c}(\mathring{a}_\ell, \mathring{\textbf{n}}_\no) + v^\transpose V + \frac12 v^\transpose\left(M-K^\transpose H^{-1} K\right) v
\end{equation}
This result is very reminiscent of the zero order Malgrange function derived in the quantum mechanical case, however, notice that here we also have a linear term, $v^\transpose V$, which is absent in quantum mechanics. Using this result, we can now write the holomorphic CLQG transition amplitude approximately as
\begin{equation}\label{eq:FullApprox}
	W^t_\mathcal{C}(H_\ell) \sim \sum_{j_\fa}\frac{1}{\sqrt{\det \textsf{Hess}}}\frac{1}{\sqrt{\det H}}\exp\left(\frac{i}{\hbar}S_\textsf{c}(\mathring{a}_\ell, \mathring{\textbf{n}}_\no) + \frac{i}{\hbar}v^\transpose V + \frac{i}{2\hbar}v^\transpose(M-K^\transpose H^{-1} K) v\right),
\end{equation}
where we deliberately left out numerical factors and normalization constants and where $\det \textsf{Hess}$ denotes the determinant of the Hessian from the propagator. An explicit expression for this complicated object can be found in~\cite{Barrett:2009,Han:2011,Kaminski:2019}.\\
There are now two questions that immediately come to mind when looking at~\eqref{eq:FullApprox}:
\begin{itemize}
	\item[a)] Under what circumstances is $H^{-1}$ defined? 
	\item[b)] Does $v^\transpose V + \frac12 v^\transpose\left(M-K^\transpose H^{-1} K\right) v$ lead to an exponentially suppressed or exponentially enhanced transition amplitude?
\end{itemize}
The first question is important to answer since $H^{-1}$ not only appears in the Malgrange function, but there is also a $\frac{1}{\sqrt{\det H}}$ factor appearing in~\eqref{eq:FullApprox}. Of course, one of the assumptions of theorem~\cref{thm:GenStatPhaseApprox} is precisely that $\det H\neq 0$. But can we be more specific than that and actually say when this is the case? The answer is yes, just as in the quantum mechanical case:\\
First of all, recall that $H$ is defined as $H:= \left.\nabla^2_y A\right|_\textsf{c}$ and that there are only two terms with potentially non-zero second order $y$-derivatives in $A$. The logarithmic term containing the normals $\mathring{\overline{\textbf{n}}}_{\no(\ell)}$ and the spinors $\textbf{z}_{\no\ell}$ does not depend on $y_\ell$ and does therefore not contribute to $H$. What remains is the purely imaginary quadratic term $i(y_\ell-s_\ell)^2t$ and the bulk (area) Regge action, $S_\textsf{Regge} = \gamma\sum_\fa j_\fa\Theta_\fa$. This action depends implicitly on~$y_\ell$ since we used the boundary spins as boundary data to find the deficit angles $\Theta_\fa$ and areas $\gamma j_\fa$ from the critical point equations. Hence, $\left.\nabla^2_y A\right|_\textsf{c}$ can be written as
\begin{equation}
	H = \left.\nabla^2_y A\right|_\textsf{c} = \left.\nabla^2 S_\textsf{Regge}\right|_\textsf{c} + 2 i t\id_{L\times L}.
\end{equation}
Since $S_\textsf{Regge}$ is a real function, $\left.\nabla^2 S_\textsf{Regge}\right|_\textsf{c}$ is a real symmetric $L\times L$ matrix. As such, it has only real eigenvalues. Hence, if $\det(\left.\nabla^2 S_\textsf{Regge}\right|_\textsf{c} + 2 i t \id_{L\times L}) = 0$ were true, this would imply the existence of the purely imaginary eigenvalue $-2 i t$ for $\left.\nabla^2 S_\textsf{Regge}\right|_\textsf{c}$, which is a contradiction\footnote{If all components of $\left.\nabla^2 S_\textsf{Regge}\right|_\textsf{c}$ are zero, $H$ can be seen to be trivially invertible.}. It follows that $H$ does not need to be assumed to be non-degenerate. It is always non-degenerate and, moreover, this also means that all assumptions of theorem~\cref{thm:GenStatPhaseApprox} are always satisfied.\\
The second question can also be answered exactly as in the quantum mechanical case. Theorem~\cref{thm:ImaginaryPart} guarantees that in a neighborhood of $v=0$ the imaginary part of $A^{(0)}(v)$ is larger than zero which in turn implies that the holomorphic CLQG transition amplitude is suppressed by the factor
\begin{equation}
	\exp\left(-\frac{1}{\hbar}\text{Im}[v^\transpose V + \frac{1}{2}v^\transpose (M-K^\transpose H^{-1} K) v]\right).
\end{equation}
This completes the definition of the approximation method. Before discussing some of its properties and subtler aspects, let us see how it performs when applied to the black hole to white hole transition.

\subsection{Application to the Black Hole to White Hole Transition}
Now that we dispose of a systematic method the evaluate CLQG transition amplitudes away from critical points, it is tempting to apply it to the black hole to white hole transition and compare the results to the computation performed in chapter~\ref{chap:CLQG}.\\
To that end, we will use the same $2$-complex as in subsection~\ref{ssec:ComputationBounceTime} and make some of the same simplifying assumptions. First of all, the $2$-complex has no internal faces which already eliminates the bulk part of the Hamilton-Regge function. That is, the first sum in~\eqref{eq:ComplexActionInNewVariables} disappears. Secondly, we shall assume the normal vectors to satisfy the critical point equations. The same is true for the areas, i.e. there will be no deviation between the area from the boundary data and the one computed from the critical point equations. With the choice of boundary data made in~\cite{RealisticObs}, this amounts to saying that the black hole transitions into a white hole of the same mass.\\
Under these assumptions, we do not need to keep track of the angles which parametrize $U_{s(\ell)}$ and $U_{t(\ell)}$. Moreover, since there is no deviation in the areas there is also no need for the parameter $s_\ell$. Hence, the vector $v$ reduces to the components $v_\ell = r_\ell = \xi_\ell-\textsf{s}\,\theta_\ell$ and, when restricted to the appropriate subspaces, the vector $V$ and the matrices $K$, $M$, $H$ assume the simple forms
\begin{align}
	V &= \left.\nabla_v A\right|_\textsf{c} = (-\gamma\mathring{a}_1, -\gamma\mathring{a}_2,\dots,-\gamma\mathring{a}_L)^\transpose\notag\\
	K &= \left.\nabla_v\nabla_y A\right|_\textsf{c} = \gamma \id_{L\times L}\notag\\
	M &= \left.\nabla^2_v A\right|_\textsf{c} = 0\notag\\
	H &= \left.\nabla^2_y A\right|_\textsf{c} = -2 i t\id_{L\times L}
\end{align}
It then follows immediately that
\begin{align}
	v^\transpose V &= \gamma\sum_{\ell\in\Gamma}\mathring{a}_\ell(\textsf{s}\,\theta_\ell-\xi_\ell)\notag\\
	\frac12 v^\transpose\left(M -K^\transpose H^{-1} K\right) v &= -\frac{\gamma^2}{4t i}\|v\|^2 = -\frac{\gamma^2}{4t i}\sum_{\ell\in\Gamma}(\textsf{s}\,\theta_\ell - \xi_\ell)^2.
\end{align}
Since $S_\textsf{c}(\mathring{a}_\ell, \mathring{\textbf{n}}_\no) = 0$ due to the above assumptions (it only contains bulk quantities), we finally get from~\eqref{eq:MainResult}
\begin{equation}
	i\,A^{(0)}(v) = i\,\gamma\sum_{\ell\in\Gamma}\mathring{a}_\ell(\textsf{s}\,\theta_\ell-\xi_\ell) -\frac{\gamma^2}{4t}\sum_{\ell\in\Gamma}(\textsf{s}\,\theta_\ell - \xi_\ell)^2.
\end{equation} 
This is exactly the same result derived in chapter~\ref{chap:CLQG} and reported in \cite{BH2WH, TimeScale, Marios_phd}! We even reproduced the $\frac{\gamma^2}{4t}$ factor of the suppression term. However, we arrived at this result in just a few lines and the applied procedure has a clear interpretation. In retrospect, we also confirm the validity of the approximation developed in~\ref{ssec:ComputationBounceTime} which had been criticized for its use of the stationary phase approximation even though there is no critical point.

\subsection{On the Range of Validity and a Mathematical Result}
Let us now return to the general expression~\eqref{eq:FullApprox} for the approximation of the holomorphic CLQG transition amplitude.\\
As we have discussed in one of the foregoing subsection, the matrix $H$ turns out to be always non-degenerate. Moreover, all assumptions of theorem~\cref{thm:GenStatPhaseApprox} are trivially satisfied. The result~\eqref{eq:FullApprox} therefore seems to be very robust and does not require strong assumptions. However, some caution is needed: To derive~\eqref{eq:FullApprox}, we needed to approximate the boundary spin sums by integrals using the Euler-Maclaurin theorem. This step is only justified, as we have seen in~\ref{ssec:ComputationBounceTime}, when the coherent boundary states are semi-classical and peaked on large areas. However, by turning the sums into integrals, we loose the periodicity of these sums. That is, we formally have
\begin{equation}
	\sum_{n_\ell = -\infty}^{\infty} \e^{-\frac{n^2_\ell}{4}t + \frac{i\gamma}{2} n_\ell\left(\textsf{s}\, \theta_\ell -\xi_\ell\right)} \approx  \int_{-\infty}^\infty \e^{-\frac{n^2_\ell}{4}t + \frac{i\gamma}{2} n_\ell\left(\textsf{s}\, \theta_\ell -\xi_\ell\right)} \dd n_\ell = 2\sqrt{\frac{\pi}{t}} \e^{-\frac{\gamma^2 \left(\textsf{s}\, \theta_\ell -\xi_\ell\right)^2}{4 t}},
\end{equation}
where the left hand side is periodic in $\textsf{s}\, \theta_\ell -\xi_\ell$ and the right hand side has lost this periodicity. We already encountered this problem in subsection~\ref{ssec:ComputationBounceTime}, where we explained that the cause of this dissonance is the fact that we only use the first term in the Euler-Maclaurin formula. That is, we neglect all the higher order terms in
\begin{equation}
	\sum_{n=a}^b f(n) = \int_a^b f(x)\dd x + \frac{f(b)+f(a)}{2}+\sum_{k=1}^{\floor{p/2}}\frac{B_{2k}}{(2k)!}\left(f^{(2k-1)}(b)-f^{(2k-1)}(a)\right),
\end{equation}
where $B_{2k}$ are Bernoulli numbers. Of course we are tempted to do so since, for $f(n)=\exp(-\frac{n^2}{4}t + \frac{i\gamma}{2} n\left(\textsf{s}\, \theta_\ell -\xi_\ell\right))$, all derivatives $f^{(2k-1)}$ with respect to $n$ are proportional to $\exp(-\frac{n^2}{4t})$ and when we take the limits $a\rightarrow-\infty$ and $b\rightarrow \infty$ we conclude that the sum on the right hand side vanishes. But this is wrong! This is one of these cases the mathematicians warned us about where taking the limit does not commute with performing the sum.\\
To restore periodicity, we therefore have to take the higher order terms into account. Alternatively, we can perform an ad hoc modification of~\eqref{eq:FullApprox} to
\begin{align}\label{eq:FullApproxModified}
	W^t_\mathcal{C}(H_\ell) \sim& \sum_{j_\fa}\frac{1}{\sqrt{\det \textsf{Hess}}}\frac{1}{\sqrt{\det H}}\exp\left(\frac{i}{\hbar}S_\textsf{c}(\mathring{a}_\ell, \mathring{\textbf{n}}_\no) + \frac{i}{\hbar}v^\transpose V + \frac{i}{2\hbar}(M-K^\transpose H^{-1} K) v\right)\times\notag\\
	&\times\prod_{\ell\in\Gamma}\vartheta_3\left(-\frac{i \pi \gamma \Delta_\ell}{t}, \e^{-\frac{4 \pi^2}{t}}\right),
\end{align}
where $\Delta_\ell := \textsf{s}\,\theta_\ell-\xi_\ell$ and $\vartheta_3$ is the third Jacobi theta function (see subsection~\ref{ssec:ComputationBounceTime} for a definition). We motivate this modification through the following exact computation:
\begin{align}
	\sum_{n_\ell = -\infty}^{\infty} \e^{-\frac{n^2_\ell}{4}t + \frac{i\gamma}{2} n_\ell\left(\textsf{s}\, \theta_\ell -\xi_\ell\right)} =& 2\sqrt{\frac{\pi}{t}} \e^{-\frac{\gamma^2 \Delta^2_\ell}{4 t}}\vartheta_3\left(-\frac{i \pi \gamma \Delta_\ell}{t}, \e^{-\frac{4 \pi^2}{t}}\right)\notag\\
	\equiv & \int_{-\infty}^\infty \e^{-\frac{n^2_\ell}{4}t + \frac{i\gamma}{2} n_\ell\left(\textsf{s}\, \theta_\ell -\xi_\ell\right)} \dd n_\ell\,\,  \vartheta_3\left(-\frac{i \pi \gamma \Delta_\ell}{t}, \e^{-\frac{4 \pi^2}{t}}\right).
\end{align}
We emphasize again that this is an ad hoc procedure, but it is the best solution we can offer for the time being. In any case, this restores the correct periodicity and it may be possible to test this modification numerically. 

A separate issue concerning the range of validity of the approximation~\eqref{eq:FullApprox} is the following: How far away can we move from the classical limit at $v=0$ and still trust that~\eqref{eq:FullApprox} (or~\eqref{eq:FullApproxModified}) accurately describes the behavior of the exact holomorphic transition amplitude? \\
This is a difficult question and we cannot provide a definitive answer. A simple strategy to get an idea of the region of validity is the following: Theorem~\cref{thm:ImaginaryPart} confirms our intuition that the holomorphic amplitude decays as soon as we move away from the classical background $v=0$. Hence, we know that we are leaving the range of validity of the approximation when the imaginary part of $A^{(0)}(v)$ becomes zero for $v\neq 0$ or even larger than zero. It follows that the range of validity is contained as a subset in the set~$\Omega_v$ implicitly defined through
\begin{equation}\label{eq:CondPos}
	\text{Im}[v^\transpose V + \frac{1}{2}v^\transpose(M-K^\transpose H^{-1} K) v] > 0.
\end{equation}
We will now show that this strategy is not viable because the above condition is satisfied for any $v$ and $\Omega_v$ is therefore completely unrestricted and devoid of useful information. In fact, we will now prove the following lemma:
\begin{mylemma}{}{}
	The condition $\text{Im}[v^\transpose V + \frac{1}{2} v^\transpose (M-K^\transpose H^{-1} K)v] > 0$ is trivially satisfied for all vectors $v$ because $\text{Im}[V] = 0$ and the $6L\times 6L$ dimensional matrix $\text{Im}[M-K^\transpose H^{-1} K]$	is real, symmetric, and positive-definite. 
\end{mylemma}
\textit{Proof}:\\
Let us start by showing that $V$ is real. From its definition, $V:=\left.\nabla_v A\right|_\textsf{c}$, we know that we need to take derivatives of the holomorphic action with respect to $r_\ell$, $s_\ell$ and the four angles $\alpha_{s(\ell)}$, $\beta_{s(\ell)}$, $\alpha_{t(\ell)}$, $\beta_{t(\ell)}$. The $r_\ell$ and $s_\ell$ derivatives produce only real terms, as can easily be checked. To compute the angular derivatives, we need to use the explicit parametrization~\eqref{eq:ExplicitParametrizationU} of the matrix $U_\no$ (see also Appendix~\ref{App:NEW}). It can then be shown that these derivatives are all zero on the critical point. Hence, $V$ is a real $6L$ dimensional vector and condition~\eqref{eq:CondPos} becomes
\begin{equation}\label{eq:CondPos2}
	v^\transpose\text{Im}[M-K^\transpose H^{-1} K]v > 0.
\end{equation}
Now observe that $\text{Im}[M-K^\transpose H^{-1} K]$ is a real symmetric $6L\times 6L$ matrix. The symmetry of $M$ and $H$ follows from their definitions (see~\eqref{eq:TaylorCLQG}) and this in turn implies the symmetry of $K^\transpose H^{-1} K$. It follows immediately that $\text{Im}[M-K^\transpose H^{-1} K]$ is diagonalizable and possesses only real eigenvalues.\\
A priori, the eigenvalues can be positive, negative or zero. Let us assume there are $N^+$ positive eigenvalues $\lambda^+$, $N^-$ negative eigenvalues $\lambda^-$ and $6L-N^+-N^-$ zero eigenvalues. Then we can rewrite condition~\eqref{eq:CondPos2} in the basis which diagonalizes the above matrix as
\begin{equation}
	\sum_{i=1}^{N^+}\lambda^+_i \tilde{v}^2_i - \sum_{j=N^{+}+1}^{N^-+N^+}|\lambda^-_j|\tilde{v}^2_j > 0,
\end{equation}
where $\tilde{v}$ is the vector $v$ expressed in the aforementioned basis. In principle, the above inequality defines a region in the vector space of $\tilde{v}$ which represents a restriction on the possible values we can consider for $\tilde{v}$. Now, let us look at the border of this region, i.e. let us consider the implicitly defined surface
\begin{equation}
	\sum_{i=1}^{N^+}\lambda^+_i \tilde{v}^2_i - \sum_{j=N^{+}+1}^{N^-+N^+}|\lambda^-_j|\tilde{v}^2_j = 0.
\end{equation}
We can solve this equation for the component $\tilde{v}_1$ and obtain
\begin{equation}\label{eq:Surface}
	\tilde{v}_1 = \pm\frac{1}{\sqrt{\lambda^+_1}}\sqrt{\sum_{j=N^++1}^{N^-+N^+}|\lambda^-_j|\tilde{v}^2_j - \sum_{i=2}^{N^+}\lambda^+_i\tilde{v}^2_i}
\end{equation}
as a function of all the other components of $\tilde{v}$, except the $6L-N^+-N^-$ components that correspond to zero eigenvalues. However, this is problematic because it means we can move freely along any of these $6L-N^+-N^-$ directions and in particular we can approach the origin as close as we want. This in turn implies that any neighborhood of $\tilde{v}=0$, no matter how small it is, always contains a part of the surface described by~\eqref{eq:Surface}. Hence, we can always find non-zero values of $\tilde{v}$ in a neighborhood of $\tilde{v}=0$ for which $\text{Im}[A^{(0)}(\tilde{v})]=0$. This is in contradiction to the statement of theorem~\cref{thm:ImaginaryPart}!\\
We conclude that $\text{Im}[M-K^\transpose H^{-1} K]$ cannot have zero eigenvalues and $\tilde{v}_1$ therefore assumes the form
\begin{equation}\label{eq:Surface2}
	\tilde{v}_1 = \pm\frac{1}{\sqrt{\lambda^+_1}}\sqrt{\sum_{j=N^++1}^{6L}|\lambda^-_j|\tilde{v}^2_j - \sum_{i=2}^{N^+}\lambda^+_i\tilde{v}^2_i}.
\end{equation}
However, here we encounter the same sort of problem. The surface described by~\eqref{eq:Surface2} is a sort of ``hyper-hyperbola'' and there are infinitely many possibilities to approach $\tilde{v}_1=0$ without making any of the other components of $\tilde{v}$ equal to zero. Hence, we find again that any neighborhood of $\tilde{v}$ contains values of $\tilde{v}$ for which $\text{Im}[A^{(0)}(\tilde{v})]=0$, which is again in contradiction to theorem~\cref{thm:ImaginaryPart}.\\
We conclude that $\text{Im}[M-K^\transpose H^{-1} K]$ cannot have negative eigenvalues. If this matrix has only positive eigenvalues,
\begin{equation}
	\tilde{v}_1 = \pm\frac{1}{\sqrt{\lambda^+_1}}\sqrt{- \sum_{i=2}^{6L}\lambda^+_i\tilde{v}^2_i}
\end{equation}
is manifestly imaginary and therefore there is no neighborhood of $\tilde{v}=0$ for which $\text{Im}[A^{(0)}(\tilde{v})]$ can become zero, in accordance with theorem~\cref{thm:ImaginaryPart}. Hence, we conclude that $\text{Im}[M-K^\transpose H^{-1} K]$ has only positive eigenvalues, which is equivalent to saying that it is a positive-definite matrix. Not only in a neighborhood of zero, but everywhere. This in turn means that condition~\eqref{eq:CondPos2} is trivially satisfied for all vectors~$v$, which proves the above lemma. 

This lemma has two implications for us: We cannot use condition~\eqref{eq:CondPos} to restrict the set of admissible values of $v$ and hence find constraints on the range of validity of the approximaton~\eqref{eq:FullApproxModified}. Secondly, we cannot start from a given background $(\mathring{a}_\ell,\mathring{\xi}_\ell, \mathring{\textbf{n}}_{s(\ell)},\mathring{\textbf{n}}_{t(\ell)})$, deform it using a non-zero perturbation vector $v$ and land on a different classical background $(\mathring{a}'_\ell,\mathring{\xi}'_\ell, \mathring{\textbf{n}}'_{s(\ell)},\mathring{\textbf{n}}'_{t(\ell)})$.\\
The meaning of this can easily be illustrated in quantum mechanics (for which this lemma is also true): Consider the phase space picture shown in Figure~\ref{fig:PSP}. Assume the classical ``background'' to be described by $(q_i,p_i;q_f,p_f)$, which in particular means there is a phase space trajectory connecting these two points. Now, use the $4n$ dimensional vector $v=(r_i,s_i,r_f, s_f)^\transpose$ to translate $(q_i,p_i)$ to $(q'_i, p'_i)$ and $(q_f,p_f)$ to $(q'_f,p'_f)$ and assume that $(q'_i, p'_i)$ and $(q'_f,p'_f)$ are also connected by a classical phase space trajectory. This provides the second ``background''. Then, the associated transition amplitude is still suppressed. The vector $v$ cannot be used to move from one classical solution to an other one. \\
Intuitively, this result is obvious since we used the propagator associated to the end points $(q_i,p_i)$ and $(q_f,p_f)$ to construct the approximated amplitude. In particular this means the matrices $M$, $K$ and $H$ depend on this choice. Since they are independent of $v$, they know nothing about the translation and consequently they only know about the first but not about  the second classical solution.

\subsection{Perspectives and Open Problems}
\label{ssec:PerspectivesAndOpenProblems}
The semi-classical expansion developed in this chapter provides a more general and mathematically more solid approach to evaluate transition amplitudes than the one presented in subsection~\ref{ssec:ComputationBounceTime} and, moreover, it is conceptually cleaner. It also sheds new light on other issues.\\
As we briefly discussed in chapter~\ref{chap:CLQG}, the classical limit of the CLQG propagator involves a sum over $2^V$ critical points. This was first observed by Barrett et al.~\cite{Barrett:2009} in the case of a single $4$-simplex, where the two critical points combine into a cosine so that one has $W_\text{CLQG}\sim \cos(S_\text{Regge})$ and this phenomenon is therefore known as \textit{cosine problem} \cite{Vojinovic:2013}. Some authors maintain that the presence of several critical points signals a pathology of the CLQG model and they seek a solution through modifications of the amplitude \cite{Mikovic:2011,Engle:2011a,Engle:2011b,Engle:2012}. Others either allude to different (more or less natural) mechanisms which should reduce the number of relevant critical points \cite{Rovelli:2005,Modesto:2005,Bianchi:2006,Bianchi:2008,Bianchi:2010b} or they fully embrace this phenomenon as a manifestation of anti-spacetime or the absence of a notion of (coordinate) time~\cite{Rovelli:2010b,Christodoulou:2012}. The results derived in the foregoing sections within the context of quantum mechanics and CLQG suggest a resolution of the cosine problem through a natural ``selection'' mechanism.\\
Indeed, we have seen that on the boundary the critical point equations coming from the spin sums imply $\textsf{s}\,\theta_\ell-\mathring{\xi}_\ell = 0$, with $\textsf{s}\,\theta_\ell \equiv \left.\PD{S_\text{Regge}}{j_\ell}\right|_{\textsf{c}}$. Hence, the boundary data selects a critical point (i.e. a sign $\textsf{s}$) and discards the second option. This is analogous to quantum mechanics where we found the equations $\left.\PD{S}{q_i}\right|_{\textsf{c}}+\mathring{p}_i = 0$. That is, the momentum $\left.\PD{S}{q_i}\right|_\textsf{c}\equiv p$ computed from the critical point equations has to match the momentum $\mathring{p}_i$ provided by the boundary data.\\
Moreover, we saw in the quantum mechanical case that the data $(\mathring{q}_i, \mathring{p}_i)$ together with the equations of motion constitute an initial value problem, in the continuum as well as in the discrete theory. Hence, the propagator evolves  $(\mathring{q}_i, \mathring{p}_i)$ along a unique classical solution. Could something similar happen in CLQG?\\
In classical GR we know that the intrinsic metric $q_{ab}$ and the extrinsic curvature $k^{ab}$ specified on a hypersurface $\Sigma$ uniquely determine the spacetime geometry via Einstein's field equations~\cite{WaldBook}. Could a discrete version of this theorem also be true? Could the data $(\mathring{a}_\ell, \mathring{\xi}_\ell, \mathring{\textbf{n}}_{s(\ell)}, \mathring{\textbf{n}}_{t(\ell)})$, which marks a point in the discrete phase space of twisted geometries, together with the equations of motion of area Regge calculus uniquely determine the discrete geometry of a spacetime region?\\
So far this is speculation, but it is encouraging that at least on the boundary the critical point equations of the holomorphic CLQG amplitude select only one critical point. Moreover, we also saw in chapter~\ref{chap:CLQG} during the computation of the black hole to white hole amplitude that the boundary data allowed only one critical point configuration to dominate while the other three were suppressed.\\
To understand whether the critical point equations supplemented by the boundary data determine a unique configuration of critical points also in the bulk, one needs to understand how the data propagates from the boundary into the bulk and how it affects the solutions. This is currently being investigated by the author. The resolution of identity in terms of heat kernel states in the twisted geometry parametrization (see end of chapter~\ref{chap:2}) is expected to play an important role here. By inserting a number of these resolutions of identity at strategic points, we can introduce auxiliary variables $(a_\ell, \xi_\ell, \textbf{n}_{s(\ell)}, \textbf{n}_{t(\ell)})$ in the bulk, just as we did in quantum mechanics. The effect may not be as profound as in quantum mechanics, but it brings us closer to a geometric formulation of the problem and it may render the equations more transparent.\\
It could also help with performing the spin sums in the bulk since the resolutions of identity introduce regulators of the form $\exp(-(j-a)^2 t)$, where $a$ is now an auxiliary area variable that has to be determined through the critical point equations. These regulators would allow to pursue a similar strategy as the one discussed in chapter~\ref{chap:CLQG} in truncating the spin sums and approximating them by integrals. This is also something that is currently being investigated and which is closely linked with the data propagation problem mentioned above.\\
The approximation method developed in this chapter has brought as a step closer towards a systematic evaluation method for holomorphic CLQG transition amplitudes away from classical solutions. However, understanding this issue of data propagation and how to do the spin sums in the bulk is crucial. Only then can we apply this approximation method to ``large'' triangulations such as the one of $I^2\times\mathcal S^2$ described in chapter~\ref{chap:MathI}.

%% file: Content/Chapter_6.tex
\chapter{Conclusion}
\label{chap:conclusion}
The recent application of CLQG to the black hole to white hole transition has catalyzed new developments related to the computation of observables and the evaluation of transition amplitudes. This thesis has contributed to it in the following ways:

At the end of chapter~\ref{chap:2}, we closed a gap in the existing literature by providing a measure for the resolution of identity in terms of Thiemann's heat kernel states in the twisted geometry parametrization. By itself this is a small result, but it is important for applications. Indeed, the measure is needed in the computation of observables as emphasized by Oeckl~\cite{Oeckl:2005, Oeckl:2018} and illustrated in this thesis in subsection~\ref{ssec:EstimationBounceTime}. Moreover, the resolution of identity in the twisted geometry parametrization could be of importance in rewriting the CLQG transition amplitude in terms of auxiliary variables which have a clear geometrical meaning. This could facilitate computations and shed new light on old issues such as the cosine problem and bring us closer to an approximation of the bulk spin sums. Work in this direction is currently ongoing.

In chapter~\ref{chap:MathI} we presented an algorithm which generates consistent simplicial triangulations for manifolds of topology $I\times\Sigma$ and which can easily be implemented on a computer. The original motivation for the development of this algorithm was the struggle encountered by the author when trying to generalize the simplicial triangulation of $I\times\mathcal{S}^2$, which we discussed in detail in subsection~\ref{ssec:AnExample}, to a triangulation of $I^2\times\mathcal{S}^2$. Regions of topology $I^2\times\mathcal{S}^2$ are naturally encountered in the Schwarzschild spacetime and they are currently considered for a modified CLQG computation of the black hole to white hole transition.\\
The algorithm has been successfully applied by the author to find a triangulation of $I^2\times\mathcal{S}^2$, however, it has not yet been used in the modified CLQG computation. The reason is two fold: First of all, the triangulation is rather large consisting of $48$ $4$-simplices, $144$ tetrahedra and $156$ triangles. It may however be possible to reduce these numbers by a series of Pachner moves\footnote{The author thanks Adam Getchell for pointing this out to him during the Bard Summer School on Quantum Gravity.}. The second reason is the lack of a systematic method to evaluate CLQG transition amplitudes which makes it a hopeless endeavor to even try to compute the black hole to white hole transition on a $2$-complex with $48$ vertices. This leads us then to chapters~\ref{chap:CLQG} and~\ref{chapter_5}.

The first sections of chapter~\ref{chap:CLQG} are devoted to deriving a path integral form of the CLQG propagator and CLQG holomorphic transition amplitude. Both were of major importance for the third section, where we presented joint work with Marios Christodoulou on an approximation method for the black hole to white hole transition amplitude~\cite{BH2WH}. This approximation rests on the stationary phase analysis of the propagator and it involves a careful treatment of the spin sums on the boundary. By exploiting the peakedness and semi-classical properties of the boundary states, it is possible to perform the spin sums and arrive at an approximate holomorphic amplitude which is suppressed. That is, the amplitude decays whenever there is a mismatch between the extrinsic angles of the boundary data and the dihedral angles computed from the stationary phase analysis of the propagator.\\
This approximation is subsequently used to compute the bounce time of the black hole to white hole transition using the measure for the resolution of identity derived in chapter~\ref{chap:2}. We also presented joint work with Marios Christodoulou~\cite{TimeScale} on the different time scales involved in the black hole to white hole transition and we clarified some conceptual issues of this model which had previously been  encountered by other authors~\cite{Ambrus:2005,Barcelo:2014b,Barcelo:2015}.

Chapter~\ref{chapter_5} finally represents the core of this thesis. There, we first took two steps back and considered the quantum mechanical holomorphic transition amplitude in the continuum theory. We studied its classical limit using a stationary phase approximation and we found that when a critical point exists, it is unique. This is not the case for the quantum mechanical propagator where one generically has to sum over several distinct critical points. The reason for this intriguing property of the holomorphic amplitude is explained in detail in subsection~\ref{ssec:BndData}.\\
We then proceeded in developing an approximation method for holomorphic amplitudes in the semi-classical regime. This approximation method allows us to move away from the critical point and we can quantitatively predict how the amplitude is suppressed when we do so. In Appendix~\ref{App:HolomorphicAmplitudes} we computed the holomorphic transition amplitudes of the free particle and the harmonic oscillator analytically and we compared them to the approximation~\eqref{eq:ApproxAmp} derived in subsection~\ref{ssec:BndData}. There is a $100\%$ agreement between the analytical and the approximate result.\\
In a first step toward CLQG, we added one layer of complexity be studying the discretized quantum mechanical holomorphic transition amplitude. The goal was to understand which results of the continuum theory continue to hold in the discrete theory and which aspects change or need to be treated with more care. We found that most results carry over to the discrete theory without alteration. In particular, we showed that also in the discrete theory the classical limit of the holomorphic amplitude is unique.\\
Finally, we discussed the holomorphic CLQG transition amplitude. We uncovered many similarities with the quantum mechanical case and we extended the approximation method developed in~\ref{ssec:BndData} to CLQG. This approximation method can be applied under rather mild assumptions since all conditions of theorem~\cref{thm:GenStatPhaseApprox} are always satisfied and otherwise one only needs to specify a classical background geometry $(\mathring{a}_\ell, \mathring{\xi}_\ell, \mathring{\textbf{n}}_{s(\ell)}, \mathring{\textbf{n}}_{t(\ell)})$ and assume the boundary coherent states to be semi-classical. The approximation~\eqref{eq:FullApprox} then captures the behavior of the holomorphic amplitude in a neighborhood of the critical point. That is, in a small region around the classical geometry. Hence, we can understand this approximation as a semi-classical expansion around a given geometry.\\
When applied to the black hole to white hole transition, under the same simplifying assumptions as in chapter~\ref{chap:CLQG}, equation~\eqref{eq:FullApprox} reproduces precisely the same result. However, the final result can be derived in a quick and systematic way and it rests on a more solid mathematical foundation. It also provides an adequate answer to the critique that has been raised with the computation of the black hole to white hole transition amplitude presented in subsection~\ref{ssec:ComputationBounceTime}.

There are several important questions and interesting research directions related to this new approximation method. One of these important questions is how to account for the periodicity in $\textsf{s}\,\theta_\ell -\mathring{\xi}_\ell$ and how to find a solid justification for the ad hoc ansatz~\eqref{eq:FullApproxModified}.\\
We also encountered several indications that the so-called cosine problem could be absent for the holomorphic CLQG amplitude. Already in subsection~\ref{ssec:ComputationBounceTime} we saw that performing the boundary spin sums leads to a suppression of three out of four critical points and that the boundary data selects the non-suppressed critical point. This result was confirmed in a more general setting in subsection~\ref{ssec:SemiClassicalHolomorphicCLQGAmplitudes} where it was shown that the boundary data selects only one critical point on the boundary. It is now important to understand how the boundary data propagates into the bulk and how it affects the solutions there.\\
In quantum mechanics it holds true in the continuum as well as in the discrete theory that the boundary data together with the critical point equations selects only one solution for the whole discretized path. This is the analogue of the classic theorem of differential equations that initial data together with second order ODE's lead to a unique solution. In classical GR there is a similar theorem roughly stating that the data $(q_{ab}, k^{ab})$ specified on a hypersurface $\Sigma$ determine the whole spacetime geometry via Einstein's field equations~\cite{WaldBook}. The question is now whether the data $(\mathring{a}_\ell, \mathring{\xi}_\ell, \mathring{\textbf{n}}_{s(\ell)}, \mathring{\textbf{n}}_{t(\ell)})$ together with the critical point equations also establish a unique discrete geometry for a given spacetime region. This question is currently being investigated.\\
Finding an answer to the data propagation problem could also help with performing the spin sums in the bulk. By strategically using the resolution of identity in terms of heat kernel states in the twisted geometry parametrization we can introduce auxiliary variables $(a, \xi, \textbf{n}_{s}, \textbf{n}_{t})$ in the bulk and, in particular, we get regulating factors of the form $\exp(-(j-a)^2 t)$. The auxiliary area variables $a$ have to be determined through the critical point equations and depend on the boundary data. If they turn out to be large, one could consider truncating the bulk spin sums and approximating them by integrals in a similar way as in subsection~\ref{ssec:ComputationBounceTime}.\\
Properly dealing with the spin sums and understanding how the boundary data propagates through the $2$-complex and what kind of solutions it determines is a necessary step if one wants to apply this semi-classical approximation method to large triangulations such as the ones generated by the triangulation algorithm of chapter~\ref{chap:MathI}. This would truly allow us to explore the semi-classical regime of CLQG in a systematic way.

%% file: Content/Appendices.tex
\appendix
 

\setcounter{equation}{0}
\renewcommand\theequation{A.\arabic{equation}}
\chapter{Simple Examples of holomorphic Transition Amplitudes}\label{App:HolomorphicAmplitudes}
\begin{center}
	\begin{figure}
		\centering
		\includegraphics[width=0.95\textwidth]{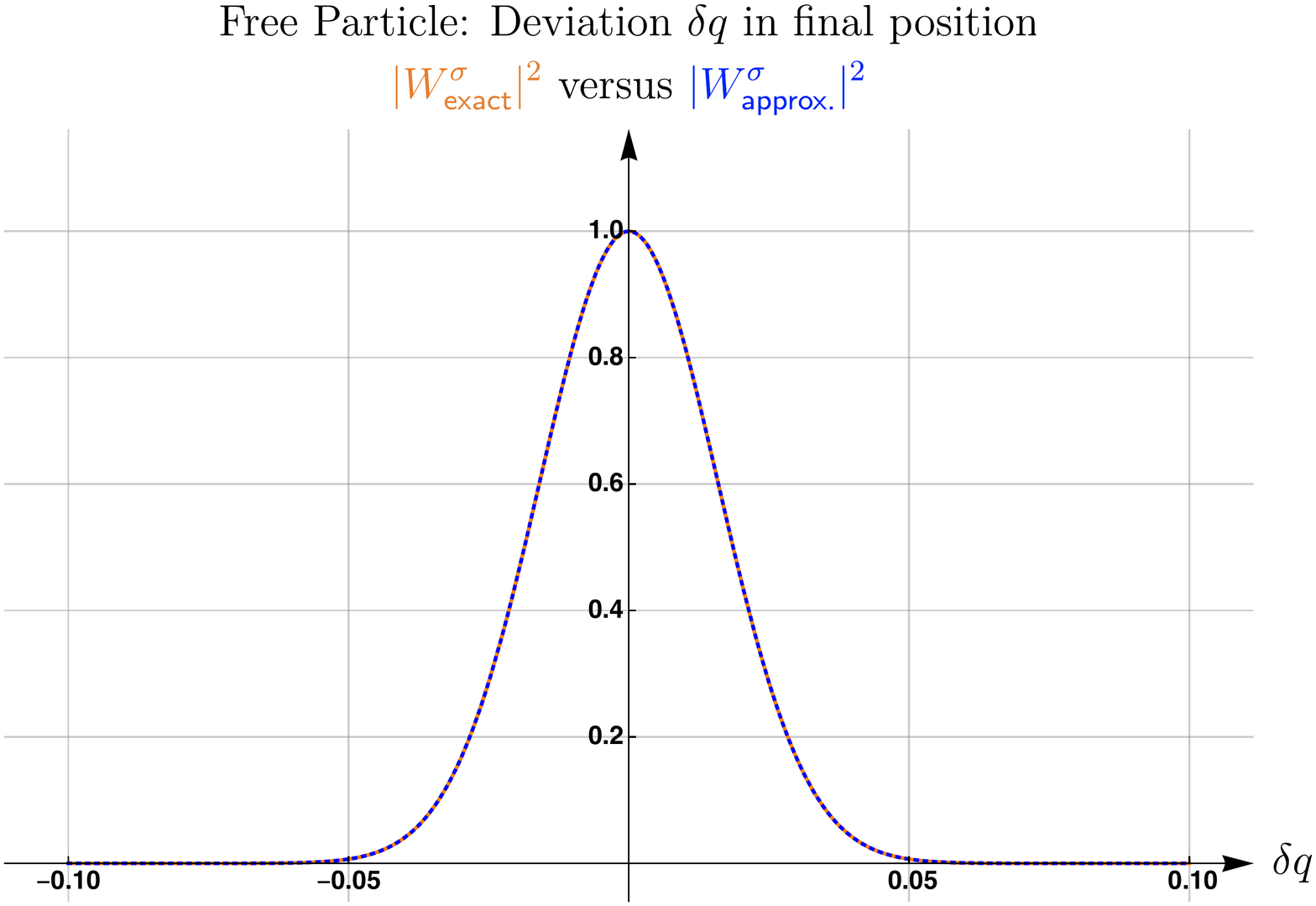}
		\caption{Comparison of the exact holomorphic transition amplitude for the free particle with the approximated amplitude. Deviation in the final position.}
		\label{fig:FP_Deviation_Position}
	\end{figure}
\end{center}

\begin{center}
	\begin{figure}
		\centering
		\includegraphics[width=0.95\textwidth]{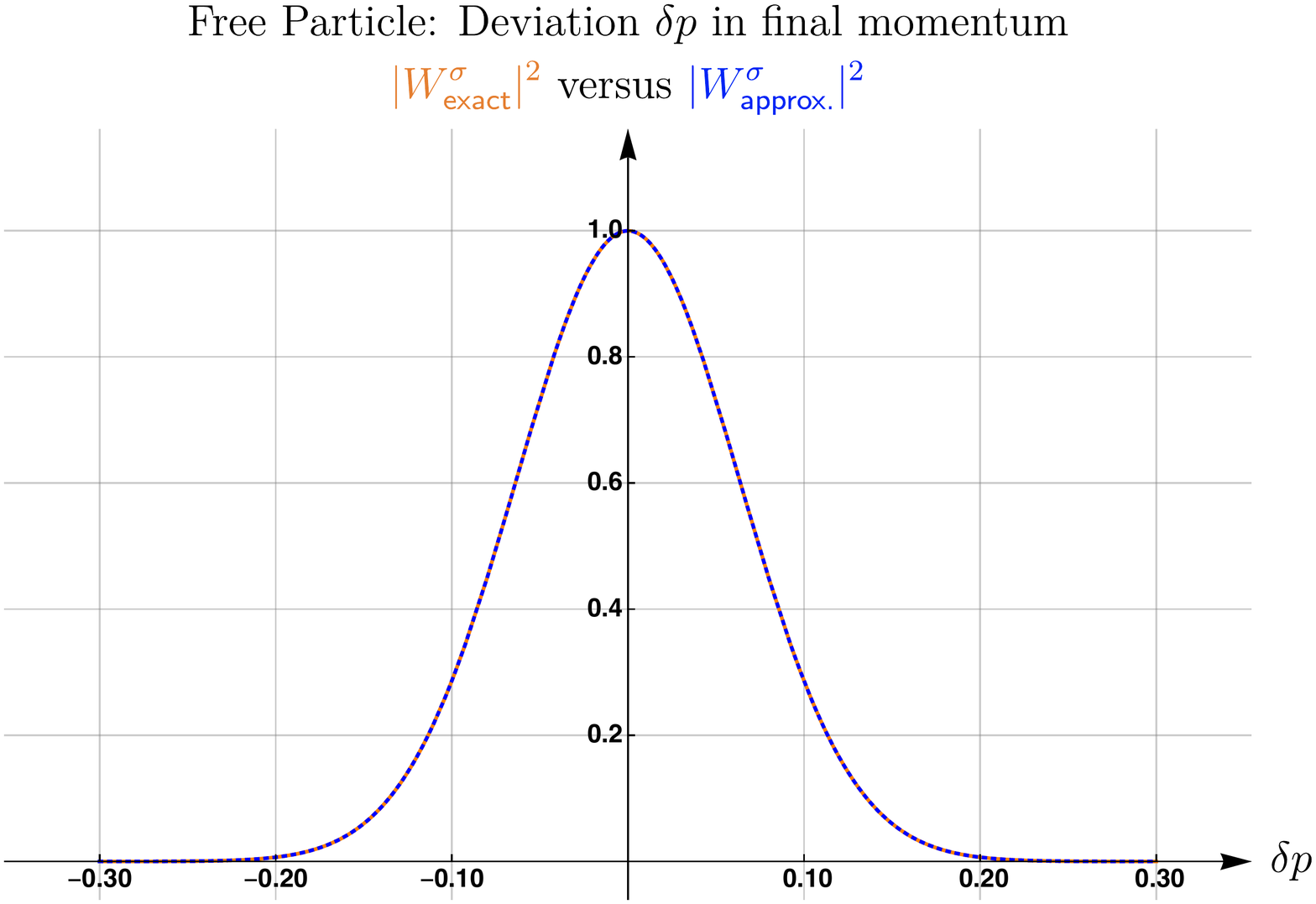}
		\caption{Comparison of the exact holomorphic transition amplitude for the free particle with the approximated amplitude. Deviation in the final momentum.}
		\label{fig:FP_Deviation_Momentum}
	\end{figure}
\end{center}

\begin{center}
	\begin{figure}
		\centering
		\includegraphics[width=0.95\textwidth]{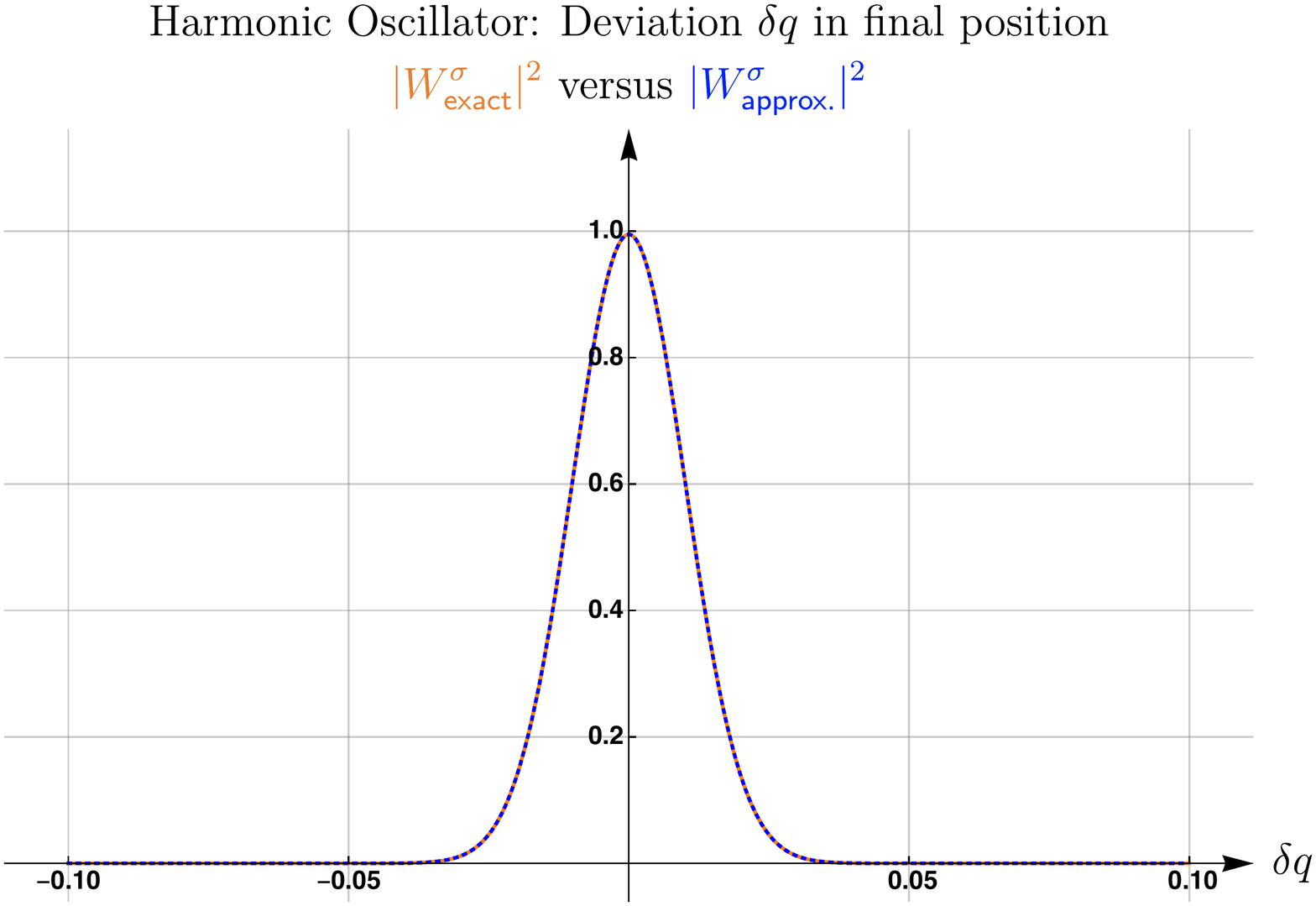}
		\caption{Comparison of the exact holomorphic transition amplitude for the harmonic oscillator with the approximated amplitude. Deviation in the final position.}
		\label{fig:HO_Deviation_Position}
	\end{figure}
\end{center}

\begin{center}
	\begin{figure}
		\centering
		\includegraphics[width=0.95\textwidth]{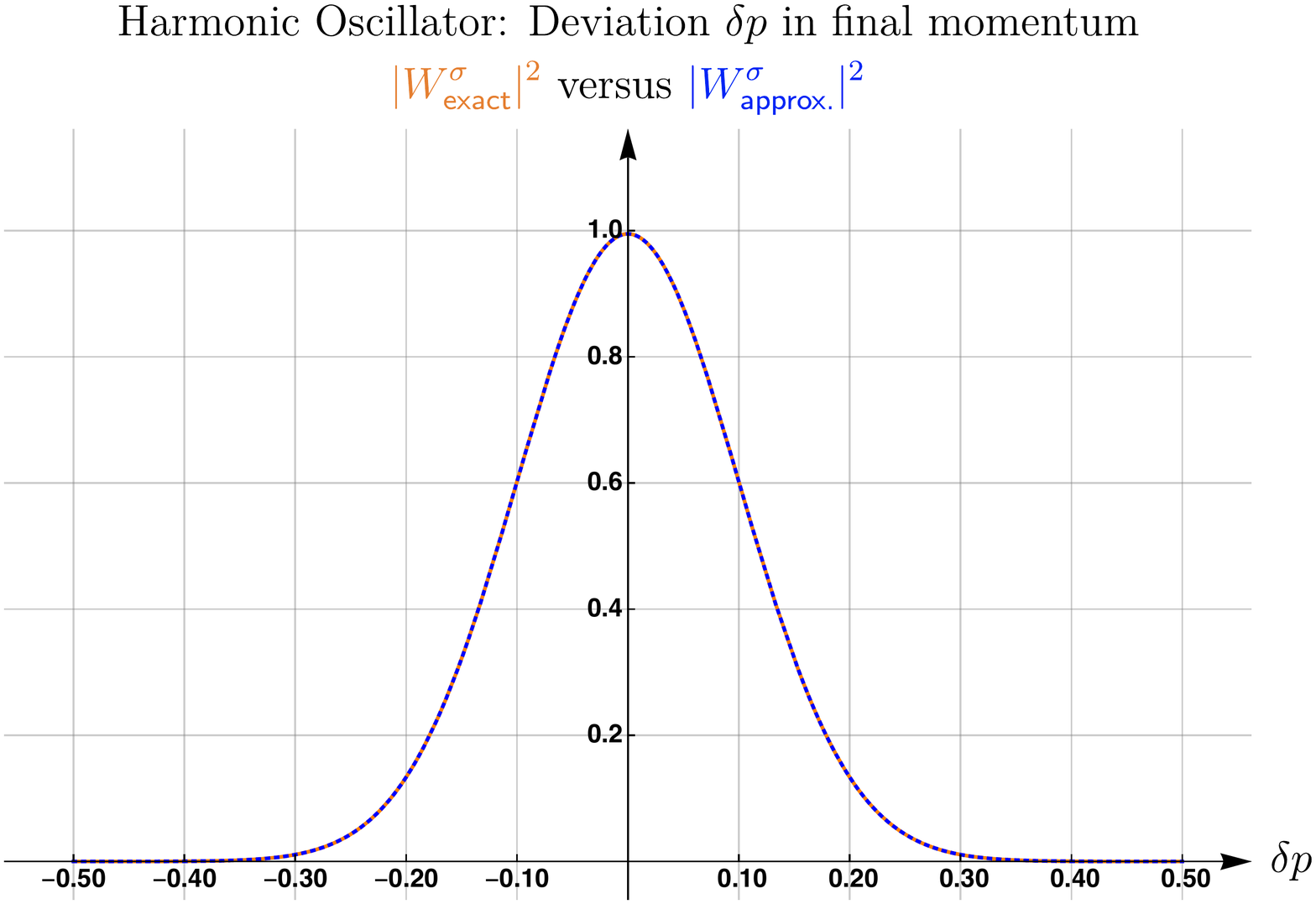}
		\caption{Comparison of the exact holomorphic transition amplitude for the harmonic oscillator with the approximated amplitude. Deviation in the final momentum.}
		\label{fig:HO_Deviatio_Momentum}
	\end{figure}
\end{center}

In this appendix we have a closer look at the holomorphic transition amplitude for the free particle and the harmonic oscillator. As it turns out, it is possible to compute these amplitudes analytically using Mathematica which gives us the opportunity to compare them with the approximate expression~\eqref{eq:ApproxAmp} derived in subsection~\ref{ssec:BndData}.\\
Let us recall that the holomorphic amplitude is defined as
\begin{equation}\label{eq:DefHolAmp}
	W^\sigma(z_i, z_f, T) := \int_{\mathbb{R}^{2n}}\dd^n x_f\dd^n x_i\,\overline{\psi^\sigma_{q_f,p_f}(x_f)}\,K(x_i, t_i; x_f, t_f)\,\psi^\sigma_{q_i,p_i}(x_i),
\end{equation}
where $K(x_i,t_i;x_f,t_f)$ denotes the propagator between $x_i$ and $x_f$ and the states $\psi^\sigma_{q_i,p_i}(x)$ and $\psi^\sigma_{q_f,p_f}(x)$ are the coherent states introduced in chapter~\ref{chap:2}. Irrespective of the dynamical system in question, these states are peaked on the data $(q,p)$ in the sense that 
\begin{align}
	\langle \psi^\sigma_{q, p} |\, \hat x\,| \psi^\sigma_{q,p}\rangle = q\quad\text{and}\quad \langle \psi^\sigma_{q, p} |\,\hat p\, | \psi^\sigma_{q,p}\rangle = p,
\end{align}
and they saturate Heisenberg's uncertainty relation. It is interesting to see what happens when these states are evolved with the propagator $K$. That is, we consider the new state
\begin{equation}\label{eq:EvolvedStates}
	\psi^\sigma_{q_i,p_i}(y,T) := \int_{\mathbb R^n}K(x,y;T)\,\psi^\sigma_{q_i,p_i}(x)\,\dd^n x.
\end{equation}
For the free particle we find that these ``evolved'' states are peaked on
\begin{equation}
	\langle \psi^\sigma_{q_i, p_i}(y,T) |\,\hat x\, | \psi^\sigma_{q_i,p_i}(y,T)\rangle = q_i + \frac{p_i}{m} T\quad\text{and}\quad \langle \psi^\sigma_{q_i, p_i}(y,T) |\,\hat{p}\, | \psi^\sigma_{q_i,p_i}(y,T)\rangle = p_i.
\end{equation}
That is, the expectation value of the position operator is the position one would classically expect for a particle starting at $q_i$ and moving with a momentum $p_i$ for a time period $T$. It also turns out that momentum is conserved.\\
A similar situation presents itself for the harmonic oscillator. There we find that the expectation values of $\hat x$ and $\hat p$ are given by
\begin{align}
	\langle \psi^\sigma_{q_i, p_i}(y,T) |\,\hat x\, | \psi^\sigma_{q_i,p_i}(y,T)\rangle &= q_i\cos(\omega T) + \frac{p_i}{m \omega}\sin(\omega T)\notag\\
	\langle \psi^\sigma_{q_i, p_i}(y,T) |\,\hat{p}\, | \psi^\sigma_{q_i,p_i}(y,T)\rangle &= p_i \cos(\omega T) - m\omega q_i \sin(\omega T).	
\end{align}
These are precisely the position and momentum one would expect from solving the classical equations of motion under the initial value conditions $q(t_i)=q_i$ and $\dot q(t_i) = \frac{p_i}{m}$. Naturally, one expects that the Heisenberg uncertainty relations for these systems are not saturated anymore. Indeed, using the evolved states~\eqref{eq:EvolvedStates}, we find for the free particle
\begin{align}
	\left(\Delta x\right)^2 &= \frac{\hbar}{2}\frac{m^2+T^2 \sigma^4}{m^2\sigma^2}\notag\\
	\left(\Delta p\right)^2 &= \frac{\hbar}{2}\sigma^2
\end{align}
which implies
\begin{equation}
	\Delta x\,\Delta p = \frac{\hbar}{2}\sqrt{1+\frac{T^2\sigma^4}{m^2}} \geq \frac{\hbar}{2}.
\end{equation}
Similarly, we find for the harmonic oscillator the following uncertainties:
\begin{align}
	\left(\Delta x\right)^2 &= \frac{\hbar}{2}\left(\frac{\cos^2(\omega T)}{\sigma^2} + \frac{\sigma^2 \sin^2(\omega T)}{m^2 \omega^2}\right)\notag\\
	\left(\Delta p\right)^2 &= 	\frac{\hbar}{2}\frac{m^2 \omega^2 + (\sigma^4 - m^2 \omega^2)\cos^2(\omega T)}{\sigma^2}
\end{align}
From this it follows that Heisenberg's uncertainty relation now reads
\begin{equation}
	\Delta x\,\Delta p = \frac{\hbar}{2}\frac{\sqrt{\sigma^8 + 6 m^2 \omega^2 \sigma^4 + m^4\omega^4-(\sigma^4-m^2\omega^2)^2\cos(4\omega T)}}{2\sqrt{2} m\omega} \geq \frac{\hbar}{2}.
\end{equation}
It can be saturated for all times if we choose $\sigma=\sqrt{m\omega}$. This choice corresponds to working with the well-known harmonic oscillator coherent states.\\
Now let us consider the holomorphic transition amplitude for the free particle. Mathematica allows us to perform all integrals in~\eqref{eq:DefHolAmp} and the answer reads
\begin{align}\label{eq:HolAmpFP}
	&W^\sigma(z_i, z_f, T) = \left(\frac{2m}{2m+i T \sigma^2}	\right)^\frac{1}{2}\notag\\
		&\times \exp\left(-\frac{2m(p_i+p_f)(q_i-q_f)T\sigma^4+(p^2_i+p^2_f)T^2\sigma^4+2m^2\left[(p_i-p_f)^2+(q_i-q_f)^2\sigma^4\right]}{2\hbar\sigma^2(4m^2+T^2\sigma^4)}\right)\notag\\
		&\times \exp\left(-i\, \frac{m(p_i+p_f)\left[4m(q_i-q_f)+(p_i+p_f)T\right]\sigma^2 - m(q_i-q_f)^2 T \sigma^6}{2\hbar\sigma^2(4m^2+T^2\sigma^4)}\right).
\end{align}
From the foregoing discussion we know that the evolved initial state is peaked on $q_i + \frac{p_i}{m}T$ and $p_i$. Hence, the holomorphic amplitude can be understood as measuring the overlap of a state peaked on $q_i + \frac{p_i}{m}T$ and $p_i$ with a second state peaked on $q_f$ and $p_f$. Given that the coherent states display a Gaussian peak in the position parameter, i.e. $\psi^\sigma_{q,p}(x)\sim \e^{-\frac{\sigma^2}{2\hbar}(x-q)^2}$, we would expect the overlap to be negligible when $q_f\neq q_i + \frac{p_i}{m}T$. Similarly, we expect the overlap to be negligible when $p_f\neq p_i$. By evaluating the holomorphic amplitude~\eqref{eq:HolAmpFP} on $q_f=q_i + \frac{p_i}{m}T$, $p_f=p_i$ and computing its modulus squared, we find
\begin{equation}
	|W^\sigma(z_i,z_f,T)|^2 = \frac{2 m}{\sqrt{4m^2+T^2\sigma^4}}.
\end{equation}
This is the probability to measure the particle which started in $q_i$ with momentum $p_i$ at $q_f=q_i+\frac{p_i}{m}T$ with $p_f=p_i$. Notice that for short time periods the probability is close to one but it decreases the larger $T$ is. This is intuitively clear: The state spreads out with $T$ and in particular its position uncertainty grows with $T$. So, even though its final position in phase space is peaked on $q_i+\frac{p_i}{m}T$ and $p_i$, the overall probability to reach this point decreases. It remains the most probable position among all alternatives, though.\\
We could also choose data $(q_f,p_f)$ which deviates from the classically expected final phase space position. If we set $p_f=p_i$ and $q_f = q_i+\frac{p_i}{m}T + \delta q$, where $\delta q$ measures the deviation between the data $q_f$ and the expected final position, we get for the modulus squred of~\eqref{eq:HolAmpFP}
\begin{equation}\label{eq:DeviationPositionFP}
	|W^\sigma(z_i,z_f,T)|^2 = \frac{2 m}{\sqrt{4m^2+T^2\sigma^4}}\e^{-\frac{2m^2\sigma^2}{(4m^2+T^2\sigma^4)\hbar}(\delta q)^2}.
\end{equation} 
The more we move away from the classically expected position, i.e. the more $\delta q$ deviates from zero, the less likely it is to find the particle at $q_f$ with $p_f=p_i$.\\
This behavior should also be described by the approximation~\eqref{eq:ApproxAmp} derived in subsection~\ref{ssec:BndData}. Hence, we can compare~\eqref{eq:HolAmpFP} to~\eqref{eq:ApproxAmp}. This comparison is shown in Figure~\ref{fig:FP_Deviation_Position} for the parameter values\footnote{These parameters were chosen such that it is possible to see something in the plots. Realistic numbers only lead to sharp lines.} $\hbar = 10^{-3}$, $m=1$, $T=10^{-2}$ and $\sigma=2$.\\
The two curves not only seem to lie on top of each other, they are actually precisely the same! That is, the expression computed from the approximation formula~\eqref{eq:ApproxAmp} yields exactly the same result as~\eqref{eq:DeviationPositionFP}. \\
It is also possible to compute the modulus squared of the holomorphic amplitude under the assumption that $q_f=q_i + \frac{p_i}{m}T$ but $p_f = p_i + \delta p$. This situation is illustrated and compared to the approximate formula~\eqref{eq:ApproxAmp} in Figure~\ref{fig:FP_Deviation_Momentum}. By carefully working out the math one can again show that the approximate formula produces precisely the same expression as the analytical computation based on the holomorphic amplitude~\eqref{eq:HolAmpFP}.
The holomorphic transition amplitude for the harmonic oscillator can also be computed analytically using Mathematica. For completeness, we report the closed analytical expression here:
\begin{align}\label{eq:HolAmpHO}
	& W^\sigma(z_i, z_f, T) = \left(\frac{2 m \omega \sigma^2}{2 m \omega \sigma^2 \cos(\omega T) + i (m^2\omega^2 + \sigma^4)\sin(\omega T)}\right)^\frac{1}{2}\notag\\
	&\times \exp\left(-\frac{\sigma^2\left[p^2_i+p^2_f+m^2(q^2_i+q^2_f)\omega^2 + 2m\omega \{p_fq_i-p_iq_f + (p_iq_i-p_fq_f)\cos(\omega T)\}\csc(\omega T)\right]}{2\hbar\left(\sigma^4 + m\omega^2 - 2 i m \omega \sigma^2 \cot(\omega T)\right)}\right)\notag\\
	&\times\exp\left(-\frac{i m \omega\left(2m(p_iq_i-p_fq_f)\omega - [p^2_i+p^2_f+(q^2_i+q^2_f)\sigma^4]\cot(\omega T) + 2\{p_ip_f + q_iq_f \sigma^4\}\csc(\omega T)\right)}{2\hbar\left(\sigma^4 + m\omega^2 - 2 i m \omega \sigma^2 \cot(\omega T)\right)}\right).
\end{align}
This long expression can again be used to test the approximation~\eqref{eq:ApproxAmp}. Figure~\ref{fig:HO_Deviation_Position} shows the modulus squared of the holomorphic amplitude and the modulus squared of~\eqref{eq:ApproxAmp} under the assumption that $q_f = q_i\cos(\omega T) + \frac{p_i}{m\omega}\sin(\omega T) + \delta q$ and $p_f = p_i \cos(\omega T) - m\omega q_i\sin(\omega T)$, where $\delta q$ acts again as a measure of deviation in the final position. Moreover, the parameter values are: $\hbar = 10^{-3}$, $m=1$, $\omega = 10$, $T=0.5$, $\sigma=3$. It can again be shown that the approximation~\eqref{eq:ApproxAmp} actually reproduces the analytical result one obtains from~\eqref{eq:HolAmpHO}.\\
Figure~\ref{fig:HO_Deviatio_Momentum} shows the same kind of plot, but now for a deviation in the final momentum. That is, it was assumed that $q_f$ agrees with the classically expected final position, but $p_f = p_i \cos(\omega T) - m\omega q_i\sin(\omega T) + \delta p$. Once again it is found that the approximation~\eqref{eq:ApproxAmp} reproduces the analytical expression. 

\setcounter{equation}{0}
\renewcommand\theequation{B.\arabic{equation}}
\chapter{An explicit Construction of $U_\no$}
\label{App:NEW}
In section~\ref{sec:SemiClassicalCLQG} we introduced the matrix $U_\no$ to measure the deviation of the spinor $\textbf{n}$ from the spinor $\mathring{\textbf{n}}$ provided by the boundary data. As discussed in subsection~\ref{ssec:SemiClassicalHolomorphicCLQGAmplitudes}, the two spinors have to be regarded as equivalent when $\textbf{n} = \e^{i\alpha}\mathring{\textbf{n}}$ because they define the same $3$d unit vector $\vec{n} = \textbf{n}^\dagger\,\vec{\sigma}\,\textbf{n}$. To ensure that $U_\no$ is a good measure for deviation, we should exclude all $U_\no\in SU(2)$ which satisfy
\begin{equation}
	U_{\no}\mathring{\textbf{n}}=\e^{i\phi}\mathring{\textbf{n}}
\end{equation}
for some $\phi\in(0,2\pi)$. It can easily be verified that this equation is solved by $U_{\no}=\exp(i\phi\,\vec{n}\cdot\vec{\sigma})$. By introducing the unit vector $\vec{w}$ which belongs to the orthogonal complement $\mathcal N^\perp:=\{\vec{u}\in\mathbb{R}^3 \, |\, \vec{u}\cdot\vec{n}=0\}$ of the linear space $\mathcal N:=\text{span}\{\vec{n}\}$ and defining $U_{\no}:=\exp(i\,\alpha\,\vec{w}\cdot\vec{\sigma})$ with $\alpha\in[0,2\pi)$, we effectively exclude all $SU(2)$ transformations, except the identity, which change the spinor $\mathring{\textbf{n}}$ only by a phase.\\
To construct such a vector and provide an explicit parametrization of $U_\no$, equip~$\mathbb R^3$ with the canonical basis $\{\vec{e}_x,\vec{e}_y,\vec{e}_z\}$ and assume that $\vec{n}$ is not collinear with $\vec{e}_z$. This allows us to introduce the vector $\vec{n}\times\vec{e}_z$ which clearly lies in the orthogonal complement $\mathcal N^\perp$. Then define
\begin{equation}
	\vec{w} := R_{\vec{n}}(\beta)\frac{\vec{n}\times\vec{e}_z}{\|\vec{n}\times\vec{e}_z\|},
\end{equation}
where $R_{\vec{n}}(\beta)\in SO(3)$ is a rotation around the axis defined by $\vec{n}$. The vector $\vec{w}$ satisfies all our requirements since it lies in $\mathcal N^\perp$, it is properly normalized, and, moreover, it can be made to point in any direction of the two-dimensional space $\mathcal N^\perp$ thanks to $R_{\vec{n}}(\beta)$.\\
Observe that the angle between $\vec{n}$ and the $z$-axis is given by $\vec{n}\cdot\vec{e}_z = \cos\theta$, and therefore we find $\|\vec{n}\times\vec{e}_z\| = \sin\theta$ with $\theta\in(0,\pi)$. The values $\theta = 0$ and $\theta=\pi$ have to be excluded from the usual $\theta$-interval $[0,\pi]$ because for these values $\vec{n}$ would be collinear with $\vec{e}_z$, which contradicts the assumption we made at the beginning. This implies that $\vec{w}$ is well-define since $\sin\theta$ cannot become zero. Using the fact that a rotation around $\vec{n}$ can be written as
\begin{equation}
	R_{\vec{n}}(\beta)\vec{v} = \vec{v}\, (\vec{n}\cdot\vec{v}) + \cos\beta \,(\vec{n}\times\vec{v})\times\vec{n} + \sin\beta\, (\vec{n}\times\vec{v}),
\end{equation}
we can further simplify the expression for $\vec{w}$. To that end, notice that $\vec{n}\cdot(\vec{n}\times\vec{e}_z)=0$. Moreover
\begin{align}
	\vec{n}\times\frac{\vec{n}\times\vec{e}_z}{\|\vec{n}\times\vec{e}_z\|} &= \frac{1}{\sin\theta}\left(\vec{n} \cos\theta - \vec{e}_z\right) = \begin{pmatrix}
		\cos\theta\cos\phi, & \cos\theta\sin\phi, & -\sin\theta
	\end{pmatrix}^\transpose 	\notag\\
	\left(\vec{n}\times\frac{\vec{n}\times\vec{e}_z}{\|\vec{n}\times\vec{e}_z\|}\right)\times \vec{n} &= \frac{\vec{n}\times\vec{e}_z}{\|\vec{n}\times\vec{e}_z\|} = \begin{pmatrix}
		\sin\phi, & -\cos\phi, & 0
	\end{pmatrix}^\transpose,
\end{align}
where we used that $\vec{n}$ is given by
\begin{equation}
	\vec{n} = \begin{pmatrix}
		\sin\theta \cos\phi\\
		\sin\theta\sin\phi\\
		\cos\theta
	\end{pmatrix}.
\end{equation}
Hence, we find that $U_\no$ can be written in terms of the two parameters $\alpha\in[0,2\pi)$ and $\beta\in[0,2\pi)$ as
\begin{equation}
	U_\no(\alpha,\beta) = \e^{i \alpha \vec{w}(\beta)\cdot\vec{\sigma}}\quad\text{with}\quad 	\vec{w}(\beta) = \begin{pmatrix}
		\sin\beta\cos\theta\cos\phi + \cos\beta\sin\phi\\
		-\cos\beta\cos\phi +\cos\theta\sin\beta\sin\phi\\
		-\sin\beta\sin\theta
	\end{pmatrix}.
\end{equation}
Relaxing the assumption that $\vec{n}$ and $e_z$ are not collinear is simple. One just chooses a different axis to define the linear space $\mathcal N$ and its orthogonal complement $\mathcal N^\perp$. All the other construction steps remain the same, but one usually finds a much more complicated vector $\vec{w}(\beta)$.